\documentclass[a4paper]{article}
\usepackage[utf8]{inputenc}
\usepackage{xcolor}
\pdfoutput=1
\title{CUPID pre-CDR}
\author{CUPID Interest Group}

\usepackage[numbers,sort&compress]{natbib}
\usepackage{graphicx}
\usepackage{sidecap}
\usepackage{subcaption}
\usepackage[nointegrals]{wasysym} 
\usepackage[left=2.5cm, right=2.5cm, top=2.5cm, bottom=2.5cm]{geometry}
\usepackage[version=3]{mhchem}
\usepackage{textcomp}
\usepackage[nottoc,numbib]{tocbibind}
\usepackage{booktabs}
\usepackage{upgreek}
\usepackage{alphabeta}
\usepackage{multirow}

\usepackage{color,hyperref,doi}
\hypersetup{colorlinks,breaklinks,
linkcolor=blue,urlcolor=blue,
anchorcolor=blue,citecolor=blue}

\usepackage{amsmath}
\usepackage{amssymb}

\usepackage{acronym}

\usepackage{lineno}

\usepackage{comment}

\usepackage{enumitem}
\setitemize{itemsep=0pt,label={--}}

\newcommand{\onbb}{$0\nu\beta\beta$}
\newcommand{\nnbb}{$2\nu\beta\beta$}
\newcommand{\thalf}{$T_{1/2}^{0\nu}$}
\newcommand{\cupid}{CUPID}
\newcommand{\cupido}{CUPID-0}
\newcommand{\cupidmo}{CUPID-Mo}
\newcommand{\cuore}{CUORE}
\newcommand{\cuoreo}{CUORE-0}
\newcommand{\cuoricino}{Cuoricino}
\newcommand{\Qbb}{$Q_{\beta\beta}$}
\newcommand{\mbb}{$m_{\beta\beta}$}
\newcommand{\Mo}{$^{100}$Mo}
\newcommand{\Te}{$^{130}$Te}
\newcommand{\Se}{$^{82}$Se}
\newcommand{\Cd}{$^{116}$Cd}
\newcommand{\Ca}{$^{48}$Ca}
\newcommand{\Ge}{$^{76}$Ge}

\newcommand{\lmo}{Li$_2$MoO$_4$}
\newcommand{\enrlmo}{Li$_2$$^{100}$MoO$_4$}
\newcommand{\cwo}{CdWO$_4$}
 
\newcommand{\teo}{TeO$_2$}

\newcommand{\Tl}{$^{208}$Tl}

\newcommand{\K}{$^{40}$K}
\newcommand{\Th}{$^{232}$Th}
\newcommand{\U}{$^{238}$U}
\newcommand{\Ra}{$^{226}$Ra}

\newcommand{\Co}{$^{60}$Co}
\newcommand{\al}{$\alpha$}
\newcommand{\be}{$\beta$}
\newcommand{\ga}{$\gamma$}
\newcommand{\ckky}{counts$/($keV$\cdot$kg$\cdot$yr$)$}

\begin{document}

\maketitle
\clearpage

\tableofcontents
\clearpage

\section*{Acronyms}


\begin{acronym}

\acro{0vDBD}[\onbb]{ - Neutrinoless Double Beta Decay}
\acro{2vDBD}[\nnbb]{ - Two-neutrino Double Beta Decay}
\acro{BI}{ - Background Index}
\acro{BM}{ - Background Model}
\acro{C-TAL}{ - CUORE Towers Assembly Line}
\acro{DM}{ - Dark Matter}
\acro{DR}{ - Diluition Refrigerator}
\acro{DU}{ - Diluition Unit}
\acro{FCS}{ - Fast Cooling System}
\acro{FCU}{ - Fast Cooling Unit}
\acro{IO}{ - Inverted Ordering}
\acro{IVC}{ - Inner Vacuum Chamber}
\acro{KIDs}{ - Kinetic Inductance Detectors}
\acro{LD}{ - Light Detector}
\acro{LNGS}{ - Laboratori Nazionali del Gran Sasso}
\acro{LSM}{ - Laboratoire Souterrain de Modane}
\acro{MC}{ - Mixing Chamber} 
\acro{NAA}{ - Neutron Activation Analysis}
\acro{NL}{ - Neganov-Luke}
\acro{NME}{ - Nuclear Matrix Element}
\acro{NTD}{ - Neutron Transmutation Doped (Ge thermistor)}
\acro{OVC}{ - Outer Vacuum Chamber}
\acro{PEP}{ - Pauli exclusion principle}
\acro{PGA}{ - Programmable Gain Amplifier}
\acro{PSA}{ - Parts storage Area}
\acro{PT}{ - Pulse Tube}
\acro{ROI}{ - Region Of Interest}
\acro{TES}{ - Transition Edge Sensors}
\acro{TSP}{ - Tower Support Plate}

\end{acronym}

\newpage

\section{Executive Summary}

The study of neutrinoless double beta decay (\onbb) is one of
the most sensitive low-energy searches for new
physics~\cite{Tanabashi:2018oca}. If observed,  
this very elusive decay would imply a lepton number violation by two
units and would prove the existence of massive Majorana
neutrinos~\cite{Tanabashi:2018oca, Elliott:2002xe, AVIGNONE2008481}. 
The decay produces a monochromatic peak in the kinetic energy
sum spectrum of the two emitted electrons.
The only necessary and sufficient signature
used to distinguish the \onbb\ decay from peaks produced by
other processes (e.g. the full energy peaks due to environmental gamma
lines) is its position at the \onbb\ transition energy. Excellent
energy resolution, large detector masses, and very small backgrounds
are of paramount importance for the discovery potential of any
experiment.

Significant experimental efforts in recent years have improved the \onbb\
sensitivity in several isotopes, and with a variety of detection
techniques~\cite{Cremonesi:2013vla, Giuliani:2012b}. The race towards discovery is
characterized by the increase in the experimental mass, 
which has led to the construction of ton-scale detectors.
With the increase in mass has also come the difficult but steady progress in removing the spurious background
events that can jeopardize the observation of a \onbb~decay.
Background reduction presents the hardest challenge.
This has led to the development of improved detection techniques and to the study of different \onbb\ candidate nuclei.
It has also driven the painstaking process of improving the radiopurity of the detector components.

Recent advances in operating large sensitive detectors  have
undoubtedly proven that cryogenic particle 
detectors (often called bolometers) are among the most attractive
devices for the construction of a large, sensitive
\onbb\ experiment. The bolometric detectors are characterized by
excellent energy resolution, which suppresses
the irreducible background from the Standard Model process of
two-neutrino double beta decay (\nnbb).  
Moreover, good energy resolution largely reduces
the risk of misidentifying a background peak for the \onbb\ one. Large detectors can be operated reliably, with low backgrounds. 
Bolometric detectors are scalable, allowing gradual, phased deployment.
Finally, these detectors have a unique feature of allo deployment of different double-beta decay isotopes in the same infrastructure. This true multi-isotope approach may prove invaluable in case of a discovery.

Over the last decade, bolometers have demonstrated excellent
operational characteristics (detector mass, resolution,
reproducibility, and background level). TeO$_2$-based bolometric
detectors such as \cuoreo~\cite{Alfonso:2015wka} and
\cuore~\cite{Alduino:2017ehq} have produced world-leading \onbb\
results. 
Moreover, bolometers containing different $\beta\beta$ candidates such as \Se,
\Mo, and \Cd\ performed equally well in dedicated
tests~\cite{Beeman:2013vda,Cardani:2013mja,Beeman:2012,Cardani:2013,Bekker:2016,Armengaud2017,Barabash:2016a}. Most
recently, \cupido~\cite{Azzolini:2018dyb,Azzolini:2019tta} at Laboratori Nazionali del Gran Sasso (LNGS) and \cupidmo~\cite{Poda:2018} at Laboratoire Souterrain de Modane (LSM) demonstrated that it is possible to improve the detector
background using active background rejection techniques that are
considered a fundamental requirement for any next generation
experiment.

\cuore\ is the largest \onbb\ detector of its kind and is the reference point for any future bolometric
\onbb\ experiment. 
With a large  cryostat able to cool a one-ton detector to
temperatures below 10 mK~\cite{DAddabbo:2018jjw,Alduino:2019xia}
and its specially designed infrastructure
ensuring a low-noise and low-radioactivity environment for its 988
crystal bolometric array, \cuore\ is searching for \onbb\ in \Te\ with
unprecedented sensitivity. \cuore\ is currently one of the most competitive
\onbb\ experiments in operation. If \onbb\ is not found at its
sensitivity scale, \cuore\ will provide fundamental information for the
next steps in this field.

A further increase in the experimental sensitivity will
require the implementation of new ideas and new technical solutions
able to surpass \cuore\ limits. At the same time, the significant
investment required to realize \cuore\ and the success of its
cryogenic and scientific operations motivates ongoing efforts towards
defining the next-generation experiment based on the same
infrastructure and know-how developed within \cuore.
A  path forward that introduces new technology capable of reducing the
backgrounds by at least two orders of magnitude while deploying up to a ton of
the isotope of interest in a few timely phases is both
scientifically motivated and plausible. 

In this document we discuss a well-defined strategy that,
starting from the present experimental configuration of \cuore, looks
toward a future experiment characterized by a substantial improvement
in \onbb\ discovery sensitivity. \cupid\, (\cuore\ Upgrade with
Particle IDentification) starts from the   
\cuore~\cite{Alduino:2017ehq} technical expertise, material selection,
background model, and excellent energy resolution,
and combines it with the capability to eliminate the backgrounds from
\al\ particles using scintillating crystal~\cite{Armengaud2017, Azzolini:2018dyb,Azzolini:2019tta}.
In addition, the high Q-value of \onbb\ in \Mo\
naturally suppresses the gamma backgrounds by over an order of
magnitude, compared to \cuore. 
We demonstrate that the scintillating bolometer technology based on
\lmo\ crystals highly enriched in \Mo\ is scalable to
ton-scale isotopic masses. An experiment deployed in the existing
\cuore\ cryostat would have a discovery potential covering the entire
inverted hierarchy region of neutrino masses. An expansion to a metric
ton of \Mo\ in a larger cryostat would enable an experiment with the
normal hierarchy sensitivity. 

We discuss strategies for deploying \cupid\ after the end of the \cuore\
scientific operations in a way that delivers results in a
timely manner, and maintains scientific leadership in \onbb\ searches
over the next decade.

\begin{figure}[h!]
  \centering
  \includegraphics[width=\textwidth]{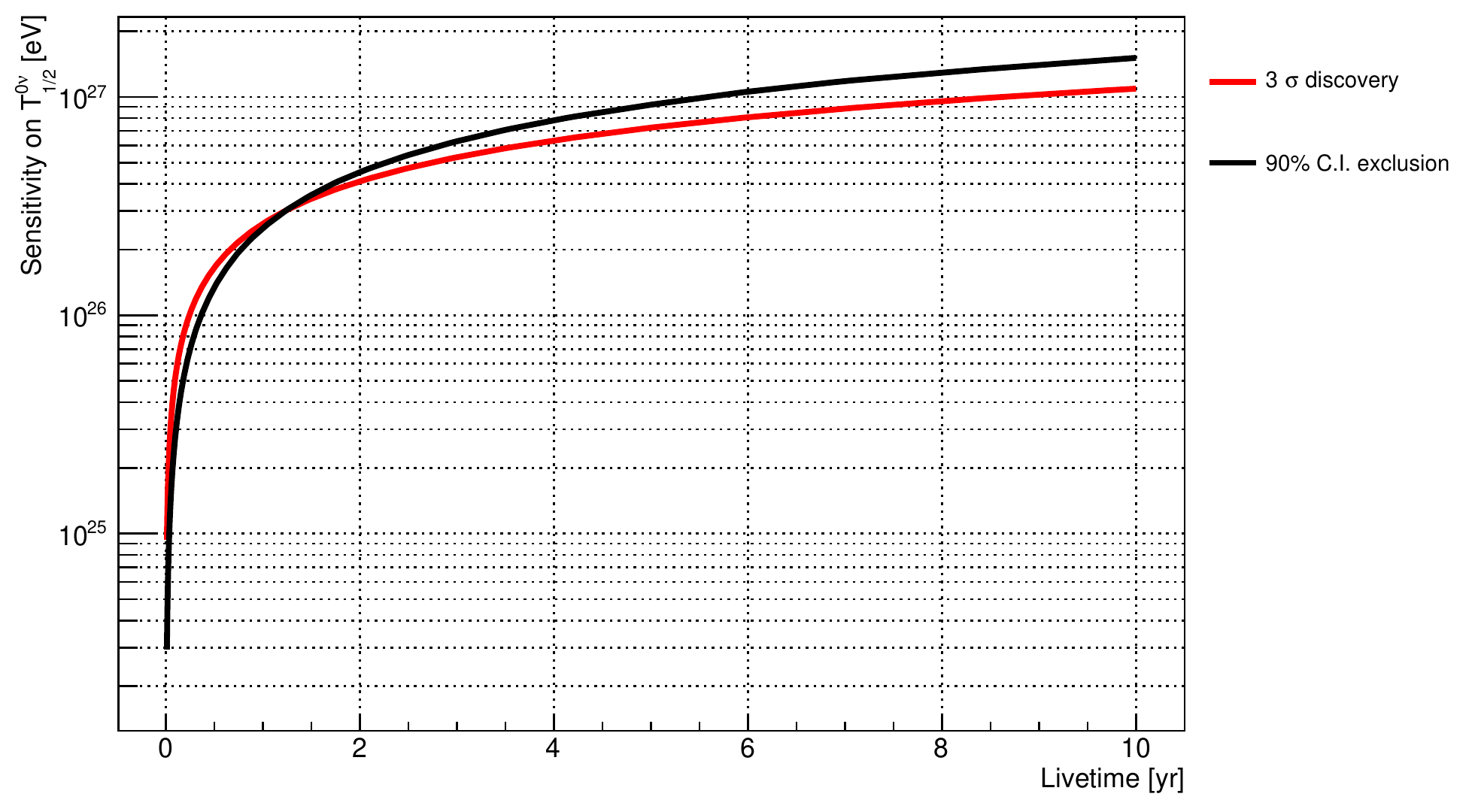}
  \caption{\cupid\ sensitivity curves as a function of the experiment livetime.
    The red curves correspond to the $3~\sigma$ discovery sensitivity,
    the black one to the $90\%$~C.I. Bayesian exclusion sensitivity.}
  \label{fig:baseline}
\end{figure}

The baseline design of the \cupid\ detector to be deployed in the \cuore\
cryostat is conservative and its sensitivity is based on mature
data-driven background model. Such a detector could be ready for
construction as early as 2021, when the technical design is
finalized. For reference, we list here the main parameters of the
\cupid\ detector in Table~\ref{tab:baseline}.
The \cupid\ discovery and exclusion sensitivity as a function of livetime are depicted in Fig.~\ref{fig:baseline}.

\begin{table}[h!]
  \caption{Main parameters of the conservative baseline \cupid\ detector design.}
  \label{tab:baseline}
  \centering
  \begin{tabular}{lr}
    \toprule
    Parameter                                   & Baseline \\
    \midrule
    Crystal                                     & \lmo \\ 
    Crystal size                                & $\oslash 50$~mm$\times$ h$50$~mm \\ 
    Crystal mass (g)                            & 308 \\
    Number of crystals                          & 1534 \\ 
    Number of light detectors                   & 1652 \\ 
    Detector mass (kg)                          & 472 \\ 
    $^{100}$Mo mass (kg)                        & 253 \\ 
    Energy resolution FWHM (keV)                & 5 \\ 
    Background index (\ckky)                    & $10^{-4}$ \\ 
    Containment efficiency                      & 79\% \\ 
    Selection efficiency                        & 90\% \\ 
    Livetime                                    & 10 years \\
    Half-life limit sensitivity (90\%) C.L.     & $1.5\times10^{27}$ y \\
    Half-life discovery sensitivity ($3\sigma$) & $1.1\times10^{27}$ y\\ 
    \mbb\ limit sensitivity (90\%) C.L.     & $10-17$\ meV \\
    \mbb\  discovery sensitivity ($3\sigma$) & $12-20$ meV\\ 
    \bottomrule
  \end{tabular}
\end{table}


\section{Motivations} 

\subsection{Lepton number violation and neutrinoless double beta decay}

Double beta decay is the rarest observed nuclear transition.
It is a weak process that takes place between two even-even isobars when the decay
to the intermediate nucleus is energetically forbidden due to the pairing interaction
and is possible for 35 nuclei. The \nnbb\ decay --- a second order weak transition ---
conserves the lepton number and was detected in eleven nuclei with half-lives in the range $10^{18}$ -- $10^{21}$~yr.
The neutrinoless process, corresponding to $(A,Z) \to (A,Z+2) + 2e^-$,
may be induced by a number of mechanisms beyond the Standard Model. 

The detection of \onbb\ decay would be a major breakthrough~\cite{Tanabashi:2018oca, Elliott:2002xe, AVIGNONE2008481},
proving that neutrino, unlike all the other fermions, is a Majorana rather than a Dirac particle:
the neutrino would be the only spin-$\frac{1}{2}$ particle coinciding with its own antimatter partner,
a possibility naturally set by its neutrality.
In this context, a new mechanism of mass generation, besides the Higgs mechanism,
could occur and naturally explain the smallness of light-neutrino masses.
In addition, the matter-antimatter asymmetry in the Universe
could be accounted for by CP violation in the neutrino sector.
In spite of these major implications on neutrinos, we remark however that \onbb\ decay
is much more than a neutrino physics experiment.
It is a powerful, inclusive test of lepton number violation,
which creates electrons according to the process $2n \to 2p+2e^-$,
implemented in the nuclear matter.
From a beyond-Standard-Model perspective, lepton number violation is as important as baryon number violation.
Therefore, the experimental search for \onbb\ decay must be pursued
with the highest possible sensitivity, regardless of the related neutrino-physics scenario.

In spite of the richness of the physics connected to \onbb\ decay (Sec.~\ref{sec:sensitivity}),
the so-called mass mechanism, which represents a minimal extension of the Standard Model, is particularly important.
In this scenario, the virtual particles mediating \onbb\ decay
are the three massive light neutrinos observed so far and which undergo flavor oscillations.
The mass mechanism neatly relates the \onbb\ decay rate
to the mixing angles and mass splittings of the three known neutrinos
(partially accessible through other searches), sets well-defined experimental targets,
and provides a metric that allows us to compare the current results
and the future sensitivities of experiments studying different isotopes
and adopting different technical approaches. 

The parameters involved in the mass mechanism appear in Eqs.~(\ref{eq:newphysics})
and (\ref{eq:majorana-mass}) of Sec.~\ref{sec:sensitivity},
where they are related to the main output of the experimental searches
consisting of the constrained (or measured, if \onbb\ decay is detected) half-life \thalf.
These parameters are the phase space $G_{0\nu}$, the axial vector coupling constant $g_A$,
the nuclear matrix element $\mathcal{M}$ relevant for the light-neutrino exchange,
and the effective Majorana mass $m_{\beta\beta}$ defined
in Eq.~(\ref{eq:majorana-mass}) of Sec.~\ref{sec:sensitivity},
which contains the neutrino-physics information delivered by \onbb\ decay investigation.
The experimental results in terms of \thalf\ can be translated to bounds
(or values) for $m_{\beta\beta}$, exploiting the knowledge about $G_{0\nu}$, $\mathcal{M}$ and $g_A$.
While $G_{0\nu}$ is exactly calculable, $\mathcal{M}$ is known to within a factor 2--3
depending on the nuclear model. The value of $g_A$ is usually taken as 1.25,
close to that measured for the free neutron.
However, values much lower than that, even by a factor 2, have been proposed.
This $g_A$ quenching would have a dramatically negative impact on the rate prediction,
and is under extensive examination and discussion in the theoretical nuclear physics community.

\begin{SCfigure}
  \includegraphics[width=0.75\textwidth]{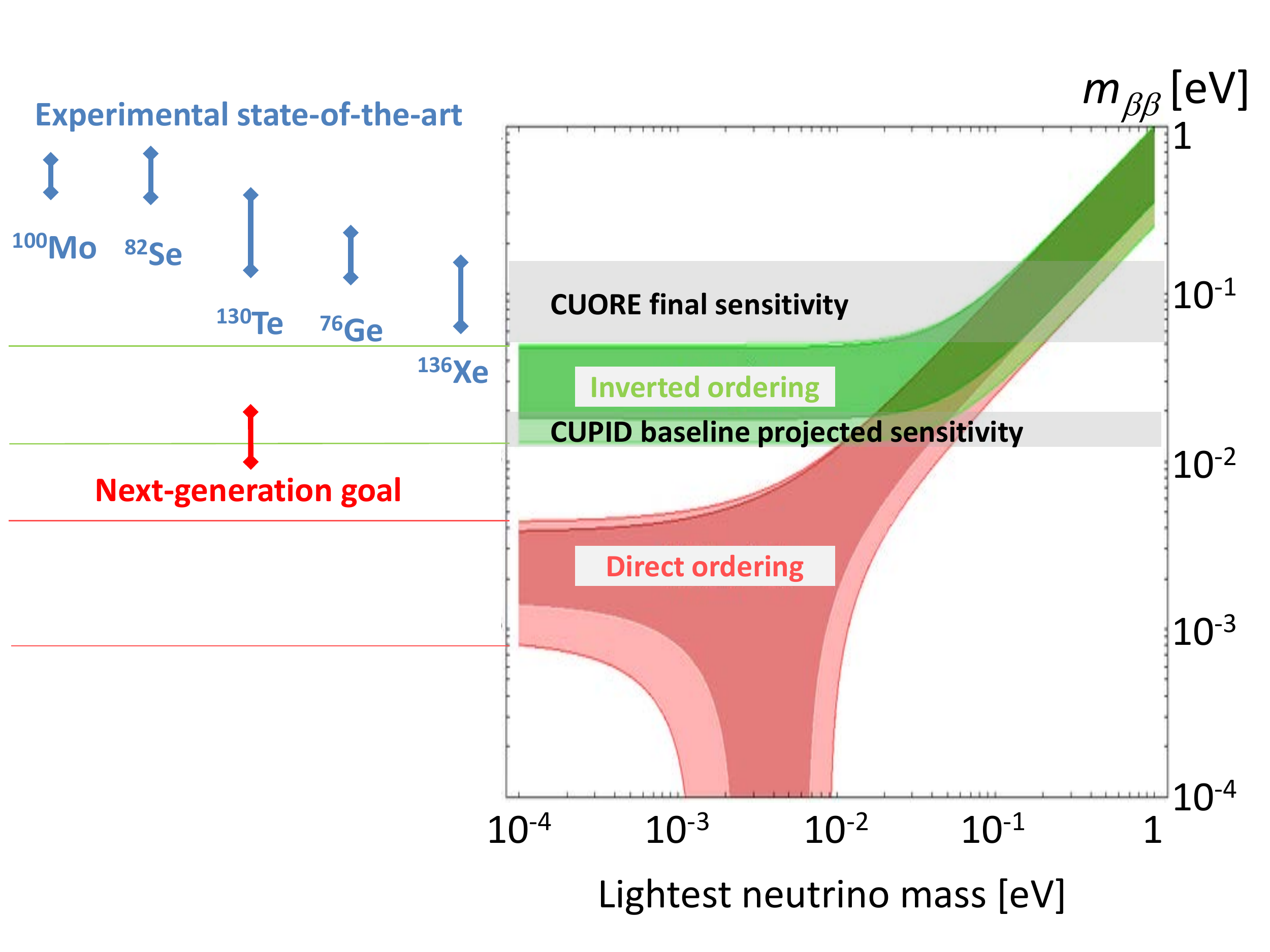}
  \caption{The effective Majorana mass $m_{\beta\beta}$ as a function of the lightest neutrino mass
    provides the parameter space typically used to compare \onbb\ decay experiments.
    The experimental state-of-the art and the goal of next-generation experiments are shown on the left.
    The final sensitivity of \cuore\ and the projected sensitivity of \cupid\ baseline are also reported.}
  \label{fig:status-prospects}
\end{SCfigure}

In this introduction, we will use the sensitivity to $m_{\beta\beta}$, which spans an interval of values
due to the uncertainties on $\mathcal{M}$, as a metric to present the state-of-the-art of the field.  It will also be used to
describe future scenarios and the role \cupid\ plays in these developments, as depicted in Fig.~\ref{fig:status-prospects}.
A thorough discussion is carried out in Sec.~\ref{sec:sensitivity}.

\subsection{Isotope selection and summary of the current searches}

\begin{SCfigure}
  \includegraphics[width=0.67\textwidth]{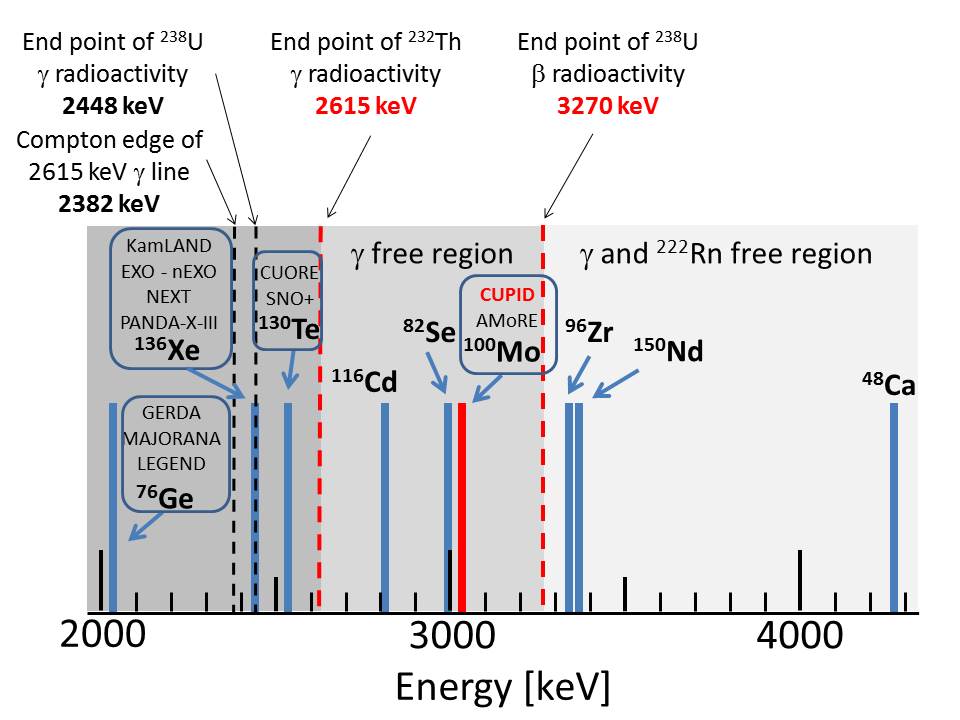}
  \caption{The positions of the expected signals in a two-electron sum-energy spectrum
    for the nine most favorable \onbb\ decay isotopes.  These are compared
    with background energy markers related to the maximum $\gamma$ energies
    of the $^{238}$U and $^{232}$Th chains and the maximum $\beta$ energy of the $^{238}$U chain.
    The main experiments are also mentioned in relation with their selected nucleus.
    For essentially technical reasons, most searches investigate the three least favorable isotopes.
    The isotope selected for the \cupid\ baseline is \Mo, while \Te\ is an alternative option.}
  \label{fig:nine-isotopes}
\end{SCfigure}

The \onbb\ decay signal is a distinctive monochromatic peak in the sum-energy spectrum
of the two emitted electrons at the Q-value (\Qbb) of the reaction.
Given the long expected lifetime ($> 10^{25-26}$ y), the \onbb\ decay search requires large sources,
containing tons of the isotope of interest.
Required detector characteristics are high energy resolution, high efficiency,
and low background~\cite{Cremonesi:2013vla,Giuliani:2012b}.
High \Qbb\ values are a crucial advantage, as this feature positively affects 
both the phase space $G_{0\nu}$ and the background level.
As a consequence, at the moment only nine isotopes are experimentally relevant
thanks to their large \Qbb\ (see Fig.~\ref{fig:nine-isotopes}).
Two energy markers are especially relevant for the expected background~\cite{Giuliani:2012b}:
the 2615~keV marker, representing the end-point of the $\gamma$ radioactivity,
and the 3270~keV marker, corresponding to the $\beta$ decay Q-value of $^{214}$Bi, which belongs
to the radon ($^{222}$Rn) progeny. 

A first group of three candidate \onbb\ isotopes ($^{76}$Ge, $^{130}$Te, and $^{136}$Xe) features \Qbb s
above 2~MeV but below both markers, and therefore have to cope
with the natural $\gamma$ background and with the radon-induced one.
However, superb detection technologies can be employed for these nuclei:
germanium diodes (GERDA~\cite{Agostini:2018tnm} and MAJORANA~\cite{Aalseth:2018}),
xenon gaseous and liquid detectors (EXO-200~\cite{Auger:2012}, NEXT~\cite{Martin-Albo:2015rhw},
and other projects), large liquid-scintillator volumes
incorporating the candidate nuclei (KamLAND-Zen~\cite{Gando:2016} and SNO+~\cite{Andringa:2015tza}),
and \teo\ bolometers (\cuore~\cite{Alduino:2017ehq}).
Thus, it is not surprising that the currently most sensitive experiments study these three nuclei
($^{136}$Xe in KamLAND-Zen, $^{76}$Ge in GERDA and $^{130}$Te in \cuore),
with limits on $m_{\beta\beta}$ of 61--165~meV~\cite{Gando:2016}, 120--260~meV~\cite{Agostini:2018tnm},
and 110--520~meV~\cite{Alduino:2017ehq} respectively.
The current results barely approach the onset of the inverted ordering (IO) region of the neutrino mass pattern,
which extends in the range 15--50~meV for a vanishing lightest-neutrino mass.
These experiments, even at the completion of their experimental program,
are far from fully exploring this region.
In addition to these three ongoing projects, there are other searches based on one of the same isotopes
that are under commissioning or fully funded, or will take data soon.
In this context, we will mention: SNO+~\cite{Andringa:2015tza},
which will study $^{130}$Te dissolved in liquid scintillator in the SNO set-up;
LEGEND-200~\cite{Myslik:2018vts}, based on the GERDA and MAJORANA experiences
and exploiting at maximum the current GERDA set-up for the study of $^{76}$Ge;
and NEXT-100~\cite{Martin-Albo:2015rhw}, a high pressure electro-luminescence TPC studying $^{136}$Xe.
Even these experiments, however, do not have the sensitivity required to cover the IO band.

Continuing our survey of the most promising isotopes, we remark that the three candidates
$^{48}$Ca, $^{96}$Zr, and $^{150}$Nd are in the best position to carry out a background-free experiment,
since their \Qbb s are higher than the $^{214}$Bi $\beta$ endpoint and
therefore are not affected by $\gamma$ radioactivity.
Unfortunately, these are currently ruled out as being part of a viable and competitive experiment
because they have a very low isotopic abundance; at this time, large-scale enrichment is impossible or prohibitively expensive.

The remaining three candidate isotopes (\Se, \Mo, and \Cd) feature an expected \onbb\ decay signal
out of the bulk of the $\gamma$ environmental background,
but radon-related contamination may still be influential.
These nuclei can be embedded in a number of compounds that allow the growth of large scintillating crystals.
Radon radioactivity is not a severe issue in this case because the active part of the detector
is not a fluid that can be contaminated by gaseous radon emanation.
This technology is currently applied to \Se\ in \cupido~\cite{Azzolini:2018dyb,Azzolini:2019tta}
and to \Mo\ in both \cupidmo~\cite{Poda:2017d} and AMoRE~\cite{Alenkov:2019jis}.
In scintillating bolometers, each event provides a thermal and a scintillation signal.
Their simultaneous detection enables the discrimination of $\alpha$ particles,
which present a generally lower light yield than that of $\beta$ particles of the same energy~\cite{Pirro2006}.
Another possibility is to perform pulse-shape discrimination in the scintillation signal~\cite{Azzolini:2018tum}
and for some crystal events in the thermal signal~\cite{Arnaboldi2011}.
These methods can reject energy-degraded surface $\alpha$s,
which are expected to be the dominant background source above 2.6~MeV.
\cupid\ will follow this experimental approach, as discussed in detail in Sec.~\ref{sec:CUPID-intro}.

\subsection{Proposed next-generation searches}

As already mentioned, the current objective of the experimental search for \onbb\ decay
is to explore deep into the IO region.
Some of the proposed next-generation projects can in principle fully cover the allowed IO region
and detect \onbb\ decay even in case of direct ordering, provided that the lightest neutrino mass is larger than 10~meV. This is the main physics goal of \cupid.

We will briefly discuss here the main next-generation projects~\cite{giuliani:2018}.
A quantitative analysis of their physics reach, in comparison with \cupid,
is presented in Sec.~\ref{sec:sensitivity} and graphically in Fig.~\ref{fig:DiscSensMbb}.

Future projects can be broadly classified into two categories:
experiments using a fluid-embedded \onbb\ source (featuring large, sensitive masses and easy scalability)
and experiments using a crystal-embedded \onbb\ source (featuring high energy resolution and efficiency).
In the first class we have Xe-based TPC projects like nEXO (evolution of the closed EXO-200),
NEXT-1.5t (evolution of the imminent NEXT-100), and Panda-X-III~\cite{Chen:2017}.
This also includes experiments which dissolve the source in a large liquid-scintillator matrix
exploiting existing infrastructures like KamLAND2-Zen (evolution of the current KamLAND-Zen-800)
and SNO+-phase-II (evolution of the imminent SNO+-phase-I).
In the second class we have experiments based on germanium diodes like LEGEND-1t
(evolution of the planned LEGEND-200) and those which exploit the bolometric technique,
like the multi-step AMoRE program (AMoRE-I and AMoRE-II,
which represent the evolution of the current AMoRE pilot),
and finally \cupid, which is based on the large experience gathered by \cuore\
and the demonstrators \cupidmo\ and \cupido.
The most prominent projects in this rich scenario are nEXO, LEGEND-1t, NEXT-1.5t,
and \cupid, which have a 3~$\sigma$ discovery sensitivity that,
at least for some matrix element calculations, reaches below 20 meV for $m_{\beta\beta}$ (see Fig.~\ref{fig:DiscSensMbb}). 

\subsection{\cupid: the next-generation bolometric search for \onbb\ decay} \label{sec:CUPID-intro}

Bolometers are ideal detectors to perform sensitive \onbb\ searches~\cite{fiorini:1984,Pirro2006,giuliani:2012a}.
They can provide an energy resolution at the few per-mil level, a total efficiency at the 70\%--90\% level,
and extremely low background thanks to the high radiopurity achievable in the crystals used as detectors~\cite{Poda:2017c}.
Moreover, several high-\Qbb\ candidates (\Ca, \Ge, \Se, \Mo, \Cd, and \Te) can be studied
with the same cryogenic apparatus.
 
The active part of a bolometer consists of a crystal coupled to a thermal sensor, which converts temperature variations into a voltage or current pulse,
depending on the sensor technology.
The typical masses for crystals used in \onbb\ decay search are in the 0.1--1~kg range.
Bolometers of this size must be operated at very low temperatures ($< 20$~mK)
so that the crystal heat capacity is low enough to provide high-amplitude signals
for energy depositions in the range of interest ($\sim$MeV-scale for \onbb\ decay).
In the \cupid\ baseline, the thermal sensor will be a Neutron Transmutation Doped (NTD) Ge thermistor,
already adopted in \cuoricino, \cuore, \cupido, and \cupidmo.
NTDs are mainly sensitive to thermal phonons and essentially act as a thermometer
(see Sec.~\ref{sec:ntds} for further details). 

\begin{figure}[htbp]
  \centering
  \includegraphics[width=1.0\textwidth]{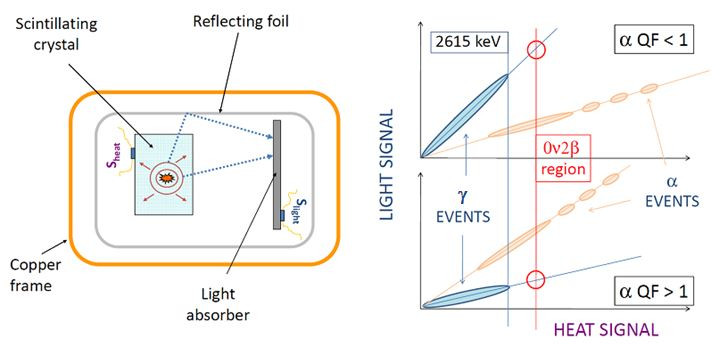}
  \caption{Left: the main elements of a scintillating bolometer consist of two phonon sensors.
    The heat signal is read from the crystal containing the \onbb\ candidate,
    and the light signal from the optical bolometer collecting the scintillation light.
    Right: concept of \al particle rejection exploiting the simultaneous measurement
    of heat and light for the same event.
    Top-right: the common situation where \al s produce less scintillation light than \be s
    and have a Quenching Factor (QF) $<1$. This is an example case of \lmo\ and \cwo\ crystals.
    Bottom-right: the opposite case with QF~$>1$. The only crystal known to follow this behavior is ZnSe.}
  \label{fig:scintillating-bolometer-scheme}
\end{figure}

The environmental $\gamma$ flux represents one of the main sources of background for most of the present experiments.
As mentioned above, the highest energy-relevant $\gamma$ line in natural radioactivity
is the 2615~keV full-energy peak of \Tl, belonging to the \Th\ decay chain.
The energy region above $\sim 2.5$~MeV is dominated by \al\ events induced by surface radioactive contaminants,
as shown by the results of \cuoricino, \cuoreo, and \cuore~\cite{Alduino:2017}.
Exploiting the luminescence of some detector materials, it is possible to distinguish between \al\ events
and signal-like events.
As the light yield for energy depositions induced by \al\ and \be\ particles  of the same energy is different,
the simultaneous detection of light and heat, and the comparison of the respective signal amplitudes,
leads to an effective rejection of the $\alpha$ background (Fig.~\ref{fig:scintillating-bolometer-scheme}).
This approach was extensively studied and proved successful in many independent
demonstrator experiments~\cite{Pirro2006,giuliani:2012a,Artusa:2014}, and is the method chosen for \cupid.
Particle IDentification (PID) is only possible with the installation of a light detector,
which in the \cupid\ design is also a bolometer facing the main crystal (Sec.~\ref{sec:LightDetectors}).
Medium-scale demonstrators like \cupido~\cite{Azzolini:2018dyb,Azzolini:2018tum}
and \cupidmo~\cite{Poda:2017d,Poda:2018}, involving tens of scintillating bolometers,
show that this technique is viable, robust and reproducible. 

The material selection for the \cupid\ crystals is based on the extensive experience
collected by the previous projects MIBETA, \cuoricino, \cuore, BOLUX, LUCIFER, ISOTTA, and LUMINEU.
The copious results achieved so far on more than 10 compounds,
joined with the preliminary results from the \cuore\ background model (Sec.~\ref{sec:bkg}),
indicate that the optimal choice for \cupid\ is lithium molybdate (\lmo)~\cite{Bekker:2016,Armengaud:2017},
containing the candidate \Mo\ with \Qbb$=3034$~keV (Fig.~\ref{fig:nine-isotopes}).
Section~\ref{sec:EnrPurCryst} describes the enrichment, the purification, and the crystallization methods
allowing the development of \enrlmo\ crystals with the required features.
The \lmo\ light yield is compatible with the desired \al-background rejection,
while the demonstrated radiopurity of the crystals, which are grown from enriched materials,
satisfies the \cupid\ requirements.

In its baseline design, \cupid\ will consist of an array of 1534 \lmo\ crystals,
grown from molybdenum enriched at $\geq95\%$ in \Mo.
The single-crystal mass will be approximately 308~g, corresponding to a total \Mo\ mass of about 253~kg.
\cupid\ will be housed in the present cryogenic facility of \cuore,
located in the Laboratori Nazionali del Gran Sasso of INFN, Italy,
and will benefit from its infrastructure and operation procedures (see Sec.~\ref{sec:overview}).
The expected background index (BI) is about $10^{-4}$~\ckky\
in the \Mo\ region of interest (ROI) (see Sec.~\ref{sec:bkg}).
These parameters will allow \cupid\ to fully explore the IO region
of the effective Majorana neutrino mass (see Fig.~\ref{fig:status-prospects})
and to have a high discovery potential if the lightest neutrino mass is larger than 10~meV,
regardless of the neutrino-mass ordering.
For a detailed discussion on \cupid\ sensitivity and physics reach,
see Sec.~\ref{sec:sensitivity} and Fig.~\ref{fig:DiscSensMbb}.

One of the aforementioned advantages of the bolometric technology is
its possible application to several favorable candidates.
Competitive experiments can be performed using crystals of ZnSe (candidate \Se),
\lmo\ (candidate \Mo), \cwo\ (candidate \Cd), and \teo\ (candidate \Te).
These multiple searches can be performed just changing the crystals
and keeping the same detector configuration, assembly procedures, cryogenic infrastructure,
readout electronics, data-acquisition system, and analysis tools~\cite{Artusa:2014,giuliani:2018a}. 

The \cuore\ background model clearly shows the current superiority of the \Mo\ option,
essentially because of its favorable \Qbb\ (see Fig.~\ref{fig:nine-isotopes}).
However, \Te\ embedded in \teo\ crystals is maintained as a possible alternative.
The \Te\ option is motivated by the excellent technical performance of the \teo\ bolometers,
whose fabrication and operation were successfully tested on hundreds of elements~\cite{Arnaboldi:2008ds,Alduino:2017ehq},
and by the uniquely high natural isotopic abundance of \Te\ ($34\%$),
which makes its enrichment especially affordable.
The PID in \teo\ requires an outstanding optical bolometer capable of detecting the feeble Cherenkov light emitted by a \onbb\ decay event~\cite{TabarellideFatis:2009zz},
which is an order of magnitude lower than the scintillation light from \lmo.
The related light-detector technology is discussed in Sec.~\ref{sec:LightDetectors},
while the overall tellurium option is presented in Sec.~\ref{sec:Te}. 

\newpage


\section{Overview of the experiment} \label{sec:overview}

\subsection{\cupid\ detector concept}

In \cupid, the main bolometer crystals will be grown from \enrlmo.
The Mo component will be enriched to $\geq95\%$ in \Mo.
At this stage of the conceptual design, we envision the use of cylindrical crystals
with 50~mm diameter and 50~mm height, corresponding to a mass of $\sim308$~g.
We selected such a mass  so that pile-up of \Mo\ ordinary \nnbb\ decay events
in a single crystal will contribute to background in the ROI at a level
compatible with the \cupid\ target BI of 10$^{-4}$ \ckky\ (see Section~\ref{sec:bkg-2nu}). 

\begin{SCfigure}[][h]
  \includegraphics[width=0.72\textwidth]{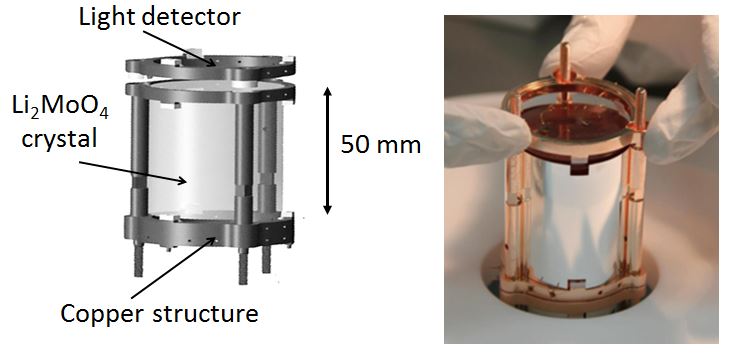}
  \caption{Left: schematic view of the \cupid\ single module according to the baseline design.
    Right: photo of a \cupido\ scintillating bolometer, prefiguring the \cupid\ single-module structure.
    The reflective foil surrounding the crystal is visible.}
  \label{fig:CUPID-single-module}
\end{SCfigure}

As in \cuore, a NTD Ge thermistor will be glued to the crystal on a flat surface,
in order to provide the thermal signal.
In the baseline design, the curved surface of each crystal are surrounded
by a light-reflecting foil to maximize light collection,
and the flat surfaces of the crystals are exposed to bolometric light detectors
fabricated from Ge wafers with 5~cm diameter.
The Ge wafers are also instrumented with an NTD as a thermal detector.
The structure of a single \cupid\ scintillating bolometer is shown in Fig.~\ref{fig:CUPID-single-module}.
The crystals and the light-detector wafers are supported by PTFE elements
connecting them to round copper frames. These frames are stacked by means of copper columns,
forming a detector tower conceptually similar to those of \cupido\ and \cupidmo.
Light detectors are placed between successive crystals,
thus each light detector will serve two crystals at the same time.
With this design, we anticipate 1534 crystals in the full array,
corresponding to about 253~kg of $^{100}$Mo.
Fig.~\ref{fig:CUPID_simulation} of Section~\ref{sec:bkg-BBcupid} shows the CUPID array
hosted inside the \cuore\ cryostat. The basic unit geometry is
a cylindrical main bolometer with reflective foil and Ge light detector disc.  This
has been used with great success in \cupido~\cite{Azzolini:2018tum}
(see Fig.~\ref{fig:CUPID-single-module}, right panel)
and \cupidmo~\cite{Poda:2018}. The array will be operated at 10-20~mK.

In the rest of this section, we describe the results achieved in \cupidmo\ and its preparation measurements.
\cupidmo\ is placed in the context of the LUMINEU project~\cite{Armengaud:2017,Poda:2017d,Poda:2018},
which shows the maturity reached by the proposed technology
and the high standard of the \lmo\ detectors in terms of energy resolution,
$\alpha$/$\beta$ rejection capabilities, internal radiopurity, and overall reproducibility of the results.
The single module of \cupidmo\ consists of a crystal of \enrlmo\ enriched at more than $\sim96\%$ in \Mo.
The crystals are cylinders with 44~mm diameter and 45~mm height coupled to NTD Ge thermistors.
At least one of the flat surfaces of each crystal is exposed to a light detector,
consisting of an NTD-instrumented Ge wafer with a diameter of 44~mm
and coated with a 70-nm-thick SiO layer on both sides to maximize light absorption.
In the tests preceding \cupidmo, we operated modules with and without reflecting foils,
achieving a satisfactory $\alpha$/$\beta$ separation in both cases.
In \cupidmo, we opted for a conservative approach and used the reflective foil in all the 20 modules installed.
The results achieved in the tests preceding \cupidmo\ and confirmed by \cupidmo\ itself are summarized below:

\begin{itemize}
\item We routinely obtained energy resolutions of $\sim5$~keV FWHM at 2615~keV.
\item The heat-light readout yields an $\alpha$ rejection at the level of 99.9\%
  with nearly full acceptance of the \be\ and \ga\ events.
\item The strict crystal production protocol guarantees internal contamination $\lesssim 5~\upmu$Bq$/$kg
  for both \Th\ and \U, and $\lesssim5$~mBq/kg for \K.
\item The heat channel features rise-times (from 10\% to 90\% of the signal maximum)
  of $\sim 15$--$30$~ms and decay-times (from 90\% to 30\% of the signal maximum)
  of $\sim 100$--$500$~ms, depending on the detector and on the operation temperature.
\end{itemize}

The size of the main crystal and of the light detector, their mechanical and geometrical arrangements,
and the readout approach adopted in both these tests and in \cupidmo\
closely resemble the baseline solution proposed for \cupid. 
Hence, we expect very similar performances for both the heat and the light detectors.

An alternative option will be considered for the crystal shape and size.
We will investigate the possibility of operating cubic \enrlmo\ crystals with a side of 45~mm,
corresponding to a mass of $\sim280$~g.
This geometry allows to arrange the crystals in tightly-packed towers
with 4 crystals per floor, as in \cuoricino, \cuoreo, and \cuore.
In this design, the amount of passive material between crystals is minimized,
substantially increasing the possibility of suppressing surface background by anti-coincidence.
This is of special importance for the \be\ component that cannot be rejected by \al\ discrimination.
Such a configuration would allow to eliminate the scintillating foil
and operate the crystals in a fully open structure.
Dedicated tests will be required to check that $\alpha$/$\beta$ separation maintains a high enough efficiency
in spite of the reduced light collection efficiency.

\subsection{Infrastructure introduction}

\cupid\ will benefit significantly from the existing infrastructure developed for \cuore.
This includes: the unique \cuore\ cryogenic system, capable of cooling $\sim$1000~kg payloads
$<10$~mK temperatures; internal (cryogenic) and external (room temperature) calibration systems;
underground cleanrooms for detector assembly and parts storage;
a mature detector assembly line; dedicated R\&D dilution refrigerators
both underground and on the surface for testing components;
and well-established procedures and facilities for the surface-cleaning of detector parts.
In the following sections, we describe the existing infrastructure in detail
and provide a conceptual overview of the upgrades envisioned for \cupid.

\subsection{Cryogenics}

%

\subsubsection{The \cuore\ cryostat}

\begin{figure}[t]
  \centering
  \includegraphics[height=0.454\textheight]{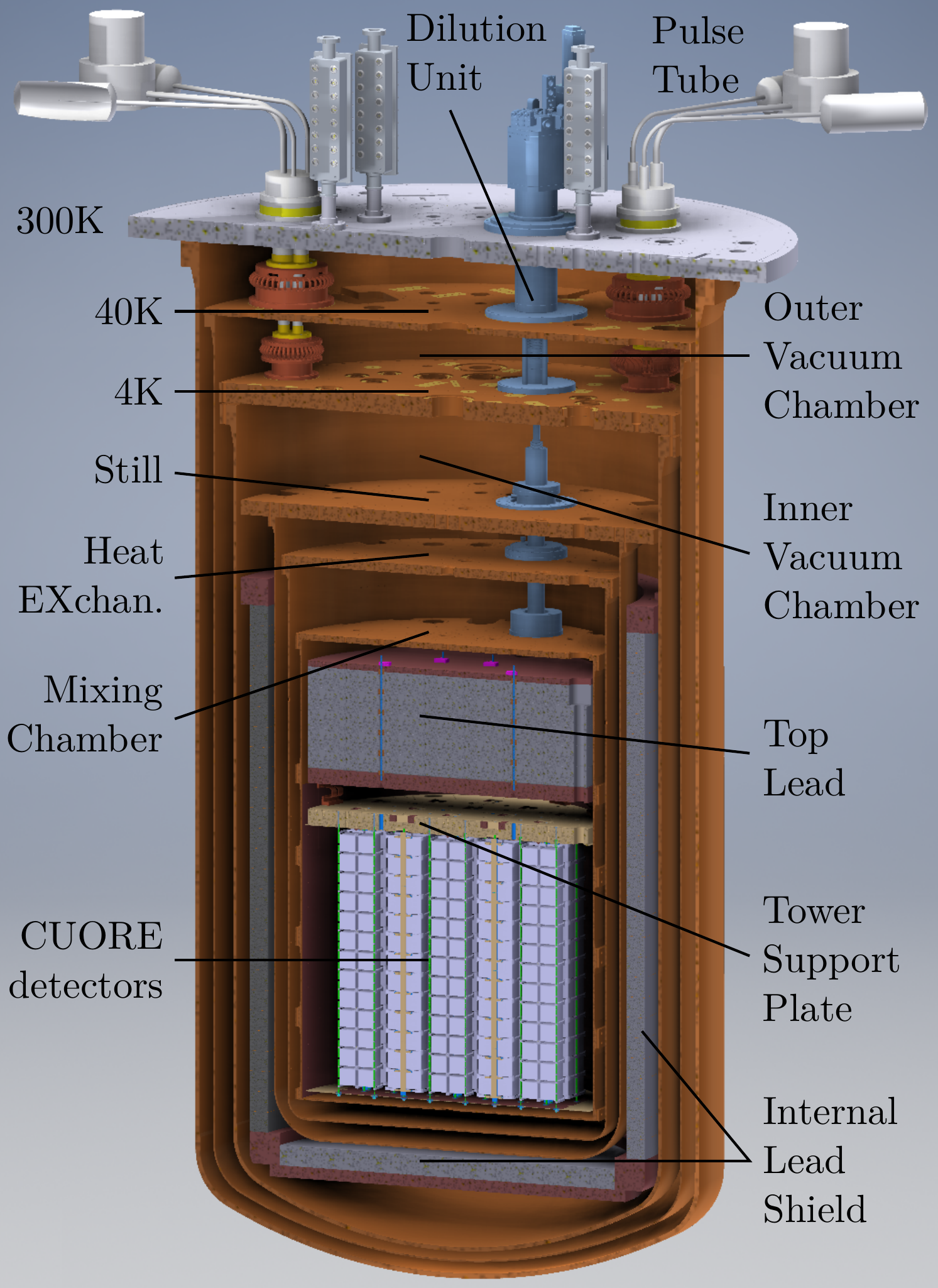}\quad
  \includegraphics[height=0.454\textheight]{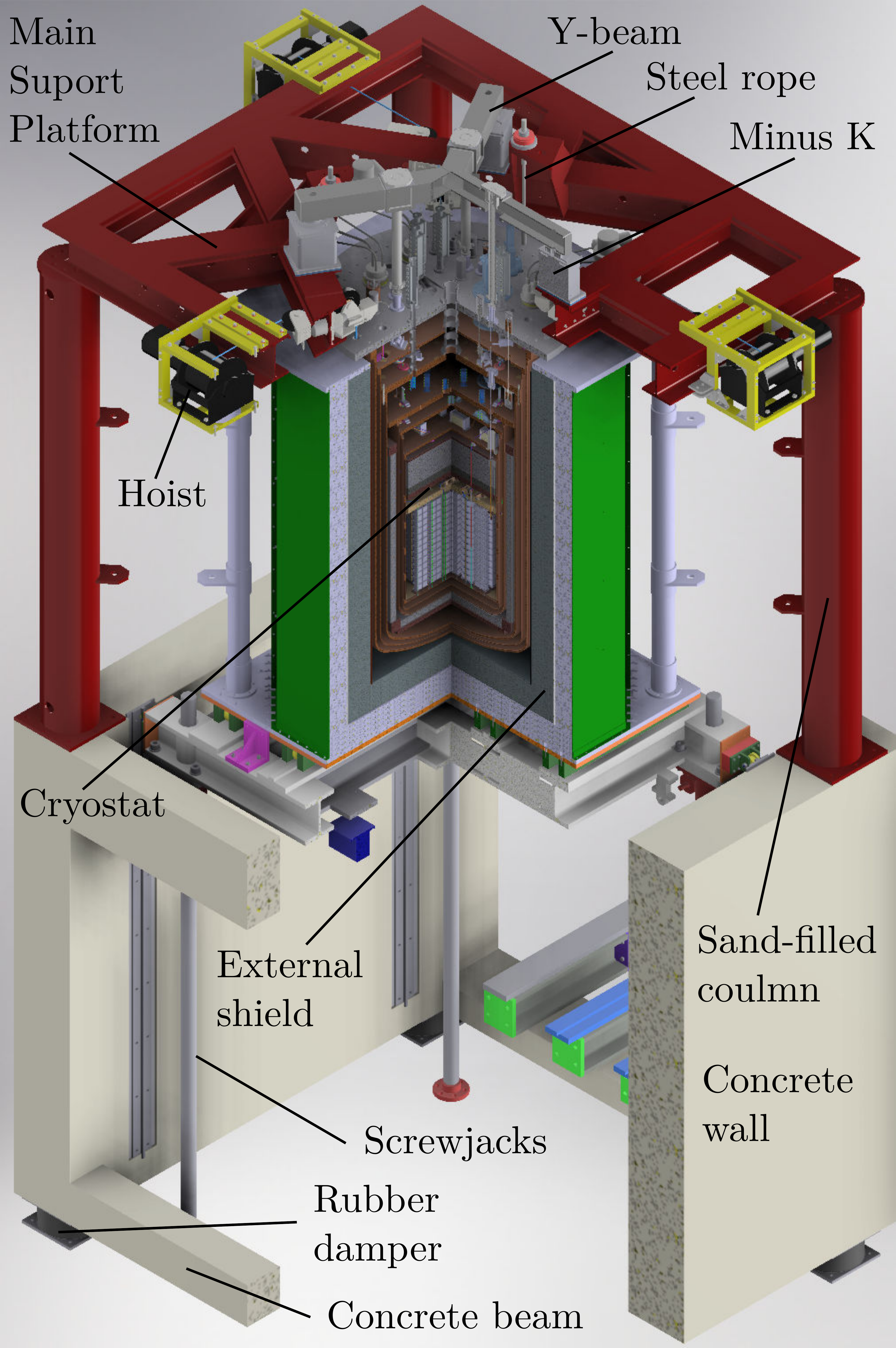}
  \caption{Left: the \cuore\ cryostat with the different thermal stages, the vacuum chambers,
    the cooling elements, the lead shields, and the detector indicated.
    Right: rendering of the cryostat support structure.  
    The external shield sits on a movable platform that can be lifted
    to surround the cryostat during the detector operation.
    When raised, the external shield is at the level of the cryostat cleanroom.}
  \label{fig:cryostat_schematic}		
\end{figure}

Figure~\ref{fig:cryostat_schematic} shows  a schematic of the \cuore\ cryostat~\cite{ Alduino:2019xia}
consisting of six nested vessels, with the innermost containing the experimental volume of about 1~m$^3$.
The different stages are at decreasing temperatures of about 300~K, 40~K, 4~K, 800~mK (also called the still),
50~mK (HEat Exchanger, HEX), and 10~mK (Mixing Chamber, MC).
The 300~K and the 4~K vessels are vacuum-tight and are called
the outer vacuum chamber (OVC) and inner vacuum chamber (IVC), respectively.
Inside the inner vacuum chamber, two lead shields protect the detectors from external radioactivity sources.
The internal lateral lead shield stands in between the still and the HEX stage, while
the top lead is positioned below the MC plate.
The \cuore\ detector is attached to the Tower Support Plate (TSP) placed right below the top lead.

The cryostat is provided with different cooling units.
The 40~K and 4~K plates are cooled down through 5 Pulse Tube refrigerators (PTs).
The still, HEX, and MC plates are cooled down by a \ce{^3He/^4He} Dilution Unit (DU) integrated in the system.

A standalone structure (Fig.~\ref{fig:cryostat_schematic}),
supports and mechanically decouples the cryostat itself from the rest of the world
to minimize the vibrational noise.
The support structure basement consists of two reinforced concrete walls
connected to its foundation through four rubber dampeners with high damping coefficients.
This solution allows for an effective decoupling of the seismic structure from the ground.
On top of the walls, four tubular sand-filled steel columns are installed around the perimeter. 
The columns in turn support the the main support platform,
a grid of steel beams from which the cryostat hangs by means of three ropes connected to the 300~K plate.

\begin{SCfigure}[][t]
  \includegraphics[width=0.65\textwidth]{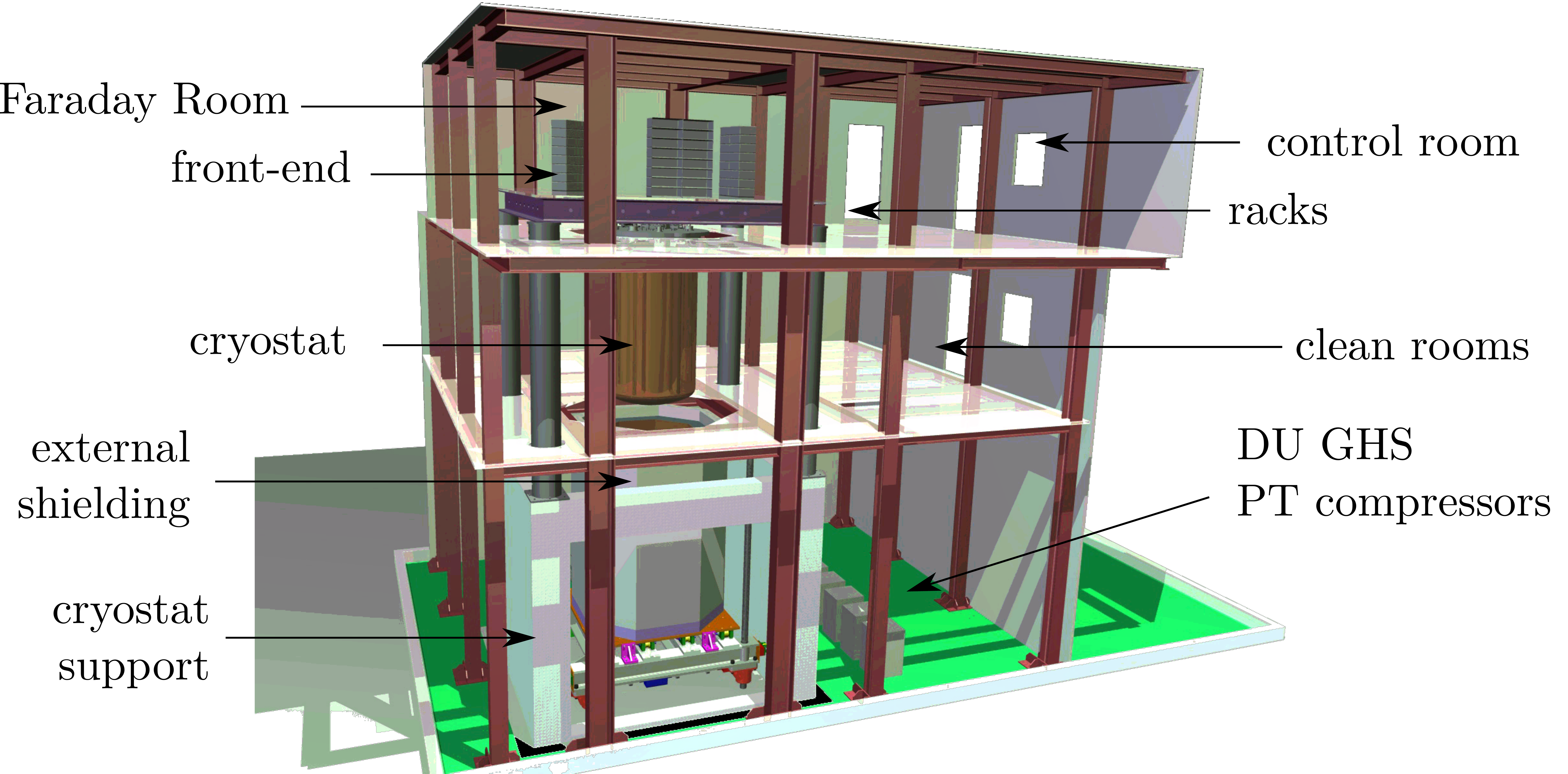}
  \caption{Rendering of the \cuore\ hosting building. The ground floor hosts
    the cryostat and shielding support structure, and most cryogenic components.
    The middle floor consists of 5 cleanrooms for the detector mounting and hosts the cryostat.
    The top floor is dedicated to the data acquisition system and to the electronics,
    which is located inside a Faraday room.}
  \label{fig:cuorehut}
\end{SCfigure}

The 300~K plate holds directly or indirectly all cryostat components
through a cascade of dedicated bars, with three bars per plate to ensure
a precise vertical alignment.
In particular, the 300~K plate holds the 40~K, 4~K and still plates;
the HEX plate is held by the still plate and supports in turn the MC plate.
Each plate holds the corresponding vessel, which acts as thermal radiation shield.
The internal lateral shield is mechanically supported by the still plate, but thermalized to the 4~K one, 
while the top lead directly hangs from the 300~K plate and is thermalized to the HEX plate. 
The TSP support is similar to that of the top lead. However, the former is mechanically decoupled from the cryostat
and is held by a Y-beam which lies on three vibration isolators directly anchored to the main support platform.
The cryostat is located inside a cleanroom, while the Y-beam is placed in a Faraday room
hosting the readout electronics (Fig.~\ref{fig:cuorehut}).

\subsubsection{Low-vibration environment}

The \cuore\ cryostat must provide low-noise environment suitable for a bolometric detector.
Noise abatement had to be extensively considered from the early design down to the commissioning phases.
The first step in this direction was to cope with mechanical vibrations,
which generate power by means of micro-friction.
When this power is dissipated on the coldest stages,
it can prevent the experimental stage from reaching a stable operating temperature.
The impact on the bolometers is even more severe.
Namely, the crystals are weakly coupled to the cooling unit 
and are more sensitive to temperature changes.
Even when the effect is not relevant for temperature stabilization,
vibrational noise still contributes to the overall noise
and directly impacts the energy resolution of the detectors.

We adopted different solutions in order to mitigate the problem,
protecting the detector from any vibration source.
On one hand, the entire cryostat support structure is intended
for decoupling the system from the surrounding environment.
On the other hand, the PTs themselves are a large source of vibrations,
hence we installed a series of devices to dissipate a significant fraction
of this power outside the cryostat.

In particular, the choice for a remote motor option for the \cuore\ PTs,
with the rotating valve separated from the PT head,
reduces the amount of transmitted vibrations and allows us to separate the valve grounding from the main one,
electrically decoupling the PT and the cryostat from the external world.
Furthermore, the PTs are not directly anchored to the cryostat plates,
as depicted in Fig.~\ref{fig:cryostat_schematic}. 
At the level of the 300~K plate, a polyurethane ring compensates
for the horizontal displacement of the PTs due to the thermal contractions at the cold stages.
Inside the cryostat, flexible thermalizations (copper braids) link the cold heads of each PT
to the cryostat flanges at the 40~K and 4~K stages, 
therefore making a softer connection between the PTs and the cryostat plates.
A substantial effort has also been made to avoid any mechanical contact
between the cryostat or main support platform and all the vibrating PT elements.
The rotating valves and the buffer volumes are suspended by means of bungee cords.
At the same time, the combination of custom rigid and flexible high pressure lines have been routed
from the compressors in order to vertically drop on the rotating valves.

The PT compressors are located 15--20~m away from the 300~K plate
and are supported by a metallic structure connected to the ground via elastomers.  The incoming/outgoing \ce{He} flexlines pass through a sandbox rigidly connected to the ground,
thus absorbing a significant fraction of the generated vibrations.
The flexlines are covered with neoprene to reduce acoustic noise,
while at the entrance of the Faraday room the presence of ceramic electric insulators
ensures that the long lines do not act like antennas.

In addition, instead of relying on the micro-stepping driver by default embedded in the compressors,
the PT rotating valves in \cuore\ are driven by 
devices specifically intended to minimize the motor vibrations.
These elements are fundamental to control the relative phases of the PT pressure oscillations
and fix them in a configuration that minimizes
the vibration transmission to the detector~\cite{Daddabbo:2018}.

The \cuore\ detector is mechanically decoupled from the cryostat.
The TSP hangs from the steel Y-beam, which is positioned on top of three Minus~K vibration isolators 
directly anchored to the main support platform (Fig.~\ref{fig:cryostat_schematic}).
The isolators behave like soft spring when subjected to small displacements
and act as a low-pass filter with a cut-off frequency of $\sim0.5$~Hz.

\begin{SCfigure}[][hp]
  \includegraphics[width=0.7\textwidth]{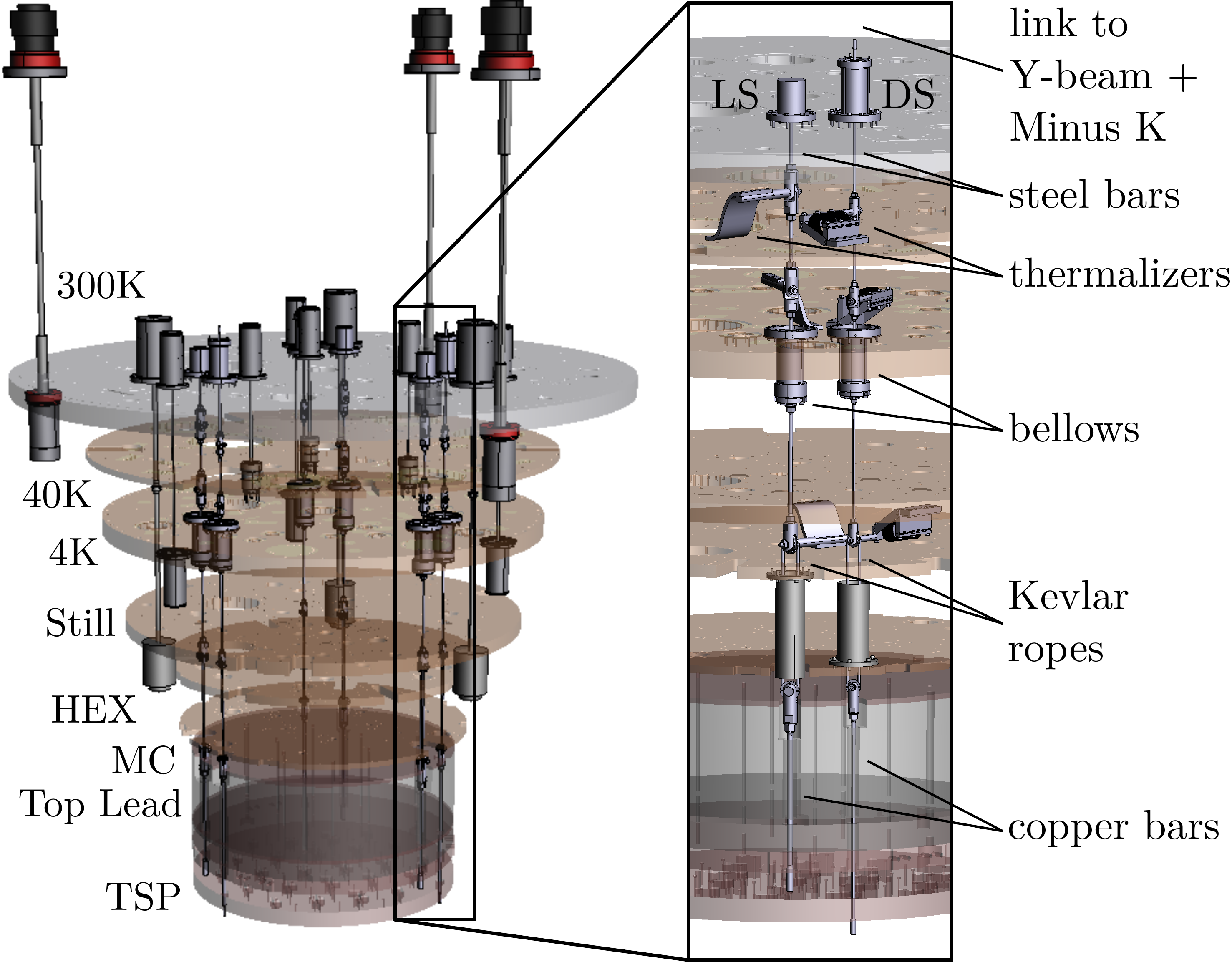}
  \caption{Rendering of the cryostat the supporting elements.
    The zoom shows the details of the detector and lead suspension systems.}
  \label{fig:cryo_structure}
\end{SCfigure}

The connection between the TSP and the Y-beam is made by means of the detector suspension system.
This, as well as the top lead suspension system, consists of three composite bars (see Fig.~\ref{fig:cryo_structure}).
The top part, between the Y-beam (300~K plate for the lead suspension system) and the the still level,
is made of segmented steel rods, which are thermalized at each thermal stage.
The middle part, between the still and the MC, consists of
a double Kevlar rope to minimize the heat input at the coldest stages.
Finally, the lowest part is made by high purity copper rods, chosen for theirs low radioactivity content.

\subsubsection{Low-radioactivity environment}

The \cuore\ cryostat must not only satisfy the requirements related to the cryogenics,
mechanics and detector performances, but also to the radioactive content
of its components, resulting in a strict selection of both the materials and the production techniques.
The use of copper was preferred for the largest masses,
i.e. for plates and vessels, resulting in a total mass of $>6$~tons.
The only exception was made for the 300~K plate and the upper part of the vessel flange,
which are made of austenitic stainless steel 
to guarantee a better mechanical stability and vacuum tightness given the huge load they sustain.

In particular, all the vessels and plates down to the HEX stage are made of Oxygen-Free Electrolytic copper,
which presents a radioactive content $<6.5\cdot10^{-5}$~Bq/kg and $<5.4\cdot10^{-5}$~Bq/kg
for \ce{^{232}Th} and \ce{^{238}U}, respectively. 
The MC flange and plate, the TSP, and the crystal holders are made of Cu NOSV,
a special copper alloy suitable for cryogenic use produced by Aurubis.
Cu NOSV features a high conductivity at low temperatures, with a residual resistance ratio  $>400$,
a low hydrogen content, and bulk radioactivity contents even lower those of Oxygen-Free Electrolytic copper.
All the copper was stored underground and extracted only for the machining operations
to avoid recontamination by cosmic radiation.

The lead shielding, inserted to protect the detector from the external radiation,
underwent a similar material selection process.
The internal lateral shield consists of a 6~cm thick lead vessel
made of radiopure lead of archaeological origin~\cite{Alessandrello:1998ats}.
The top lead consists of five 6~cm thick disks of pure commercial lead
piled and sandwiched between two copper plates. 
The choice for this material is less critical since the top lead
is separated from the detectors by more than 9~cm of copper.
High purity cooper has been used for the support parts:
Oxygen-Free Electrolytic copper for the internal lateral shield rings and bottom plate, and Cu NOSV for the top lead plates.

The entire cryostat is installed inside a cleanroom (see Sec. \ref{sec:cleanroom}).
The high cleanliness standards reduced the risk of recontaminating the various parts
during the commissioning phase and allowed a safe installation of the detectors.
In particular the 300~K plate lies at the level of the cleanroom ceiling
and separates the cleanroom environment from the standard one
(see Fig.~\ref{fig:cuorehut} for comparison).

\subsubsection{Cooldown system and thermal performance}

Given its extraordinary size and mass, \cuore\ utilizes a customly designed cryogen-free cryostat
instrumented with multiple PMTs and a custom \ce{^3He}/\ce{^4He} dilution refrigerator.
The total mass to be cooled down is about $13.7$~tons, of which $12.7$~tons to below 4~K.
The heat content to be extracted from the system is very large:
the inner vacuum chamber enthalpy difference between room temperature (295~K) and 40~K,
which represents more than $95\%$ of the total, is $\sim 6.9\cdot10^8$~J.
The PT cooling power is $>100$~W at 300~K and quickly decreases at lower temperatures, down to  $<50$~W at 100~K.
Moreover, PTs are designed to work at high temperatures only for short periods.
Therefore, we designed and realized a dedicated apparatus, the Fast Cooling System (FCS),
to reduce the cool down time by directly injecting cold \ce{He} gas
inside the inner vacuum chamber and forcing its circulation through a dedicated cooling circuit (see Sec \ref{sec:fcs}).
The FCS drives the initial phase of the cool down and is turned off
when the system is at $\leq$ 100 K at all thermal stages.
Meanwhile, the PTs are turned on.
The PTs, first supporting the FCS and then alone,
eventually bring the 40~K and 4~K stages to their base temperatures while the inner stages follow the 4~K one.
The cool down is completed by the DU, which reaches and maintain the coldest stage temperatures,
keeping a steady detector temperature of $\sim10$~mK.

The \cuore\ cryostat utilizes five PT415-RM by Cryomech.
In principle, 3~PTs would be enough to guarantee the 40~K and 4~K stage base temperatures. 
However, we decided to keep a spare unit in case one fails and,
since an inactive PT creates a connection between the 300~K, 40~K, and 4~K stages,
thus representing a thermal load for the system, we installed 2 extra PTs.
We characterized the individual PTs in order to quantify the available cooling power,
which is $\sim80$~W and $\sim2$~W at 35~K and 3.5~K, respectively,
for the corresponding thermal stages and with 4 active PTs.

The \cuore\ DU is a high-power Joule-Thomson custom CF-3000, manufactured by Leiden Cryogenics,
and is a standalone element with respect to the cryostat.
As for the PTs, the DU is provided with two incoming mixture lines for redundancy,
so that it is possible to continue stable operations in case one of the two lines fails.
The presence of two independent condensing lines also yields a higher cooling power during cooldown,
reaching flows $>8$~mmol/s.

Aiming at operating the \cuore\ cryostat at base temperature of $\sim 10$~mK,
we set very stringent requirements on the DU cooling power
and performed several tests both at the construction site (Leiden, The Netherlands) and at LNGS.
The cooling power is $2$~mW at $100$~mK and $4~\upmu$W at $10$~mK,
while stable operating temperatures of $\sim$ 6~mK were achieved during the cryostat commissioning.
The obtained values exceeded the expectations.

\subsubsection{\cuore\ cooldown}

\begin{SCfigure}
  \includegraphics[width=0.7\textwidth]{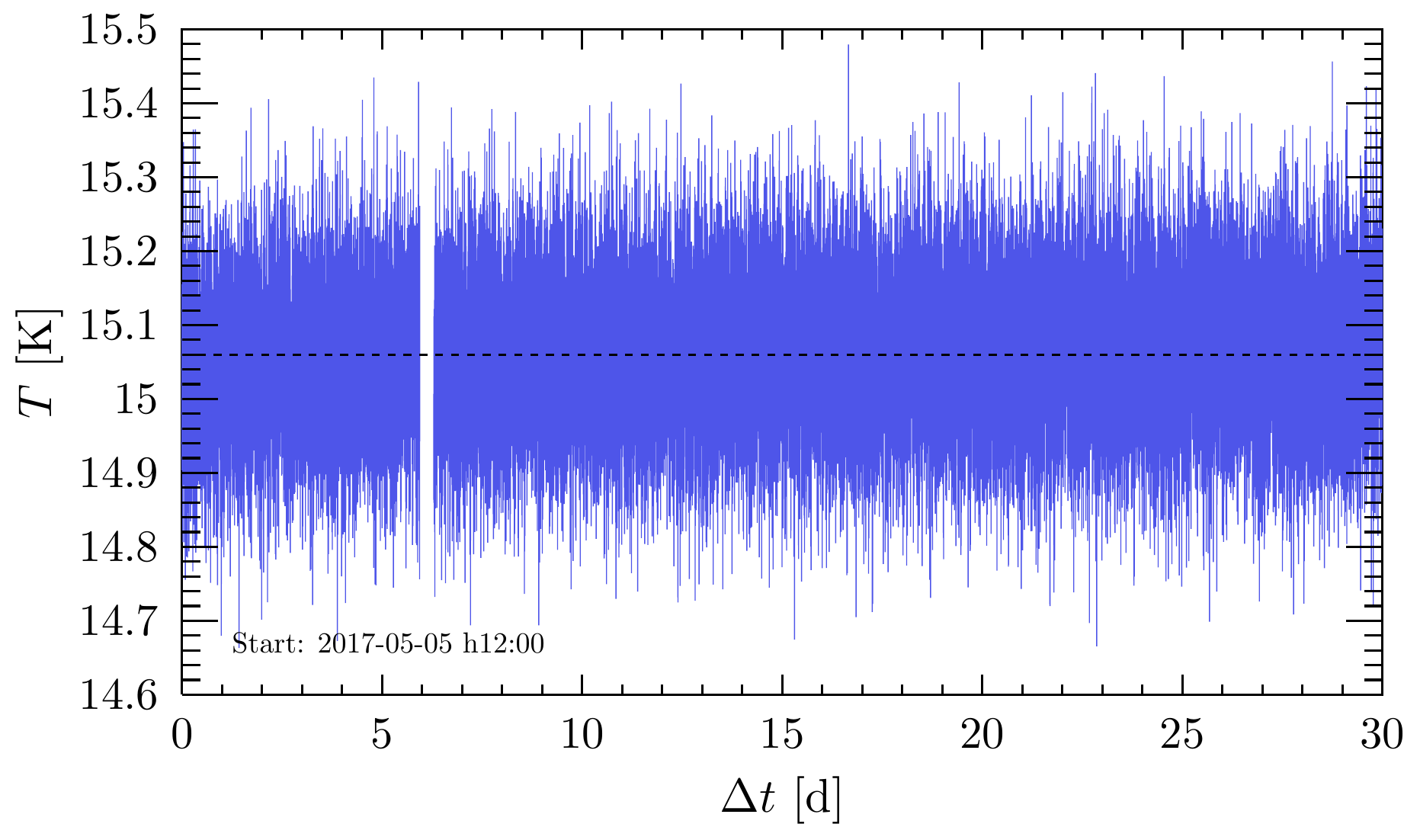}
  \caption{Base temperature stability over a thirty days period
    during the initial phase of the \cuore\ data taking. 
	The MC temperature is $15.06\pm0.10$~mK. The dashed line indicates the average value.
	The gap on day 6 is due to hardware operations on the cryogenic system.}
  \label{fig:R5_TBase}
\end{SCfigure}

The \cuore\ cryostat is a very complex and, despite the large masses involved, a very delicate machine. 
Custom design and construction were required for most of its components
and numerous tests had to be performed to verify the correct functioning of the various parts.
At the same time, special attention was required to ensure that any addition or modification
to the apparatus  did not compromise the overall performance.
The commissioning of the cryogenic system proceeded by steps and involved a period of about four years.
The ultimate test for the cryogenic system could only be demonstrated
by the cooldown and operation of the full \cuore\ detector itself.	

As expected, it took about 20 days in order to bring the 40~K and 4~K stages to their base temperatures.
Once the DU was turned on, it took less than 4 days to reach the temperatures 0.89~K, 55~mK and 8~mK for the still, HEX, and MC stages, respectively.

An initial characterization of the detector identified 15~mK as a suitable operating temperature.
We thus performed a first science run at this temperature.
Fig.~\ref{fig:R5_TBase} shows the stability of the system for over one month of operation,
which ultimately confirms the success of the \cuore\ cryogenic infrastructure.

\subsubsection{Lessons learned}

In the design of the \cuore\ cryogenic system, we considered many requirements.
Building a system capable of cooling down a ton of detectors to mK temperatures
in a seismically safe environment with ultra-low vibrations and radioactivity levels
is an extremely complex challenge.
These requirements are often in conflict, thus we gave the priority to the most critical aspects,
i.e. the thermal performances of the system. 
As an example, we kept large safety margins in the design of the thermal conductances
between the PTs and the cryostat resulting in a lower temperature of the 40~K and 4~K stages
at the price of a lower mechanical decoupling between the PTs and the cryostat plates.

The \cuore\ cooldown has been successful and the cryogenic system performances exceeded the expectations.
On the other hand, the vibration isolation of the detectors is far from optimal
and we had to adopt several corrections in order to decrease the detector noise. 
The main sources of vibrational noise measured on the \cuore\ detectors
turned out to be the PTs themselves, which are too strongly coupled to the cryostat structure.
For this reason we introduced various modifications in the structure outside the cryostat.

With the knowledge accumulated in \cuore\ we can now
refine the cryogenic system acting on the interior part of the cryostat and improving the vibration induced noise.

\subsubsection{Upgrades for \cupid}

We envision two main upgrades of the \cuore\ cryogenic system oriented towards the vibrational noise suppression.
First, we plan to reduce the mechanical coupling between the PTs and the cryostat with two dedicated R\&Ds:
\begin{itemize}
\item the search, design and characterization of copper conductances with high residual resistance ratio
  that can guarantee a comparable thermal conductivity with a lower mechanical coupling with respect to the present arrangement;
\item the development of innovative heat exchangers based on gaseous helium.
\end{itemize}
The latter type of coupling is more difficult to implement in the \cuore\ cryostat
but could bring relevant improvements.
A similar idea was already developed by Cryoconcept.
If successful, the gas heat exchangers could bring the possibility of running the cryostat
with just 3 PTs an additional benefit.

The second planned improvement is to increase the rigidity of the cryostat internal structure.
Indeed, we verified that the \cuore\ detectors suffer from low frequency noise
generated by tiny power dissipation originated by friction of moving parts.
These  movements are induced by the PTs vibration.
Stiffening the cryostat internal structure will limit the impact of the PTs vibrations on the detectors,
keeping them at frequencies higher than the signal bandwidth. 

\begin{figure}[ht!]
\centering
\includegraphics[width=0.99\textwidth]{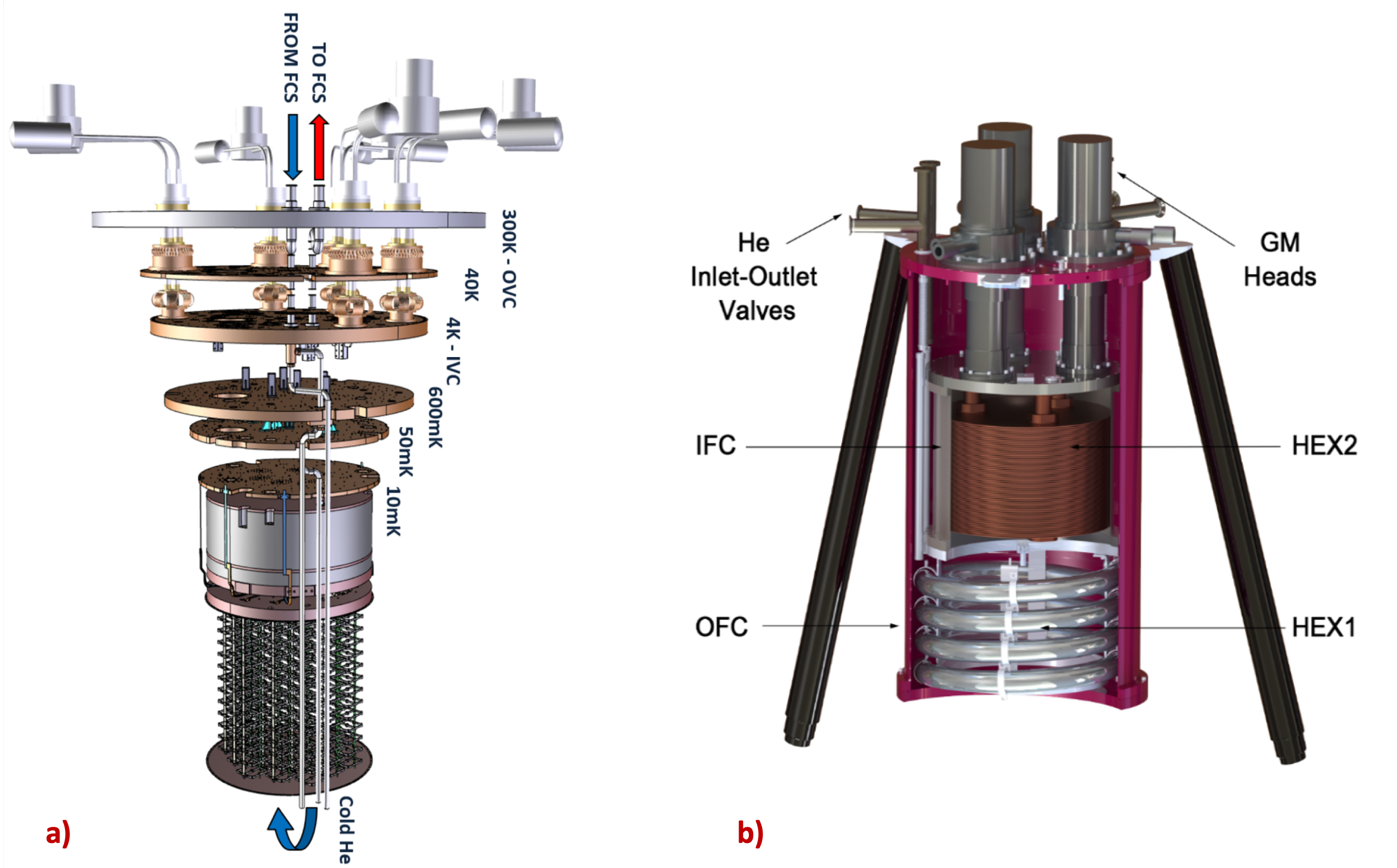}
\caption{Helium circulation inside \cuore/ cryostat. Cold $^{4}$He enters from
the valve located on the top of the 300K plateand  goes directly to the bottom, exchanges heat; b) Cross section view of the Fast Cooling Unit.}
\label{fig:cep-1}
\end{figure}
\subsubsection{\cuore\ precooling stage: the Fast Cooling System}\label{sec:fcs}

The CUORE FCS is composed of the five PT refrigerators and an external system that flows cold $^{4}$He inside the IVC to improve the heat exchange, as shown in Fig.~\ref{fig:cep-1}.a. This external system involves a Fast Cooling Unit (FCU), which is a cryostat containing two heat exchangers, a gas blower, double-walled flexible pipes, a filtering system and several other components and sensors. 
\begin{figure}[hb!]
\centering
\includegraphics[width=0.73\textwidth]{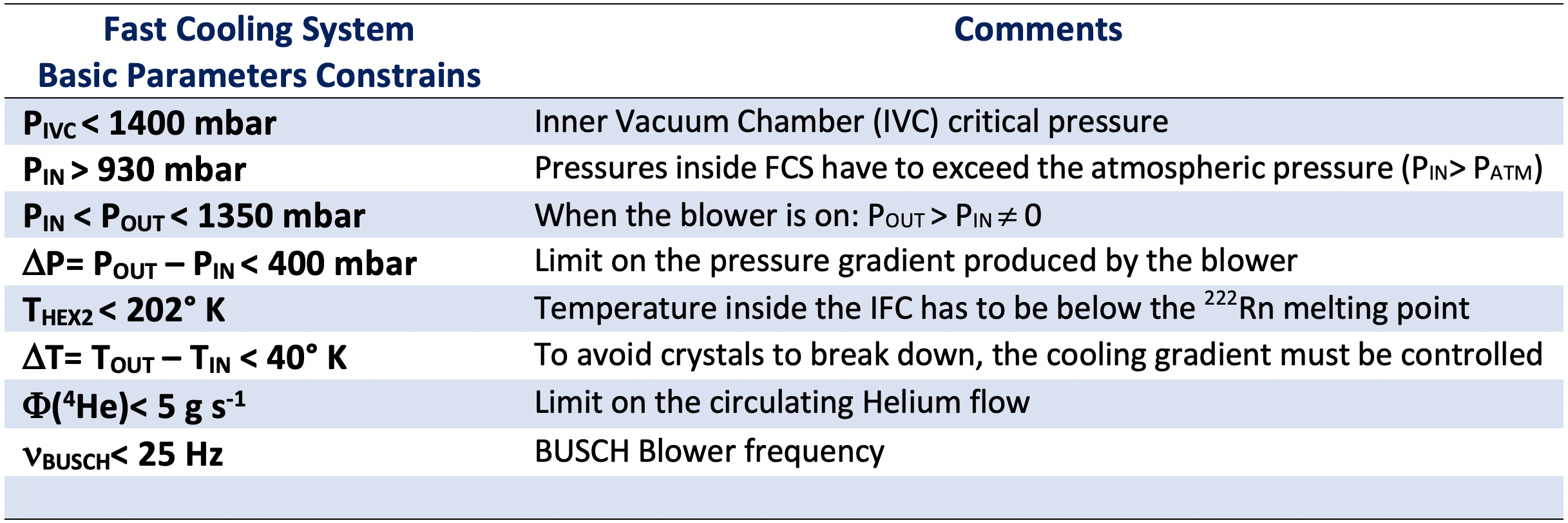}
\caption{Fast Cooling System operational parameter constrains.}
\label{fig:FCS-Table}
\end{figure}
The FCS helium circuit is designed to have two main temperature stages: stage one, upstream of the FCU, has helium gas at room temperature, where the He blower forces the gas moving; stage two, downstream of the FCU, has helium at low temperature, where the gas goes into \cuore\ cryostat for cooling purpose. This procedure allows us to use a normal helium blower (no cryopumps are needed for gas circulation), while the temperatures of the helium, entering the cryostat, can go down to $\sim$ 30-40 K. 

The entire precooling process has several critical issues that have to be taken into account. For example, to avoid damage to the crystals, the temperature gradient needs to be $\Delta T<$ $40$ K and the pressures of the He gas entering ($P_{\text{out}}$) or exiting ($P_{\text{in}}$) from the cryostat has to be limited within a specific range ($\Delta P\equiv$ $P_{\text{out}}$ – $P_{\text{in}}<$ 350 mbar). Another important aspects is that the helium flux $\Phi_{\text{He}}$ cannot exceed 5 g s$^{-1}$. Moreover, to prevent possible $^{222}$Rn radioactive contamination, the entire gas circuit pressures need to be kept above the external pressure and the He temperatures inside the heat exchanger, largely below the $^{222}$Rn melting point of $\sim$ 202 K. Fig.~\ref{fig:FCS-Table} shows a list of the FCS basic parameter constrains as adopted during \cuore\ precooldown operations.
\begin{figure}[]
\centering
\includegraphics[width=0.95\textwidth]{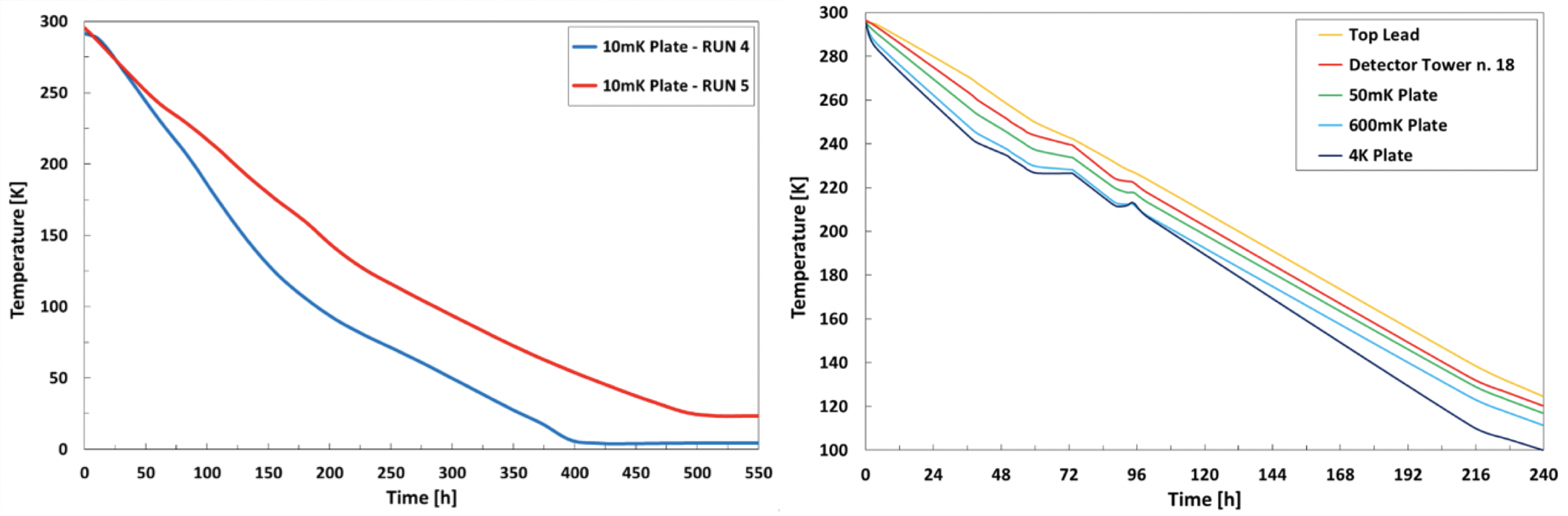}
\caption{Left Panel -- Comparison of the FCS performances on the 10 mK plate between Run 4 and Run 5 (CUORE Run 1) cooldown (see text for Run details). During the Run 5 cooldown, for technical reasons, only three out of five PTs were used to reach ~25 K; the remaining PTs and the DR have been switched on later. Right Panel -- Physics Run: cryostat temperature trend cooling down with the FCS.}
\label{fig:FCS-Plots}
\end{figure}



During the first \cuore\ precooldown in December 2016, the FCS successfully brought the large mass of \cuore’s cryostat and crystals down to 100 K in 10 days (see Fig.~\ref{fig:FCS-Plots}). 


\noindent
A peak cooling speed value of 1.09 $\pm$ 0.02 Kh$^{-1}$ was measured inside the IVC, while having a mean value around 0.90 $\pm$ 0.02 Kh$^{-1}$ decreasing to 
0.49 $\pm$ 0.02 Kh$^{-1}$ at the end of the precooling process. 
Fig.~\ref{fig:FCS-Plots} shows the results achieved during the FCS Run 4 (no detector installed, only a mini test tower) and the FCS Run 5 that correspond to \cuore\ Run I, the first physical run. The lower performances achieved during Run I are mainly due to three factors: for technical reasons, PT refrigerators were switched on later; only three out of five PTs were used; and interruptions were experienced during the cool down due to external factors such as blackouts.

An increase of $\sim$\,20-30\,\% of cooling efficiency and speed can be easily reached if the system runs smoothly, without too many changes in the gas circulation rate, 
and at least 4 of the 5 PTs are used.

\subsection{The \cuore\ Cryogeny Monitor and Control System}

The \cuore\ Cryogeny Monitor and Control System is an important part of the whole \cuore\ Slow Control System. It consists of about hundred LabVIEW based Virtual Instruments, organized in a more general expandable architecture, dedicated to monitor and control different instrumentation and related variables. It runs on a dedicated 17 slots NI PXIe located at the second floor of the \cuore\ hut. It acquires all the required information coming from all the different instrumentation and controls.  This can be done manually or automatically for most of the vital aspects of the experiment such as the HV quality, the UPS status and performances, the water cooling complex, and all the Cryogenic equipment.

\begin{figure}[ht!]
\centering
\includegraphics[width=0.45\textwidth]{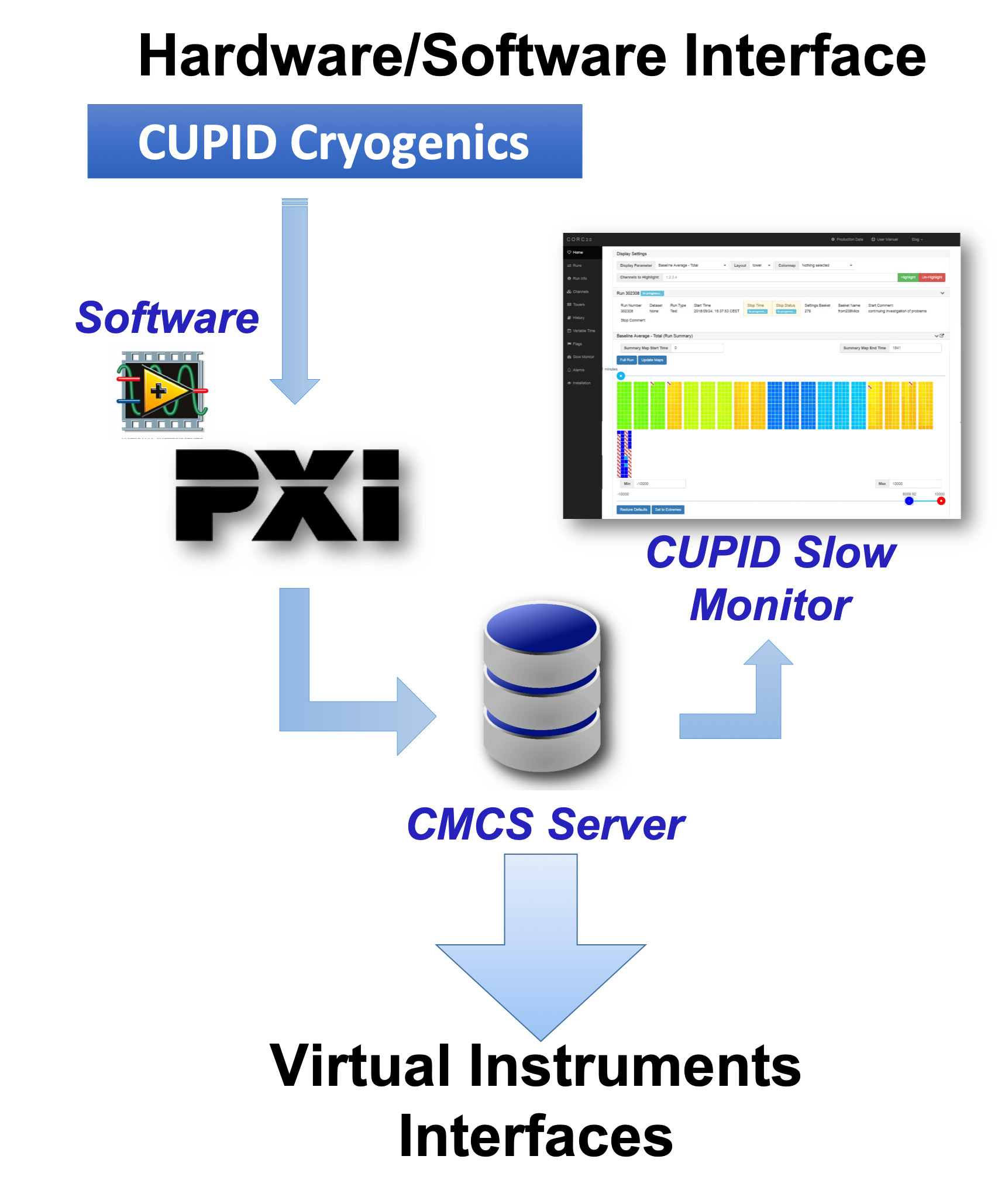}
\includegraphics[width=0.50\textwidth]{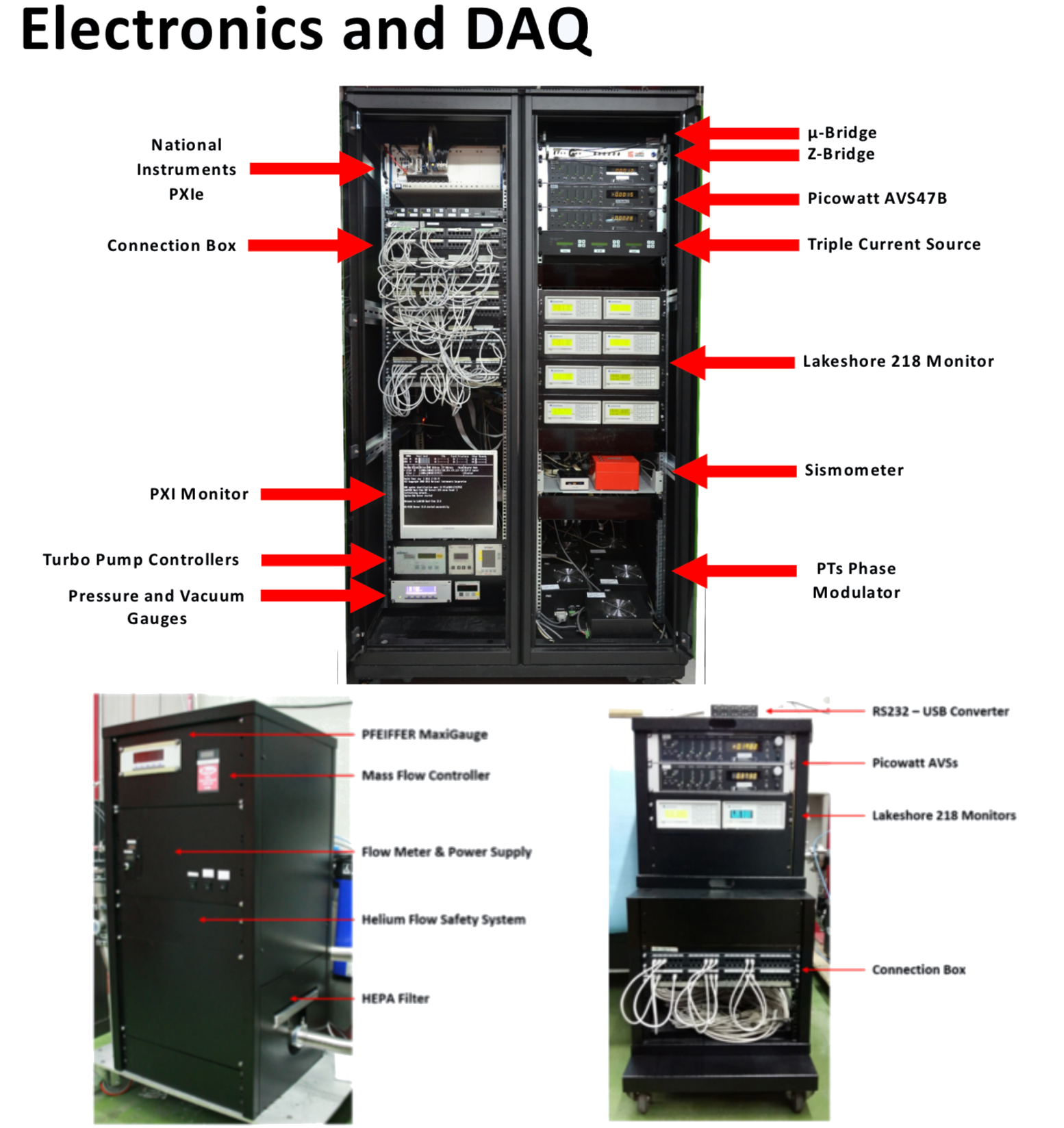}
\caption{Left Panel -- \cupid\ Cryogenic Monitor and Control System Hardware/Software Interface. Right Panel -- Electronics and DAQ for the cryogenic system as it's at the present for \cuore\ Experiment.}
\label{fig:CMCS-4-5}
\end{figure}


\subsection{Electric demand and infrastructures}

The maximum power load that LNGS can deliver to the \cuore\ experiments and to the associated infrastructures is $P_{\text{max}}$= 162 kW.
The delivered power can be shared between normal service, with power P$_{\text{NS}}$,
and the UPS Service with power P$_{\text{UPS}}$.
Without modifying the present electrical network, 
the following constrains apply:
$P_{\text{NS}} + P_{\text{UPS}}$ $<$ 162 kW 
and 
P$_{\text{UPS}}$ $<$ 140 kW.
Presently, \cuore\ uses two UPS Systems, called UPS-1 and UPS-2.
UPS-1 electrical loads are the water cooling, the PTs, the DU and related pumps,
the main vacuum pumps, the backup pumps, the PLC rack, 
the cleanroom, and the radon abatement system.
UPS-2 electrical loads are the electronics, DAQ, and all equipment located in the Faraday room.
The total maximum load size, under UPS is 96~kW. 
The existing \cuore\ electric infrastructure allows us to have an extra UPS system
with a maximum power load of 50~kVA (40~kW), leaving more than enough power (25~kW)
for normal electric service available for other needs.

Based on a measurement performed on \cuore\ during standard running conditions, we measured a
UPS-1 electrical peak power of $P_{\text{peak}}^{\text{UPS-1}} \sim 41.8$~kW 
($\sim 50 \% P_{\text{max}}^{\text{UPS-1}}$)
with an average electrical power of 
$P_{\text{AVG}}^{\text{UPS-1}} \sim 17.2$~kW.
Similarly, UPS-2 delivered
$P_{\text{peak}}^{\text{UPS-2}} \sim 10.2$~kW
($\sim 50\,\% P_{\text{max}}^{\text{UPS-2}}$), with
$P_{\text{AVG}}^{\text{UPS-2}} \sim 5.2$~kW.
These measurements confirm that there is enough room for adding small cryogenic devices,
an emergency chiller, and an extra UPS
without modifying or expanding the present electric network.


\subsection{Water cooling system}\label{wcs}

Each \cuore\ PT is connected to a water-cooled PT1010 Cryomech compressor
that needs to dissipate, in water, a maximum power of  12~kW 
for a total 60~kW for the full PT system.
Each compressor works with an average water flux of 11.5~l/min. 
As a comparison, the DU needs to dissipate a power of just $\sim 2$~kW.
The present \cuore\ water cooling system is twofold.
Under normal running conditions, the circuit that cools down
all the critical components of the \cuore\ cryogenic system
passes through a heat exchanger directly  connected to the Hall A conduit
of the primary laboratory water circuit.

An emergency condition state occurs when the laboratory-chilled water cannot reach
the \cuore\ heat exchanger.  In such a situation, the secondary water circuit is isolated from the laboratory
primary  circuit and directly connected to a small chiller (McQuay, M4AC 120 CR), located outside the Hall A.
This chiller has a buffer of 300~L of cold water and can deliver 10$^{\circ}$C water
for a maximum integrated power of 40~kW.
At present, only three out of five PTs, plus the DU, can be operated during an emergency condition.
The size of the chiller was originally not intended to work for the \cuore\ cryogenic system,
as it was inherited from the much smaller experiment Cuoricino.
During precooling, a larger number of Cryomech compressors (7-8) run at the same time and require the dissipation of $\sim90$--100~kW of total heat power.
We plan to double the size of the external chiller
and put the chilling system under UPS to allow stable operations
even when there are temporary external problems.


\subsection{Detector assembly and installation} 

In this section we describe the assembly procedure and facilities used for \cuore.
At this stage of the conceptual design we expect many of these procedures will be similar for \cupid.
We highlight lessons learned and areas which will differ and will need technical R\&D to develop procedures. 

The assembly process for \cuore\ consisted of four steps:
\begin{itemize}
\item gluing, i.e., fixing sensors to the crystals;
\item mechanical assembly of the copper frames, crystals and CuPEN readout strips into towers;
\item wirebonding, i.e., forming the electrical connection between sensors and the readout strips;
\item tower closing, extraction and storage.
\end{itemize}
Each \cuore\ tower is composed of 13 identical floors with 4 bolometers each.
These are held in position by a composite structure formed by pure copper frames
providing the necessary mechanical support and PTFE holders
that provide the cleanest possible contact to prevent crystal recontamination.

\subsubsection{Underground cleanroom facilities} \label{sec:cleanroom}

\begin{figure}
  \centering
  \includegraphics[width=.7\textwidth]{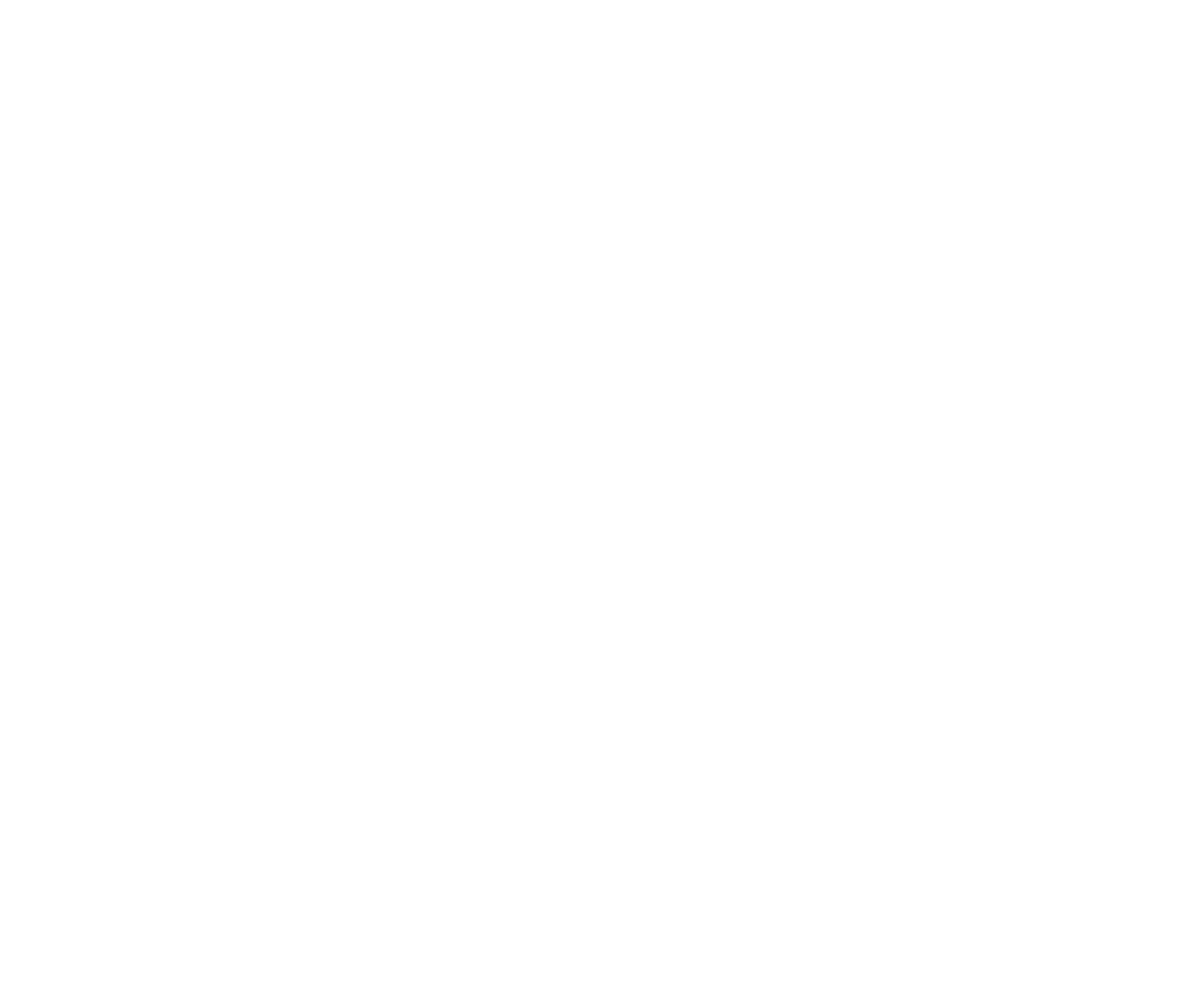}
  \caption{Map of the \cuore\ cleanroom. It is a class 1000 cleanroom with approximately 100~m$^2$ floor space divided into five sub-volumes.}
  \label{fig:CleanRoomMap}
\end{figure}

The four main assembly steps took place in the \cuore\ cleanroom
on the first floor of the \cuore\ hut underground at the LNGS (Fig.~\ref{fig:cuorehut}).
The cleanroom is a class 1000 with approximately 100~m$^2$ floor space,
divided into five sub-volumes (called CR1, CR2, CR3, CR4 and CR5, Fig.~\ref{fig:CleanRoomMap}).
CR1 is an entrance area, divided in sub-areas by curtains in order to separate the vestibule
from the entrance to the other rooms (except CR4, which is accessible from CR3 only).
The use of the other rooms changed according to the different phases
of the \cuore\ construction, in particular during the detector assembly and cryostat commissioning phase,
and the detector installation phase.

In the construction phase, CR2 was dedicated to the semi-automatic gluing system.
Namely, CR2 has a feedthrough towards the ground floor of the hut,
where the pump and the main unit of the robotic arm of the gluing system are located.
It is served with water lines for the gluing system chiller.
The \cuore\ gluing proceeded independently from other activities in the hut. 

The mechanical assembly and wirebonding took place in CR3 and CR4
through the \cuore\ Towers Assembly Line (C-TAL).
Dedicated glove boxes were used to assemble and wire the several components in a clean environment,
preserving the radiopurity of the detector during the construction.
An ancillary glove box was also used to prepare the sub-assemblies for the operations in the main ones.
Once completed, the \cuore\ towers were stored in dedicated areas of CR3 and CR4
in special boxes flushed with nitrogen.

The heavy mechanical work of constructing the cryostat took place in CR5.
In reality, CR5 could not be maintained as a cleanroom until the end of the cryostat commissioning.
CR5 personnel used the same entrance to the hut as personnel doing critically clean work in CR2, CR3, and CR4,
leading to difficulties maintaining cleanliness of CR5 itself.
Therefore, the installation of the \cuore\ detector into the cryostsat
took place in a special soft-wall cleanroom (called CR6),
designed to fit into CR5 and supplied with radon-free air.
The introduction of CR6 was not the only modification
of the \cuore\ cleanroom in view of the tower installation.
CR1 was re-organized in order to guarantee a clean path between CR3/CR4,
where the stored towers were located, and CR6;
CR2 became a second vestibule; CR4 became a storage area.

Some modifications of the logistical environment will be required
of the \cuore\ assembly line in order for it to be used for the \cupid\ detector assembly.
Primarily, the wire bonding operations must be decoupled from the detector mounting.
Wire bonding is a delicate task that can only be performed by trained personnel,
and can interfere with the detector mounting if the two operations are performed in the same space.
Considering that both the procedures must be completely revised for \cupid,
this is an opportunity to properly decouple these tasks
and relocate the bonding equipment outside C-TAL.
Moreover, completely decoupling CR5 from the other rooms is desirable
to preserve the cleanliness of the other cleanroom areas.

\subsubsection{Other support facilities}

We also performed many core assembly tasks in other facilities provided by LNGS.
We cleaned and processed the small parts such as heaters, thermistors, PTFE supports,
and tools for assembly using cleanrooms, wet benches, and laminar flow hoods in the above-ground chemical labs.
LNGS also provided lab space for electrical testing the CuPEN readout strips.
We used the underground parts storage area, adjacent to the \cuore\ hut,
to store parts prior to assembly, as well as to store glued crystals prior to integration into towers.
We also employed a fume hood and a cleanroom for cleaning and reprocessing
of thermistors and heaters that were not successfully glued to the crystals.
We expect \cupid\ will require similar support.  

\subsubsection{Gluing}

\begin{SCfigure}
    \includegraphics[width=.7\textwidth]{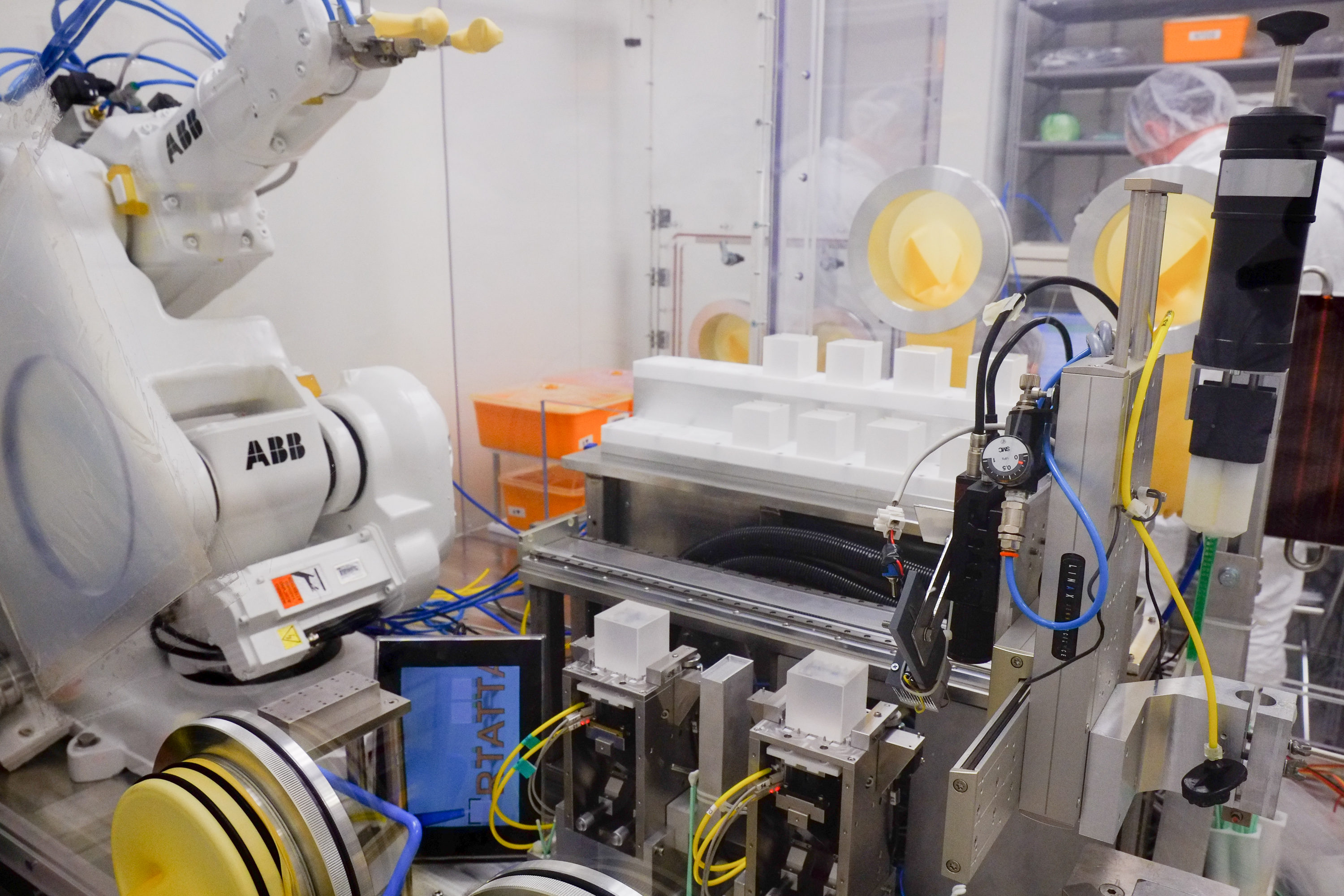}
    \caption{Front view of the \cuore\ semi-automatic gluing tool, with
      the robotic arm for crystal handling, the crystal and sensor staging,
      the two z-positioners (hosting a crystal each), and the robotic arm for the glue deposition.}
    \label{fig:gluing}
\end{SCfigure}

As noted earlier, we performed the gluing in CR2.
We carried out the entire operation in a nitrogen-flushed glovebox
to prevent any radon recontamination of the crystals.
The system consisted of a station for crystal loading and inspection:
a robotic arm moved the crystals from the inspection station to a staging position in the glovebox.
Precision positioners were designed to hold the sensors
(thermistors and heaters, see Sec.~\ref{sec:LightDetectors}) in a fixed position
relative to the crystal during gluing.
A human operator mounted the sensors on the positioner
while a second operator prepared glue for the upcoming operation.
Once ready, a second robotic arm printed a matrix of glue dots on the sensor.
The operators then instpected the deposited glue matrix
with the aid of high resolution cameras and imaging processing software.
If the glue matrix passed quality inspection, the robotic arm moved a crystal
from the staging area to the positioner allowing the crystal face to come into contact with the glue matrix.
The glue was allowed to cure for 50 minutes.
After curing, the robotic arm repositioned the glued crystal to the staging area.
At the end of the operation, the operators inspected the cured glue matrix
with the aid of high-resolution cameras.
A semi-automatic GUI interface controlled the entire procedure,
which required a team of three technicians and one physicist supervisor.
A picture of the \cuore\ gluing tool is displayed in Fig.~\ref{fig:gluing}.

For \cupid, new sensor positioners are needed because of the cylindrical geometry
of the \lmo\ crystals and Ge wafer light detectors.
In particular, the latter can be designed starting from the solution
adopted during the gluing of the \cupido\ detector.
In this case the Ge wafer was encapsulated in a dedicated PTFE holder,
allowing us to handle them as ``normal'' crystals \cite{Azzolini:2018tum}.

The robotic positioning arm and glue dispensing arm requires reprogramming
to account for the different geometries and glue matrix patterns for each sensor-detector combination.
In \cuore, one tower worth of crystals were glued every two weeks,
with operations mostly confined to a single 8-hour shift on weekdays.
The time was dominated by the cure time of the glue,
i.e. the time that had to elapse before moving on the next crystal.
The system had approximately 20\% downtime for routine and non-routine maintenance. 

\subsubsection{Mechanical assembly}

\begin{figure}
  \centering
  \includegraphics[height=0.22\textheight]{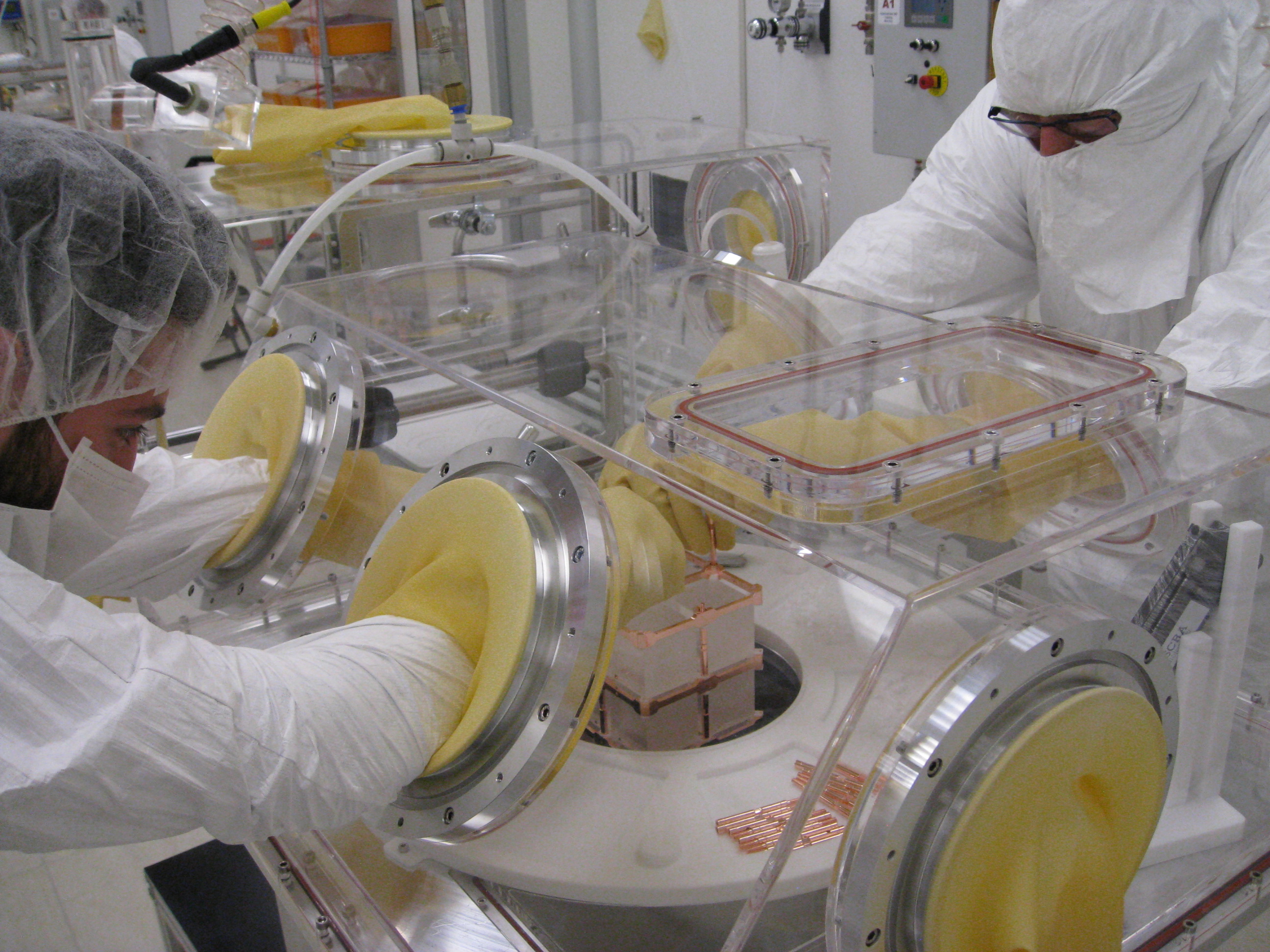}\quad
  \includegraphics[height=0.22\textheight]{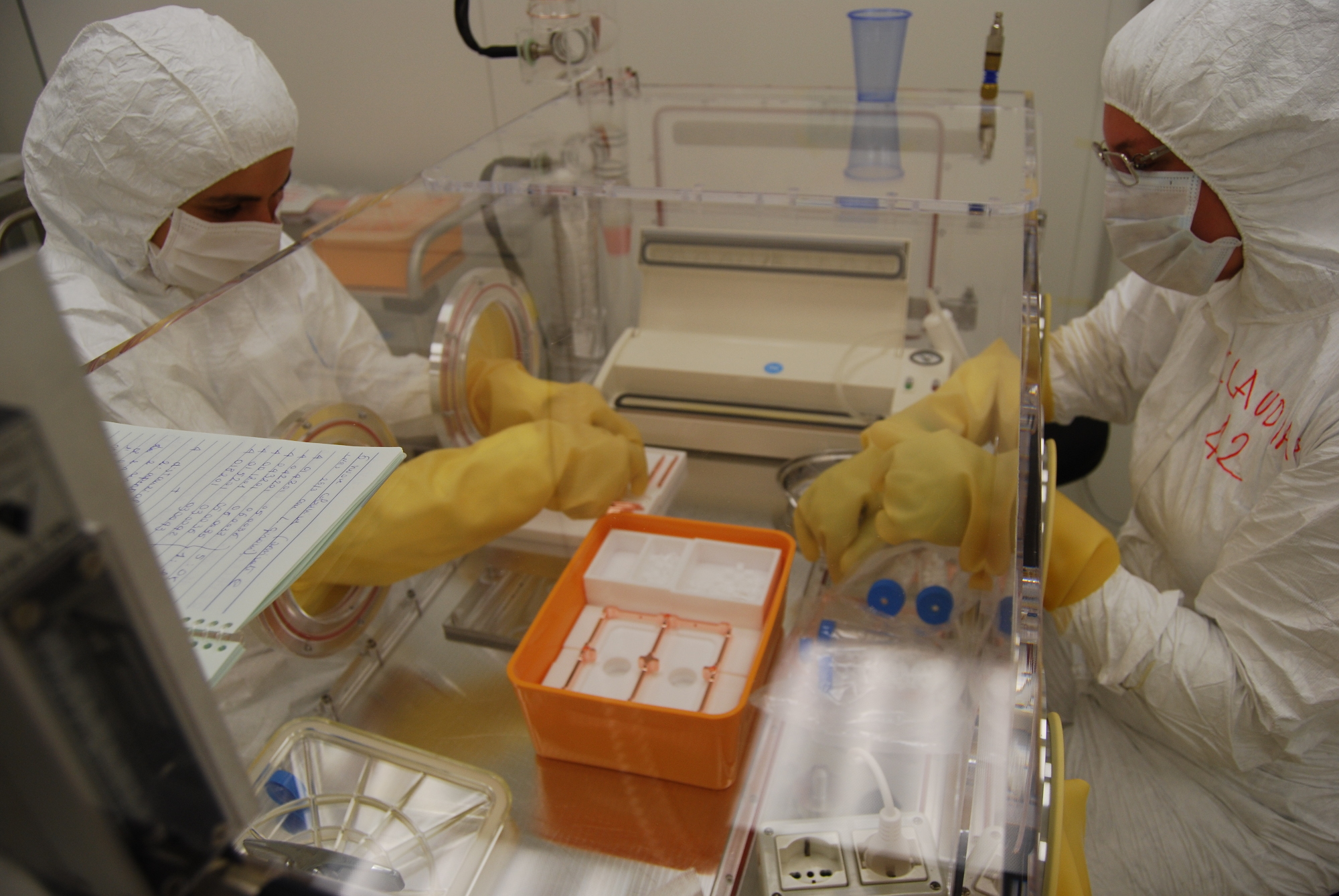}
  \caption{Left: A \cuore\ tower under construction into the Mech Box. Right: People working at the End Of the World Glove Box.}
  \label{fig:gloveboxes}
\end{figure}

The mechanical assembly of the glued crystals and copper frames into towers took place in CR3 and CR4.
We performed all steps in specially designed, task-specific,
nitrogen-flushed gloveboxes~\cite{Buccheri:2014bma} (the C-TAL).
The operations include:
\begin{itemize}
\item unpacking and quality checking of all copper and PTFE pieces,
  running-in of all threaded components to avoid gauling during the assembly;
\item assembly of copper frames, PTFE holders, and crystal into towers;
\item gluing the left and right pack of CuPEN strips to their respective backing copper frame
  and allowing the glue to cure;
\item attaching the CuPEN strip assemblies to the tower.
\end{itemize}
These operations involved three different glove boxes (Fig.~\ref{fig:gloveboxes}):
one for the mechanical assembly,
one for the preparation of the wirings on related cable trays,
and one for the installation of the cable trays on the tower.
The main workstation was mounted on a stainless steel air-tight nitrogen flushed chamber
equipped with an automated system able to lift and rotate the tower under-construction, as needed.
After the tower was assembled, it was removed and stored in a nitrogen-flushed PMMA chamber,
and parked inside dedicated racks equipped with nitrogen lines (Fig.~\ref{fig:cabling}).
The assembly of one tower required a team of four technicians and one week of time.

\subsubsection{Wire bonding} 

\begin{figure}
  \centering
  \includegraphics[height=0.2235\textheight]{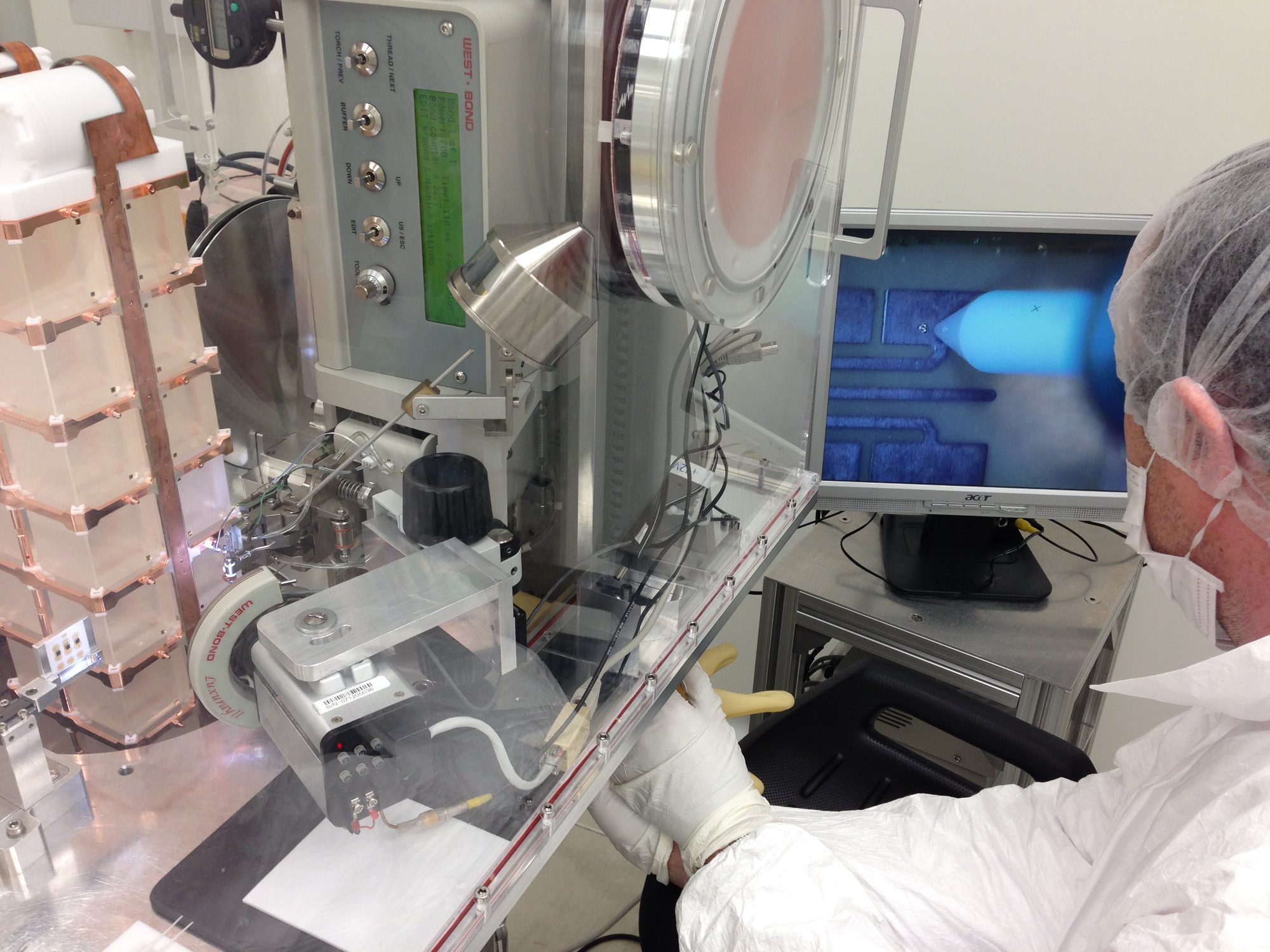}\quad
  \includegraphics[height=0.2235\textheight]{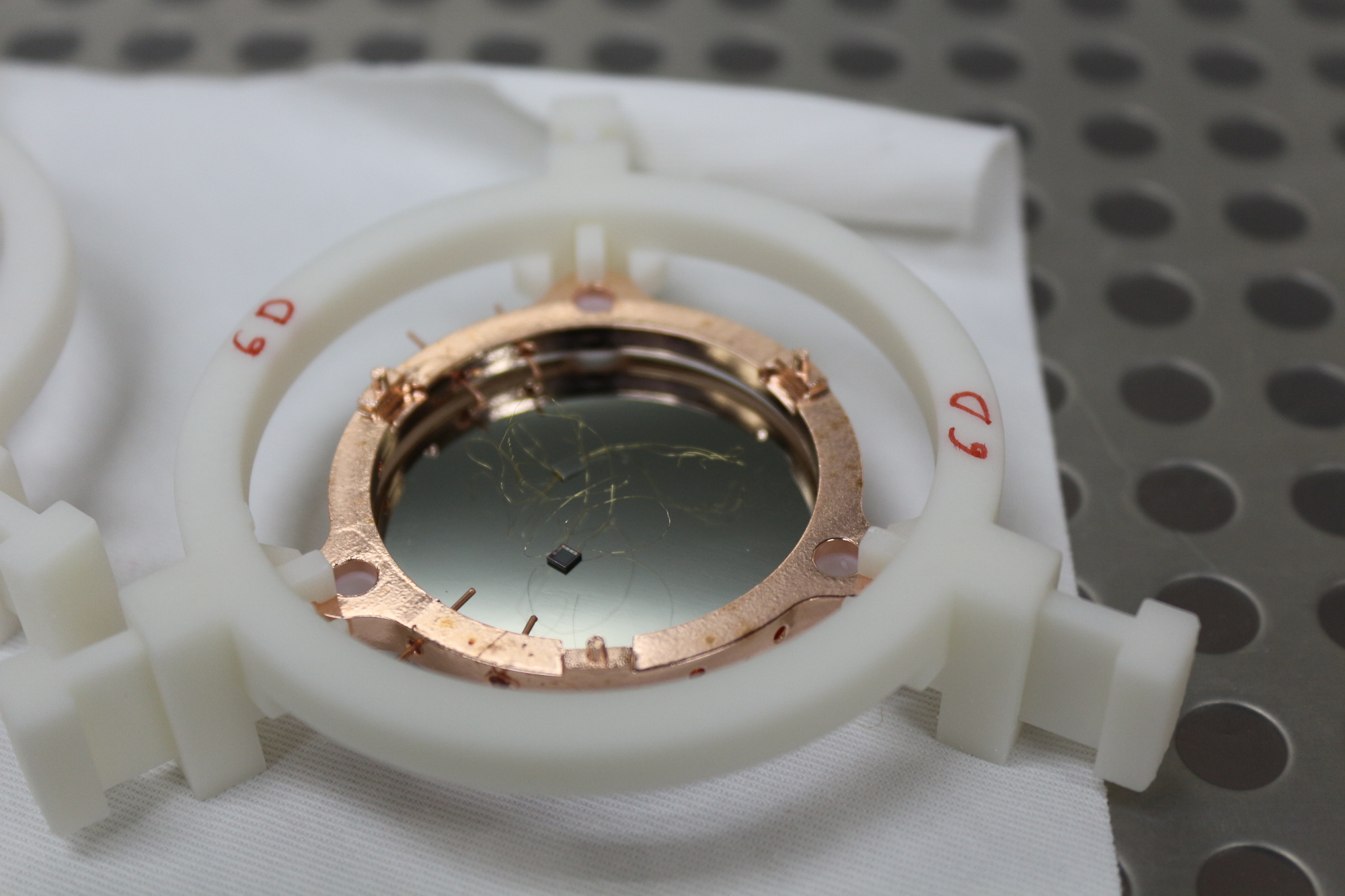}
  \caption{Left: bonding of a \cuore\ tower. Right: a \cupido\ light detector pre-assembled in its 3D printed housing.}
  \label{fig:preAssembledLD}
\end{figure}

In \cuore, the tower design was such that all bonding pads
on the sensors and CuPEN strips were accessible from the side
of the tower after mechanical assembly.
This allowed wirebonds connecting the sensor pads to the readout to be formed at any time.
We modified the wire bonding machine to work in a horizontal mode
rather than the typical vertical mode and
mounted it on travelling rails to increase its range of motion.
This allowed for course-positioning of the bonding head relative to the target pad.
The final fine-positioning was achieved in the standard way
using the integrated manipulator of the bonding machine.
The entire setup was enclosed in a nitrogen flushed glovebox
to avoid radon contamination of the tower (Fig.~\ref{fig:preAssembledLD}).
Each sensor was instrumented with four $\sim1$~cm long wires.
The operation was performed in 12 hours shifts by two operators: one operating the wirebonder,
and one controlling the course positioning of the tower and bonding machine.
The time required to bond a full tower was about one week.
Assembled towers were typically bonded in batches of three.

At this stage of the conceptual design,
is unlikely that the same bonding protocol can be followed for {\cupid};
technical R\&D on a new protocol is needed.
The \lmo\ crystals will have a cylindrical geometry
and the curved outer surface will be wrapped in light-reflecting foil.
This constrains the sensors to be attached on one of the flat faces of the crystal.
Sensors will not be accessible for bonding once assembled into a tower.
Similar considerations apply to the Ge wafer light detectors,
which will be interposed between adjacent floors of the tower.
A possible approach is to build stand-alone 4-crystal floors
or modules and 4-wafer floors that can be stacked together into a tower.
Each module could be instrumented with a set of readout strips
that connect bonding pads near the sensor location
to other bonding pads accessible from the side of the tower.
The sensors could be bonded after assembly of the module,
and bonded modules could be stacked into towers later.
Once the floors are stacked into towers, the externally accessible pads
could be wirebonded to the final readout. This would allow most of the wire bonding
to proceed in parallel with gluing.
The tower assembly protocol would need to include measures to protect delicate wire bonds
on each module during assembly.

This approach of having stand alone modules of pre-assembled light detectors
was successfully used for the construction of the towers of the \cupido\ detector.
In particular, thanks to dedicated 3D printed holding tools (see Fig.~\ref{fig:preAssembledLD}),
it was possible to avoid the introduction of an additional copper support frame
to complete the light detector modules.  This allowed adjacent floors
to share a single frame as it was for the \cuore\ towers \cite{Azzolini:2018tum}
.

\subsubsection{Closing and storage}

\begin{figure}
  \centering
  \includegraphics[height=0.278\textheight]{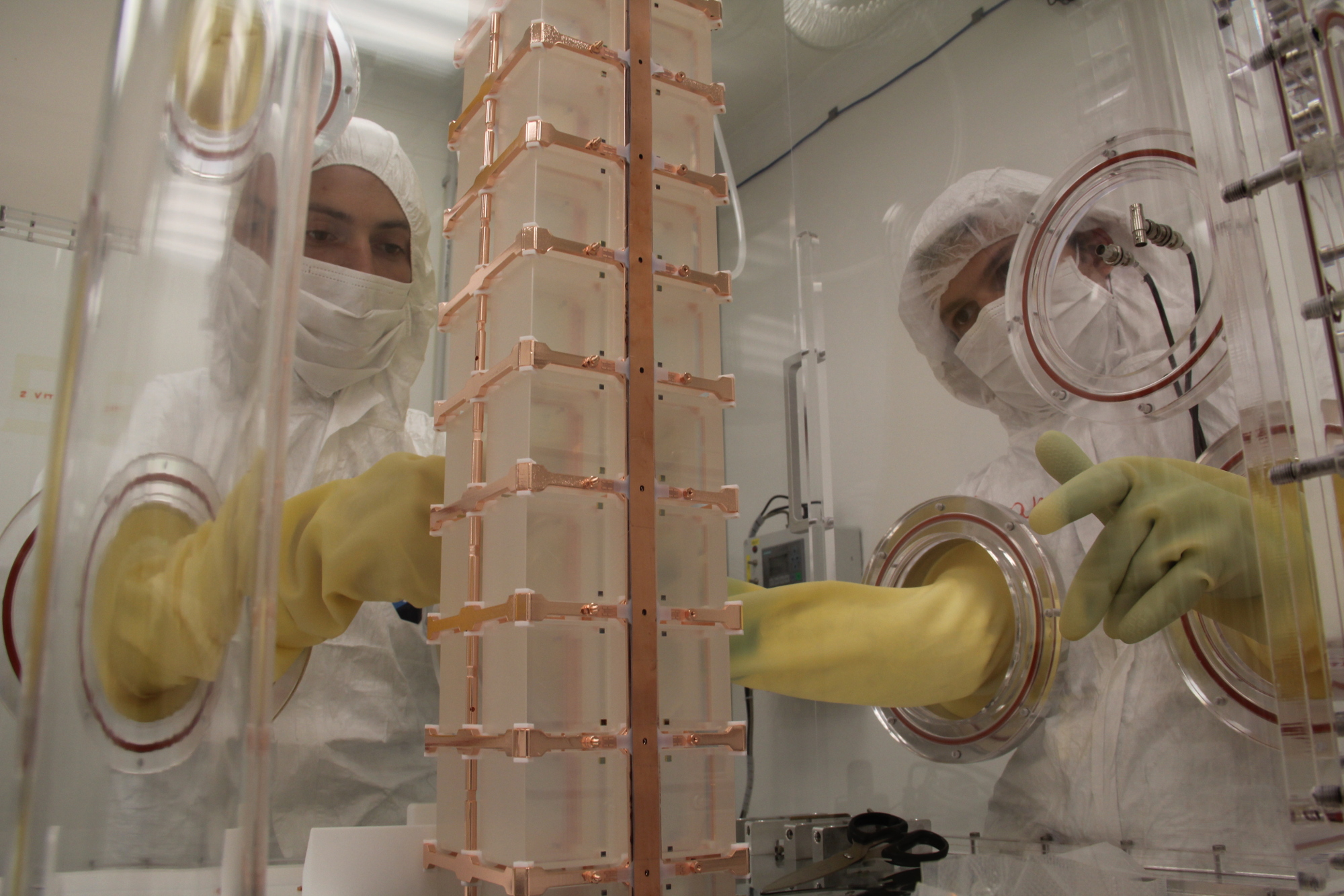}\quad
  \includegraphics[height=0.278\textheight]{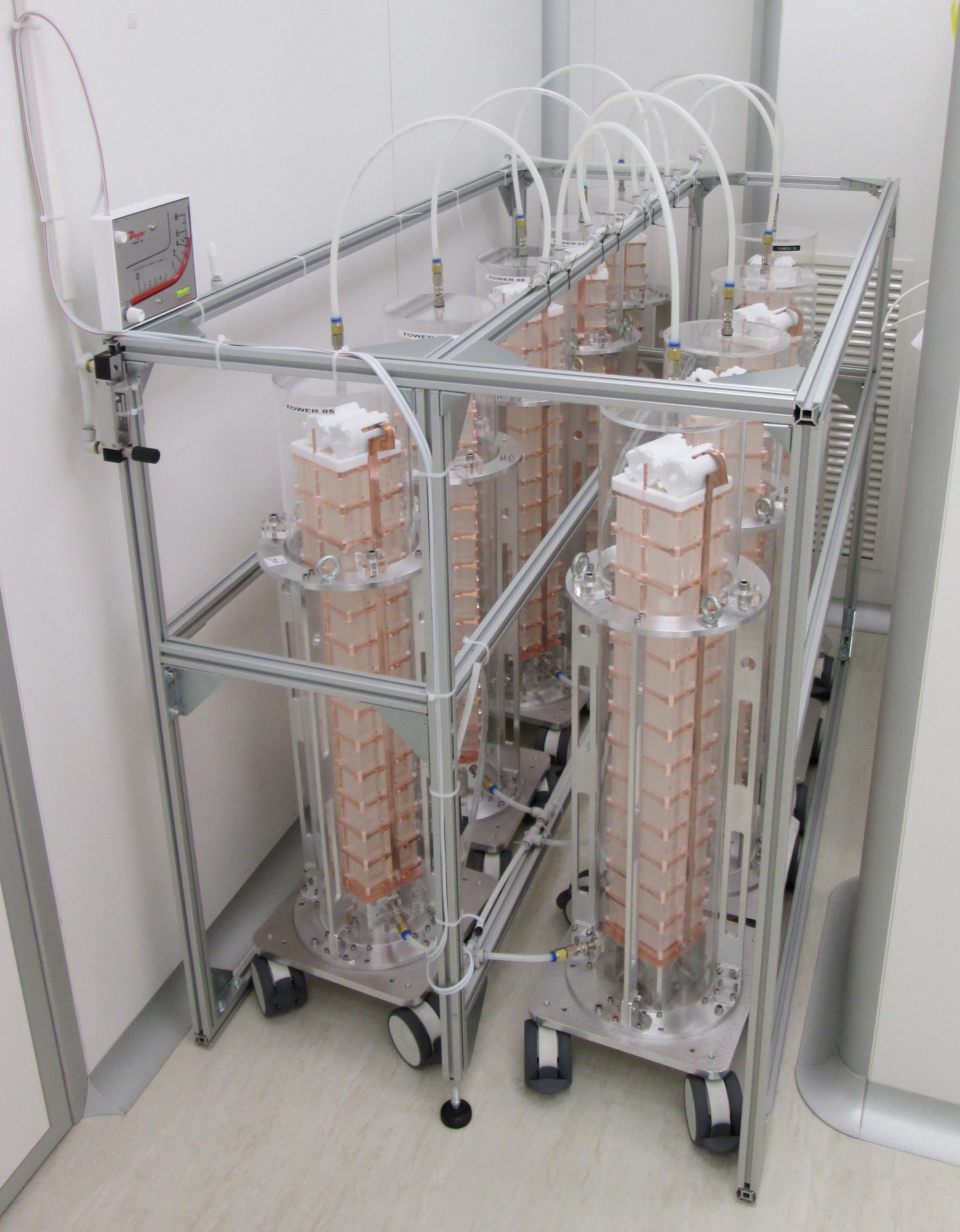}
  \caption{Left: tower cabling. Right: \cuore\ towers inside their LN$_2$ storage box.}
  \label{fig:cabling}
\end{figure}

The final step of \cuore\ assembly after wirebonding
was to place protective copper covers over the CuPEN strip assemblies (Fig.~\ref{fig:cabling}).
These covers held to strip assemblies tightly against the frame
so that they could not freely vibrate. The completed towers were then removed
from the workstation and stored in individual nitrogen-flushed boxes
in CR4 (see Fig.~\ref{fig:cabling}).
We anticipate similar storage arrangement will be adequate for \cupid. 

\subsubsection{Detector installation}

\begin{figure}
  \centering
    \includegraphics[height=0.2285\textheight]{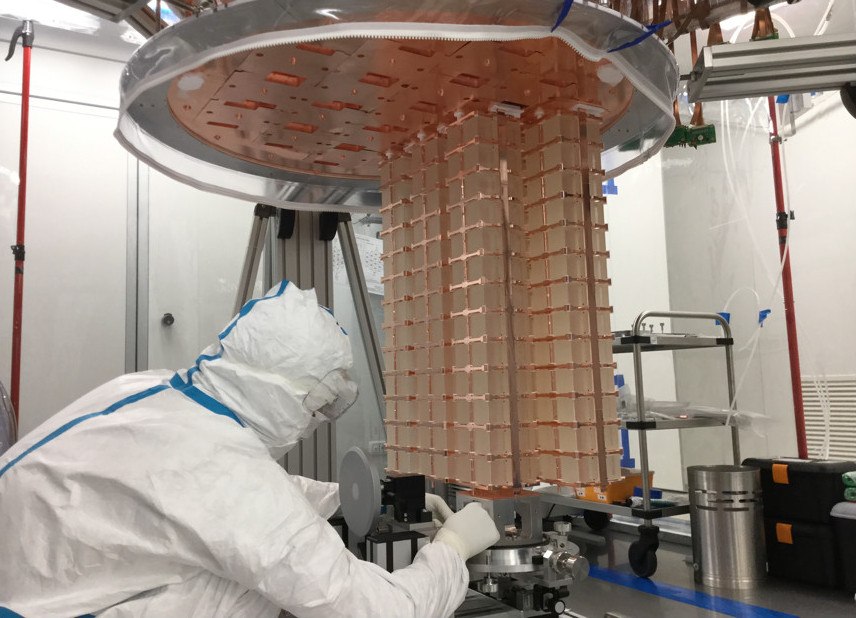}\quad
    \includegraphics[height=0.2285\textheight]{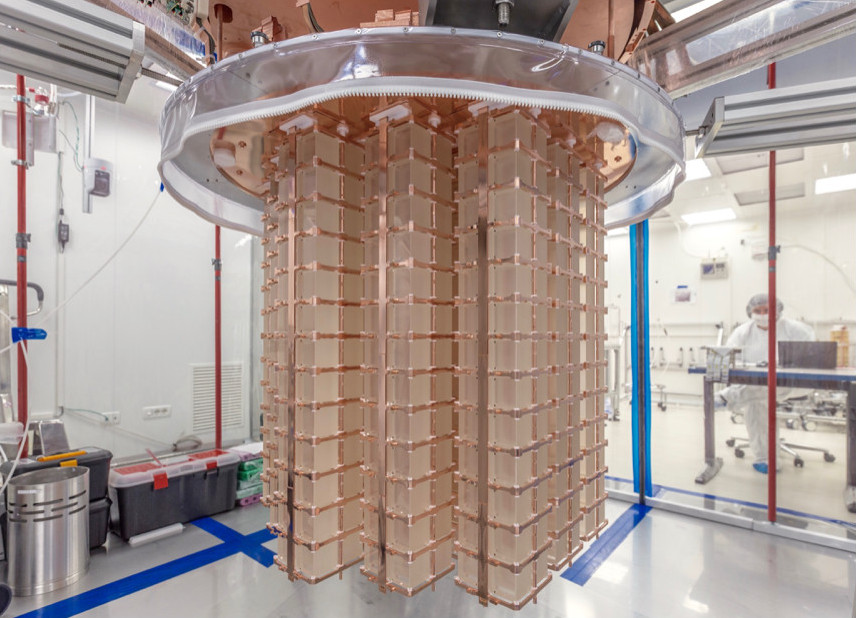}
    \caption{Installation of the \cuore\ towers. The top part of the plastic radon bag
      is visible around the tower support plate.
      Left: an operator positions a tower using the cart. Right: tower installation completed.}
  \label{fig:install}
\end{figure}

In \cuore, the detectors hang from the TSP, which is in thermal contact with the 10~mK stage.
We used a custom-designed cart, which provided fine adjustment of the XYZ position and tilt of the tower,
to position the tower under the assigned spot beneath the TSP
and raise it into position (Fig.~\ref{fig:install}, left).
The tower was then fixed to the TSP and the readout strips connected.
The operations were carried out in a specially designed soft-walled cleanroom (CR6),
which was supplied with radon-free air.
The towers were installed by a team of three people at a rate of one per day.
For redundancy, a cylindrical plastic barrier could be quickly positioned
around the bottom of the cryostat, covering the mounted towers.
This bag was put in place and continuously flushed with nitrogen
when personnel were not working in CR6. and when personnel entered or exited the room.
The pace of installation was set by several factors:
the time needed for the radon level in CR6 to reduce to acceptable levels
after entering or exiting the area;
the time needed to re-clean the working area and tools before the next operation;
and the desire not to overexert the operators performing the critical and stressful installation task.
We anticipate a similar setup and protocol will be adequate for {\cupid}.

\subsubsection{Radon abatement}

The storage and assembly procedures for \cupid\ envision that parts
will not be exposed to air, and consequently radon, following their final surface cleaning.
This will be achieved by packing and storing parts in a pure nitrogen environment
and subsequently assembling parts into detectors in nitrogen-flushed glove boxes.
However, for the safety of personnel carrying out the operation,
the attachment of detector towers to the 10~mK stage of the cryostat will be done in air.
To reduce the radon in the air to an acceptable level during this installation step,
we will reuse the radon abatement system
developed for and used by \cuore~\cite{1748-0221-13-01-P01010}.
The system process is as follows.
Intake air is pressurized $\sim9$~atm, passed through an oil vapor separator and microfilters,
and dried so that the dew point is below $\sim-70$\textcelsius.
The air is filtered again to remove liquid and dust after which it is cooled to $\sim-55$\textcelsius,
and flushed through two large activated carbon filters (in series) to trap radon.
The cooling is necessary to increase the radon absorption efficiency of the charcoal.
The output (radon-free) air is then past through a series of HEPA filters and heated to room temperature.
The $^{222}$Rn activity at the output is reduced to $<5$~mBq/m$^3$ (from $\sim30$~Bq/m$^{3}$ ambient at LNGS),
and the system can produce $\sim120$~m$^{3}$/h.
In the current conceptual design, we envision that the \cupid\ tower installation
will be very similar to our experience in \cuore, and the existing radon abatement system, with standard maintenance, will be adequate for the needs of \cupid\ installation.  


\subsection{Front-end electronics}\label{sec:electronics}

In view of \cupid, we are developing several improvements to the front-end system,
aimed at obtaining higher channel density, lower noise for light detectors,
and improved DAQ resolution and sampling speed.
These upgrades regard the back-end electronics and DAQ system,
the very front-end, the power supply, and the stabilization pulser.

\subsubsection{The back-end electronics system stage: anti-aliasing and DAQ}

The back-end electronics of \cuore\ performs analog filtering and digitization.
It is based on a custom anti-aliasing board with four cut-off frequency settings
(15, 35, 100, and 120~Hz)~\cite{Arnaboldi:2018:frontend} and commercial DAQ boards from National Instruments,
operating at a sample rate of 1~kHz per channel~\cite{DiDomizio:2018:DAQ}.
If the present configuration used in \cuore\ were to be adopted for \cupid\ without modifications,
a channel increase of at least a factor 3 would be necessary,
leading to high power consumption, occupied space, and cost.
Apart from the increased number of channels,
there will also be the necessity to operate at higher sampling rate for optimal pile-up rejection.
This will require the re-design or modification of the anti-aliasing filters
to allow a wider bandwidth of both the heat and light channels.

For the \cupid\ back-end we will adopt an upgraded anti-aliasing filter with integrated digitization circuitry.
The new anti-aliasing board will offer the capability to digitally select
the cut-off frequency in a range from 24~Hz to 2.5~kHz, with a resolution of 10 bits.
This will allow independently tuned cut-off frequencies for heat and light channels,
and a larger degree of flexibility in the arrangement and customization of each channel.
The digitization will be performed by 24-bit delta-sigma ADCs, with a maximum sampling rate of 25~kHz per channel.
The data transfer from the board to the DAQ storage will be managed by an FPGA
through UDP protocol on inexpensive 1~Gbit/s Ethernet interfaces.
The power consumption of this solution is up to a factor 5 lower
than the previous solution adopted for \cuore, while the space occupied is half.
Another advantage of this configuration is that the storage infrastructure
can be placed farther from the DAQ (either above ground or in a less crowded underground space).

\begin{figure}
  \centering
  \includegraphics[height=0.2691\textheight]{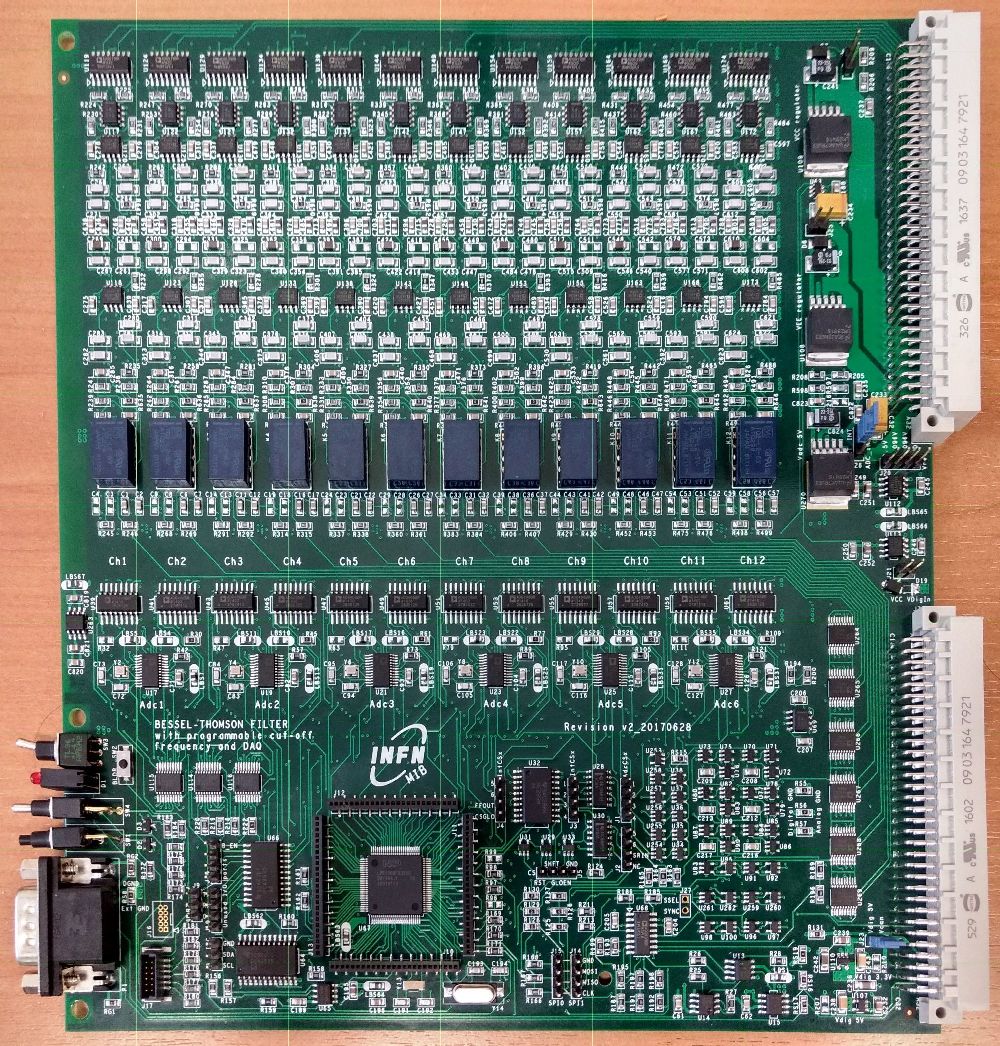}\quad
  \includegraphics[height=0.2691\textheight]{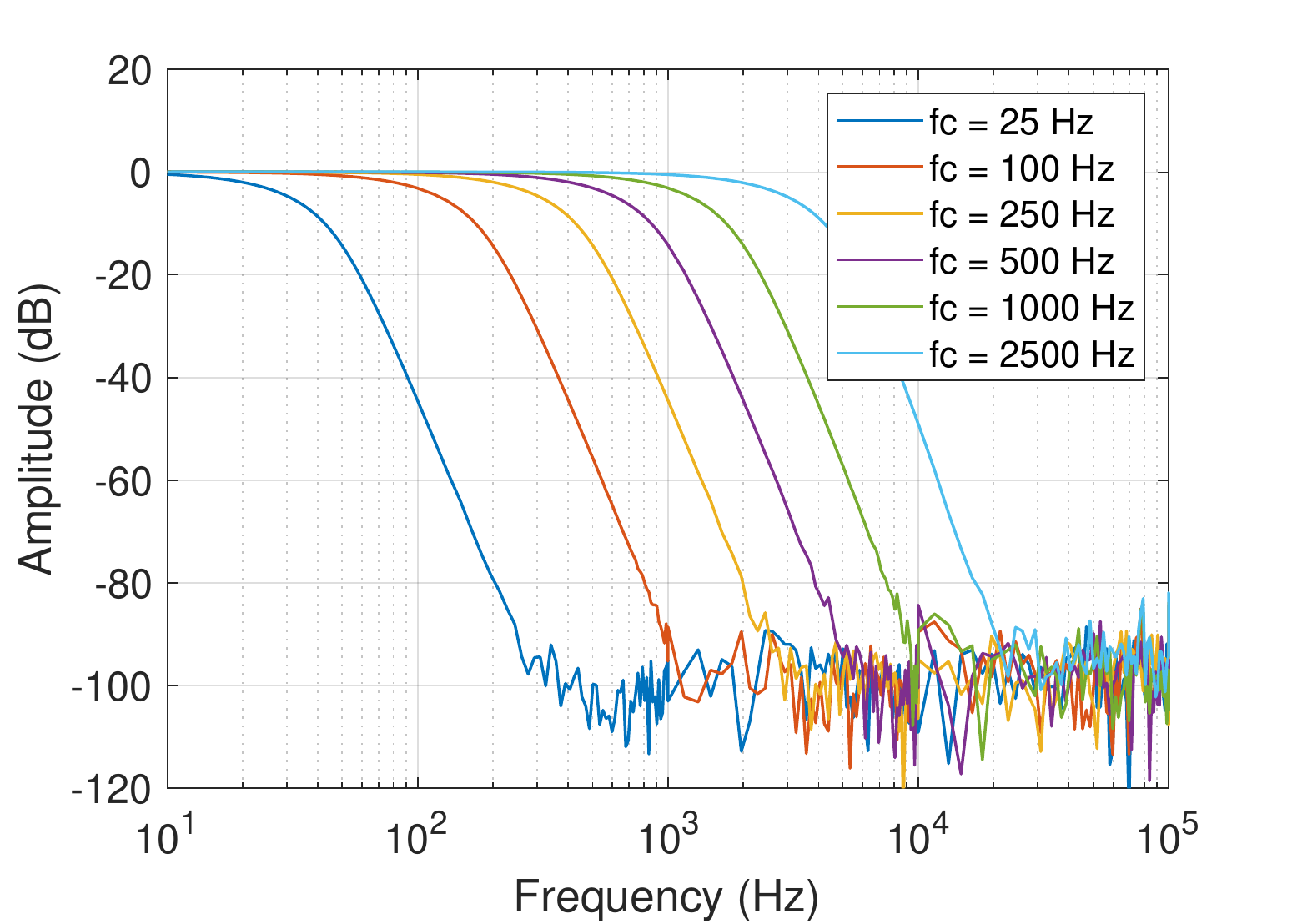}
  \caption{Left: photo of a prototype of the new anti-aliasing filter with integrated DAQ.
    Right: transfer function of the filter at different cut-off frequency settings (from 25~Hz to~2.5 kHz).}
  \label{fig:FilterDAQ}
\end{figure}

A photograph of the first prototype of the board is shown in Fig.~\ref{fig:FilterDAQ}.
The figure also shows the transfer function of the board at different cut-off frequency settings.

Another important feature foreseen for the new anti-aliasing filter is a Programmable Gain Amplifier (PGA).
Moving the PGA stage away from the very front-end boards
allows to save space at the top of the cryostat, where the very front-end boards reside.

As a backup solution, the newly designed anti-aliasing board
can be operated in a fully analog mode with an external DAQ system,
making it compatible with the present \cuore\ readout scheme.

\subsubsection{Very front-end and Faraday room}

We intend to operate the very front-end at room temperature,
and locate it on the top of the cryostat, as in \cuore.
In the \cuore\ scheme, the front-end of each channel consists of an input preamplifier,
a second stage PGA, and the detector biasing system \cite{Arnaboldi:2018:frontend}.
The Faraday room, already in place, has shown the expected shielding performances~\cite{Bucci:2017a:faradayroom}.
In order to save space inside the Faraday room, the PGA stage will be moved outside,
at the input of the anti-aliasing filter.

\begin{SCfigure}
  \includegraphics[width=.7\textwidth]{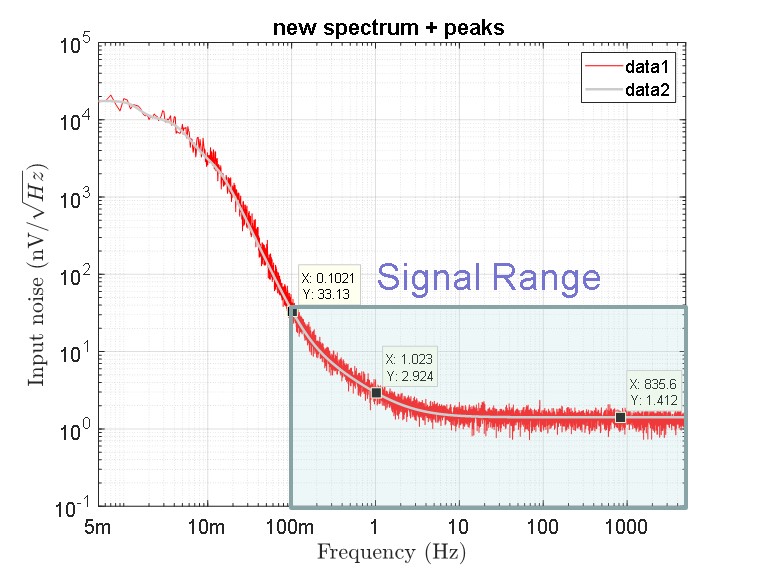}
  \caption {Series input noise of the new pre-amplifier in the low series noise and high power configuration.}
  \label{fig:Pre_Noise}
\end{SCfigure}

Power dissipation inside the Faraday room is a sensitive quantity to be considered in the design phase.
For \cupid, we intend to minimize it in two ways.
First, we selected a new family of low noise commercial operational amplifiers with lower supply current;
this will be used in most of the analog circuitry.
Furthermore, we intend to consider different preamplifiers for light and thermal channels,
characterized by a different level of series input noise
and inversely proportional to their operating current.
When operated at about 150~mW, the preamplifiers show an input series noise below 1.4~nV/$\sqrt{\text{Hz}}$,
(see Fig.~\ref{fig:Pre_Noise}).
If the power is reduced to 70~mW, their input series noise increases to about 3~nV/$\sqrt{\text{Hz}}$.
The total expected power dissipated by the front-end electronics,
which will be up to 3000 channels, inside the Faraday room (dimensions $6 \times 6 \times 2.5$~m$^3$),
is about 600~W, or 6.7~W/m$^3$ on average.
No additional cooling system is foreseen with respect to \cuore.
The differential voltage-sensitive front-end preamplifiers
show also small parallel noise and, with the new generation of JFETs we recently selected,
the expected current at each input gate is about 100~fA,
or 0.13~fA/$\sqrt{\text{Hz}}$. The input capacitance is below 20~pF.

The general scheme of the biasing circuitry for the detectors will be maintained as in \cuore,
i.e. the detectors will be biased with a constant current
injected through very large value resistors (tens of G$\Omega$ range).
The value of the resistors will be optimized depending on the expected impedance
of the detectors at their optimal working temperatures.

\subsubsection{Power supply and stabilization pulser}\label{sec:pulser}

The power supply of \cuore\ was designed with 3 stages:
AC/DC, DC/DC~\cite{Arnaboldi:2015a:powersupply}, and linear~\cite{Carniti:2016a:linear} regulators.
By design, it is floating with respect to earth and to shows exceptionally low thermal drift and noise.
The AC/DC and DC/DC regulators are located outside the Faraday room,
while the linear regulators are located inside the front-end racks.
We intend to maintain the same approach in \cupid.

The periodic calibration of the detector array with known radioactive sources
is expected between physics runs.
However, to stabilize the operation of the detectors during each physics run, two techniques are used.
First, since the front-end amplifiers are DC-coupled to the detectors,
and are designed to show a negligible temperature drift,
the baseline of each detector is proportional to its temperature.
The baseline can therefore be monitored and used to stabilize the detector response
over long data collection periods.
Second, for \cuore\ we developed a series of pulsers having
a stability $\lesssim1$~ppm$/$\textcelsius~\cite{Alfonso:2018a:pulser},
which are used to simulate particle hits by sending short voltage pulses
on heating resistors glued to each detector crystal (see Sec.~\ref{sec:heaters}).
The height of the pulser signals,
which are triggered periodically by the DAQ system and properly tagged,
is used to track any temperature or gain variation of the detectors during each run. 
The same strategy and design will also be used for \cupid.

\subsection{Computing infrastructure: data acquisition, storage, online processing, and monitoring}

The \cupid\ experiment will handle an amount of data about a factor of $\sim$10 larger than in \cuore.
This rough estimate comes from considering a larger signal bandwidth, about 3000 channels instead of 1000,
and a further factor of $\sim$2 for a conservative estimation.
\cuore\ is producing about 100~GB/day of raw data,
so a reasonable estimation for \cupid\ is 1~TB/day.
In this section we briefly summarize how we plan to manage these data
in terms of data acquisition and storage, and of the overall computing infrastructure of \cupid.

As already stated in Sec.~\ref{sec:electronics}, a simplification
of the data acquisition hardware is foreseen for \cupid.
It consists in replacing the commercial digitizers used in \cuore\
with custom boards featuring a higher channel density, serving both as anti-aliasing filters and digitizers.
This improvement will make the readout system significantly more compact
and will remove the limitations on the operating system of the readout computers,
imposed in \cuore\ by the commercial drivers of the digitizer boards.
The \cuore\ data acquisition system~\cite{DiDomizio:2018:DAQ} has a modular structure,
so it can be easily adapted to the different digitizer hardware and to the larger number of channels of \cupid.
Another simplification in the data acquisition system is likely to take place in \cupid.
The \cuore\ DAQ system not only took care of acquiring and saving the continuous bolometer waveforms,
but also of searching for signal triggers and building the corresponding events.
However, in \cupid\ the larger amount of data will make it preferable
to move the process of triggering and event-building offline.
This change of paradigm will make the data acquisition system simpler and more scalable,
and will allow more sophisticated trigger algoritms as long as enough computing resources are available.

In \cuore, a multi-site approach is being pursued for storing the acquired and processed data.
The primary storage site is located above ground at the LNGS,
with two other mirror storage sites at CNAF (Italy) and NERSC (USA).
This approach proved to be robust against data losses and we plan to replicate it for \cupid.
Some care must be taken to make sure that the bandwidth of the data links
between underground and above-ground sites at LNGS,
and from LNGS to the secondary storage sites,
are still large enough to deal with the larger amount of data expected in \cupid.

Apart from some minor upgrades, we plan to use the same computing infrastructure that we are using in \cuore.
It consists of services and resources distributed over several physical sites.
The core services, whose availability is compulsory for the operation of the data acquisition system,
are located underground at LNGS in order to minimize the downtime due to malfunctioning of the network.
The other non-critical services required for the operation of the experiment are located above-ground at the LNGS.
The day-by-day data processing, whose main purpose is to assess the quality of the acquired data,
is performed on a small-size computing cluster above ground at the LNGS~\cite{ulite:arXiv:1212.4658}.
Monte Carlo simulations and heavier data processing,
usually performed with longer latency (on the month to a few months timescale),
are carried out at remote clusters (mostly CNAF but also NERSC) where more computing resources are available.
The amount of computing resources used in \cuore,
which involved a few hundred parallel cores,
is small compared to the typical needs of high energy physics experiments,
and it is unlikely that the transition to \cupid\ will have a significant impact in this regard.
However, in the event that more parallel and efficient data processing algorithms
will be exploited in \cupid, the migration to a more efficient computing paradigm (e.g. GRID-based) will be considered.

In summary, the data acquisition, storage, and computing model of \cupid\
will be heavily based on what was already put in place for \cuore.
None of the planned upgrades is expected to imply significant risks for the successful outcome of the experiment.

\subsection{Calibration systems }



\subsubsection{\lmo\ calibration}

An accurate calibration is mandatory for \cupid,
in particular in the 3~MeV region where the  \onbb\  signal of \Mo\ is expected.
A  $^{232}$Th source emits a \ga\ line at 2615~keV and is an obvious choice
since this natural radionucleide is present in many common materials.
A $^{56}$Co source emits \ga\ lines up to 3612~keV,
offering the possibility to calibrate at 3 MeV and above.
Due to the relatively short half-life (77 days) the $^{56}$Co source
can only be used for a few months after its production.

\begin{figure}
    \centering
    \includegraphics[height=0.2542\textheight]{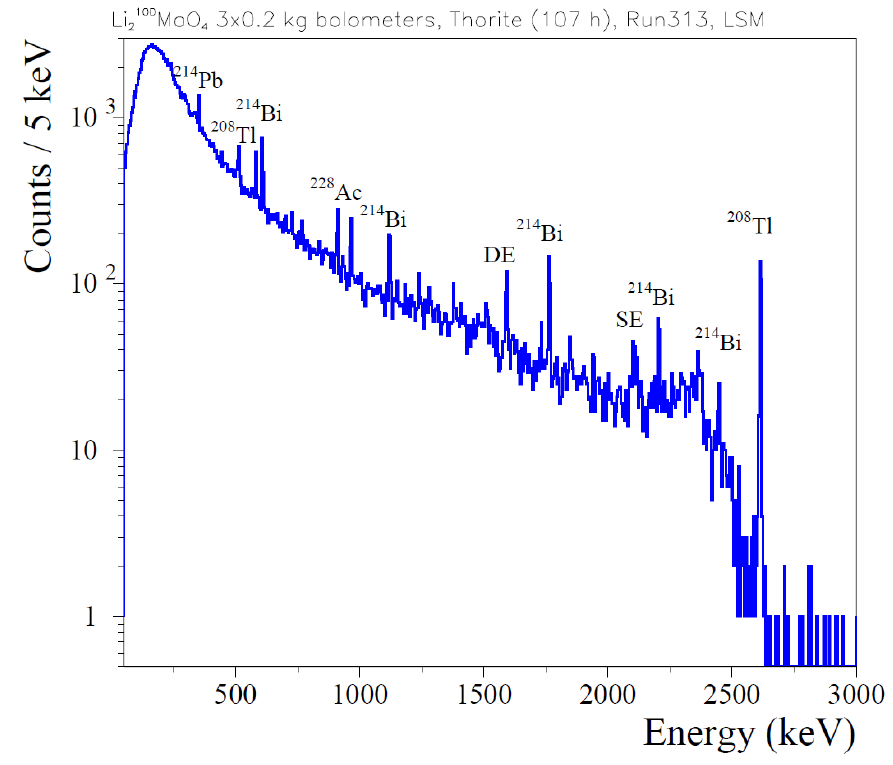}\quad
    \includegraphics[height=0.2532\textheight]{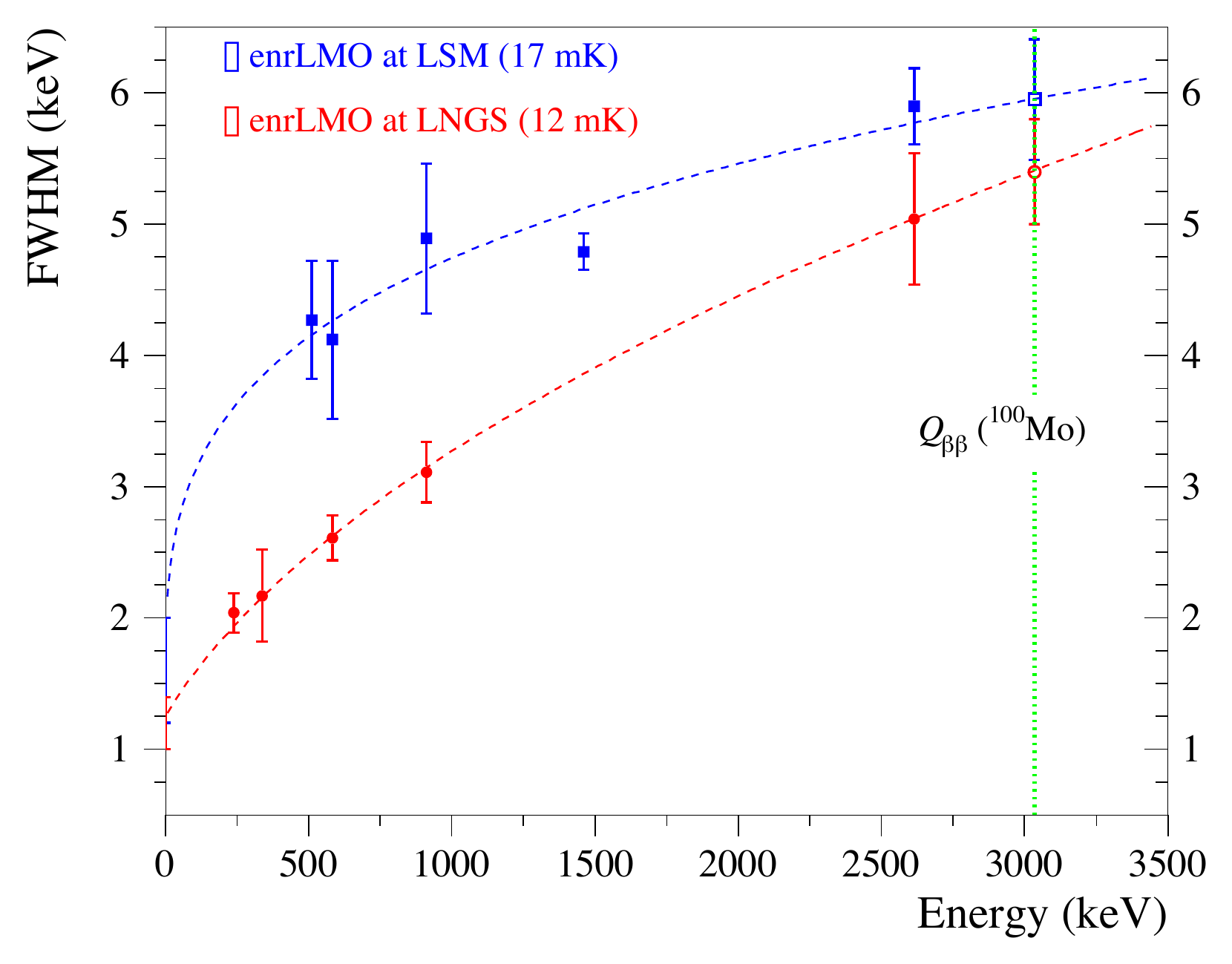}
    \caption {Left -- Energy spectrum of \cupidmo\ towers for 2.7 kg$\cdot$d exposure calibrated with a $^{232}$Th source. Right -- FWHM resolution scaling and extrapolation to the ROI (hollow markers) at different working points at LSM (blue) and LNGS (red).}
    \label{fig:heat_calibration}
\end{figure}

Currently the \cuore\ detector can be calibrated by an internal calibration system,
which deploys low-activity thoriated strings from room temperature into the cryogenic volume,
and by an external calibration system that deploys high activity $^{232}$Th and $^{60}$Co sources
in the volume between the outside of the OVC and inside of the external lead shield.
The activities of the internal and external calibration sources are optimized by Monte-Carlo simulations
in order to avoid a large number of pile-up events while also maintaining
a reasonable duration of the calibration run.
Figure~\ref{fig:heat_calibration} (left) shows the calibration of the \cupidmo\ towers using a $^{232}$Th source.

\subsubsection{Light detector calibration}

\begin{figure}
  \centering
  \includegraphics[height=0.2745\textheight]{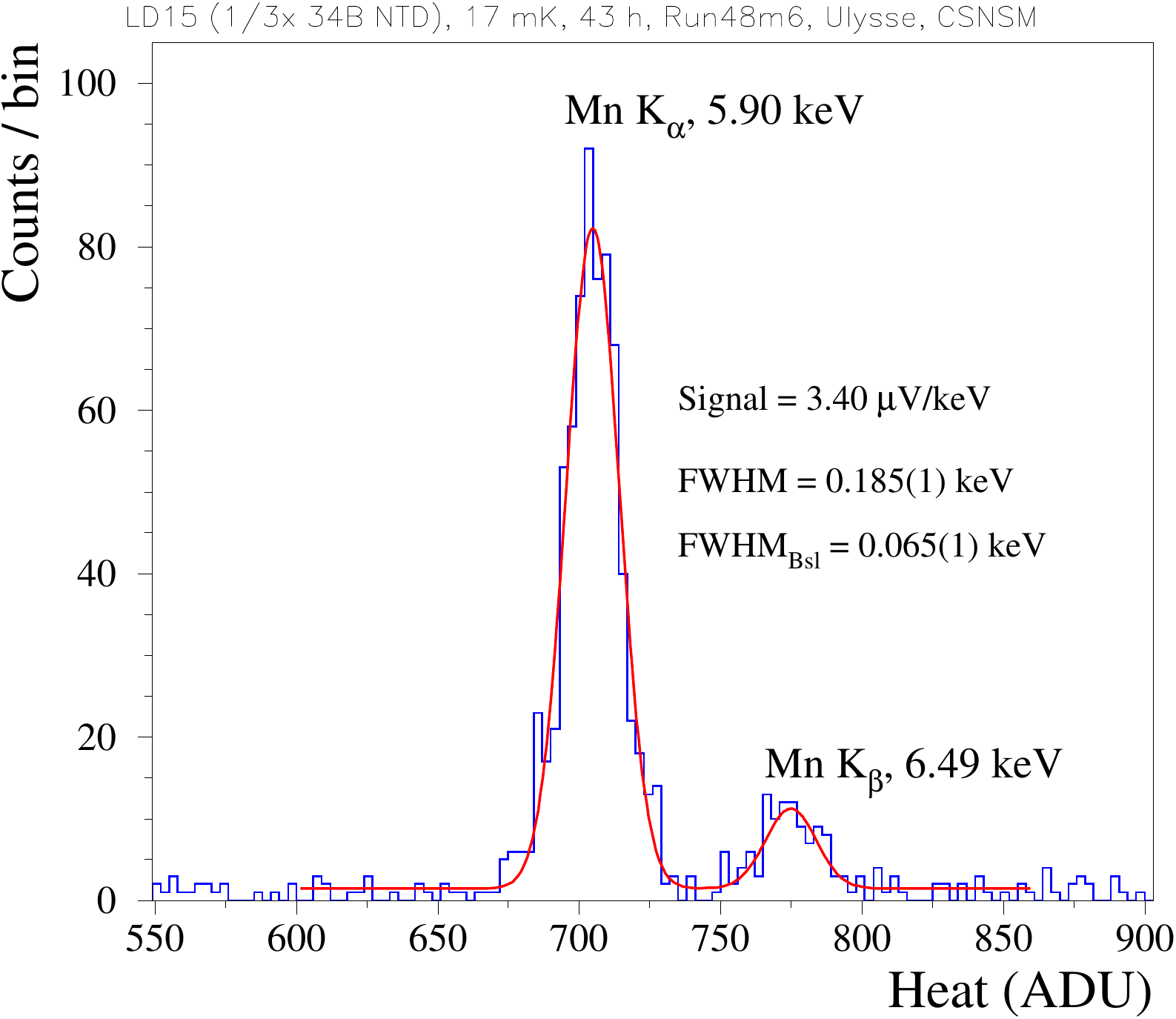}\quad
  \includegraphics[height=0.2745\textheight]{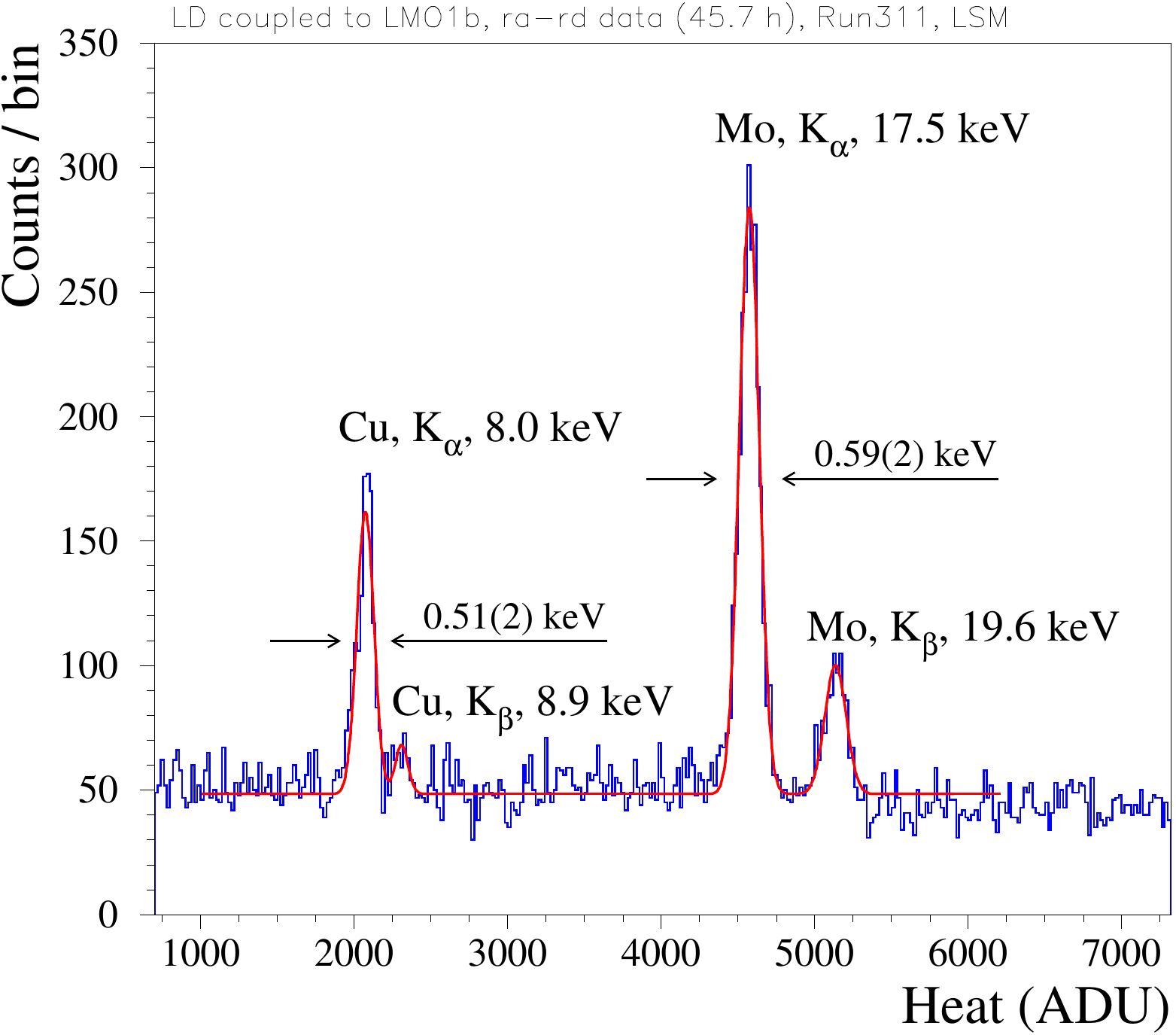}
  \caption{Left -- Energy spectra of X-rays accumulated by 44.5~mm diameter Ge
    light detectors instrumented with NTDs~(see Sec.~\ref{sec:LightDetectors}).
    Right -- The measurements were performed in the above-ground cryogenic laboratory of CSNSM (left)
    and at LSM. The energy scale is given in analog-to-digit units (ADU).}
  \label{fig:LD_NTD_calibration}
\end{figure}

A calibration of the light detectors is impossible using \ga\ sources
of the kind used for the calibration of massive detectors.
Thus, it is common to use an X-ray $^{55}$Fe source to calibrate light detectors (Fig.~\ref{fig:LD_NTD_calibration}, left).
The use of a $^{55}$Fe source might be problematic in an ultra-low background environment.
Hoever, the light detector calibration is optional because it does not strongly affect
the $\alpha$/$\gamma$ separation capability.
In order to overcome the issue, we can use a high intensity $\gamma$ source
(e.g. $\sim$100~Bq $^{60}$Co source~\cite{Berge2018})
to stimulate X-ray fluorescent of materials (mainly, \lmo\ crystal),
which are close to light detectors.
An example of such calibration is illustrated in Fig.~\ref{fig:LD_NTD_calibration}, right~\cite{Poda:2017d}.

\subsection{Muon tagger}
\label{sec:MuonTagger}

Atmospheric muons produce two types of backgrounds in a bolometric detector:
direct energy deposits by muons passing through the crystals, and muon-induced spallation. 
Interactions from the muons in the rock or the external lead shield
cause spallation products that can lead to fast neutron backgrounds. 
Muons can be effectively tagged in \cupid\ if they go directly through multiple crystals;
however, they can cause irreducible backgrounds if they clip corners
or interact in the surrounding materials to create secondaries
that can produce particles mimicking \onbb-like signal.
Neutrons produced by cosmic rays are energetic and  difficult to shield.
Photons emitted in (n, n'$\gamma$) or (n, $\gamma$) reactions
can appear near \Qbb.

\begin{figure}[htbp]
  \begin{center}
	\includegraphics[width=0.95\textwidth]{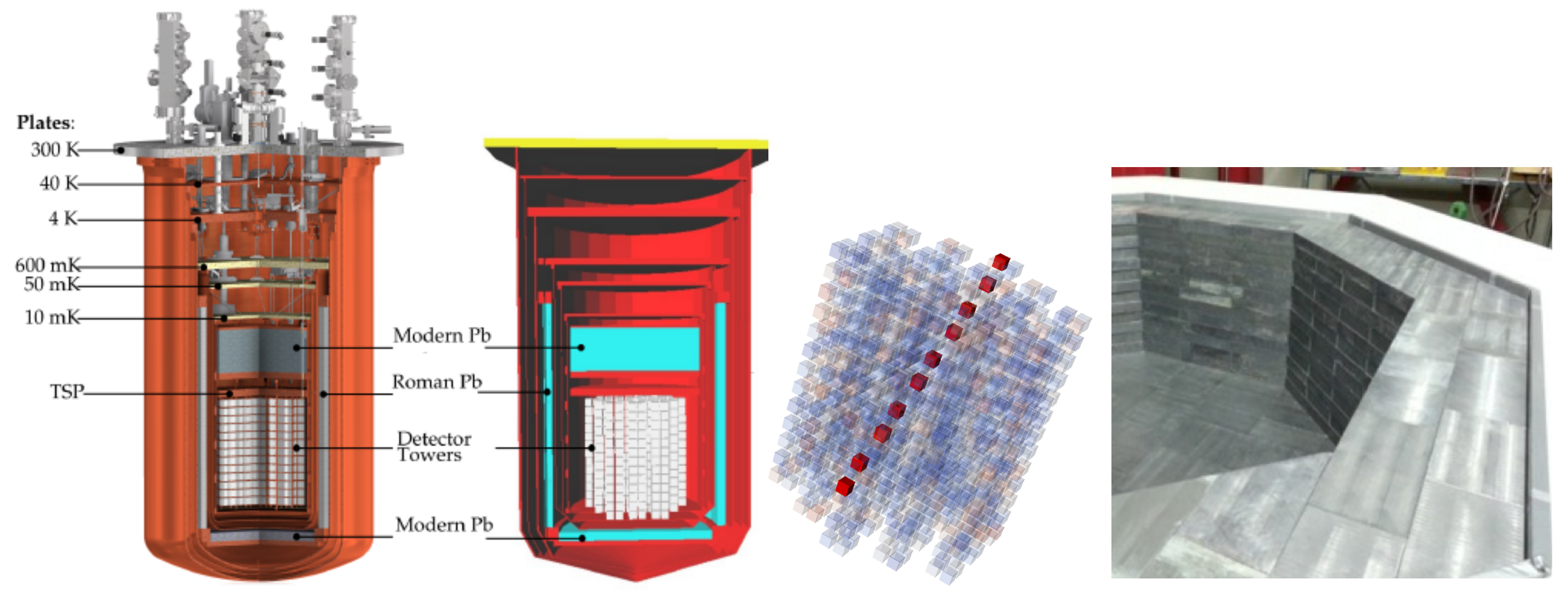}
	\caption{Left: 3D view of the \cuore\ apparatus and geometry implemented in Monte Carlo simulations.
      Center: Simulated muon event in \cuore.
      The solid red color indicates crystals hit by muons, the transparent colors
      indicate the simulated base temperature of each crystal.
      Muons that go through multiple crystals will be easy to tag.
      Right: \cuore\ Pb and polyethylene shielding outside the cryostat.}
	\label{fig:muonCUORE}
  \end{center}
\end{figure}

The muon flux at Gran Sasso National Laboratory is
$\sim2.6 \cdot 10^{-8}$~cm$^{-2}$s$^{-1}$, as reported by MACRO and
BOREXINO~\cite{Cecchini:2009ug,Bellini:2011yd,Mei:2005gm}.
The muon-induced backgrounds in \cuoricino\ have also been measured and projected onto
what we expect in \cuore\ and \cupid~\cite{Andreotti:2009dk,Bellini:2010}.
As summarized in Sec.~\ref{sec:bkg-BBcupid},
the total background rate expected from muons in the \cupid\ detector is $\sim 10^{-4}$\ckky.
This background can be reduced by at least an order of magnitude by a dedicated muon tagger/veto system.
The system to be developed for \cupid\ will use the design of the muon tagger
system developed for \cuore\ (Fig.~\ref{fig:muonCUORE})
and the experience of the similar systems built for \cuoricino\ and \cupido. 

The muon tagger consists of a set of vertical scintillator panels
to be arranged around the \cupid\ cryostat
and a set of horizontal panels above or below the cryostat.
The large scintillator panels will be read out with optical fibers
arranged in an x-y grid over the panels to provide good coverage
and x-y positioning at economical costs.
The dimensions and segmentation of the tagger is being optimized
to provide tagging and crude tracking capabilities.
The tagger will allow us to identify muons that produce
single-site energy deposits in the \cupid\ crystals,
or miss the inner crystal array but enter the nearby lead shielding.
The geometry of the system will be customized to fit the available space around the cryostat.

Simulations provide estimates for the the expected event rates:
1.8~muons/hour in the \cupid\ crystals, 7.0~muons/hour in lead shield.
With the preliminary tagger design, we estimate that
for the muons leaving a single-site energy deposit near the \Qbb\ ROI ($2950-3100$ keV),
$(77\pm1)$\% would be detected in the vertical tagger panels,
and $(62\pm1)$\% would be detected in the tagged by the horizontal panel.
Combined, the two panels provide over 99\% rejection of the single-site muon-induced events. 



\subsection{Cryogenic testing infrastructure}

Even if \cupid\ is not in an R$\&$D phase, there are still some issues that need to be tested in order to define the final layout of the detector (e.g. the mechanical structure) as well as to assure the quality of its components (crystals, sensors, etc.).
In this section we briefly describe the underground and above-ground facilities where those measurements can be conducted.

\subsubsection{Underground cryostats at LNGS}

INFN has several test cryostats located deep underground in the LNGS.
All of them are already equipped with electronics and DAQs in order to operate large arrays of scintillating bolometers.

The largest one is the one in the Hall-A of the LNGS that hosted Cuoricino~\cite{Arnaboldi:2008ds}, \cuoreo~\cite{Alfonso:2015wka}, and is presently hosting the CUPID-0~\cite{Azzolini:2018dyb} experiment.
It consists of a completely custom wet Oxford 1000 dilution unit, able to reach a base temperature of 
$\sim$6.5~mK with a cooling power of 1~mW @ 100~mK. The available experimental volume is approximately a cylinder of 178~mm diameter and a length of 880 mm, resulting in a total experimental volume  of $\sim$ 22~L. This volume is surrounded by a 1.4-cm-thick Roman lead cylindrical shield. A Roman lead disc, 10-cm-thick and 17~cm in diameter, is placed  below the Mixing Chamber to shield against the intrinsic radioactivity of the dilution refrigerator.
The cryostat is externally shielded by two layers of lead with a minimum thickness of 20 cm.   
The external lead shielding is surrounded by an acrylic glass anti-radon box flushed with nitrogen gas to reduce Rn contamination. Finally, a 10-cm-thick layer of borated
polyethylene is used as neutron shielding. The entire apparatus is enclosed in a Faraday cage to minimize electromagnetic interference.
The cryostat is instrumented with 136 twisted pairs of NbTi wires from room temperature down to the Mixing Chamber. Presently we have 66 acquisition channels, the same type of electronics used in \cuore.
The detectors can be mounted into the cryostat within a dedicated cleanroom, located just above the cryostat, in which the dilution unit can be raised.  This guarantees a clean 
environment with a relatively low radon content ($\sim$ 5~Bq/m$^3$).
Due to its peculiar characteristics, this facility is dedicated to large scale demonstrators that need a very low $\gamma$ background environment. The present achieved background level at 3~MeV is of the order of $\sim$3.6$\times10^{-3}$ \ckky. The cryostat needs a $^4$He liquid Helium refill ($\sim$ 60~L) every two days, that takes roughly 40 minutes to complete. The total mean dead time induced by the refilling procedure is about 90 minutes.
The time needed to reach the base operational temperature, starting from room temperature, is $\sim$10-12 days.

The Hall-C CUPID-0 R\&D test facility  consists of  custom Oxford TL 200 wet dilution cryostat.
The base temperature is 7.5~mK with a cooling power of 0.2~mW~at~100~mK. The experimental volume is smaller with 
respect to the one in Hall-A: a cylinder of 182~mm in diameter and a height of 310~mm, totaling $\sim$ 8~L. 
The radioactive shielding is less effective with respect to the Hall-A cryostat, consisting of a Roman lead disc, 5-cm-thick and 17.5~cm in diameter, placed just below the Mixing Chamber. The cryostat is then externally shielded by two layers of lead with a minimum thickness of 15~cm. The external lead shielding is surrounded by an acrylic glass anti-radon box
flushed with nitrogen gas to reduce Rn contamination. Just outside, a 1-cm-thick shield of CB$_4$ acts as neutron catcher.  The neutrons are thermalized by an 8-cm-thick layer of polyethylene.
The entire apparatus is enclosed in a Faraday cage to minimize electromagnetic interference.
Using the 2615 keV $\gamma$-line of $^{208}$Tl as an evaluation metric, the background level that can be achieved in this cryostat is about 6 times larger with respect to the one in Hall-A. 
The cryostat is equipped with different type of readout electronics:
\begin{itemize}
    \item 10 COLD electronics channels~\cite{Arnaboldi:2006mx} framed {\em inside} the cryostat with Load Resistor of 27+27 G$\Omega$;
    \item 10 COLD electronics channels framed {\em inside} the cryostat with Load Resistor of 2+2 G$\Omega$;
    \item 12 standard electronic channels, similar to the \cuore\ ones;
    \item 24 additional twisted wires from room temperature down to the Mixing Chamber;
    \item 2 400 kHz DC squids (anchored at the 1.2~K thermal stage).
\end{itemize}

A Helium re-liquefier continuously fills the cryostat, keeping the He level constant, resulting in an almost 100\% duty cycle.
The time needed to reach the base operational temperature, starting from room temperature, is roughly 5-6 days.

The IETI test-cryostat~\cite{Barucci:2009} is a Pulse Tube dry dilution refrigerator, located in the same hut as the 
CUPID-0 R\&D facility, just few meters away. This is a completely custom cryostat based on a MKN-CF-500 dilution 
unit produced by Leiden Cryogenic, with a cooling power of 0.5~mW at 120 mK and a base temperature of the order of 7~mK.
The experimental volume is a cylinder of 32~cm in diameter and a height of 17~cm totalling 13~L.

The cryostat is equipped with 12 standard electronic channels, similar to the one of \cuore, and 52 twisted pairs of NbTi wires from room temperature to the Mixing Chamber. 
This cryostat can be shielded with an outside layer of 10~cm of lead, while the internal shielding is of 3~cm of lead below the MC. This facility is, therefore, mainly devoted for testing the performances of the detectors, thermometry, and NTD characterization.
The time needed to reach the base operational temperature, starting from room temperature is, about 3-4 days.

\subsubsection{Cryostats outside LNGS}

Beside the LNGS underground cryostats, there are several above-ground cryostats available among the various institution of the CUPID interest group (e.g. at Berkeley, CEA, Milano-Bicocca, MIT, and Rome). These can be used for all the measurements that don't require a low-background environment, such as sensors characterization, test of detectors performances, thermometry, etc. 
Underground facilities are also available.
\subsubsection{Edelweiss-II cryostat at LSM}

CUPID-Mo is presently installed in the EDELWEISS-II cryostat located in the Laboratoire Souterrain de Modane (LSM) where the rock overburden of 4800 m w.e. reduces the cosmic muon flux down to 4 $\mu$/m$^2$/day. 
   The detectors are placed in the 10 mK chamber and arranged on plates supported by three vertical bars. Below the 10 mK plate, at 1K, a 14 cm thick lead plate shields the detectors from the gamma-rays induced by the radioactivity in the cold electronics, the dilution unit, and other cryogenic parts.  Five thermal copper screens at 1~K, 4.2~K, 40~K, 100~K, and 300~K complete the cryostat. Resistors together with electrical connectors are installed at the 1K stage below the lead shielding. At the 100K stage the cold JFETs are connected on an extractable panel. The electronics to bias the JFETs, the DACs to bias the detectors, the final amplification, the anti-aliasing filter, and the digitization are all integrated in a single room-temperature module, called the Bolometer Box, which is attached to the cryostat. 

An 18-cm-thick outer layer of modern lead shields the cryostat against ambient gamma background. A 2-cm-thick inner Roman lead layer has been cast directly on the modern lead. An outer 50-cm-thick polyethylene shielding protects the detector against ambient neutrons. The lead and polyethylene shielding is mounted on a mild steel structure with rails allowing the opening of the two halves of the shielding structure.
In addition, a 100 m$^2$ plastic scintillator active muon veto surrounds the polyethylene.

\subsubsection{CROSS cryostat at LSC}

The CROSS cryostat was installed in April 2019 at the Laboratorio Subterraneo de Canfranc, at 2500 m.w.e. The cryostat was fabricated by Cryoconcept and has a unique method to reject vibrations induced by the Pulse Tube: it utilizes a thermal machine, which
guarantees the precooling down to 4 K of the 3 He- 4 He mixture employed in the dilution refrigeration. This
method avoids a direct mechanical coupling between the cold stages of the Pulse Tube and the refrigerator
and it is not available in systems produced by other companies.
The experimental volume (a cylinder d= 300 mm, h= 605 mm) is limited upwards by a copper floating plate (at 10 mK) that is suspended by a system of wires and
springs to reduce further vibrations. The detectors are connected to this plate and
are surrounded by a thin copper screen at 10 mK as well. The experimental volume is shielded from
above by a fixed 13-cm-thick low-radioactivity internal lead shield. 
Laterally and downwards, the experimental volume is surrounded by the aforementioned copper screens and then by the external lead shield of 25 cm.


\newpage
\section{Enrichment, purification, and crystallization} \label{sec:EnrPurCryst}

The success of \cupid\ will strongly depend on the of the \enrlmo\ crystals quality
in terms of energy resolution, light output, and radioactive contamination.
In this section, we describe the available sources of \Mo\-enriched Mo,
the material purification process, and the available crystal growth techniques.

\subsection{Procurement of \Mo-enriched molybdenum}
\label{sec:Mo100_procurement}

\Mo-enriched molybdenum is produced using gas centrifuges.
First, molybdenum is converted to MoF$_6$ gas by fluorinating natural molybdenum:
\begin{center}
$^{\text{nat}}$Mo + 3F$_2$ $\rightarrow$ $^{\text{nat}}$MoF$_6$,
\end{center}
\noindent and then enriched with a cascade of centrifuges.
At the next stage, the $^{100}$MoF$_6$ gas is converted into $^{100}$MoO$_3$
using the wet chemistry processes:
\begin{align*}
  ^{100}\text{MoF}_6 + \text{H}_2\text{O} + \text{HNO}_3 &
  \rightarrow \text{(precipitation and calcination)}
  \rightarrow ^{100}\text{MoO}_3 \quad ,\\
  ^{100}\text{MoO}_3 + \text{H}_2\text{O} + \text{NH}_3 &
  \rightarrow \text{(precipitation and calcination)}
  \rightarrow ^{100}\text{MoO}_3\quad,
\end{align*}
\noindent  which produces $^{100}$MoO$_3$ in the form of a fine powder.
If necessary, $^{100}$MoO$_3$ is reduced to metallic powder $^{100}$Mo.
Thus, enriched molybdenum is supplied to the customer in the form of MoO$_3$ powder, or metal powder.
Enrichment is usually in the 90--99\% range.

Several Russian companies can produce stable isotopes by centrifuging.
In order to successfully enrich molybdenum, a company must master a procedure
for converting molybdenum to the gaseous phase and vice versa.
Due to the difficulty of these processes, enriched molybdenum is currently produced
by only one enterprise, the Electro-Chemical Plant (ECP) in Zelenogorsk, Russia~\cite{ecp},
although the \Mo\ production can in principle be organized at other companies too.
ECP is mainly engaged in the production of enriched uranium for nuclear power plants.
The production of stable isotopes at this company started in the 1970s.
Eventually, a dedicated production line was set for the extraction of stable isotopes
using the SVETLANA plant with an independent gas centrifuge complex.
Currently, a large number of isotopes are produced here for medicine,
electronics, biology, chemistry, and physics.
In recent years, the interest in $^{98}$Mo has increased, from which $^{99}$Mo is then obtained.
This is due to the widespread use of the $^{99}$Mo isotope in medicine
(more precisely, $^{99m}$Tc, which is the product of $^{99}$Mo $\beta$ decay).
In this case, $^{99}$Mo can be obtained from \Mo\ on accelerators.
In this regard, we can expect an increased request of significant quantities (tens of kg)
of $^{98}$Mo and \Mo\ from nuclear medicine in the near future.

In the Soviet Union, and then in Russia, \Mo\ had been produced in large quantities ($>1$~kg) since the 1980s.
The production was mainly associated with \onbb\ decay experiments.
A summary of the main \Mo\ production batches is reported in Table~\ref{tab:enrhistory}.
To date, $\sim$145~kg of \Mo\ has been produced in the world.
The first contacts with ECP representatives showed that the production of 300~kg
of \Mo\ with $>95\%$ enrichment is feasible in a 3--5~year framework.

\begin{table}[h]
  \centering
  \caption{\Mo\ production history. The last three columns report the measured
  contamination levels of the material after enrichment and prior to crystal growth.}
  \label{tab:enrhistory}
  \begin{tabular}{lcccccc}
    \toprule
    Date & Amount & Enrichment & Current & \multicolumn{3}{c}{Contamination [mBq/kg]} \\
         & [kg]   & [\%]       & Owner   & $^{228}$Th & $^{226}$Ra & $^{40}$K \\
    \midrule
    1980s   & $\sim1$   & 99.5   & INR, Kiev (Ukraine)   & $\sim2.5$ & $\sim10$ & $\sim100$ \\ 
    1980s   & $\sim1$   & 98.5   & ITEP, Moskow (Russia) & $\sim2.5$ & $\sim10$ & $\sim100$ \\
    1990s   & $\sim12$  & 95--99 & NEMO-3 collaboration  & 1--5      & 1--6     & 50--140   \\
    2016-19 & $\sim120$ & 96     & AMoRE collaboration   & $<2$      & $<4$     &           \\
    \bottomrule
  \end{tabular}
\end{table}

\subsection{Purification of materials for crystal growth}
\label{sec:pur-cryst-growth}

\subsubsection{Enriched \Mo\ purification} \label{sec:mo-pur}

The purity level of commercially available molybdenum does not
satisfy the radiopurity requirements for the growth of high quality \lmo\ crystals.
Indeed, the scintillation performance of ZnMoO$_4$ crystals, which are very similar to \lmo,
was substantially improved after some additional
molybdenum purification~\cite{Beeman:2012,Chernyak:2013,Berge:2014,Chernyak:2015}.
Table~\ref{tab:mo-purity} reports impurity values
for commercial 5N grade molybdenum oxide and enriched \Mo.
Roughly, the purity level of the enriched molybdenum after production
should be improved by one order of magnitude in order to obtain
high quality scintillating crystals.
The purity levels of the enriched isotope \Mo\ as delivered by the producer
are even worse.

\begin{table}[htb]
  \caption{Purity level of commercial 5N grade molybdenum oxide and of enriched \Mo.}
  \centering
  \begin{tabular}{lccccccc}
    \toprule
    Material                                       & \multicolumn{6}{c}{Impurity concentration [ppm]} \\
                                                   & Na    & Si        & K     & Ca    & Fe    &  W \\
    \midrule
    Commercial 5N grade MoO$_3$ used                         & \multirow{2}{*}{24} & \multirow{2}{*}{9} & \multirow{2}{*}{67} & \multirow{2}{*}{15} & \multirow{2}{*}{$<18$} & \multirow{2}{*}{96} \\
    for ZnMoO$_4$ crystals~\cite{Nagornaya:2009,Gironi:2010} &       &           &       &       &       &  \\
    \midrule
    Samples of enriched \Mo\ used in \cite{Belli:2010}        & \multirow{2}{*}{10} & \multirow{2}{*}{50--360} & \multirow{2}{*}{$<30$} & \multirow{2}{*}{40--50} & \multirow{2}{*}{10--80} & \multirow{2}{*}{200} \\
    (values after production)                                &       &           &       &       &       &  \\
    \bottomrule
  \end{tabular}
  \label{tab:mo-purity}
\end{table}

The NEMO-3 collaboration observed 
high radioactive contamination of enriched \Mo\ samples
using low-background HPGe \ga\ spectroscopy.
The radioactive contamination of the samples was on the level of $\sim0.1$~Bq/kg
for $^{40}$K, $^{234}$Th, and $^{235}$U,
and of a few mBq/kg for $^{208}$Tl, $^{228}$Ac and $^{214}$Bi~\cite{Arnold:2001}.
Similarly, the poor optical quality and high radioactive
contamination of ZnMoO$_4$ and \lmo\ crystals produced
from commercially available high purity grade
molybdenum~\cite{Nagornaya:2009,Gironi:2010,Dubovik:2010,Barinova:2010,Cardani:2013}
with natural isotopic composition
confirms the need for additional material purification
prior to \lmo\ crystal growth.

We developed a two stage technique for molybdenum purification~\cite{Berge:2014}.
The procedure consists of a two-stage sublimation with addition of zinc molybdate,
and a double recrystallization
from aqueous solutions of ammonium para-molybdate by
co-precipitation of impurities on zinc molybdate sediment.
The addition of high purity ZnMoO$_4$ (up to 1\%) substantially reduces
the tungsten contamination, whose impurity is hard to remove
because of its chemical affinity to molybdenum.
During the sublimation at high temperature the following exchange reaction occurs:
\begin{center}
  ZnMoO$_4~+~$WO$_3~\rightarrow~$ZnWO$_4~+~$MoO$_3$.
\end{center}
In addition to inducing a substantial decrease of tungsten,
the sublimation reduces the concentration of radioactive impurities
(K, Th, U, Ra, Pb) and other contaminants,
e.g. transition metals that affect the \lmo\ crystals optical quality
(Ca, Cu, Fe, Mg, Na, Si, Zn, etc.).
The sublimate is annealed in air atmosphere to obtain yellow color stoichiometric MoO$_3$.
The analysis of the obtained sublimate by atomic emission spectrometry
showed an improvement of the MoO$_3$ purity level by at least one order of magnitude.
The efficiency of molybdenum purification by
sublimation is presented in Table \ref{tb:subl-pur}.

\begin{table}[htb]
  \caption{Efficiency of molybdenum oxide purification by sublimation.}
  \centering
  \begin{tabular}{lcccc}
    \toprule
    Material               & \multicolumn{4}{c}{Impurity concentration (ppm)} \\
                       & Si        & K         &  Fe   & W \\
    \midrule
    Initial MoO$_3$         & 600       & $100-500$ & 6     & $200-500$ \\
    After $1^{\text{st}}$ sublimation  & $100-500$ & $10-50$   & $2-6$ & $100-200$ \\
    After $2^{\text{nd}}$ sublimation  & 70        & $1-8$     &  $<1$ & $30-40$ \\
    \bottomrule
  \end{tabular}
  \label{tb:subl-pur}
\end{table}

After the sublimation process, the molybdenum is further
purified by double recrystallization of ammonium molybdate in aqueous solutions.
Molybdenum oxide is dissolved in ammonia solution at room temperature.
To improve the recrystallization process efficiency,
zinc oxide is dissolve at the level of $1-2$~g/L in the ammonium para-molybdate solution
at a pH$>6$ to initiate precipitation.
Ammonia is then added to the solution to reach a pH$=7-8$.
The precipitation of zinc molybdate occurs after several hours.
The ZnMoO$_4$ sediment absorbs the impurities from the solution.
A further pH increase leads to the precipitation of contaminants in the form of hydroxides.
It should be stressed that the basic solution with a pH$\sim 8-9$
provides the most favorable conditions for Th and U precipitation.
After the separation of the sediment, the solution is evaporated to 70\%
at a temperature near its boiling point.
Ammonium oxalate is then added to the solution as a scavenger
for the residual iron impurities which are thus bound together and removed from MoO$_3$.

The purity levels of the so-obtained MoO$_3$ (Table~\ref{tb:final-pur}) fully satisfy the \cupid\ requirements.
Natural and enriched \lmo\ crystal tested at LSM and LNGS 
were characterized by $^{228}$Th and $^{226}$Ra contamination levels $<10~\upmu$Bq/kg~\cite{Armengaud:2017}.

\begin{table}[htb]
  \caption{MoO$_3$ purity levels before and after the purification by sublimation and recrystallization.}
  \centering
  \begin{tabular}{lcccc}
    \toprule
    Material       & \multicolumn{4}{c}{Impurity concentration (ppm)} \\
    ~           				& Si 	& K  			& Fe  	& W \\
    \midrule
    Initial MoO$_3$                   & 60                  & 50                  & 8                  & 200                  \\
    Recrystallization from            & \multirow{2}{*}{30} & \multirow{2}{*}{20} & \multirow{2}{*}{6} & \multirow{2}{*}{220} \\
    aqueous solutions                 & & & & \\
    \midrule
    Sublimation and recrystallization & \multirow{2}{*}{30} &\multirow{2}{*}{10} & \multirow{2}{*}{5} & \multirow{2}{*}{130} \\
    from aqueous solutions            & & & & \\
    \midrule
    Double sublimation and recrystalizzation & \multirow{2}{*}{--} & \multirow{2}{*}{$<10$} & \multirow{2}{*}{$<5$} & \multirow{2}{*}{$<50$} \\
    from aqueous solutions            & & & & \\
    \bottomrule
  \end{tabular}
  \label{tb:final-pur}
\end{table}

\subsubsection{Lithium purification}
\label{sec:li-pur}

The chemical affinity of lithium with potassium resulted
in a considerably high \K\ contamination of $\sim 0.1$~Bq/kg
in one of the first \lmo\ crystals tested~\cite{Barinova:2009swc}.
Similarly, the affinity of lithium with radium
and the observation of $^{226}$Ra in some \lmo\ samples~\cite{Armengaud2017},
imposes strict requirements on the purity of lithium carbonate (Li$_2$CO$_3$)
employed for the \lmo\ crystal growth.
The 99.99\% purity level of the commercially available high purity Li$_2$CO$_3$
is high enough to produce high quality, radiopure \lmo\ crystals~\cite{Armengaud:2017,Grigorieva:2017}.
Nevertheless, an R\&D for additional Li$_2$CO$_3$ purification 
based on the washing of precipitate and recrystallization is underway.

\subsection{Chemical processing and synthesis of \lmo}

The production of \lmo\ crystals can be performed starting from molybdenum of different forms --
e.g. metal ingots, metal foils, or oxide -- according to the following procedure.
The highly purified molybdenum oxide and commercial high purity
lithium carbonate is annealed at temperature $\sim(350$--$400)$\textcelsius,
weighted in the required stoichiometry ratio, and mixed.
The powder is then placed directly into the platinum crucible
used for the crystal growth, where the following reaction of \lmo\
solid-state synthesis takes place:

\begin{equation*}
  \text{MoO}_3 + \text{Li}_2\text{CO}_3 \rightarrow \text{Li}_2\text{MoO}_4 + \text{CO}_2.
\end{equation*}

\noindent
The synthesis is carried out in three stages:
\begin{itemize}
\item fast heating until the synthesis begins;
\item slow heating for 5--10~hours at $\sim 450$\textcelsius,
  until the release of CO$_2$ gas has terminated;
\item fast heating to the melting temperature.
\end{itemize}

 \subsection{Crystal growth by the low-thermal-gradient Czochralski technique}
 \label{sec:LTG-Cz}

\begin{SCfigure}[][ht]
  \includegraphics[width=0.7\textwidth]{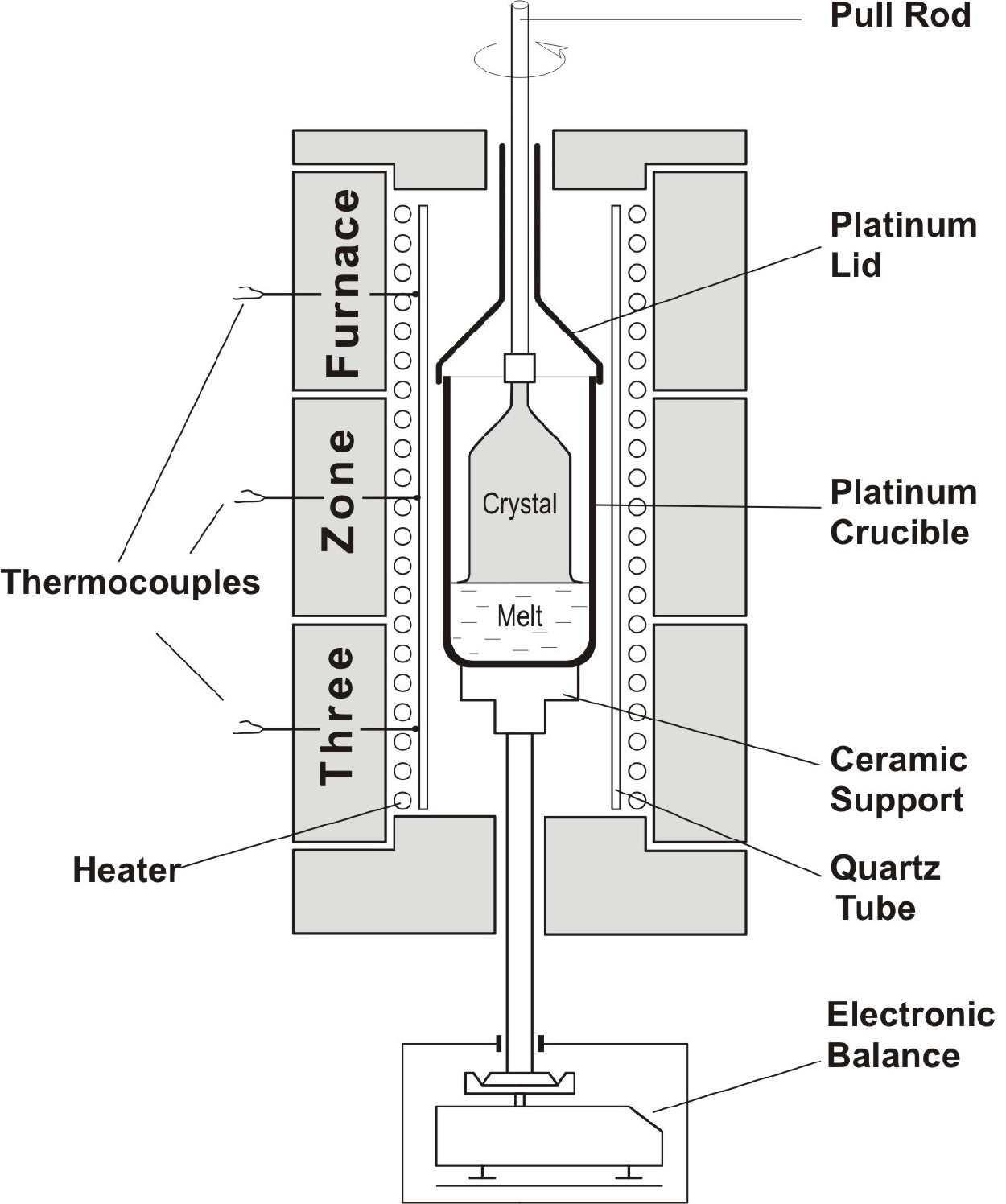}
  \caption{Scheme of the furnace for crystal growth with thelow thermal gradient
    Czochralski technique.}
  \label{fig:growing-set-up}
\end{SCfigure}

Typically, \lmo\ crystals are grown in air atmosphere from the
synthesized powder with the low-thermal-gradient Czochralski
technique~\cite{Pavlyuk:1993,Borovlev:2001}
using a platinum crucible of 70~mm diameter and 130~mm height.
The platinum crucible is placed into a three-zone resistance furnace
with low thermal conductivity bottom and top thermal insulation.
The scheme of the growing set-up is shown in Fig.~\ref{fig:growing-set-up}.
The crucible is covered by a platinum lid with a pipe socket through which the pull rode with
the crystal holder is introduced into the inner space.
During the entire growth process the crystal remains inside the crucible.
The heater and control system allow to keep the axial and radial
thermal gradients within $0.05-1.0$~\textcelsius/cm.
This crystal growth technique is fully controlled by the weighing method
at all stages including the seeding, as any visual inspection is impossible.

The growth rate varies from 0.5~mm/h to 0.75~mm/h.
The rotation velocity is about 3~rounds/min.
A typical duration of the growth process is 10--12~days.

A second-crystallization crystal growth is applied in order to have better quality crystals (higher crystal perfection and lower level of impurities). The second-crystallization crystal growth is carried starting from the
crystals grown in the first crystallization,
and from the scraps obtained from the scintillation elements production.
The following materials are used in a typical second-crystallization process:
\begin{itemize}
\item larger parts of the crystals after the first crystallization;
\item scraps of the second crystallization crystals after the cuts
  to produce the scintillation elements (see Sec.~\ref{sec:cut}).
  The scraps are etched by ultra-pure water to a $\geq10~\upmu$m thick remove surface layer.
\end{itemize}
Tens of Li$_2$MoO$_4$ and Li$_2$$^{100}$MoO$_4$ crystals have been
grown by the described technique \cite{Grigorieva:2017}.

\subsection{Scintillation elements production and surface treatment}
\label{sec:cut}

Grown crystal boules are subject to a visual inspection
performed in a dark room using a collimated light source
(laser pointer) to localize scattering centers.
With the goal of minimizing the presence of scattering defects, this allows evaluation of the geometry
and position of the scintillation element that can be cut from the ingot.

As a first step of crystal processing,
the crystal boule is fixed mechanically on a holder
without using any resin or other gluing material
in order to avoid dissolving the resin or the glue in the organic liquids
used in the cutting process.  This also minimizes the
pollution of the scraps after the cutting process.

Crystal boules are cut with strip tape with
a diamond layer applied at its end using high purity kerosene as a coolant.
The coolant liquid with crystal  sludge is collected in a
plastic tub to save as much of the enriched material as possible.

The obtained crystal bars are glued on a glass plate with a thin layer of organic resin.
Cylindrical shapes are obtained by cutting the crystal with a tubular drill
with a $<1$~mm thick diamond tip,
while parallelepiped shapes are cut with a wire-saw.
In all cases, kerosene is used as a coolant liquid.

The shaped crystals are further polished using an emulsion of silicon carbide powder and vacuum oil as a slurry.
The sludge produced at this stage is highly contaminated, but of limited amount.
Therefore, it is not reprocessed.

The final polishing is carried out using radiopure silicon
oxide abrasive powder and vacuum oil. 
Polishing is carried out on synthetic fabric.
A layer of several tens of micrometers is removed.
Losses at this stage are almost negligible.  Nevertheless,
even in this case residuals could be collected for the recovery of enriched Mo.

\begin{SCfigure}
  \includegraphics[width=0.5\textwidth]{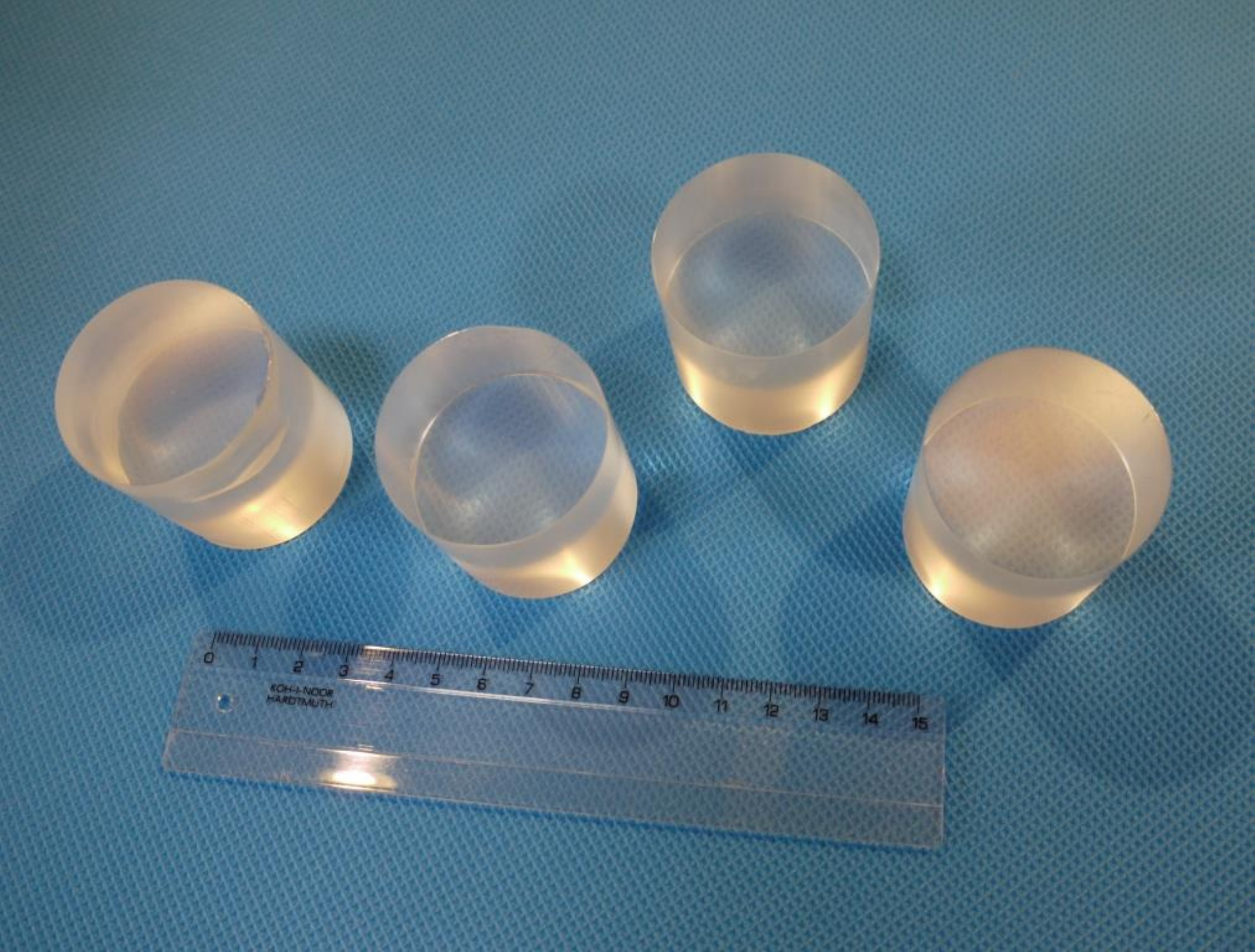}
  \caption{\enrlmo\ crystal 
    produced for the \cupidmo\ pilot experiment.}
  \label{fig:scint-elements}
\end{SCfigure}

An example of \enrlmo\ crystals
produced for \cupidmo\ is shown in Fig.~\ref{fig:scint-elements}.

\subsection{Enriched $^{100}$Mo recycling}

Some methods for the recovery of isotopically enriched molybdenum
both from the crystal growth residues and the mechanical treatment scraps
have been developed in the last few years.
Large crystal residues and trimmings are grinded before being placed
into the crucible of the growing set-up.
The scraps obtained from machining are fired
to remove traces of organic compounds (kerosene, resin, etc),
and then dissolved and filtered.
The solution is evaporated to obtain a solid precipitate.
The obtained \lmo\ powder is decomposed by acid to molybdic acid (H$_2$MoO$_4$).
Molybdic acid is then annealed to molybdenum oxide (MoO$_3$).
The molybdenum oxide is purified with the procedure described in Sec.~\ref{sec:mo-pur}.
The sublimation process is repeated two-three times depending on the MoO$_3$ contamination level.
The recycling process recovers $>90\%$ of the molybdenum from scraps of different origin.

\subsection{Other possible crystal growth methods}

Alternative crystal growth techniques are under investigation
for the large scale production of \lmo\ crystals for \cupid.
These include the ordinary Czochralski crystal growth
and the Bridgman method.

\subsubsection{Ordinary Czochralski crystal growth}
\label{sec:Cz-ordinary}

The French ANR-funded R\&D project CLYMENE aims at exploiting
the ordinary Czochralski crystal growth for the production
of scintillating crystals to be operated as bolometers in rare events searches.
A \lmo\ crystal with a mass of 230~g was successfully grown
with this method in a platinum crucible under an air atmosphere.
The crystal characterization demonstrates promising radiopurity levels (Table~\ref{tab:OCgrowth})
an optical transmission $\alpha_{\text{ABS}}(589\text{nm})\sim0.05~\text{cm}^{-1}$),
and good thermal properties, with $\theta_{D}\sim765$~K and no phase transition down to 2 K~\cite{Velazquez:2017}.
Above-ground low temperature tests of the material as a scintillating bolometer~\cite{Velazquez:2017,Buse:2017}  
showed an energy resolution of few keV and a relative light yield of $\sim0.97$~keV/MeV,
possibly yielding a highly efficient $\alpha$/$\gamma$ separation.
The R\&D is continuing with the goal of developing crystal boules of $>1$~kg
with improved properties by means of combined numerical
simulations and well-designed implementations of the \lmo\
crystals Czochralski growth process \cite{Stelian:2018}.
A 820~g \lmo\ crystal has already been pulled at a rate of
$\sim 2$~mm/h and will be operated as a bolometer in the near future.

\begin{SCtable}
  \centering
  \caption{Radioactive contamination levels of the first \lmo\ crystal
    grown with the ordinary Czochralski method.}
  \label{tab:OCgrowth}
  \begin{tabular}{lc}
    \toprule
    Isotope & Contamination [mBq/kg] \\
    \midrule
    \K & $<40$ \\
    \Ra & $<0.37$ \\
    \Th & $<0.21$ \\
    $^{228}$Th & $<0.27$ \\
    \bottomrule
  \end{tabular}
\end{SCtable}

In addition, \lmo\ crystals of up to 2'' in diameter, grown by the Czochralski technique,
were produced by Radiation Monitoring Devices Inc. (Watertown, MA, USA).
The crystal compounds were synthesized in a separate step prior to crystal growth.
Transparent, colorless crystals were produced when materials of suitable purity were used.
The bolometric test of the material will be realized soon at CSNSM (France).

\subsubsection{Crystal production by Bridgman method}
\label{sec:Bridgman}

Large-size \lmo\ single crystals have been recently grown with the vertical Bridgman method
in the Ningbo University (China)~\cite{Chen:2018} in Shanghai SICCAS High Technology Corporation (China).
Commercial compounds with 3N purity of both Li$_2$CO$_3$ and MoO$_3$ powders
were used without additional purification for the crystals production.
The grown crystal boules exhibit some residual green-yellow coloration,
which is most probably due to a high contamination level of transient metals.
We investigated the transmittance spectra, scattering spectra,
laser-stimulated luminescence spectra, and luminescence decay time of these \lmo\ crystals,
and performed an above-ground bolometric test of a $20\times20$~mm-diameter sample at CSNSM (France),
obtaining satisfactory results
in terms of the performance and radiopurity.
A low temperature test of a $45\times45\times45$~mm sample is also foreseen.
Further R\&D for the growth of radiopure \lmo\ crystals using the Bridgman method is ongoing.


\subsection{Quality assurance: detector prototypes and tests}

Given the \cupid\ background goals, the \lmo\ crystal radiopurity is crucial.
Even if a dedicated protocol is defined for the quality control of the crystal production process,
only a cryogenic test can determine if the final crystals are suitable for the experiment.
Therefore, following the \cuore\ example \cite{Alessandria:2012},
cryogenic measurements will be perform to test random samples
from each batch of the \lmo\ crystal production.
Each test will be performed using a mounting scheme as similar as possible
to a floor of the final \cupid\ detector, and operated for several weeks
in order to test both the bolometric performance and the compliance of the crystals
to the required radiopurity specifications.


\newpage


\section{Phonon and light sensors} \label{sec:LightDetectors}

In this section we discuss the sensors for both the heat and light detection.
In CUPID, the signals consists of a phonon pulse induced in either the \lmo\ crystal
or the Light Detector (LD) adsorber, and read-out by dedicated sensor.
The most mature sensor technology is represented by Neutron Transmutation Doped (NTD) germanium thermistors,
consisting of a small Ge crystal whose resistance rises sharply as the temperature decreases.
Possible alternatives are Transition Edge Sensors (TES), in which a superconductive film
is kept within the normal-to-superconducting transition, or microwave Kinetic Inductance Detectors (KID),
which measure the change of the kinetic inductance of a superconductive element
following the absorptions of athermal phonons.

The sensor baseline for CUPID, both for the \lmo\ crystals and the light detectors,
consists of Ge thermistors,
although TESs and KIDs are under study as possible light detectors
for their superior signal-to-noise ratio and speed.

Based on the experience gained in the framework of the CUPID R$\&$D projects,
we selected NTDs for the readout of both the heat and light channel.
The LDs consist of a thin Ge wafer acting as an absorber, coupled to an NTD.
The features and the performances of such LDs satisfy the requirements for CUPID,
with the advantage of exploiting the same electronics and the wiring system
already implemented in CUORE. Several lines of R\&D are still underway to develop
NTDs with Neganov-Luke amplification, TES, and KIDS, for their higher sensitivity and faster response.
To complement the light detectors, R\&D on the optimization
of the light collection and of the mechanical supports are being carried out.

In this section we first describe the NTDs and the heaters used for the offline stabilization of the recorded data.
Second, we detail the various LD technologies. Finally, we illustrate the optimization of the light collection
and of the LD coupling to the thermal bath.

\subsection{Neutron transmutation doped Ge thermistors}\label{sec:ntds}

NTDs have resistance values in the 1--100~M$\Omega$ range at their operation point,
which is tuned in the 10--20~mK range for CUPID.
They are biased at constant current and respond with a voltage pulse to a temperature change.  

Nuclear transmutation induced by a thermal neutron flux is the best technique
to achieve uniform doping in the bulk of a semiconductor.
The target neutron dose for the CUPID bolometers is about $4.3 \times 10^{18}$~ n/cm$^{2}$.
We envision a specific NTD production for CUPID, which will exploit the experience developed in CUORE and its precursors.
We will irradiate several Ge wafers in a nuclear reactor using slightly different doses to fine tune the NTD response.
Hundreds of individual thermistors can be typically extracted from each wafer.
We will select the NTDs with the correct resistance-temperature behavior
through a campaign of low-temperature characterization.
This can be performed in several above-ground facilities available to the CUPID collaboration.
The baseline geometry for the heat-channel NTDs is $3 \times 3 \times 1$~mm,
with two metalized $3 \times 1$ sides separated by 3~mm for wire bonding.
The mass of this sensors is about 50~mg.
NTDs with these dimensions were already adopted in CUORE, CUORE-0, and Cuoricino.
As for the LDs, smaller NTDs are required to increase the sensitivity.
In fact, the heat capacity of the \lmo\ crystal dominates that of the NTD,
while the opposite is true in the case of LDs, due to the small mass ($\sim$1~g) of the Ge wafer.
In this case, therefore, the mass of the NTD plays an important role.
Based on the lessons learned in CUPID-0, CUPID-Mo, and related R\&D,
we plan to employ NTD sensor for the LDs with a mass of about 5-10 mg
and the same contact distance as the NTDs for the heat channel,
but the contact cross section modified accordingly.      

\subsection{Heaters}\label{sec:heaters}

A critical issue in operating bolometric detectors over long periods of time
is keeping their response stable despite the unavoidable temperature fluctuations of the cryogenic setup.
In CUORE, CUPID-0, and CUPID-Mo this is achieved using a pulser
that periodically delivers a fixed amount of energy (through the Joule effect)
into the bolometer by means of a resistor thermally coupled to it,
generating a pulse as similar as possible to a signal due to particle interactions \cite{Alessandrello:1998bf}.
The off-line study of the variation of the detector response for the same energy deposition,
due to the cryogenic instabilities, can be used to correct their effects.

In CUORE the heating element used to deliver the Joule heat pulses consists of a resistive device, called simply a heater.
It is made up of heavily doped semiconductor material (well above the metal-to-insulator transition).
Square voltage pulses are injected into the devices with a programmable pulse generator,
tuning the amplitude and the time width so as to develop a few MeV thermal energy in the heater (see Section \ref{sec:pulser}).
The details of the fabrication and characterization of the CUORE heaters can be found in Ref.~\cite{Andreotti:2012zz}.
We envisage a similar production for the CUPID experiment.

\subsection{Baseline technology: Ge light absorbers with NTD sensors}
\label{sec:Ge_LD_NTD}

\begin{figure}
  \centering
  \includegraphics[width=\textwidth]{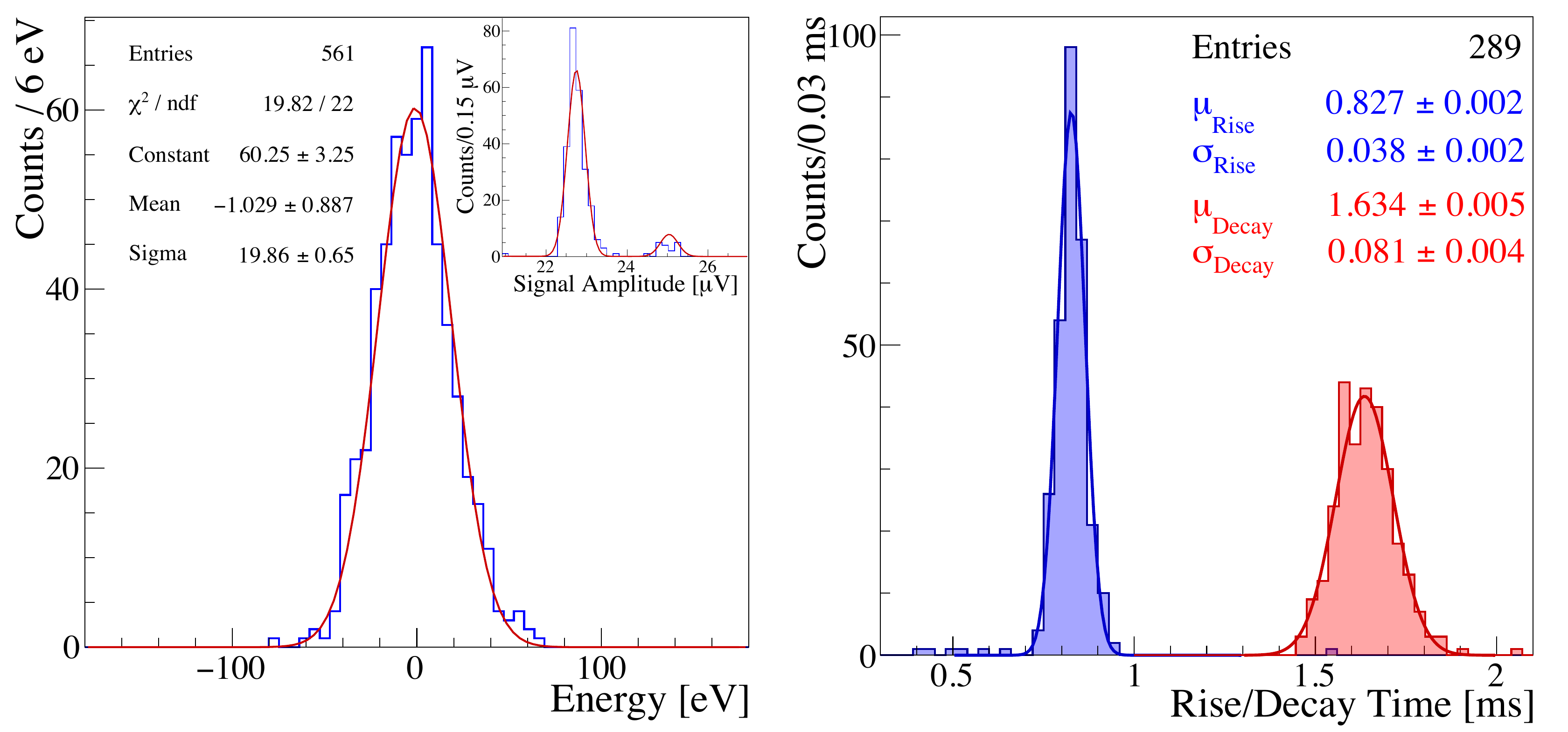}
  \caption{Performance of a Ge-LD tested at LNGS in the framework of the CUPID R$\&$D program~\cite{Barucci:2019ghi}.
    Left: energy distribution of the random sampled noise. The width of the distribution ($\sigma = 20$~eV)
    represents the baseline energy resolution of the Ge-LD.
    The upper inset shows a $^{55}$Fe calibration spectrum.
    The x-axis unit is in absolute voltage drop across the thermistor.
    The RMS resolution on the 5.9~keV and 6.5~keV X-ray peaks is 59 eV.
    Right: rise and decay times distributions for the $^{55}$Fe X-rays.}
 \label{fig:LD_NTD_fastGe}
\end{figure}

Germanium wafers instrumented with NTDs
are one of the most well-established technologies for the detection of scintillation light
in cryogenic experiments~\cite{Bellini:2018,Pirro:2017,Poda:2017c}.
The NTD readout of the LDs allows us to take advantage of the electronics
and the wiring system already implemented in  CUORE.
The long-standing experience acquired in the framework of the bolometric \onbb\ decay  experiments
LUCIFER~\cite{Beeman:2013b}, which is now renamed as CUPID-0~\cite{Artusa:2016,Azzolini:2018tum},
LUMINEU~\cite{Armengaud:2017,Poda:2017d}, and its extension CUPID-Mo~\cite{Poda:2018},
allows us to reach very high performance in terms of energy resolution, stability, reproducibility, and reliability.
The performance of the LDs developed by the mentioned collaborations are summarized in Table~\ref{tab:LDGeNTD_performances}.
The recent CUPID R\&D results on development of a fast Ge-based optical bolometer \cite{Barucci:2019ghi} are also listed in Table~\ref{tab:LDGeNTD_performances} and illustrated in Fig.~\ref{fig:LD_NTD_fastGe}.

\begin{table}
  \centering
  \caption{Performances of NTD Ge-instrumented LDs based on $44.5 \times 0.175$~mm Ge wafers.
    The results are from CUPID-0 (currently running), LUMINEU (completed), and CUPID-Mo (recently started).
    The most recent results achieved in the framework of the CUPID R\&D program are quoted as well.
    All values correspond to the harmonic mean.
    The volume of all NTDs is comprised between 0.5 and 2.8~mm$^3$, and their maximum length
    between 2.2 and 3~mm.}
  \begin{tabular}{lcccccc}
    \toprule
    ~ & \multicolumn{2}{c}{{\bf CUPID-0}} & {\bf LUMINEU}                               & \multicolumn{2}{c}{{\bf CUPID-Mo}} & {\bf CUPID R\&D}	\\
    ~ & \cite{Artusa:2016} 	              & \cite{Azzolini:2018tum} & \cite{Poda:2017d}	& commissioning & upgraded 	         &\cite{Barucci:2019ghi} \\
    \midrule
    SiO coating       & \multicolumn{2}{c}{Single side}											& Single side								& \multicolumn{2}{c}{Double side}  &None\\
    Operated detectors                & 6    & 31     & 4$^a$   & 17$^b$  & 17$^c$ & 1   \\
    Measurement time [yr]             & 0.1  & 2      & 0.3     & 0.2     & 0.3    & 0.1 \\				
    \midrule
    Temperature [mK]                  & 20.0 & 17.5   & 17.0    & 20.5    & 20.0   & 15  \\
    NTD $R_{\text{work}}$ [M$\Omega$] & 0.8  & 4.2    & 0.7     & 1.2     & 0.8    & 1.5 \\	
    \midrule
    Rise time [ms]                    & 1.8  & 3.5    & 2.0     & 0.8     & 1.2    & 0.8 \\				
    Decay time [ms]                   & 5    & 7      & 13      & 6       & 13     & 1.6 \\				
    Signal [$\mu$V/keV]               & 1.3  & --$^d$ & 1.5$^e$ & 1.2     & 0.96   & 3.9 \\				
    Baseline RMS [eV]                 & 41   & --$^d$ & 35$^e$  & 167$^f$ & 58     & 20  \\	
    \bottomrule
  \end{tabular}
  \label{tab:LDGeNTD_performances}

\end{table}

The baseline design for the CUPID light detectors is a disk-shaped pure Ge wafer
with double-sided SiO anti-reflecting coating, instrumented with NTDs.
Ge substrates, provided by UMICORE Electro-Optic Material (Geel, Belgium),
are double side polished wafers with a 50~mm diameter and an average thickness of 175~$\upmu$m,
with an impurity concentration below 2~$\times$~10$^{10}$ atoms/cm$^3$.
Prior to the light detectors construction, each wafer will be coated on both sides
with a 70~nm SiO anti-reflecting layer~\cite{Mancuso:2014,Azzolini:2018tum},
which enhances the light collection by a factor of $\sim2$~\cite{Mancuso:2014}.
A small NTD, with a mass of 10 mg or less, will be glued on the Ge surface
using the same gluing tools as the main \lmo crystals.
To minimize the $1/f$ noise in sliced NTDs faced over the CUPID-Mo commissioning,
it will be necessary to perform a dedicated polishing of thermistors as done in CUPID-0.
The use of $\sim 1$~mg NTDs would yield a faster response,
further increase the signal amplitude, and reduce the baseline noise~\cite{Coron:2004,Armengaud:2017}.
The baseline design for the Ge wafer holder consists of PTFE clamps and a copper housing.
The thermal link to the heat sink is provided by 25-$\upmu$m-diameter gold bonding
wires, which also provide the electrical connection with the NTD. 


\subsection{Alternative Options}

\subsubsection{Neganov-Luke-assisted NTDs}\label{sec:NL-NTD}

\begin{figure}
  \centering
    \includegraphics[height=0.2525\textheight]{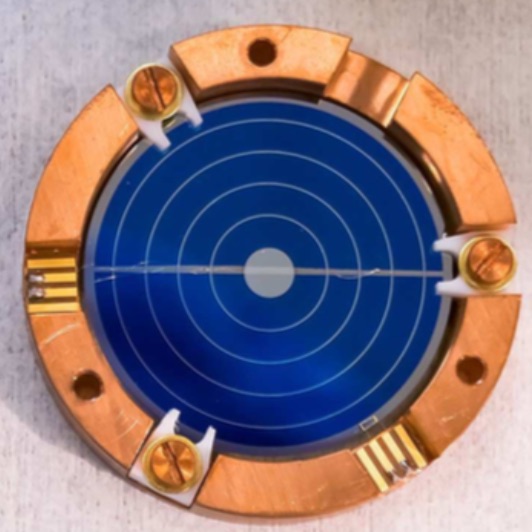}\quad
    \includegraphics[height=0.2525\textheight]{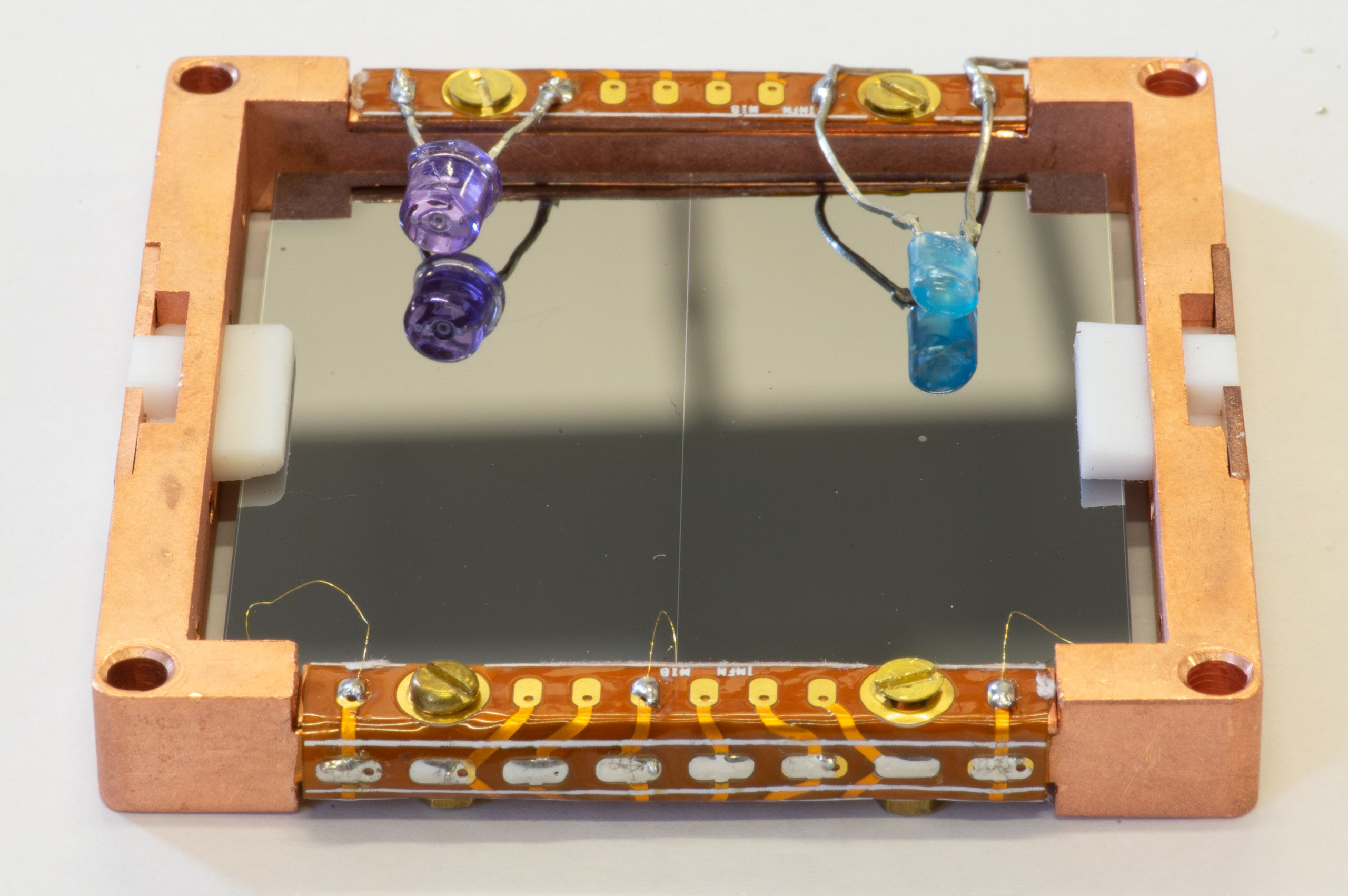}
  \caption{Left -- picture of a NL-NTD base germanium light detector. The absorber is realized with a 44~mm-diameter, 175~$\mu$m-thickness high purity germanium wafer. Concentric annular electrodes are deposited on the surface, which is then coated with a thin SiO (blue-ish) layer to enhance the photon absorption in the visible wavelength.
  Right -- a picture of a NL-NTD base silicon light detector. The absorber is realized with a 50$\times$50 mm, 625~$\mu$, high-resistivity silicon. Electrodes of different geometries can be realized by ion implantation to realize ohmic contacts.}
  \label{fig:NL-NTD}
\end{figure}

Standard NTD-equipped Ge or Si-absorber-based light detectors can be upgraded to exploit the so-called Neganov-Luke amplification mechanism \cite{Luke:1988} by adding a set of electrodes on the absorber surface(s)(NL-NTD, Fig. \ref{fig:NL-NTD}), in order to apply a static electric filed to the semiconductor. The realization relies on well-assessed evaporation, sputtering, and ion implantation techniques.

Using this amplification mechanism, typical performances reached are of 10-20~$\mu$V/keV responsivity and 10-20~eV FWHM baseline resolution under grid bias up to 70~V (300~V), corresponding to a factor $\sim$10 (50) improvement in the signal-to-noise ratio with respect to the standard Ge (Si) detectors. Several NL-NTD prototypes of about 15~cm$^2$ (up to 25~cm$^2$ with Si) detection surface have been realized and operated several time in above-ground experiments and in underground facilities to perform R\&D measurements where a specific low energy detection was required, well beyond the limit one could get with standard NTDs equipped light detectors. 

The technique of the NL-NTD light detectors is now mature enough to be deployed in a large-scale experiment, allowing us to reach detection thresholds of about 10 times smaller than the standard NTD light detector.  This comes with a minimal price to pay of 2 additional wires, required to bias the electrode sets.

\subsubsection{Transition edge sensors}\label{sec:tes}

A TES is a superconducting sensor that has been used in multiple experiments since their development \cite{STP3, PolarBear-TES, SuperCDMS-TES}. This technology is a mature option that can be used in cryogenic experiments with requirements for low noise, fast rise-times, and good energy resolution. TES devices have very small footprints compared to NTD thermistors and as a result have a significantly lower heat capacity, which leads to a shorter thermal time constant. TES time constants are further enhanced by negative electrothermal feedback \cite{Irwin2005}, making these devices quite capable of meeting pile-up rejection requirements for a Mo-based CUPID design. In particular IrPt bi-layer TESes, utilizing the proximity effect \cite{Holm1932}, have shown reproducible and controllable transition temperatures \cite{IrPtTc_arxiv}.


Unlike NTDs, TES devices are low-impedance ($R_n \sim$ 540\,m$\Omega$), allowing for a complete electrothermal characterization of the TES by measurement of the complex impedance using varying AC input signals. This is useful for determining the TES heat capacities, thermal conductances, thermal response, and theoretical energy resolution. For the test TES devices at present, we observe rise times at 33\,mK of $\sim$ 100\,$\mu s$ with decay times of $\sim$ 5\,ms. The current energy threshold of these devices is $\sim$ 10 \,eV. By sending 20\,$\mu s$ voltage pulses across the Pt resistor we can deposit a known amount of energy into the Si wafer and observe the TES response. We can use the observed TES response to compute the energy dissipated through the TES via electrothermal feedback and compare to the known energy deposition. At present there appears to be a factor of $\sim$ 50 discrepancy between these two. A possible reason for this is that the TES is not thermalized to the Si wafer strongly; this is an ongoing area of investigation. This can perhaps be corrected via the addition of small amounts of Au to thermally anchor the TES more strongly to the Si substrate and will be investigated. If this is found to be the case, it will lead to an improved energy threshold and energy resolution for the device.

\begin{figure}
    \centering
    \includegraphics[height=0.2142\textheight]{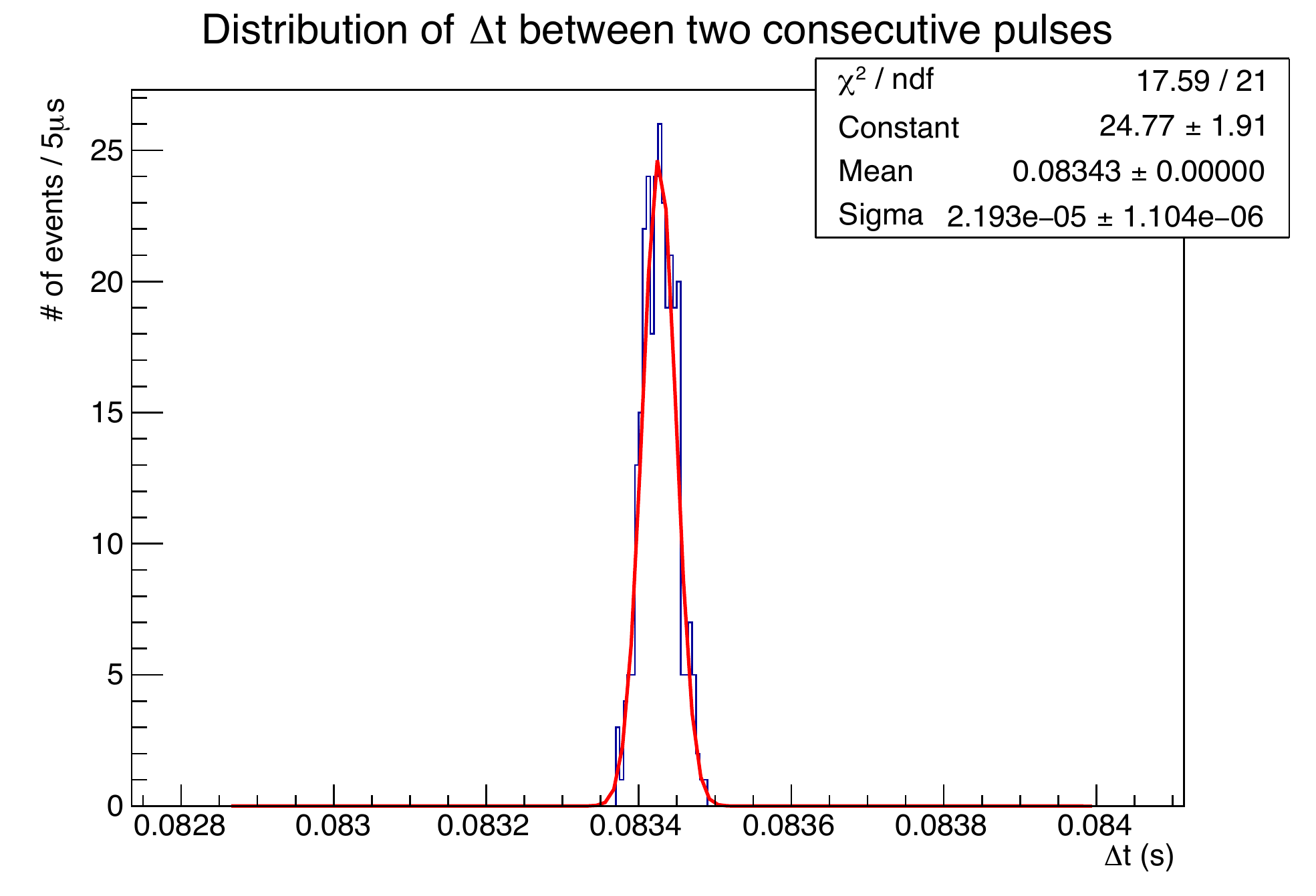}\quad
    \includegraphics[height=0.2142\textheight]{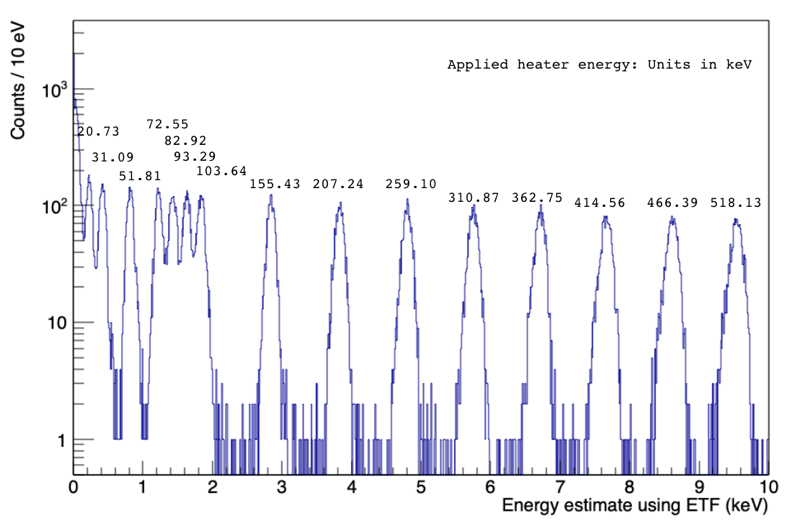}
    \caption{Left -- timing resolution of an IrPt sensor at $T$ = 32\,mK. Using the deposited Pt resistor, pulses of constant energy with a fixed separation time are deposited into the Si wafer. A reconstruction of the time between the pulses shows a timing resolution of $\Delta T \sim$ 22\,$\mu$s. Right -- reconstructed energy response by computing the energy removed through the TES via electrothermal feedback. This is proportional to the integral of the TES pulse. Each peak is labeled with the corresponding energy deposited into the Si wafer by the Pt resistor.}
    \label{fig:TES-timing-etf}
\end{figure}

TES devices require cold-stage amplifiers, with SQUIDs being the preferred devices to use for this purpose. SQUIDs introduce an extra heat load via the power dissipation during their operation ($\sim$ 1\,nW per SQUID) along with associated cabling. In order to minimize the total cooling power required, frequency division multiplexing of the TES devices can be implemented. This requires the addition of an inductor and capacitor in series with a TES (as close as possible to avoid parasitic inductance). By tuning the values of L and C for each TES, a network of RLC filters is formed with each particular RLC filter allowing only a specific AC bias frequency through. This technique has been employed successfully with CMB experiments with active R\&D to implement multiplexing factors $>100$ \cite{CMB-Multiplex}. For CUPID, a more modest multiplexing factor of $\sim$ 10 is enough. At present the development of both operating and readout electronics for the frequency division multiplexing are proceeding.
TES devices are a mature technology used in rare event searches for dark matter and in CMB experiments. Additionally, high multiplexing factors are already demonstrated in CMB experiments. This, combined with the intrinsically fast time constants, low noise, and good energy resolution, make TES an appealing alternative.
\label{sec:TES}

\subsubsection{Kinetic inductance detectors}\label{sec:MKID}

KIDs have been successfully applied in astrophysics searches providing superb energy resolution and natural multiplexed readout~\cite{Day:2013}. Because the active surface of these devices is limited to few mm$^2$ they can be operated as cryogenic light detector in $0\nu\beta\beta$ search only exploiting the phonon-mediated approach, proposed by Swenson et al.~\cite{Swenson:2010yf} and Moore et al.~\cite{Moore:2012au}.

\begin{figure}
    \centering
    \includegraphics[height=0.215\textheight]{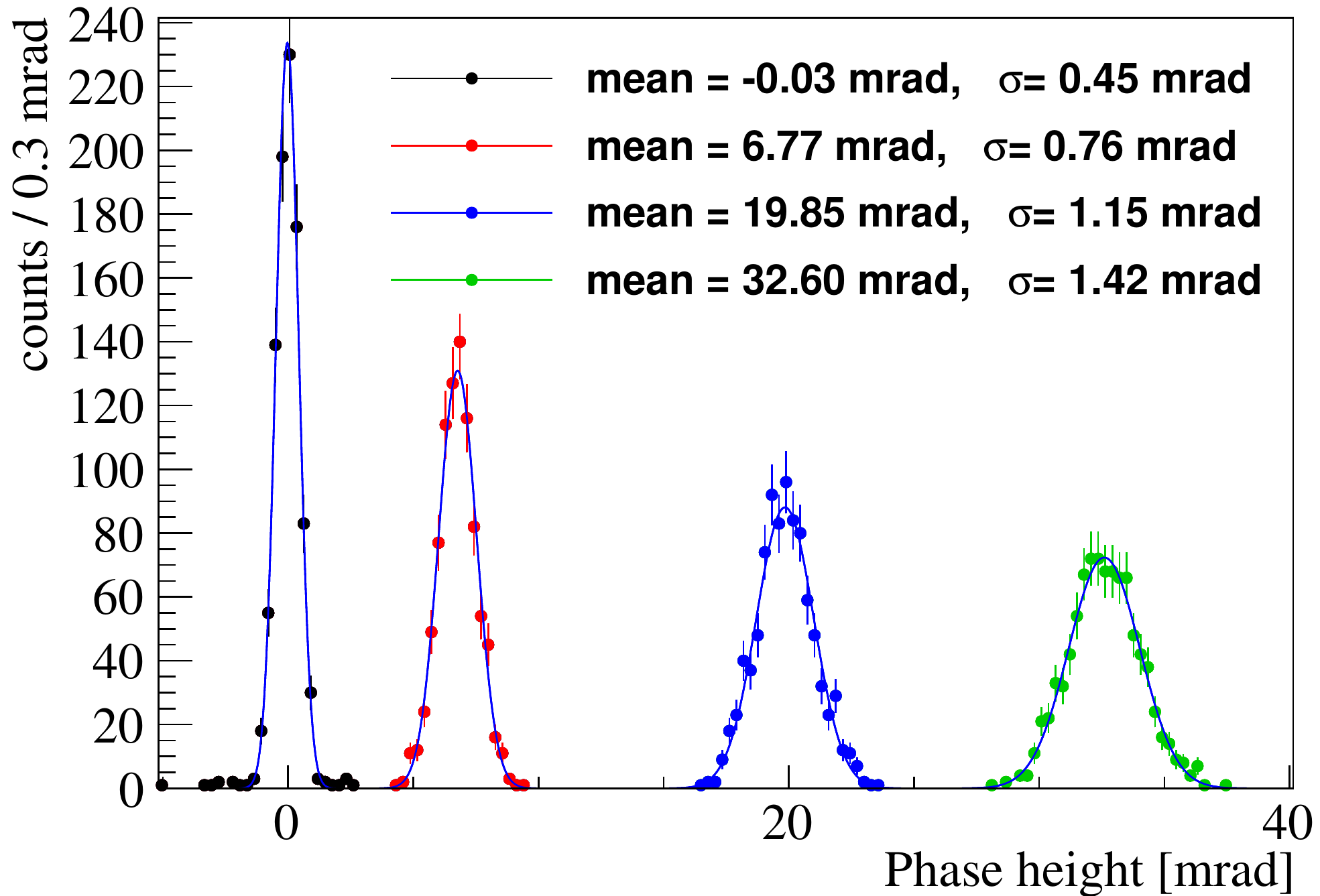}\quad
    \includegraphics[height=0.215\textheight]{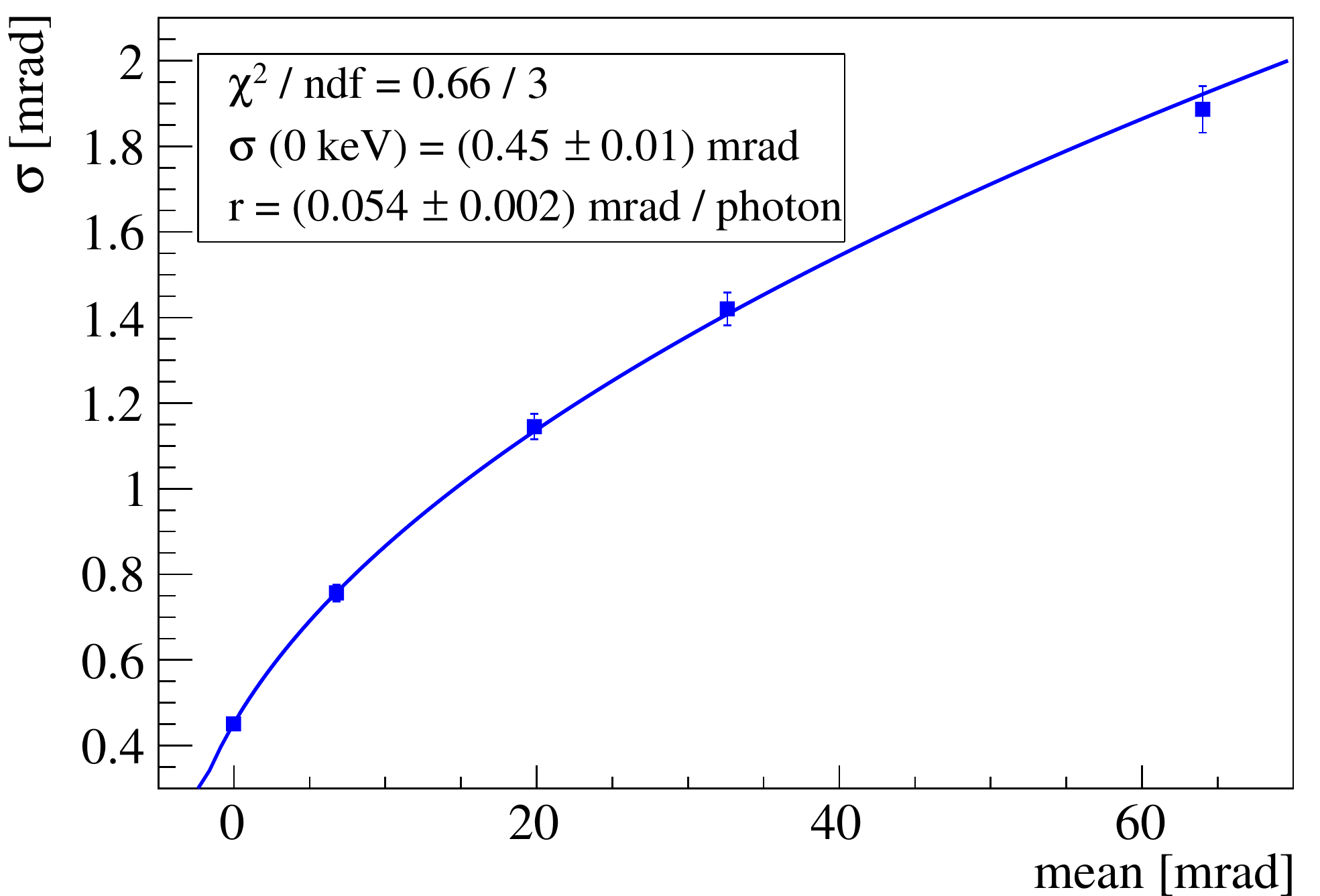}
    \caption{Left: histograms of pulse heights in the phase readout for light signals of different energies.
    Right: standard deviation $\sigma$ versus mean m of the Gaussian distributions shown in the left panel. The plot includes one more point at 65 mrad not shown in the left panel to not compress the scale for the other distributions. The curve is well described by the Poisson statistics of photons. The resolution at zero amplitude corresponds to 25~eV~RMS. See Ref.~\cite{Cardani:2018krv} for more details.}
    \label{fig:KIDResult}
\end{figure}

In such an approach, KIDs are deposited on a large insulating substrate featuring a surface of several cm$^2$. Then, the light emitted by cryogenic calorimeters interact with the substrate producing phonons, which travel within the material until a small fraction of them (5-10~\%) are absorbed by a KID.

\begin{SCfigure}
    \includegraphics[width=0.7\textwidth]{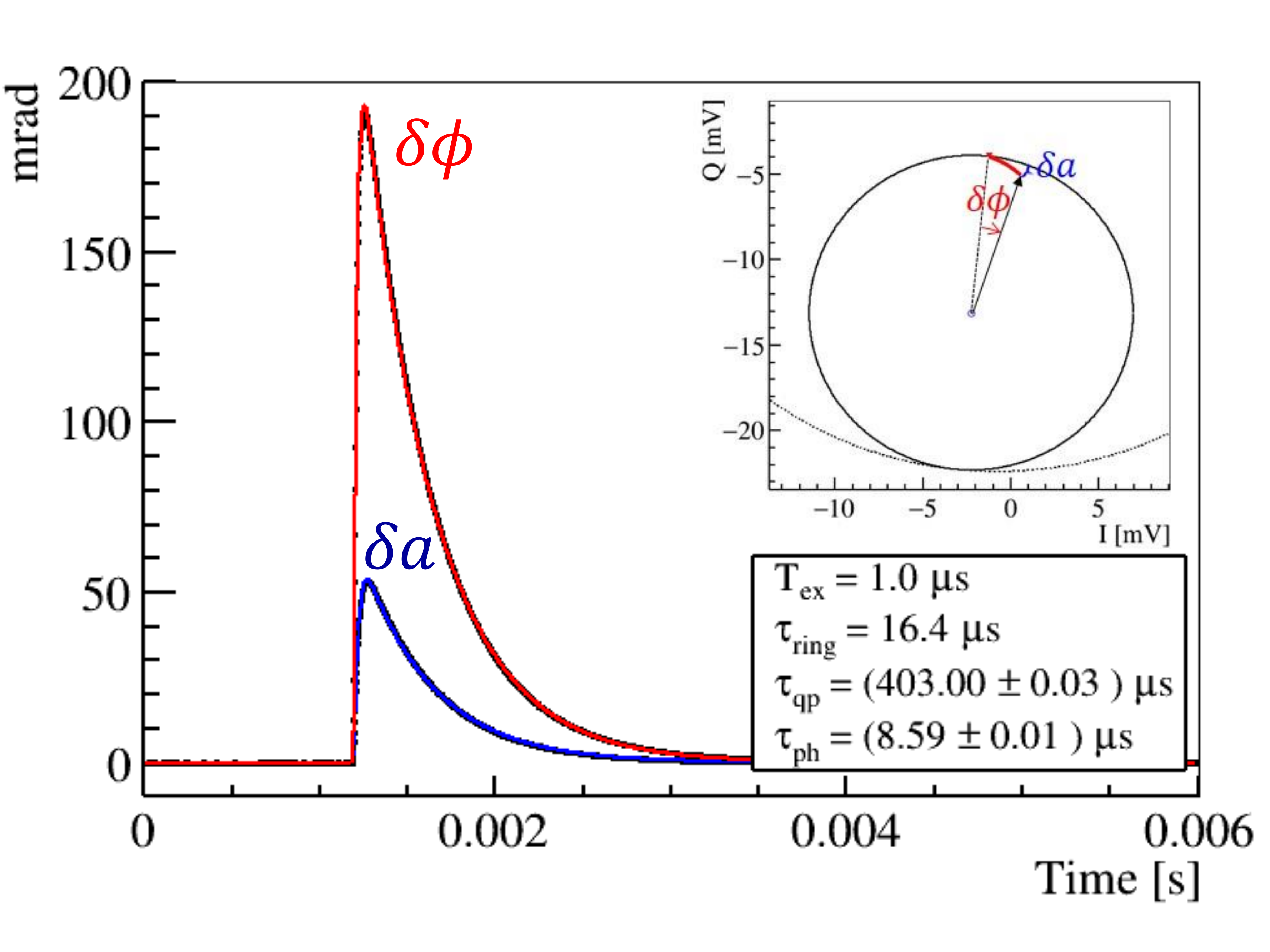}
    \caption{Phase (red) and amplitude (blue) variation measured from the center of the resonance loop induced by optical pulses produced with a 400~nm LED lamp located at room temperature and routed inside the cryostat by an optical fiber. The photon burst had a time duration of 1~$\mu$s.}
    \label{fig:KIDpulse}
\end{SCfigure}

Exploiting the phonon-mediated technique, the CALDER project~\cite{Battistelli:2015vha} realized a cryogenic light detector using a $2\times2$~cm$^2$ Si substrate 300~$\mu$m thick monitored with one aluminum KID. The detector showed a baseline energy resolution of about 80~eV~RMS, constant in a temperature range starting from 10~mK up to 200~mK~\cite{Bellini:2016lgg}. Moving from Al to more sensitive superconductors, such as AlTiAl multi-layer, the baseline energy resolution has improved up to 25~eV~RMS~\cite{Cardani:2018krv}. The high energy resolution of this prototype allowed us to obtain a solid estimation of the detector performance exploiting an absolute energy calibration based on the Poisson statistics of the photons absorbed into the substrate as shown in Fig.~\ref{fig:KIDResult}.

\begin{figure}
    \centering
    \includegraphics[height=0.3039\textheight]{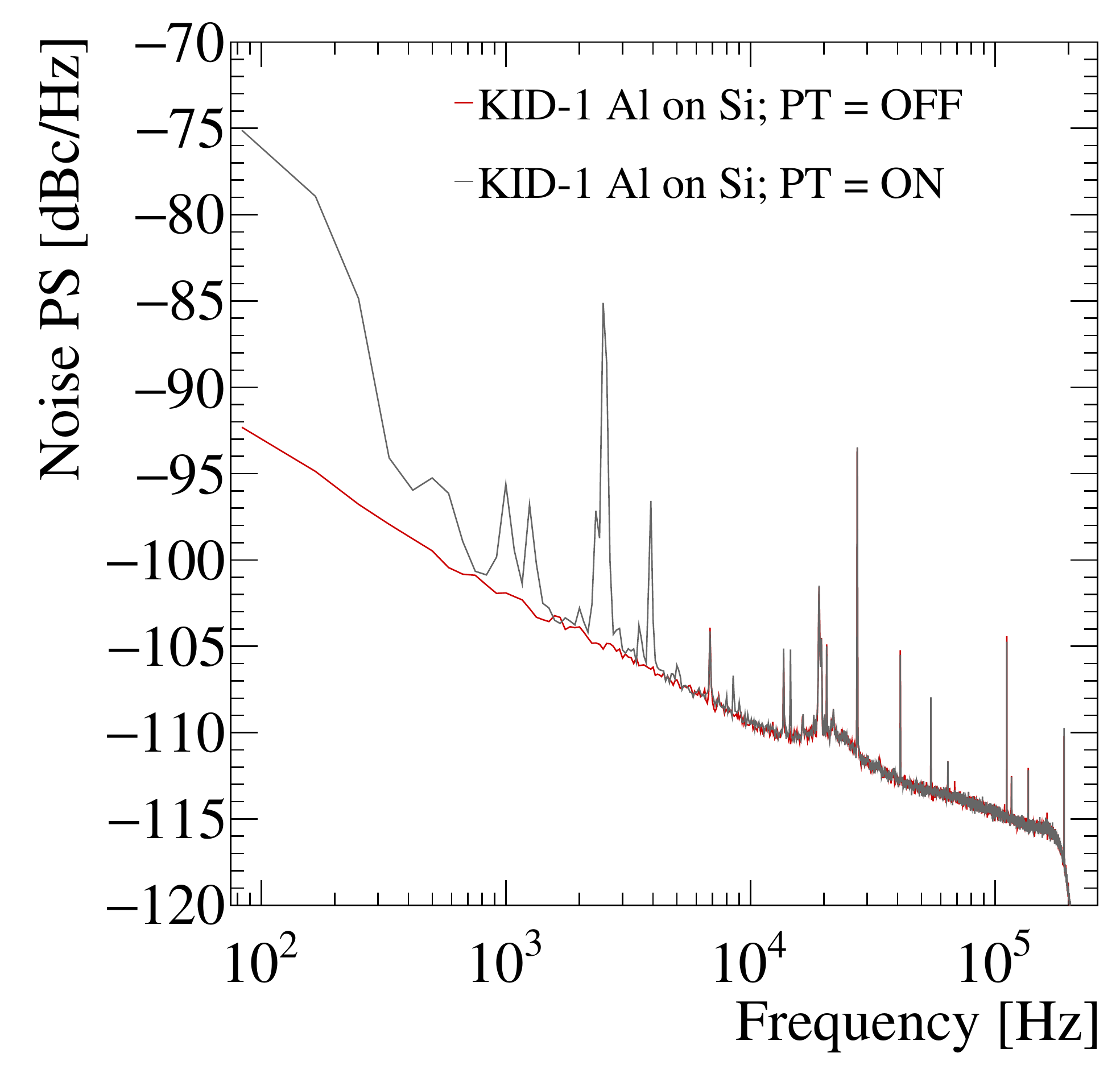}\quad
    \includegraphics[height=0.3039\textheight]{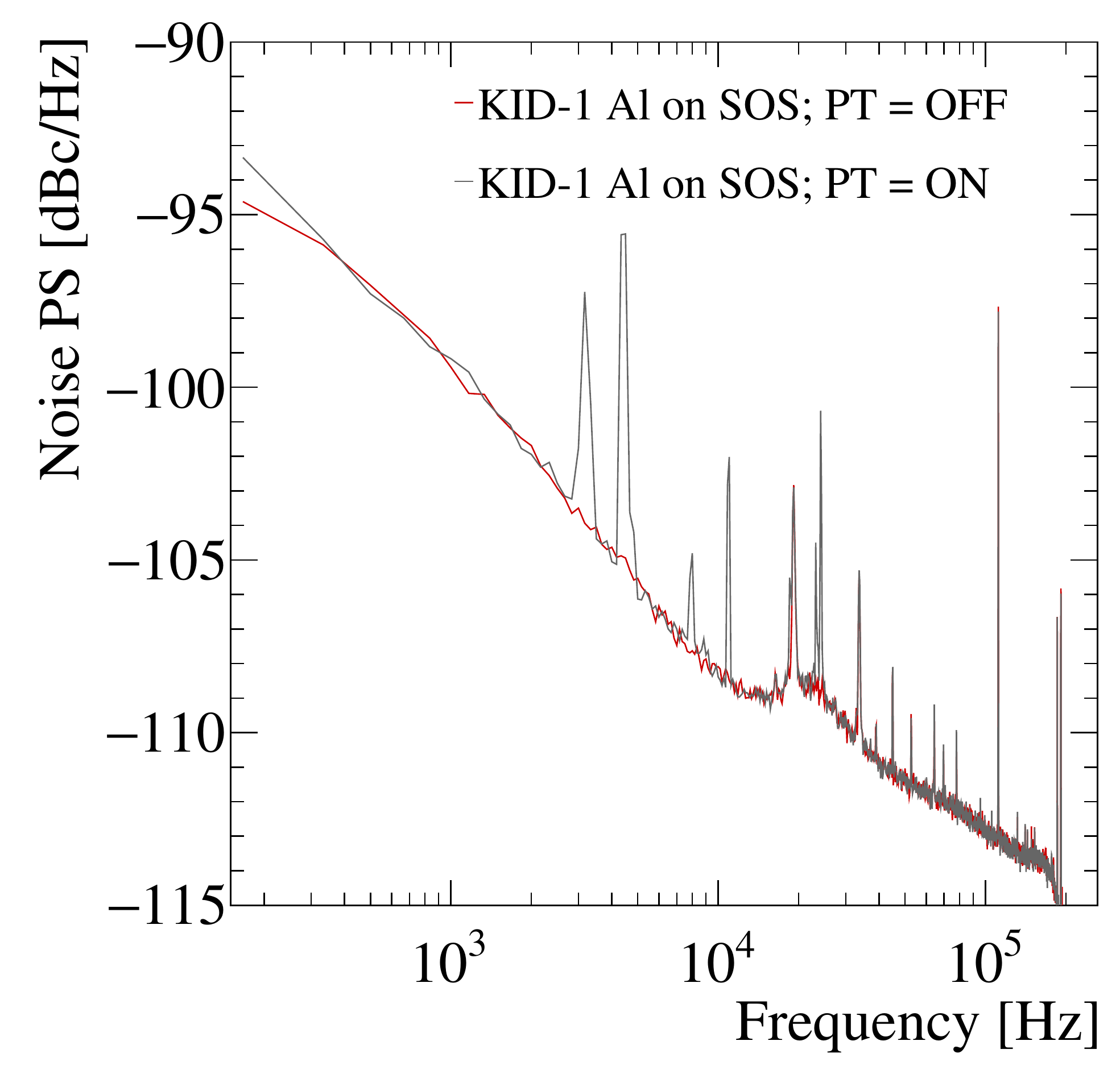}
    \caption{Noise power spectra measured for the same Al resonator evaporated on Si (left) and silicon on sapphire, i.e. SOS, (right) substrates with pulse tube off (red line) and on (black line). The noise increase at low frequency caused by the vibrations induced by the pulse tube is visible only for Si substrates.}
    \label{fig:KIDNoise}
\end{figure}

A very important feature of KIDs is their fast time response as shown in Fig.~\ref{fig:KIDpulse}. The typical rise and decay time are $\sim10^2$ times faster than NTD Ge thermistors, substantially increasing the capability to reject pile-up events. This would allow us to further suppress the background coming from the $2\nu\beta\beta$ pile-up events in $^{100}$Mo-enriched cryogenic calorimeters. 
Furthermore, exploiting sapphire substrates instead of silicon, KIDs results are barely affected by the mechanical vibrations induced by pulse tube refrigerator. This is shown in Fig.~\ref{fig:KIDNoise} where an increase of the detector noise at low frequency is visible when the pulse tube is operating only for Si substrates. Although this effect was observed in all the sapphire substrates measured, a clear explanation of this experimental evidence has not been yet formulated.
In summary, in addition to offering a sensitivity competitive with the NTDs, KIDs provide a stronger reduction of the background events coming from pile-up thanks to the faster time response and overcome the problems caused by the mechanical vibrations induced by pulse tube refrigerators that, in NTD- and TES-based detectors, could lead to a worsening of the baseline energy resolution.

\subsection{Light collection optimization}
In order to relax the constraints
on the light sensors, there is an active R\&D program for the improvement of the light collection efficiency by means of geometry \cite{Danevich2014}, reflective housing \cite{Casali2015}, and anti-reflective coatings of the light detectors \cite{Hansen2017, Mancuso2014}.  The R\&D also focuses on careful study of the optical characteristics of the crystals \cite{Casali2017}. In general, the bolometric technique poses a stringent constraint on the detector design since a bolometer can only be coupled by a weak thermal link and one cannot employ refractive index matching to guide the light to its detector. Hence, with the  interface between the cryostat vacuum and the crystal, light trapping due to total internal reflection can lead to a significant fraction of signal loss. Furthermore, an improvement of the light collection has to be weighed against drawbacks like the loss of coincidence information for short-range particles with a passive reflective foil between the crystals or the possibility of extra backgrounds from radioactive contamination in the additional passive material.

The light yield of large size  Li$_2$MoO$_4$ crystals (\diameter\ 44 mm  $\times$  h 40 mm) with 3M RMF VM2000/VM2002 or Enhanced Specular Reflector Films and a SiO antireflective coating on the Ge light detectors has been measured in several single crystal and few crystal prototype experiments \cite{Armengaud2017, Poda2017, Velazquez2017, Buse2018}. The light yields obtained for events around the Q-value of  $^{100}$Mo range from 0.7-1 keV/MeV, and are expected to be of similar magnitude for the CUPID baseline design of (\diameter\ 50 mm $\times$ h 50 mm) Li$_2$MoO$_4$ crystals. Preliminary results from twenty detectors (\diameter\ 44 mm  $\times$  h 45 mm, but one of \diameter\ 44 mm  $\times$  h 40 mm) currently operated in the CUIPD-Mo demonstrator at LSM confirm and exceed these previously achieved light yields.
As the light yield of ~1 keV/MeV is at a very comfortable level for particle discrimination, current R\&D is focused on evaluating if sufficient particle discrimination can be maintained while simplifying the detector design for CUPID. A cryogenic test with 45~mm cubic crystals with and without reflective foil is in preparation for mounting and underground operation at the LNGS. An additional effort to complement existing measurements of the optical properties of Li$_2$MoO$_4$  \cite{Bekker:2016, Barinova2014, Spassky2017}  with an absorption length measurement at cryogenic temperature is in preparation. Monte Carlo simulation codes to simulate the expected light yield of Li$_2$MoO$_4$ have been adapted from previous studies \cite{Caravaca2017a} and have already been used successfully for the simulation of the light yield in TeO$_2$. 

\subsection{Thermal and mechanical support}
The baseline energy resolutions achieved by massive low temperature calorimeters are a few orders of magnitude worse than theoretical predictions. It has been pointed out in previous work \cite{Pyle:2015arxiv} that this effect can occur when there is a mismatch between the signal and sensor bandwidths. If the coupling between the photon absorber and the thermal bath is larger than the thermal coupling between the absorber and the sensor, this can lead to sub-optimal collection of signal phonons in the sensor. This, in turn, leads to degradation of the energy resolution.

An ideal weak link should have an extremely low heat capacity and thermal conductance.  While polytetrafluoroethene (PTFE) has been extensively and reliably used over the last decade as the weak thermal link between the absorber and the thermal bath, it does add a non-negligible parasitic heat capacity to the device, affecting its performance. We are currently working towards developing suitable alternatives that can offer better bandwidth matching for next generation detectors. A single crystal sapphire (Al$_2$O$_3$) seems to be the most ideal due to its extremely low heat capacity.  One can also utilize the thermal boundary resistance between sapphire and copper (thermal bath) to further reduce the thermal coupling between the absorber and thermal bath. On the other hand, increasing the thermal coupling between sensor and the absorber using suitable number of gold wire bonds can also help tune the signal bandwidth. 

However, unlike PTFE, sapphire is not easily machinable. We are looking at the feasibility of using sapphire as a support structure for the light detectors. Further testing to optimize the mounting arrangement and device characterization is underway.

\newpage
\section{Material selection and treatment}

Materials prescreening for radiopurity and strict effective handling
and storage protocols are essential to achieve the near zero background goal of \cupid.
In this section we describe the treatment and handling of passive materials
after selection for use in the experiment, the storage protocols to be followed for all parts,
and the radioassay techniques and facilities currently available to the collaboration.

\subsubsection{Surface cleaning}
\label{sec:cleaning}

The aggressive cleaning procedures applied to the \cuore\ copper parts
have proven effective in reducing the degraded \al\ background
to the target level~\cite{ALESSANDRIA201313,Alfonso:2015wka,Alduino:2017ehq}.
At this stage of the \cupid\ conceptual design, we plan to maintain
the same aggressive surface cleaning procedure to achieve the lowest attainable \al\ background
and minimize potential surface \be\ background. 

The copper surface cleaning procedures for \cuore\ were developed and executed
at the Laboratori Nazionali di Legnaro of INFN, Italy.
We implemented three levels of cleaning, which we designate as
chemical low, chemical high, and standard \cuore\ protocol.
The applied protocol depends on the distance of the considered copper surface
to the bolometer and can be summarized as follows.

\begin{figure}
  \centering
  \includegraphics[height=0.1635\textheight]{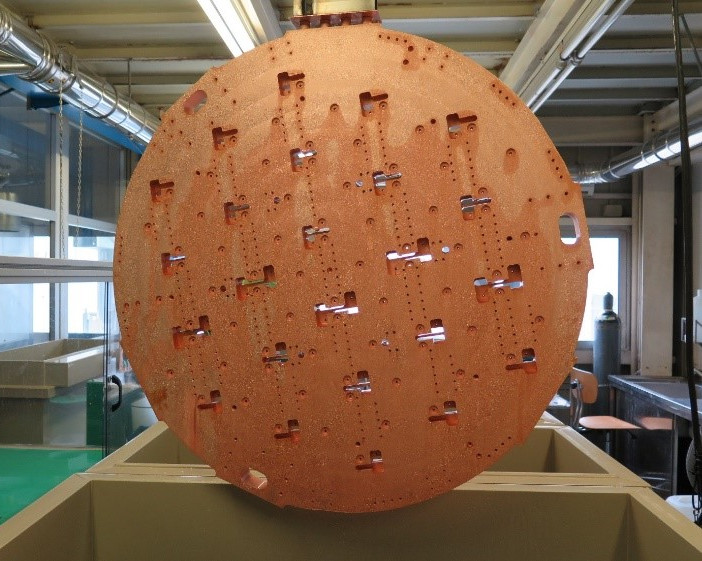}\quad
  \includegraphics[height=0.1635\textheight]{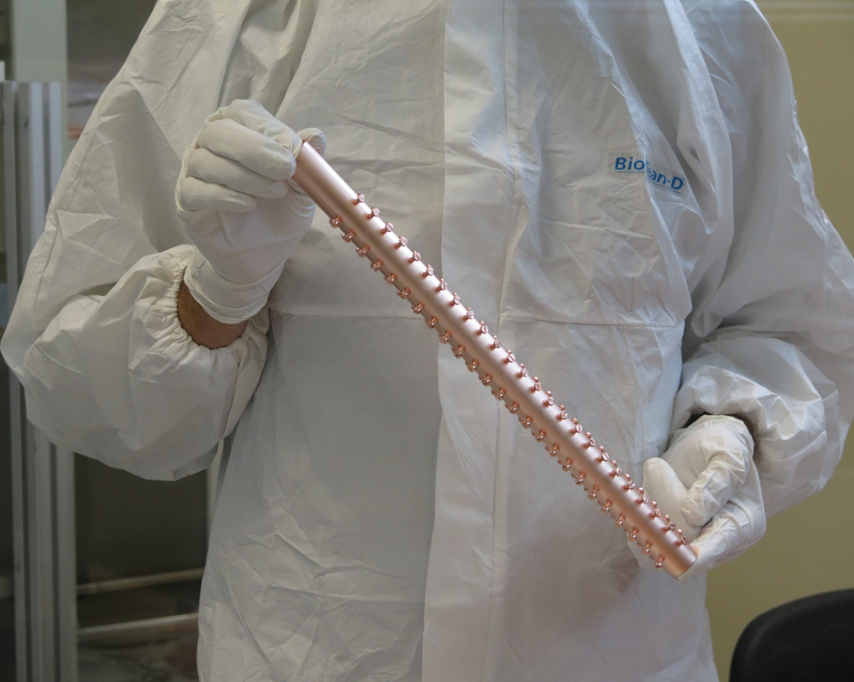}\quad
  \includegraphics[height=0.1635\textheight]{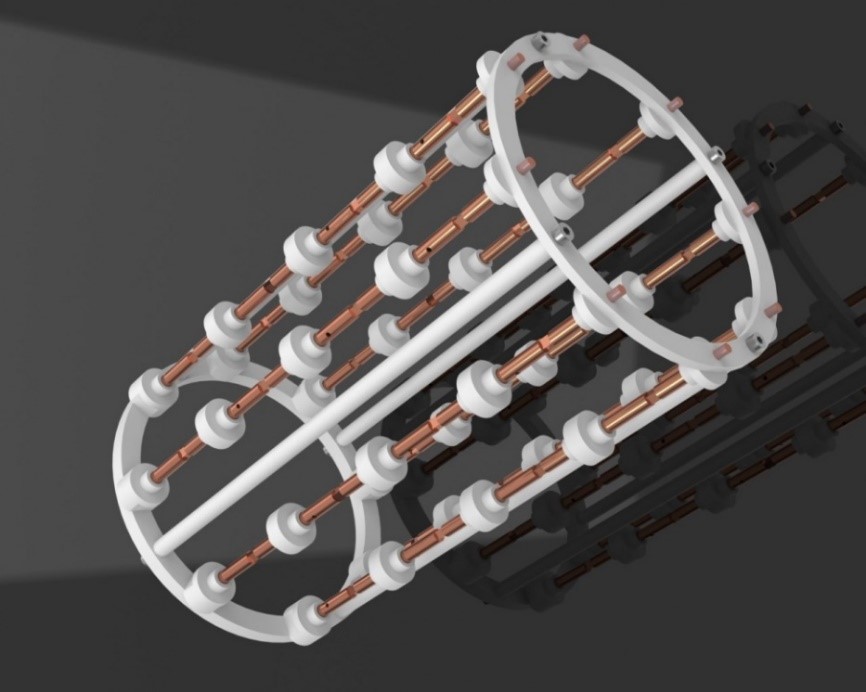}
  \caption{Left -- \cuore\ TSP plate cleaned with the chemical high protocol.
    Middle -- \cuore\ tiles screws cleaned with standard cleaning protocol after plasma cleaning.
    Right -- rendering of the sample holder used for the \cuore\ columns electro polishing,
    with the PTFE protection.}
  \label{fig:copper_parts}
\end{figure}

\begin{itemize}
\item The chemical low protocol consists of precleaning plus cold chemical etching;
\item The chemical high protocol consists of precleaning plus cold chemical etching
  followed by hot chemical etching (Fig.~\ref{fig:copper_parts});
\item The standard \cuore\ protocol consists of pre-cleaning, tumbling,
  electropolishing, and plasma cleaning. It is the highest standard of cleaning (Fig.~\ref{fig:copper_parts}). 
\end{itemize} 
 
The steps of the standard \cuore\ protocol were adapted from procedures for resonant
cavity production~\cite{doi:10.1116/1.571983,doi:10.1116/1.576420}.
The parts were extracted from the final cleaning stage in a cleanroom ISO 6
and packaged to prevent recontamination of the treated surfaces.
Ultra clean reagents were not employed since the expense of supplying
the large volumes needed for \cuore\ cryostat components was deemed unsustainable.
The procedure was designed for zero-deposition of contaminant material on the treated pieces,
therefore special attention was given to the manipulation of all the components
in order to prevent re-contamination.
The chemical plant employed for cleaning  more than 7700 \cuore\ copper components  remains in place
and can be reactivated for \cupid\ with only a partial refurbishing.
The different dimensions and structure of the \cupid\ copper components
involve a new areas of research and development of the PTFE characteristics used during cleaning (Fig.~\ref{fig:copper_parts}).  For example, care must be taken to guarantee the pieces' tolerances and the established erosion rates during the electro-chemical treatment.
We envision two possible methods for PTFE cleaning:
\begin{itemize}
\item chemical cleaning followed by an atmospheric plasma
  cleaning immediately before the installation in cleanroom;
\item chemical cleaning followed by an RF plasma cleaning in vacuum at Laboratori Nazionali di Legnaro.
 \end{itemize} 
To ensure adequate quality control, the \cupid\ cleaning protocol
will be supported by Inductively Coupled Plasma Mass Spectrometry (ICP-MS)
to obtain an immediate response on the process quality and efficacy. 

\subsubsection{Material handling and critical storage}

The storage facilities for the experiment must provide:
(i) a reduced cosmic ray flux to minimize activation of long-lived isotopes, such as \Co;
(ii) a clean, radon-free environment to prevent surface recontamination of critically cleaned parts;
(iii) a chemically compatible environment, e.g. a low-humidity environment for hygroscopic parts.
A parts database will be maintained as was done for \cuore\ to track the history of each component.

Storage facilities were available for \cuore\ at the LNGS
and at a shallower depth at Baradello, Como, in Italy.
We plan to reuse these for \cupid.
Underground storage will be reserved for raw materials, in particular copper, tellurium, and molybdenum.
The material will be transported to above ground facilities
for machining, crystal growth, and cleaning according to the need,
but particular attention will be paid to minimizing the exposure to cosmic rays.
Where international transportation is required, surface shipping will be used. 

A dedicated Parts Storage Area (PSA) at the LNGS is needed for cleaned parts.
Such a facility already exists from \cuore, and consists of a hut supplied with standard lab air.
The hut has two rooms: an insertion/extraction work area with barcode scanner
and a workstation to interface to the parts storage database.
The user can scan the barcode of the incoming/outgoing part and note all relevant information.
The timestamp is automatically logged by the computer.
The second room consists of a bank of air-tight, nitrogen-flushed storage lockers
that provide a low-humidity environment  compatible with the needs of the slightly hygroscopic \lmo.
Crystal shipping will be performed using multi-bagging and silica-gel-sealing
to ensure adequate protection against humidity.

We developed and followed this protocol  for handling cleaned parts for \cuore.
\teo\ crystals and ultra-cleaned copper parts were stored in a triple layer
of vacuum-sealed polyethylene bags. Oil-free vacuum pumps were used
to prevent oil or grease from diffusing back into the bags.
The bags were radio-assayed to prevent contamination of the enclosed components,
leading us to choose polyethylene as  it forms a good radon barrier.
The triple protection provided redundant outer layers
that could be removed without exposing the enclosed part.
The external layer was removed after transporting the part
from the PSA to the anteroom of the cleanroom.
The second layer was removed when the part was transferred from the cleanroom
to the assembly glove box, where the part was to be used.
The final innermost layer was removed when the part was used in assembly.
A unique bar code identifying the part was fixed
to the outside of each layer of plastic so that its history could be tracked.   

The storage area was flushed with  boil-off nitrogen
from a large liquid nitrogen dewar in Hall C of the LNGS.
The level in this dewar was monitored by the vendor,
with redundant monitoring by the \cuore\ collaboration,
and refilled once the level reached a predefined minimum.
This system has worked continuosly and reliably for several years. 

For \cupid, we will reuse this storage infrastructure.
Additional nitrogen-flushed storage, conceptually similar to the existing lockers,
may be added to the cleanrooms for intermediate storage of partially assembled parts.

\subsection{Radiopurity assessment}\label{sec:radio-meas}

To contain the background counts at the levels required by the \cupid\ sensitivity goal,
the material selection will be a crucial effort of the construction phase.
In this respect, the collaboration will take advantage
of the intense radiopurity campaigns already established for \cuore\ and \cupido\ experiments.
The infrastructures owned by the collaborating institutions
will constitute the basis for the needed assay programs,
while new detecting systems will be developed
when the existing apparatuses do not reach the required sensitivity.
The measurement campaigns will be organized into four main areas:
\ga\ ray spectroscopy  to inspect contaminants in the bulk of the construction materials;
Neutron Activation Analysis (NAA) for high sensitivity measurement of trace elements in the samples;
ICP-MS to assess the concentration of long lived isotopes in the materials;
and surface measurements to verify cleaning protocols and eventual radon progenies implantation.
Details are given in the following sections.

\subsubsection{HPGe \ga\ ray spectroscopy}\label{sec:hpge}

Several High-Purity Germanium (HPGe) detectors for \ga\ ray spectroscopy
with different sensitivities are available for material screening at the LNGS, Milano, and LSM.
Many of them can be used to screen and certify the materials
that will be installed relatively far away from the \cupid\ bolometrics,
while only few of them have the required sensitivities to measure and select
the most critical materials to be mounted next to the crystals.
In particular, the detectors installed in the underground low-level activity laboratory
at the LNGS can achieve very high performances for ultra-low background measurements.
At the radioactivity laboratories of LMS and Milano-Bicocca,
very sensitive detectors for \ga\ rays measurements are also available.

At LSM, one coaxial HPGe from Mirion Technologies is shared with the Edelweiss dark matter experiment.
The integrated background between in the 40--3000~keV range
is 178~counts/day and the sensitivity is of the order of 1~mBq/kg.

At Milano Bicocca, two high efficiency n-type detectors with around 100\% relative efficiency
realized by ORTEC are available for sensitive measurements.
The experimental setup is designed to permit the use of the two detectors as a single spectrometer,
increasing the detection efficiency, as well as the use of the two detectors in coincidence,
allowing a high background rejection.
The sample volume can be as large as 600~cm$^3$ without efficiency loss.
The sensitivities of this spectrometer are in the range of 0.5~mBq/kg for the U and Th chains.

Additionally, one detector of BEGe type, with 50\% relative efficiency
and specifically configured for low energy measurements in low background conditions
is also available at Milano Bicocca.
Its sensitivity is in the range of few mBq/kg for the U and Th chains,
with sample volumes up to 250~cm$^3$.
The sensitivity performances can be greatly increased by coupling this detector
with NAA, as detailed later.

\subsubsection{Neutron activation analysis}\label{sec:naa}

The Milano Bicocca research group has considerable experience in NAA measurements.
NAA was applied in the past to various materials like plastics and metals.
An instrumental NAA allowed determination of the limits on the high-purity copper
used for \cuore\ at a level $<10^{-12}$~g/g for U and Th radioactive contaminants.
In contrast, the actual achieved sensitivities on the plastic or organic materials
are $<10^{-13}$~g/g for U and Th, and of 10$^{-14}$~g/g for $^{40}$K.
The irradiation facility uses the TRIGA Mark II research reactor located
in the LENA laboratory of the University of Pavia.
The reactor is sufficiently  close to the Milano Bicocca laboratory
where the spectrometric measurements are performed.
The Bicocca radioactivity laboratory is equipped with several HPGe detectors
specifically suited for such measurements.
In particular, there is one HPGe configured for high sensitivity measurements on small samples.  There are two very low thresholds detectors with high energy resolution
for the precise measurements of the low energy \ga's emitted by the activated U and Th nuclei.  Finally, there is one system with an HPGe detector coupled to a liquid scintillator detector that,
by means of a coincidence technique,
allows a high background rejection for measurements in which many activated interference nuclei are involved.

\subsubsection{ICP mass spectrometry}\label{sec:icpms}

\cupid\ has access to mass spectrometry laboratories located at the LNGS and Milano Bicocca
for the  identification of U and Th contamination.
Both labs are equipped with two magnetic sector mass spectrometers
(Element XR at Milano Bicocca and Element 2 at LNGS, both from Thermo Fisher)
with an ultimate sensitivity of $\sim10^{-15}$~g/g,
and two quadrupole spectrometers with a sensitivity of $\sim10^{-12}$~g/g.
Both are also equipped with all the facilities
(chemical labs, distillation, and purification machines, etc)
necessary to guarantee these sensitivities.

\subsubsection{Surface measurements}\label{sec:surf}

Surface contamination on materials close to the detectors could be extremely dangerous.
To certify cleaning and handling protocols of such materials,
different measurement approaches are available, such as the 
BiPo-3 detector~\cite{Barabash:2017hst} and a few optimized large area silicon barrier detectors.

\begin{figure}[!htbp]
    \centering
    \includegraphics[scale=0.7]{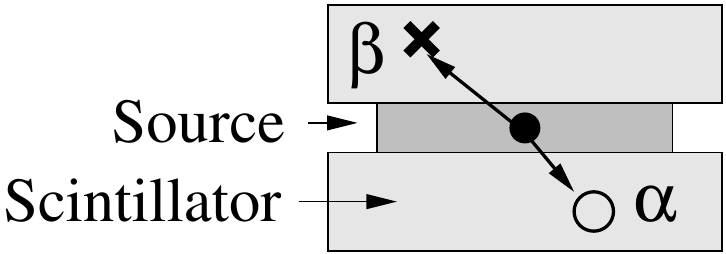}
     \hspace{1.5cm}
    \includegraphics[scale=0.5]{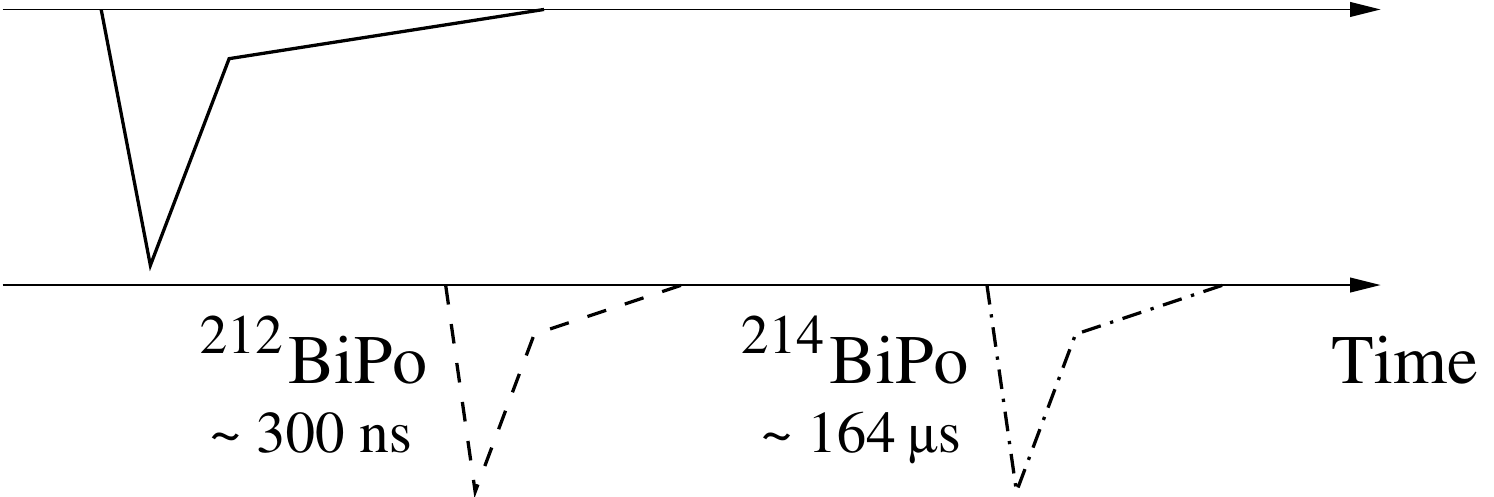}
    \caption{Schematic view of the BiPo detection principle with plastic scintillators
      and the time signals seen with PMTs for a BiPo event.
      Left -- the dot represents the contamination, while the cross and open circle
      represent energy depositions in the scintillators by the prompt beta and the delayed signal, respectively.
      Right -- the prompt \be\ and the delayed \al\ signals observed by the two scintillators
      surrounding the thin source.}
    \label{fig:BiPo-principle}
\end{figure}

Measurements of thin materials will be performed with the BiPo-3 detector,
initially developed in the framework of the SuperNEMO experiment
to evaluate the radiopurity of the SuperNEMO sources ~\cite{BiPo-JINST}.
BiPo-3, running since 2013 in the Canfranc Underground Laboratory in Spain,
can reach sensitivities of the order of $10\upmu$Bq/kg for \Tl, and $100\upmu$Bq/kg for $^{214}$Bi.
The underlying concept of the BiPo-3 detector is to observe the $^{214}$Bi-$^{214}$Po
and $^{212}$Bi-$^{212}$Po cascades, which emit an electron followed by a delayed \al\ particle.
The material under investigation is sandwiched between two low-radioactivity
thin polystyrene scintillators, which provide a very clear time-topology signature
(see Fig.~\ref{fig:BiPo-principle}).

Surface barrier detectors at Milano Bicocca,
specifically selected for low background measurement during the \cuore\ construction,
will be available for the certification of the surface radiopurity.
The sensitivities of those detectors are of the order of $10^{-7}$~Bq/cm$^2$
for the \al\-emitting isotopes contributing to the lower part of the U and Th radioactive chains,
and of the order of $10^{-6}$~Bq/cm$^2$ for the upper part of the same chains.
Dedicated R\&D developments are under study in order to further increase the achieved specific sensitivities.



\newpage
\section{Background sources and suppression}\label{sec:bkg}

\subsection{Background sources and their mitigation}\label{sec:bkg-sources}

\begin{figure}[h!]
  \centering
  \includegraphics[width=0.48\textwidth]{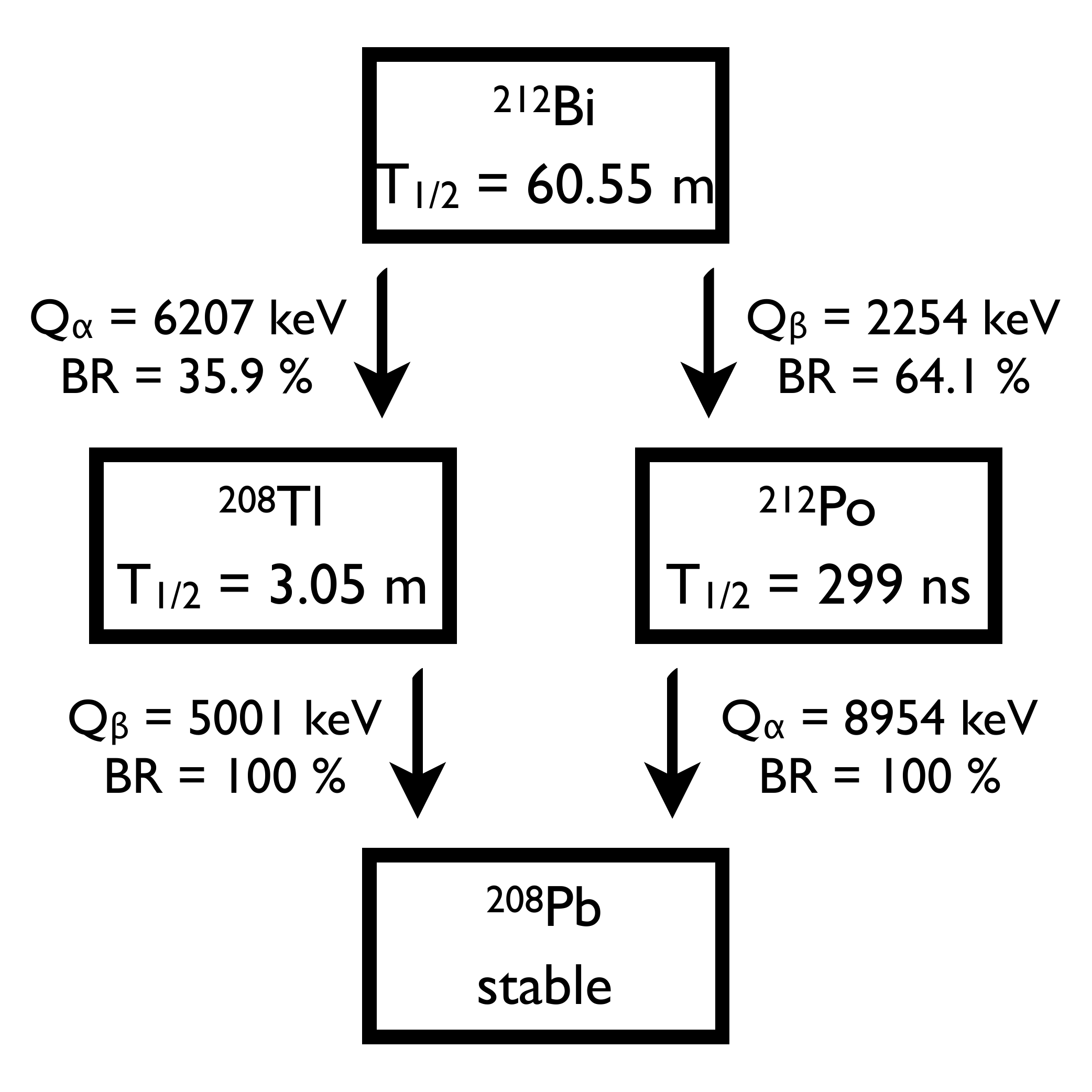}
  \includegraphics[width=0.48\textwidth]{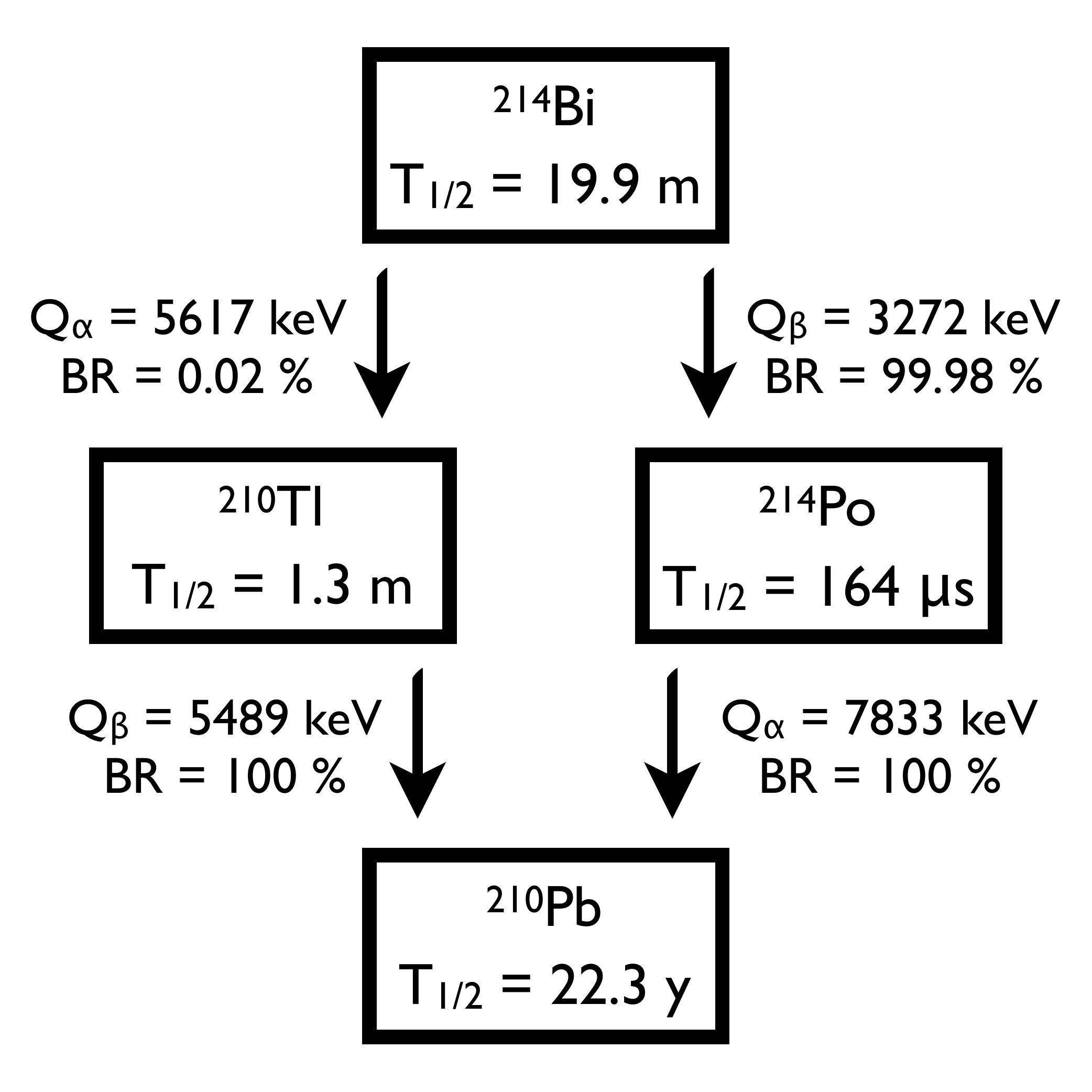}
  \caption{Decay scheme of $^{208}$Tl (left) and $^{214}$Bi (right).  Both of these isotopes decay with the emission of high energy $\beta$s. }\label{fig:DelayedCoincidences}
\end{figure}

The \onbb\ decay signature in CUPID will be a monochromatic line centered at \Qbb=3035~keV.
We  group sources that may produce events in the same energy region, hiding the expected peak, in the following categories:
\begin{itemize}
\item \textbf{External sources}: mainly environmental $\mu$s (that can't be shielded)
  and to a minor extent $\gamma$s and neutrons (that are already efficiently shielded in the CUORE set-up);
\item \textbf{Far sources}: radioactive contaminants in the cryostat
  (that is a system of nested copper vessel acting as thermal shields)
  or in the radiation shields installed inside (made of lead) and outside
  (made of lead and of polyethylene) the cryostat. In the following we will refer to these elements as the ``shields.''
  Only $\gamma$s emitted by the shields can reach the bolometers. The condition
  to mimic a \onbb\ signal is that the sum energies of the $\gamma$s produced in the decay
  are larger than \Qbb.  This takes into account the possibility of the simultaneous
  detection in a bolometer of two or more $\gamma$s emitted in the same cascade.
  In the $^{238}$U and $^{232}$Th chains only two isotopes satisfy this condition:
  $^{214}$Bi that has few $\gamma$s with energy larger than \Qbb\
  and $^{208}$Tl that contributes only through the cascade 583+2615 keV;
\item \textbf{Near sources}: radioactive isotopes with Q$>$\Qbb\ contaminating the ``detector''
  (comprising the \lmo\ crystals, the light detectors, the light reflecting foils, the copper holder,
  the PTFE stands, and the 10 mK shield containing the detector array).
  All the particles emitted by these radionuclei can reach the bolometers,
  even the short range $\alpha$s produced by $^{238}$U and $^{232}$Th
  and their daughters or the $\beta$s produced by $^{214}$Bi and $^{208}$Tl decays (Fig.~\ref{fig:DelayedCoincidences}). 
\end{itemize}


The CUPID detector will offer several passive and active technologies to reduce the background in the region of interest.
Passive techniques involve the use of copper and lead shields. It also includes the selection, cleaning, and storage of all the materials involved in the detector construction.

Active techniques, successfully exploited in CUORE (Sec.~\ref{sec:cuore bkg model}) and CUPID-0 (Sec.~\ref{sec:cupid0 bkg model}), involves the capability of operating the detector in anti-coincidence, the implementation of a time-veto, and the use of particle identification. Additionally, we anticipate the installation of plastic scintillators to act as a $\mu$ veto.

Operating the array in anti-coincidence allows us to exploit the high granularity of the device to reject multi-site events, such as $\mu$s either crossing the detector or inducing showers in the shields or, again, multi-Compton interactions induced by $\gamma$ rays. 
This technique applies, with different efficiency, to any contaminant or environmental flux, while the time-veto technique  described below can be used only in the case of $^{214}$Bi and $^{208}$Tl contaminants in the detector (since it requires one to be able to detect the $\alpha$ decay of the mother or the daughter).

The time-veto allows us to reduce the background produced by the $\beta/\gamma$ decays of $^{208}$Tl ($Q\sim$5\,MeV) by first detecting its mother, $^{212}$Bi. This isotope decays via $\alpha$ emission to $^{208}$Tl, which subsequently decays with a short half-life of 3.05 minutes (Fig.~\ref{fig:DelayedCoincidences}). As a consequence, the $\beta/\gamma$ interactions produced by $^{208}$Tl decay can be efficiently suppressed by opening a time-veto of few minutes after the detection of an $\alpha$ particle \cite{Azzolini:2019nmi}.
Another example of application of a time-veto is for $^{214}$Bi. This isotope decays with a branching ratio of 99.98$\%$ to $^{214}$Po, which subsequently $\alpha$-decays with a short half-life of 163\,$\mu$s (Fig.~\ref{fig:DelayedCoincidences}). Given the slow time-development of bolometric signals, these events sum up in a single pulse with energy well above the region of interest. Despite the low branching ratio of this decay mode (0.02$\%$), $^{214}$Bi can also decay via $\alpha$ emission to $^{210}$Tl, a $\beta$ emitter with a high Q-value of 5.4 MeV. In this case, it is possible to exploit the fast decay of $^{210}$Tl (1.3\,minutes) to reject its interactions with a time-veto.
This analysis technique has not been exploited in CUORE, as the background of this experiment is dominated by $\alpha$ interactions. In CUPID-0, on the contrary, the background is dominated by $\beta/\gamma$ interactions, as the $\alpha$ contribution can be rejected by exploiting the light yield. Thus, the implementation of the time-veto has allowed us to suppress the background by a factor 3.7 with a negligible dead-time increase. 

While the described techniques are available for all the arrays of cryogenic calorimeters, CUPID will offer an additional tool for particle identification, i.e. the simultaneous readout of heat and light. The suppression of the $\alpha$ background via particle identification was successfully demonstrated by CUPID-0, one of the demonstrators of the CUPID technology.
Furthermore, because the light signal is faster with respect to the heat signal, it will contribute to the abatement of pile-up in the region of interest (see Sec.~\ref{sec:bkg-2nu}).


\subsubsection{$\alpha/\beta$ discrimination}\label{sec:discrimination}

The discrimination capability offered by the simultaneous readout of heat and light depends on the light output of the crystal, the light collection efficiency, and the performance of the light detector. 

The CUPID experiment anticipates using cylindrical crystals surrounded by a reflecting foil and coupled to light detectors consisting of Ge disks (50\,mm diameter). Even if we do not dispose of data in the final geometrical configuration, we can rely on the results obtained from past R$\&$D activity with very similar devices: four cylindrical crystals with diameter of 44\,mm and height ranging from 40 to 46\,mm~\cite{Poda2017}. 

\begin{SCfigure}
  \centering
  \includegraphics[width=0.5\textwidth]{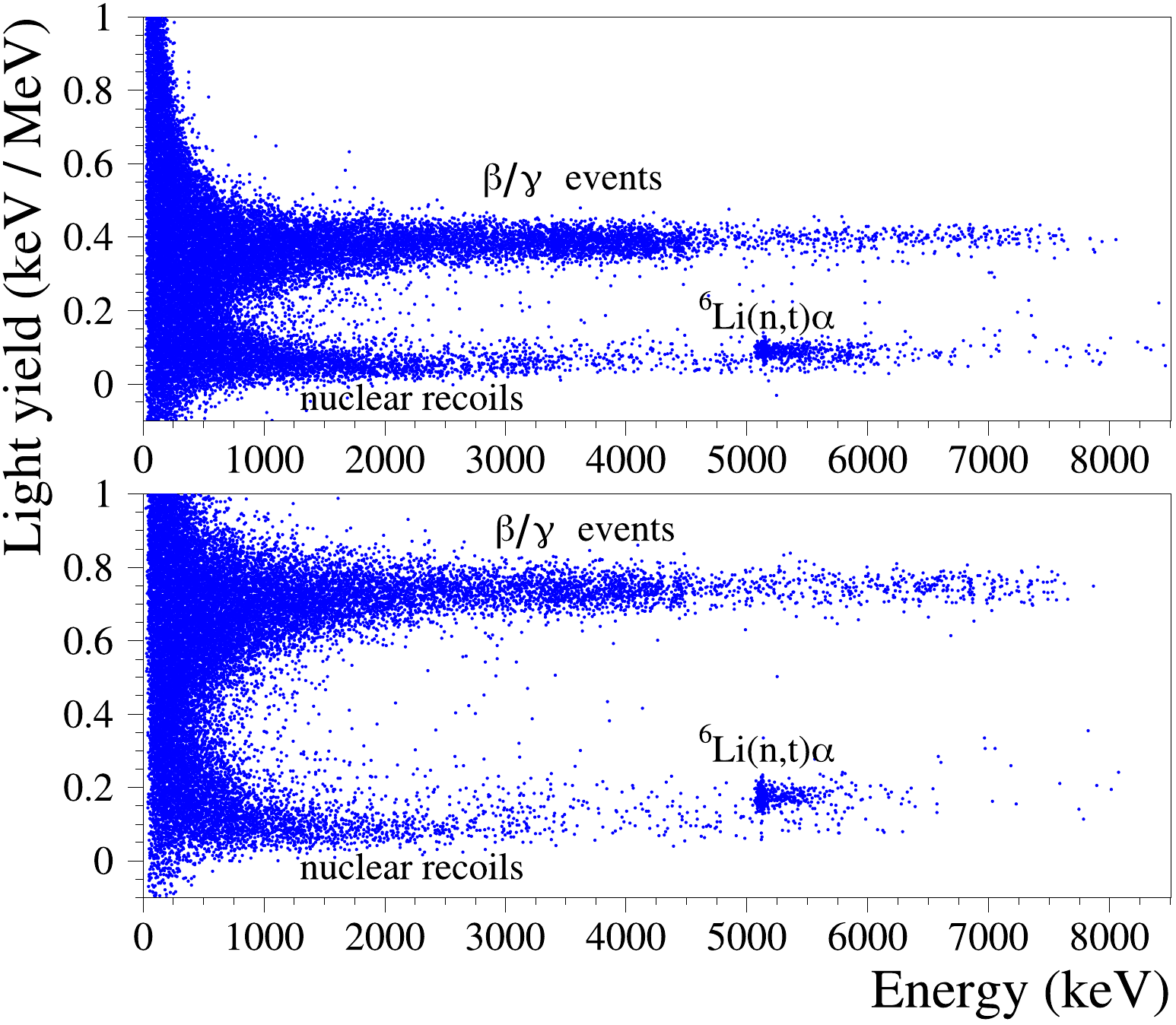}
  \caption{Light yield reported as a function of the energy released in Li$_2$MoO$_4$ crystals measured without (top) and with (bottom) reflecting foil \cite{Poda2017}. The events shown in the plot are produced by an Am:Be neutron source.}\label{fig:LightVsHeat}
\end{SCfigure}

The crystals were measured with the reflecting foil, obtaining a comparable light yield of 0.73-0.77 keV/MeV.  In a subsequent run the reflector was removed, resulting in a reduction of the light yield by a factor 2 (Fig.~\ref{fig:LightVsHeat}).

To understand the effect of this variation on the particle identification capability, it is useful to introduce the Discrimination Power ($DP$), defined as:
\begin{equation}
  DP = \frac{|\mu_{\beta/\gamma}-\mu_{\alpha}|} {\sqrt{\sigma^2_{\beta/\gamma} + \sigma^2_{\alpha}}}
\end{equation}
where $\mu$ and $\sigma$ are the average value and RMS of the $\beta/\gamma$ and $\alpha$ light distributions, respectively. Both these parameters depend on the energy and the DP in the ROI.

The requirement of a background induced by $\alpha$ particles lower than 10$^{-4}$\,\ckky\ is satisfied for $DP(E=Q_{\beta\beta})> 3.1$~\cite{Poda:2017c}. 

The test-runs with four crystals without reflecting foil featured $DP=9$, while adding the reflecting foil allowed us to reach $DP>12$.  As a consequence, in CUPID we expect the $\alpha$ background to be negligible in both cases.

The presence of the reflecting foil would be more conservative to guarantee a high discrimination capability: the CUPID light detectors have never been tested in the CUORE facility and in a pessimistic scenario they could perform more poorly than in other cryostats. Thus, the reflecting foil would ensure a higher light collection efficiency, compensating for potentially worse performance of the light detectors.

On the other hand, the reflecting foil would prevent the study of coincidences between adjacent crystals, which would help in background suppression. Moreover, according to the CUPID-0 background model, contaminations of the reflecting foils and holder surfaces contribute to the background in the region of interest for the double beta decay of $^{82}$Se at the level of 2.1$\pm0.3^{+2.2}_{-1.0}\times$10$^{-4}$ \ckky. This value would be similar in CUPID, pointing to an important source of background.

In view of CUPID, we are investigating possible alternatives for the reflecting foil in order to keep a high light collection while reducing the background produced by the reflector. 
The simplest solution consists of developing other reflecting materials, featuring contamination levels lower by about a factor 3 with respect to the one used by CUPID-0. Nevertheless, we are also considering another promising technology, in which the reflecting foil is replaced by a thin metallic coating of the crystal itself (for example, with tens of nm of ultra-pure aluminum). This technique optimizes the light collection while suppressing the background due to the presence of the reflecting foil. Furthermore, if the coating is thin enough, radioactive decays on the surface can interact in more than one crystal, enabling the study of coincidences between adjacent crystals and thus providing an additional tool for background suppression. The reproducibility and reliability of the \lmo\ coating will be tested in the next cryogenic runs in the LNGS.  The merits of this solution will then be evaluated and the baseline detector design will be updated accordingly

Finally, it is worth mentioning that the Li$_2$MoO$_4$ prototypes tested in view of CUPID showed hints of particle identification using the heat channel alone (without light detectors). This feature has not yet been deeply investigated so it is not used in the background calculation presented in this document, but it could provide an additional tool for the suppression of $\alpha$ interactions.

\subsubsection{Rejection of 2$\nu$ background}\label{sec:bkg-2nu}

The generally slow response time of bolometers can cause accidental pile-up of \nnbb\ events and/or background events to contribute to the rate in the ROI at a detectable level, thus limiting the sensitivity of an experiment. This is especially true for Mo-based bolometers, given the fast \nnbb\ decay time of $7.1 \times 10^{18}$yr for molybdenum. Following plausible estimates \cite{Artusa:2014,Chernyak2012} we get for a 300g \lmo\ crystal with 100\% enrichment a \nnbb\ pile-up rate of $3.5\times 10^{-4}$ \ckky\ $\times \Delta t /[1\rm{ms}]$, where $\Delta t$ is the minimum resolving time of the heat channel. For ZnMoO$_4$ crystals \cite{Chernyak2014}, this background can be reduced to $\sim 10^{-4}$ \ckky\ by using a high sampling rate of the data acquisition and by developing algorithms to find the origin of the signal with high accuracy.
The discrimination capability is not only a function of $\Delta t$ but also of the relative amplitudes among the two piled-up pulses and the S/N ratio of each pulse. Further improvements could be obtained by exploiting the faster time response of light detectors and the Neganov-Luke amplification that increases the S/N ratio.

The pile-up rejection efficiency is estimated using simulated light events and results about 86\% for standard light detectors and 98\% for Neganov-Luke ones \cite{Chernyak2016}. Discrimination on the Neganov-Luke light channel can bring the \nnbb\ background to a level of $6 \times 10^{-5}$ \ckky. \\

The pile-up among two radioactive background pulses or among a \nnbb\ decay and a radioactive decay can also contribute to the ROI, with an amount depending on the radioactive background rate. Assuming this to be 0.1 mHz, the resulting pile-up background in the ROI from radioactive sources and radioactive sources plus \nnbb\ is negligible~\cite{Artusa:2014}. Regarding cosmogenic isotopes, the global activation rates after 90 days of exposure at sea level plus one year of cool down, are below 0.1 nuclei/(kg day) (see sec.~\ref{sec:bkg-cosmo}). This cannot give rise to significant pile-up.

\subsection{Background in CUORE and CUPID-0}

Both the CUORE and CUPID-0 experiments have developed a background model (BM) able to describe their measured spectra in terms of 1) radioactive contamination coming from different parts of the detector and the experimental set-up and 2) the contribution of environmental muons (neutron and gamma contributions are negligible).

\begin{SCfigure}
  \begin{subfigure}[t]{.7\textwidth}
    \includegraphics[width=\textwidth]{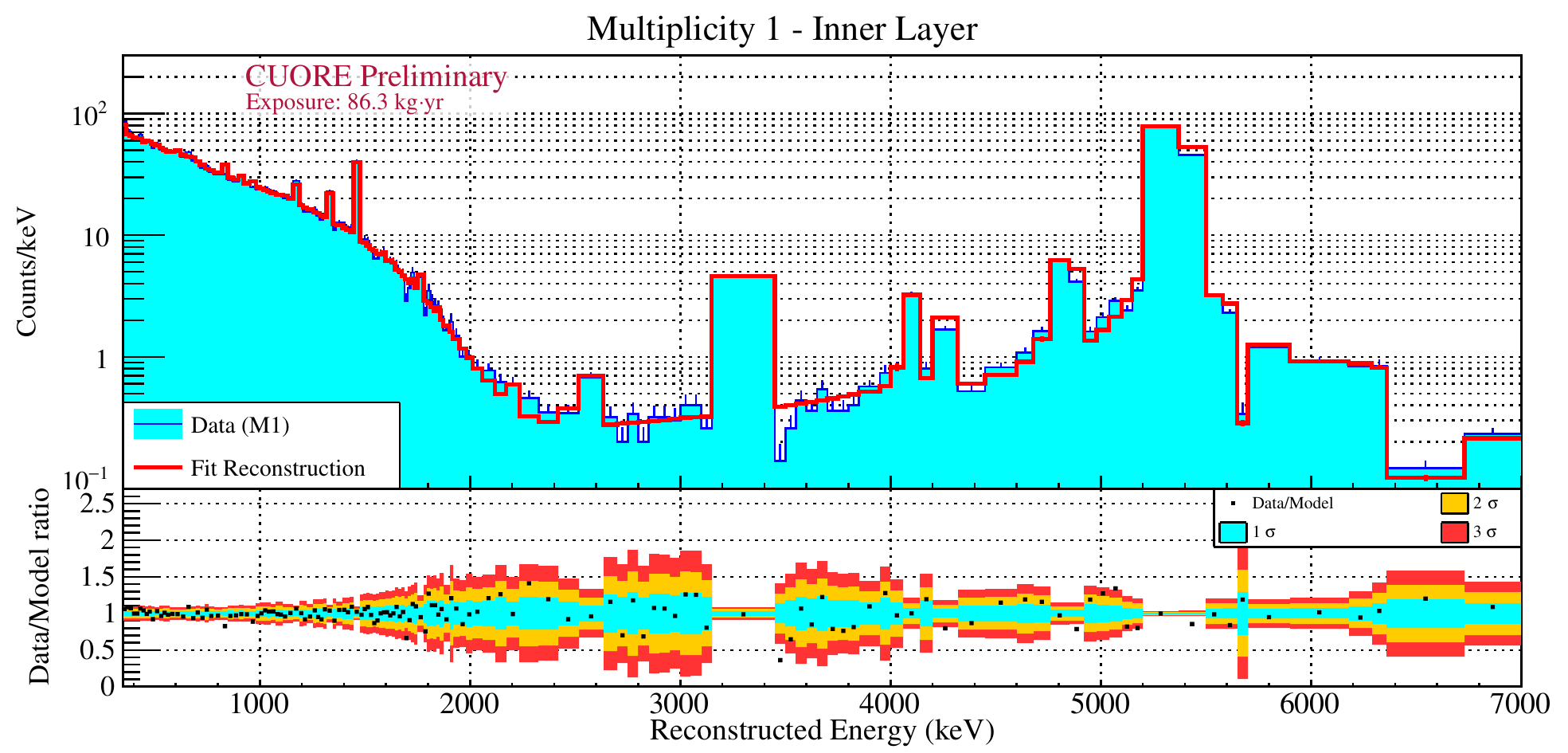}\\
    \includegraphics[width=\textwidth]{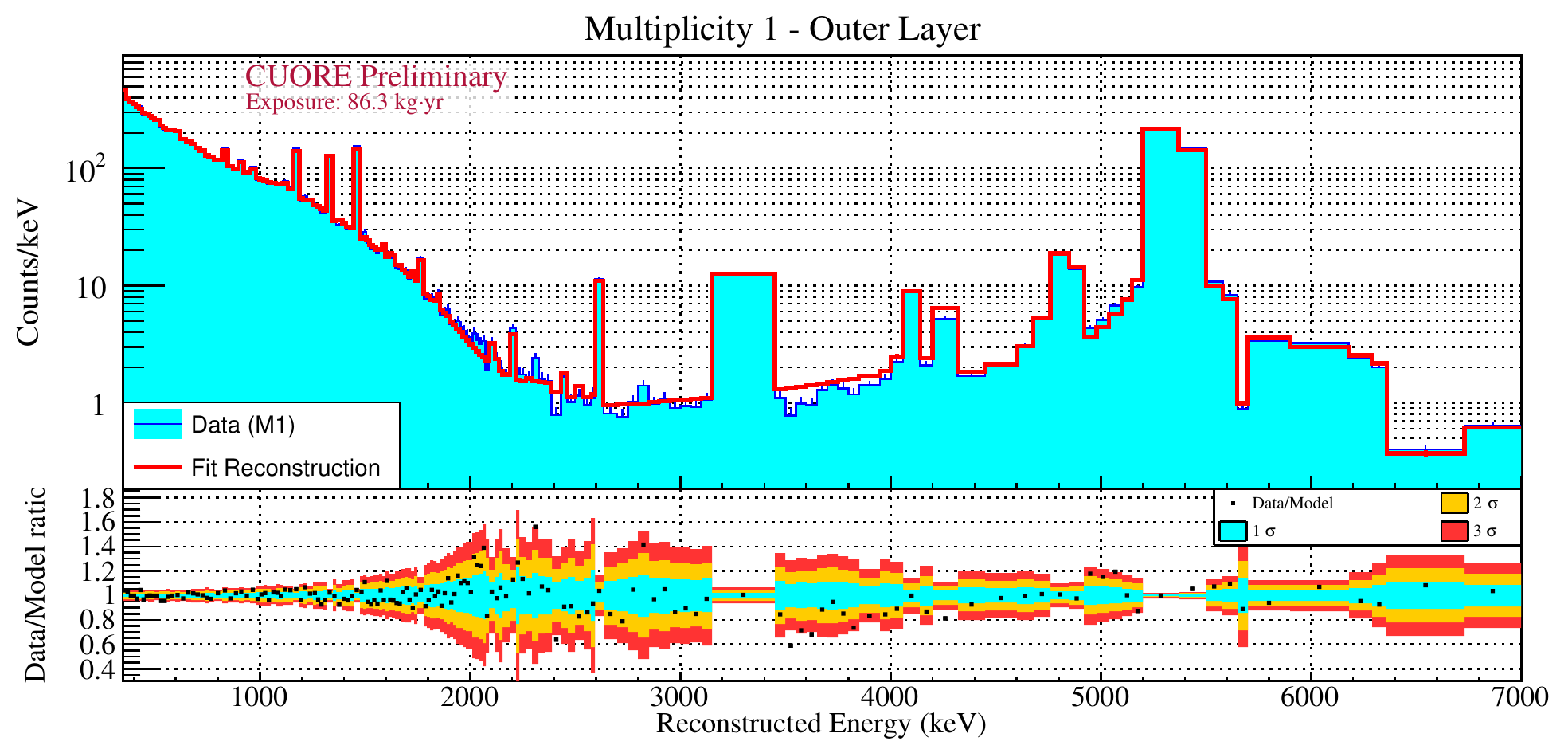}
  \end{subfigure}
  \caption{CUORE background reconstruction. The detectors are divided in two groups:
    detectors in the very core of the array (top panel) and detectors in an outer layer
    (bottom panel) with a ``thickness'' of about two crystals.}
  \label{fig:CUORE_BM_fit}
\end{SCfigure}

\begin{SCfigure}
  \begin{subfigure}[t]{.7\textwidth}
    \includegraphics[width=\textwidth]{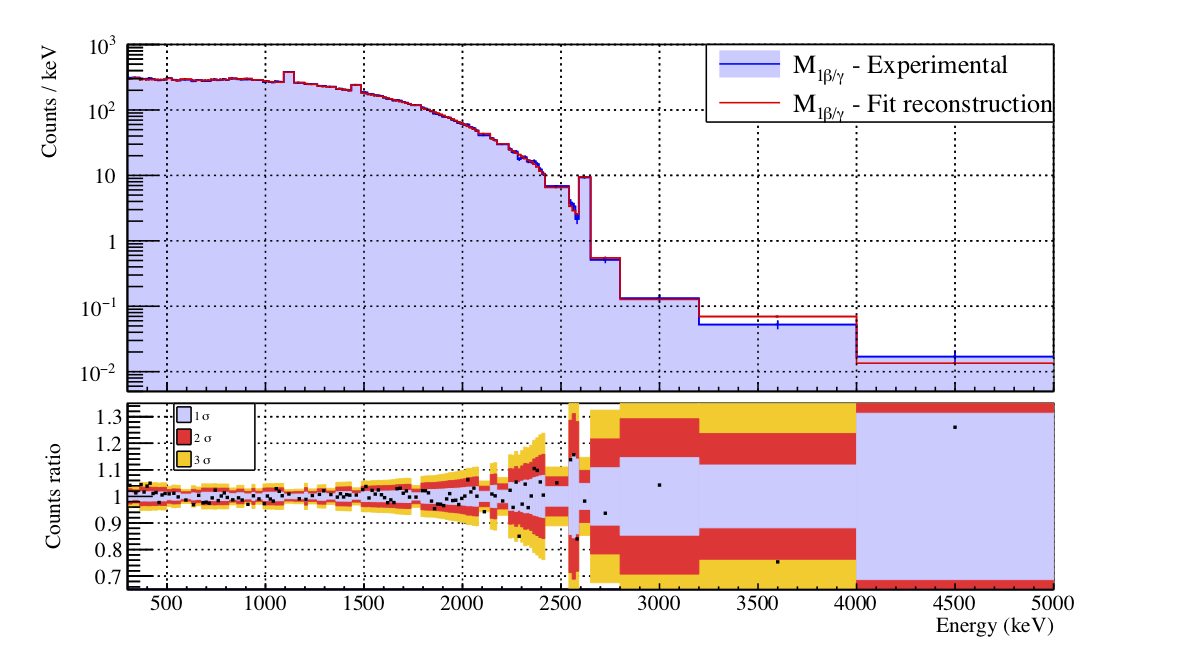}
    \includegraphics[width=\textwidth]{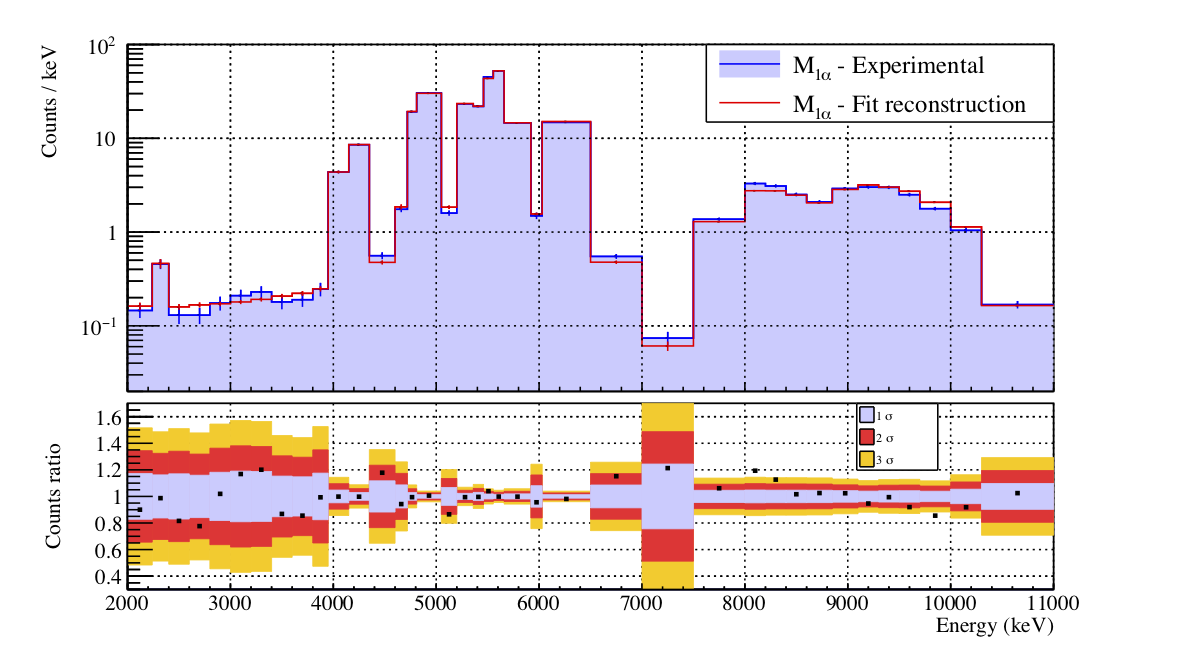}
  \end{subfigure}
  \caption{CUPID-0 background reconstruction. In the top panel, the spectrum of $\beta$/\/$\gamma$ events.  In the bottom panel, the spectrum of $\alpha$ events \cite{Azzolini:2019nmi}.}
  \label{fig:CUPID-0_BM_fit}
\end{SCfigure}

The model follows the procedure first developed for CUORE-0~\cite{Q0_2nu}. A detailed description of the experimental set-up is the basis of a Monte Carlo simulation that adopts GEANT4 for particle interaction and propagation.  Custom code is used to reproduce the detector, DAQ, and analysis features. Many simulations, describing all the possible (distinguishable) background sources, are generated with this code. They include surface contamination of the materials near the bolometers, bulk contamination of materials both near and far from the bolometers, and cosmogenic muons. 
A linear combination of these simulations is fitted to the data, having as free parameters the activity of each source. The quality of data reconstruction is illustrated in Fig.~\ref{fig:CUORE_BM_fit} for CUORE and Fig.~\ref{fig:CUPID-0_BM_fit} for CUPID.

Although still a work-in-progress for both experiments, the results obtained so far provide useful information for the construction of the CUPID background budget. It proves that there is a high degree of understanding of background sources. This is particularly important in the case of CUORE since we plan to host the CUPID detector in the same infrastructure. 

\subsubsection{CUORE background model} \label{sec:bkg-BMCuore}
\label{sec:cuore bkg model}

\begin{figure}[h!]
  \centering
  \includegraphics[height=.2145\textheight]{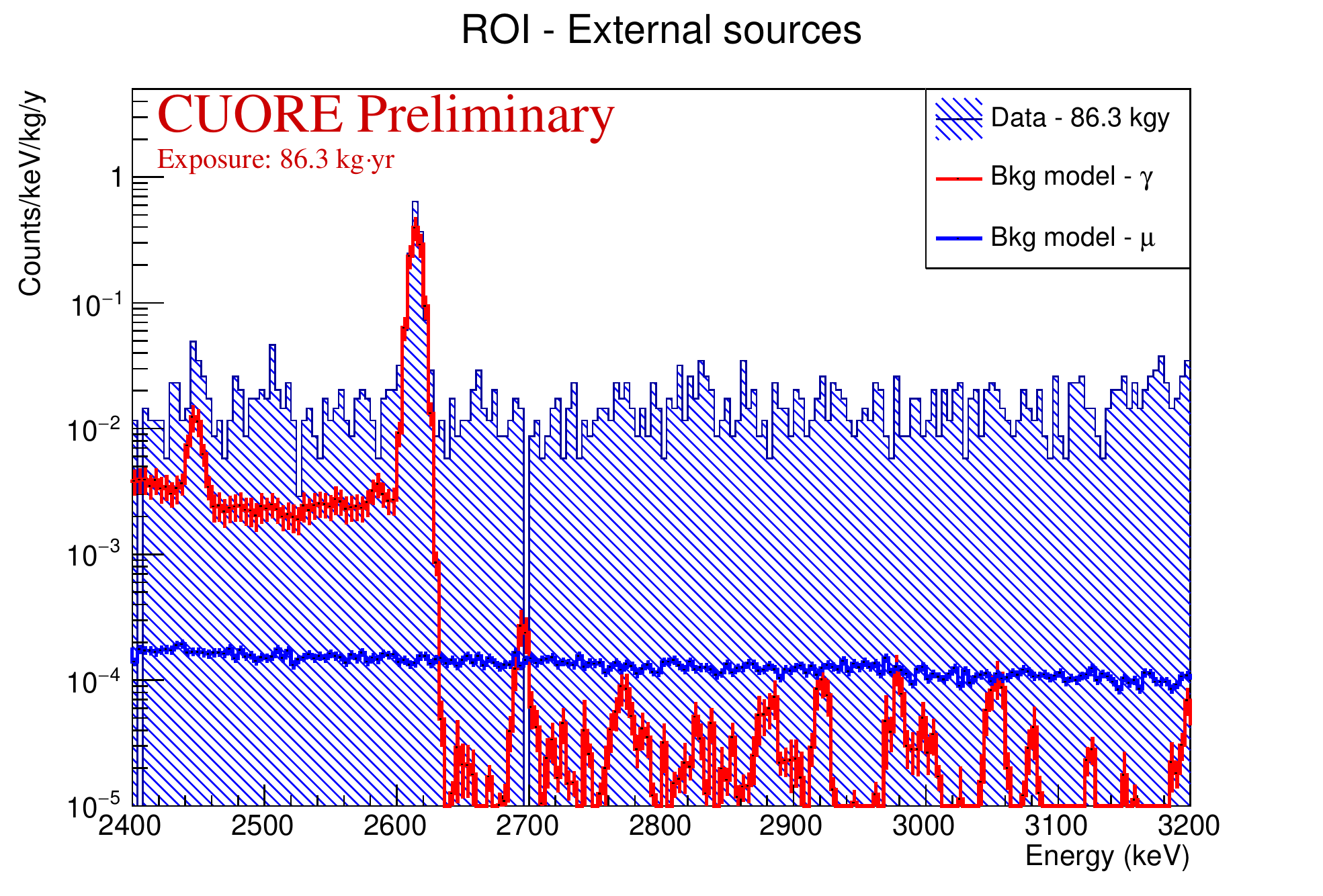}\quad
  \includegraphics[height=.2145\textheight]{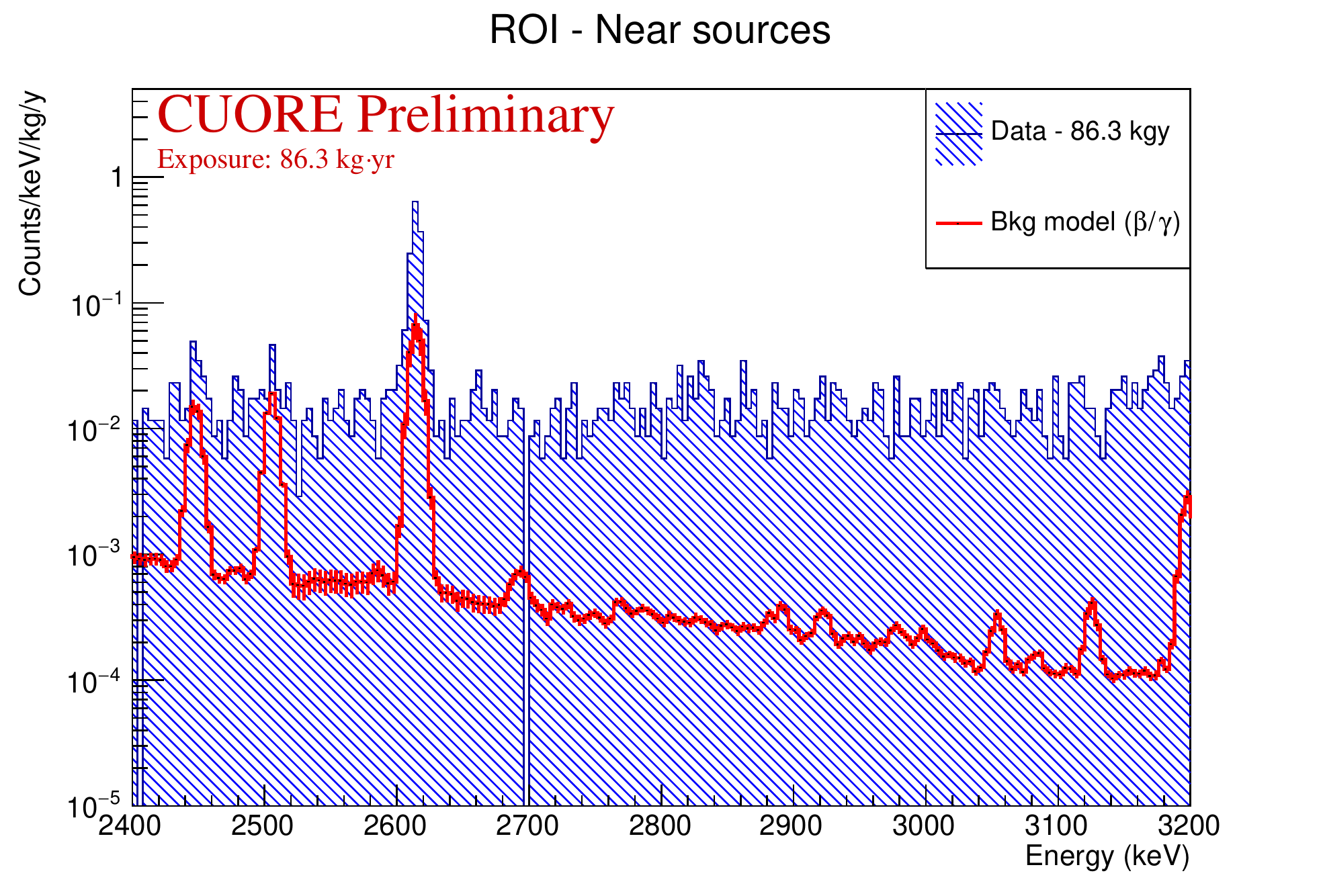}
  \caption{CUORE experimental data (filled histogram) and BM reconstructed contributions. Left -- the cryostat/shields (red) and cosmogenic muons (blue) contribute only through their $\gamma$ emission. Right -- the reconstructed  $\beta/\gamma$ contribution of the detector, mainly due to the copper infrastructure.}
  \label{fig:CUORE_BM_contributions}
\end{figure}

Here we summarize the CUORE BM results that are particularly relevant for CUPID. 
The left panel of Fig.~\ref{fig:CUORE_BM_contributions} shows the summed contributions of the cryostat thermal shields with those of the radiation shields. This is a pure $\gamma$ contribution since none of the shields directly faces the detector. CUPID will use this same infrastructure and the background counting rate will scale mainly with the detection and anti-coincidence cut efficiencies. The same is valid for the $\mu$ (induced) background.

\begin{figure}[h!]
  \centering
  \includegraphics[height=.224\textheight]{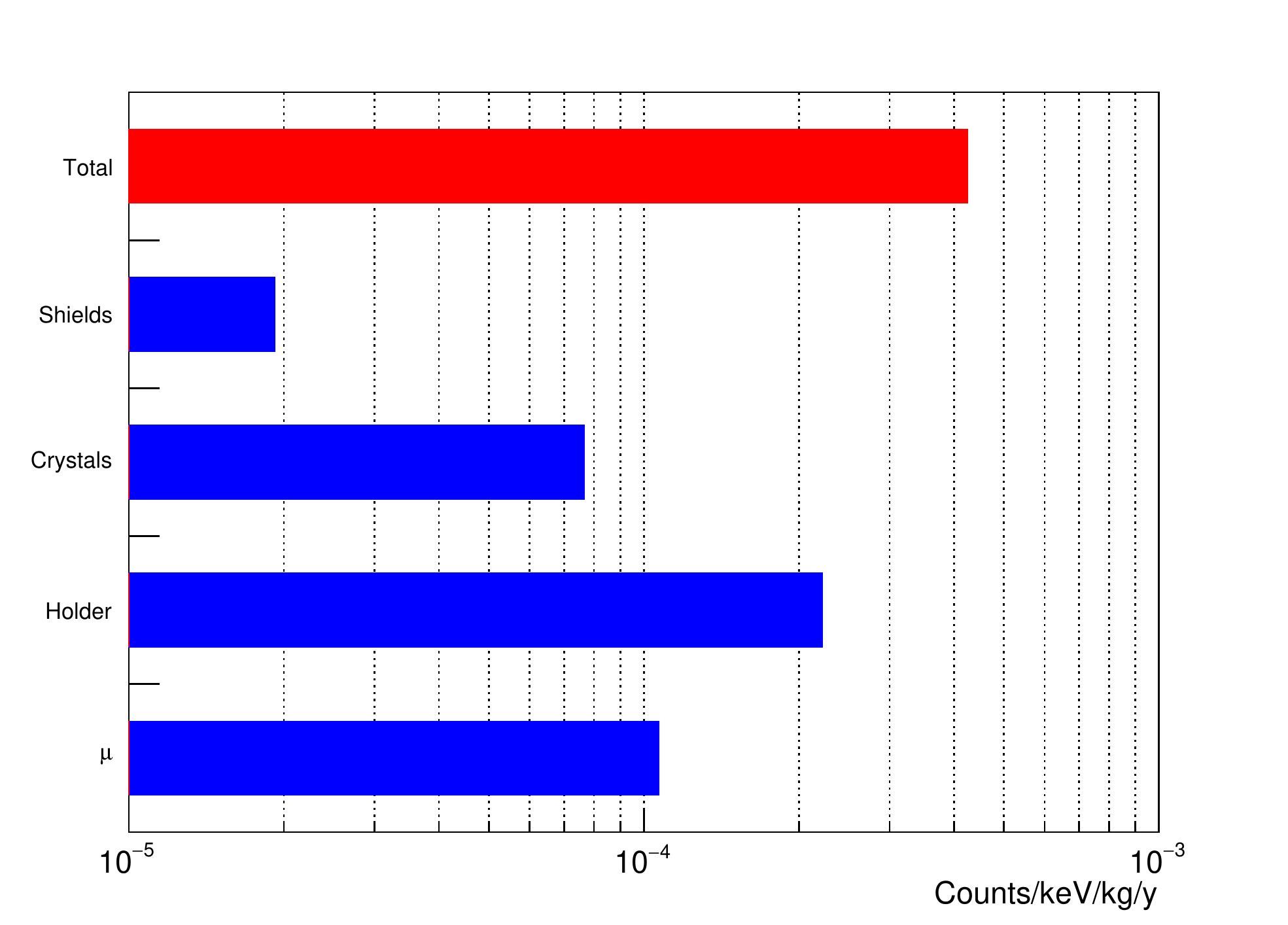}\quad
  \includegraphics[height=.224\textheight]{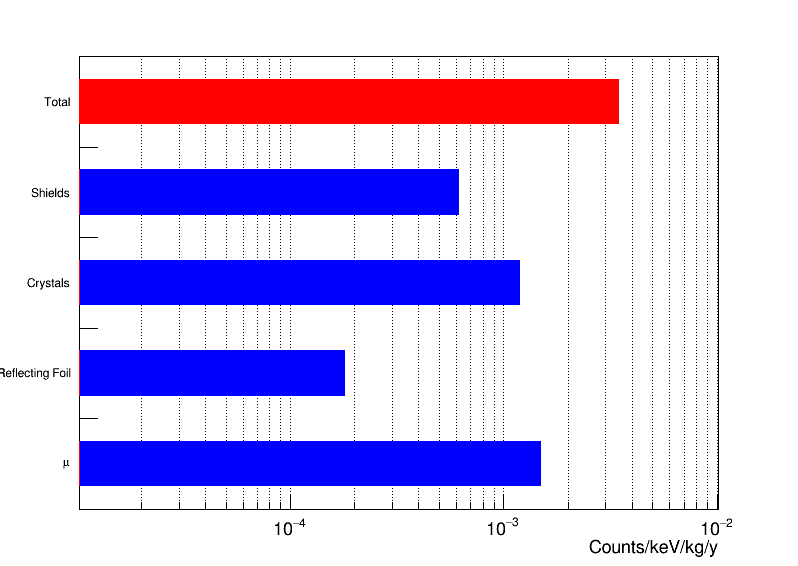}
  \caption{Breakdown of the $\beta/\gamma$ count rate in the $^{100}$Mo \onbb\ region for CUORE (left) and for CUPID-0 (right). The main contributions identified by the respective BMs are shown.  The integration region is 3000-3200 keV for CUORE and 2800-3200 keV for CUPID-0. In the case of CUPID-0, the \nnbb\ $^{82}$Se contribution is not included in crystal contribution. A time-veto cut is applied both to data and to simulations in order to reject the $^{208}$Tl induced background.}
  \label{fig:CUPID0_rate}
  \label{fig:CUORE_rate}
\end{figure}

The right panel of Fig.~\ref{fig:CUORE_BM_contributions} shows the contribution due to the detector  infrastructure, mainly the copper frames and the copper 10 mK shield surrounding the array. A cut on $\alpha$-induced events is applied to show only the background that would be observed after $\alpha$ rejection (the experimental counting rate at 3 MeV is dominated by the surface contamination of copper). In CUPID this contribution will scale as the detection efficiency and the amount of copper directly facing the crystals.
Fig.~\ref{fig:CUORE_rate} (left panel) shows the breakdown of the CUORE counting rate predicted from the BM in the 2800-3200 keV region (only $\beta/\gamma$ contribution).

\subsubsection{CUPID-0 background model} \label{sec:bkg-BMCupid0}
\label{sec:cupid0 bkg model}

\begin{figure}[h!]
  \centering
  \includegraphics[height=.18\textheight]{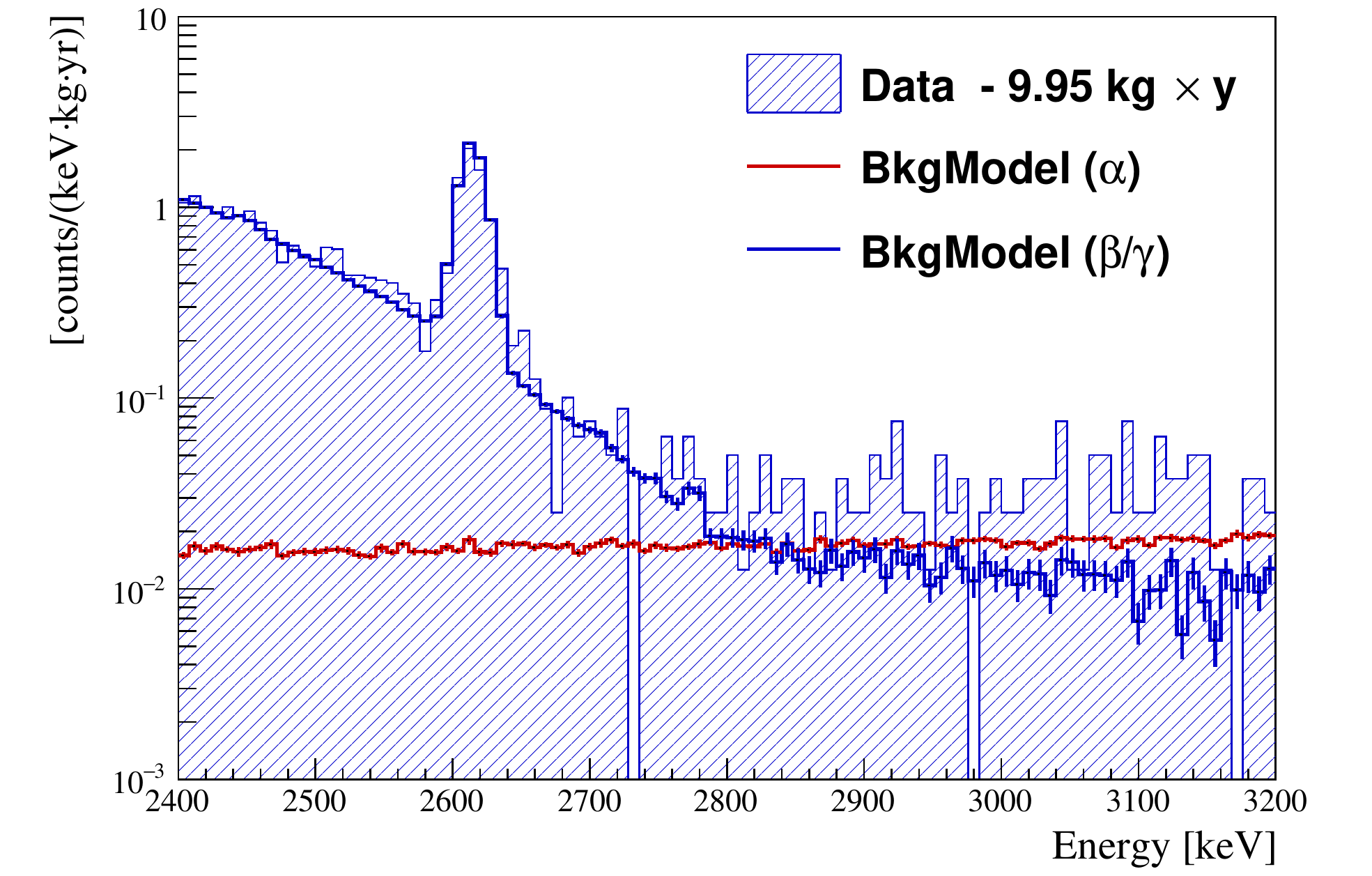}\quad
  \includegraphics[height=.202\textheight]{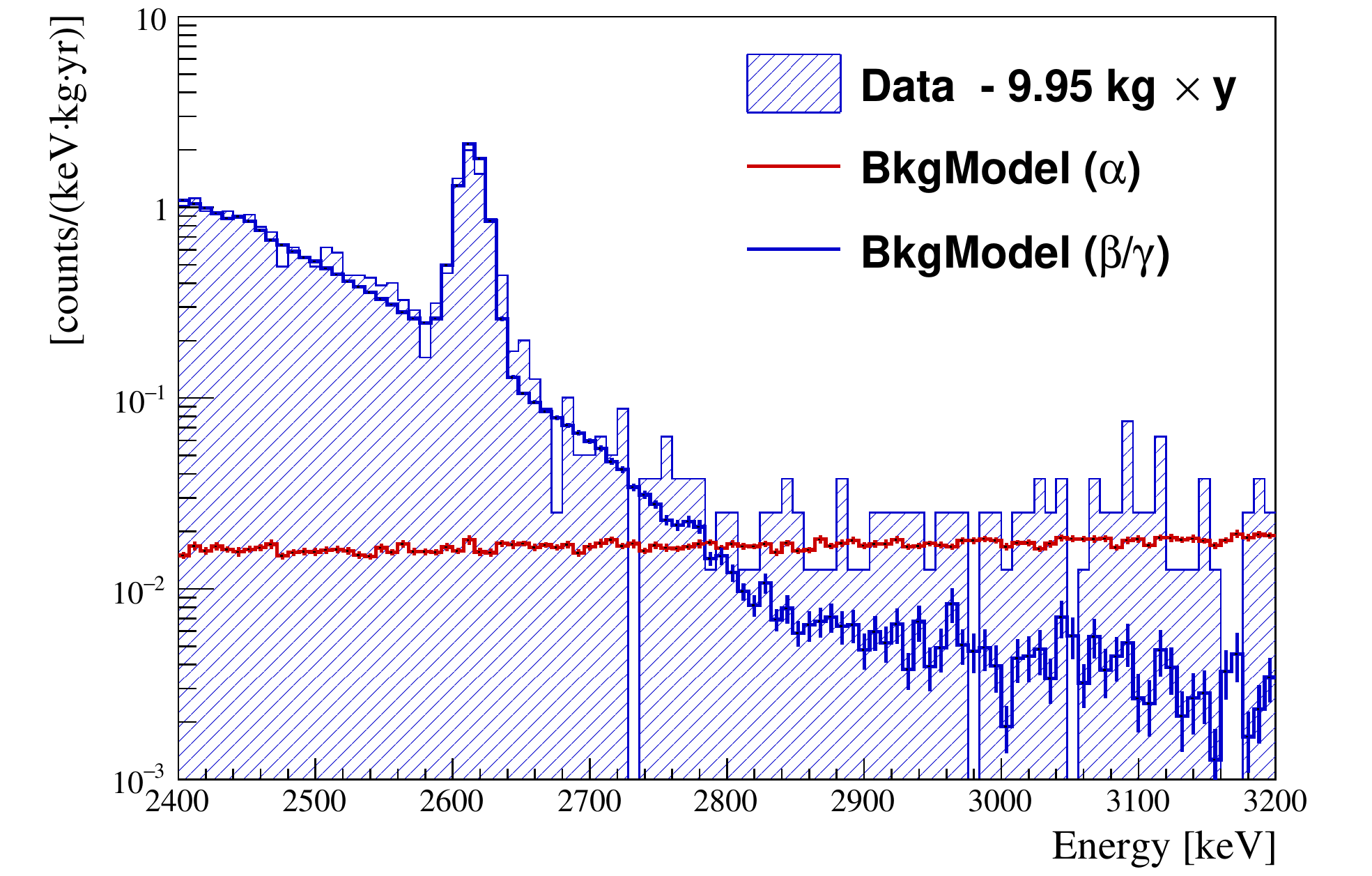}
  \caption{CUPID-0 BM reconstruction of the $\alpha$ and the $\beta$/\/$\gamma$ components before (left panel) and after (right panel) the delayed coincidence cut. Figures reprinted from \cite{Azzolini:2019nmi}.}
  \label{fig:CUPID0_BM_particleID}
\end{figure}

CUPID-0 detectors are ZnSe scintillating bolometers that allow particle identification.
In the BM, the spectra due to pure $\alpha$ and pure $\beta/\gamma$ events are separated both in the
simulations and in the data. Particle identification becomes weaker below 2~MeV, a region dominated by $\beta/\gamma$ and in particular by the 2$\nu\beta\beta$ of $^{82}$Se.
Therefore all events below 2~MeV are included in the $\beta/\gamma$ spectrum. The same kind of cut is applied on all the simulations used in the BM.
Fig.~\ref{fig:CUPID0_BM_particleID} shows the reconstructed contribution of the $\alpha$
and  $\beta/\gamma$ component (left) and  the effect of a delayed coincidence cut (right).
This cut efficiently removes the $\beta/\gamma$ contribution due to the $^{232}$Th
contamination of ZnSe crystals, which is particularly high in CUPID-0 (about 10 $\mu$Bq/kg~\cite{Azzolini:2018tum}).
Details of the CUPID-0 background model can be found in \cite{Azzolini:2019nmi}.

\begin{figure}[h!]
  \centering
  \includegraphics[height=.205\textheight]{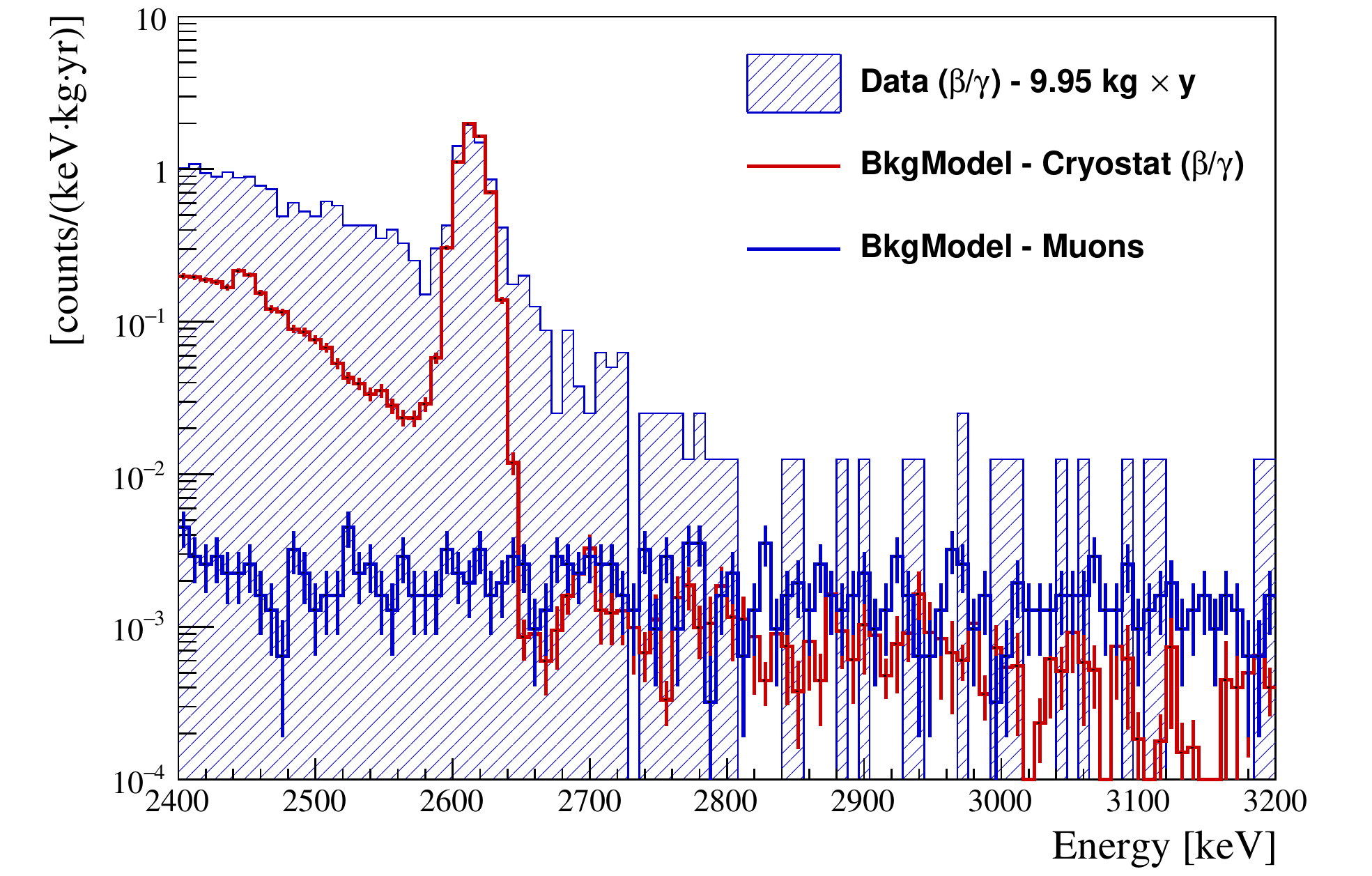}
  \includegraphics[height=.205\textheight]{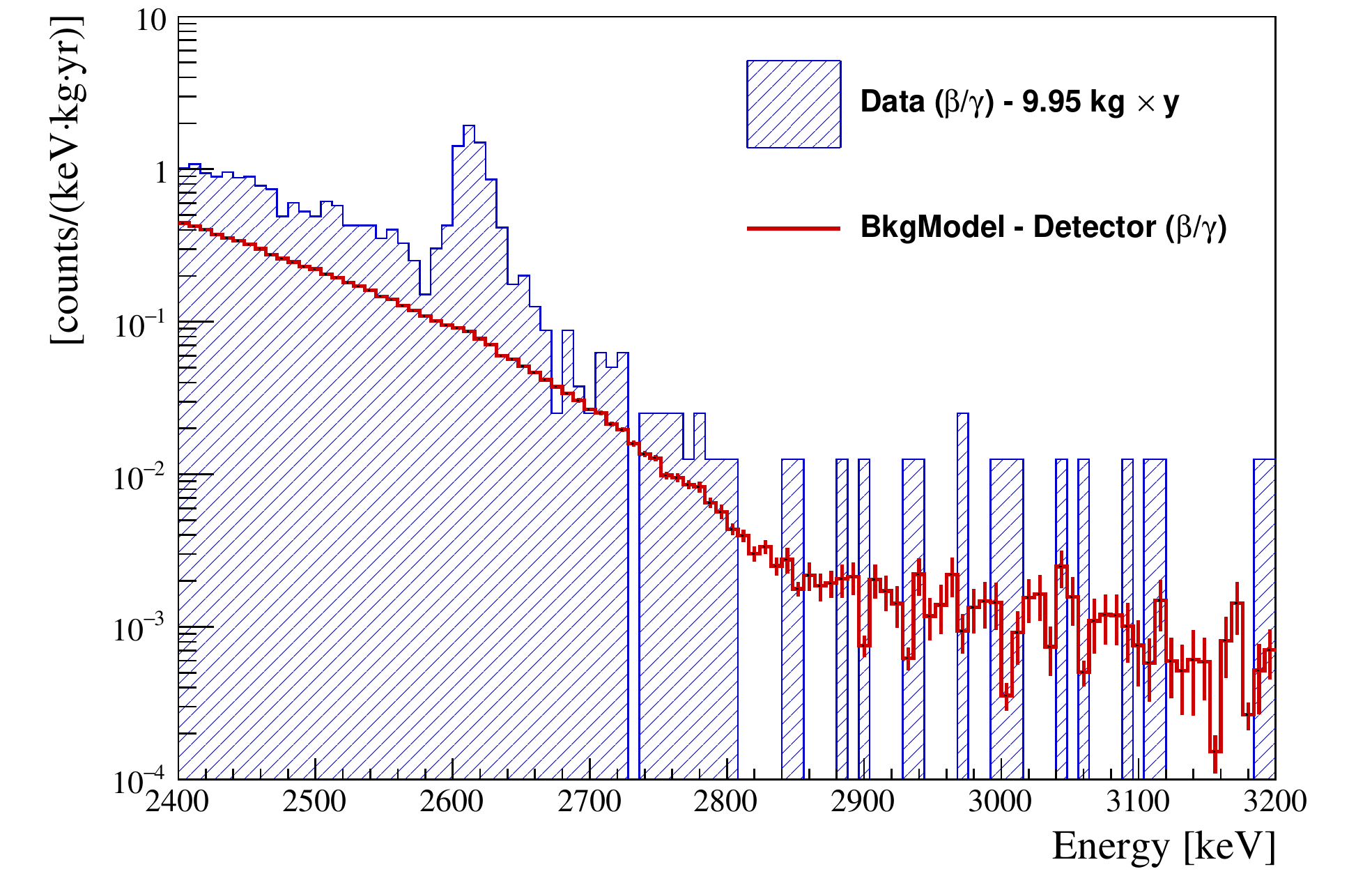}
  \caption{CUPID-0 experimental data (filled histogram) after
	$\alpha$ particle rejection and delayed coincidence cut
    The left panel shows the reconstructed contribution of
    the cryostat (red) and the cosmogenic muons (blue).
    The right panel shows the reconstructed contribution of the
    detector, mainly coming from the ZnSe crystals, reflecting foil, and holder.}
  \label{fig:CUPID0_BM_cryostat}
\end{figure}

Fig.~\ref{fig:CUPID0_BM_cryostat} shows the CUPID-0 spectrum after $\alpha$ particle rejection and delayed coincidence cut as compared with the reconstructed contribution.
Fig.~\ref{fig:CUPID0_rate} (right panel) shows the breakdown of the CUPID-0 counting rate predicted from the BM in the 2800-3200 keV region (only $\beta/\gamma$ contribution). The delayed coincidence cut is applied both to the data and to the simulations.

\subsection{Background in CUPID-Mo demonstrator}\label{sec:bkg-BMLumineu}

Multiple tests with enriched Li$_2$MoO$_4$ scintillating bolometers were performed at LSM and the LNGS. The tests allowed us to evaluate the background contribution of the crystals and set-up.  
At this stage, one of our top priorities is the evaluation
of the radioactive contamination in the Li$_2$MoO$_4$ crystals and
its expected contribution to the CUPID background.

\begin{SCfigure}
  \centering
  \includegraphics[width=0.7\textwidth]{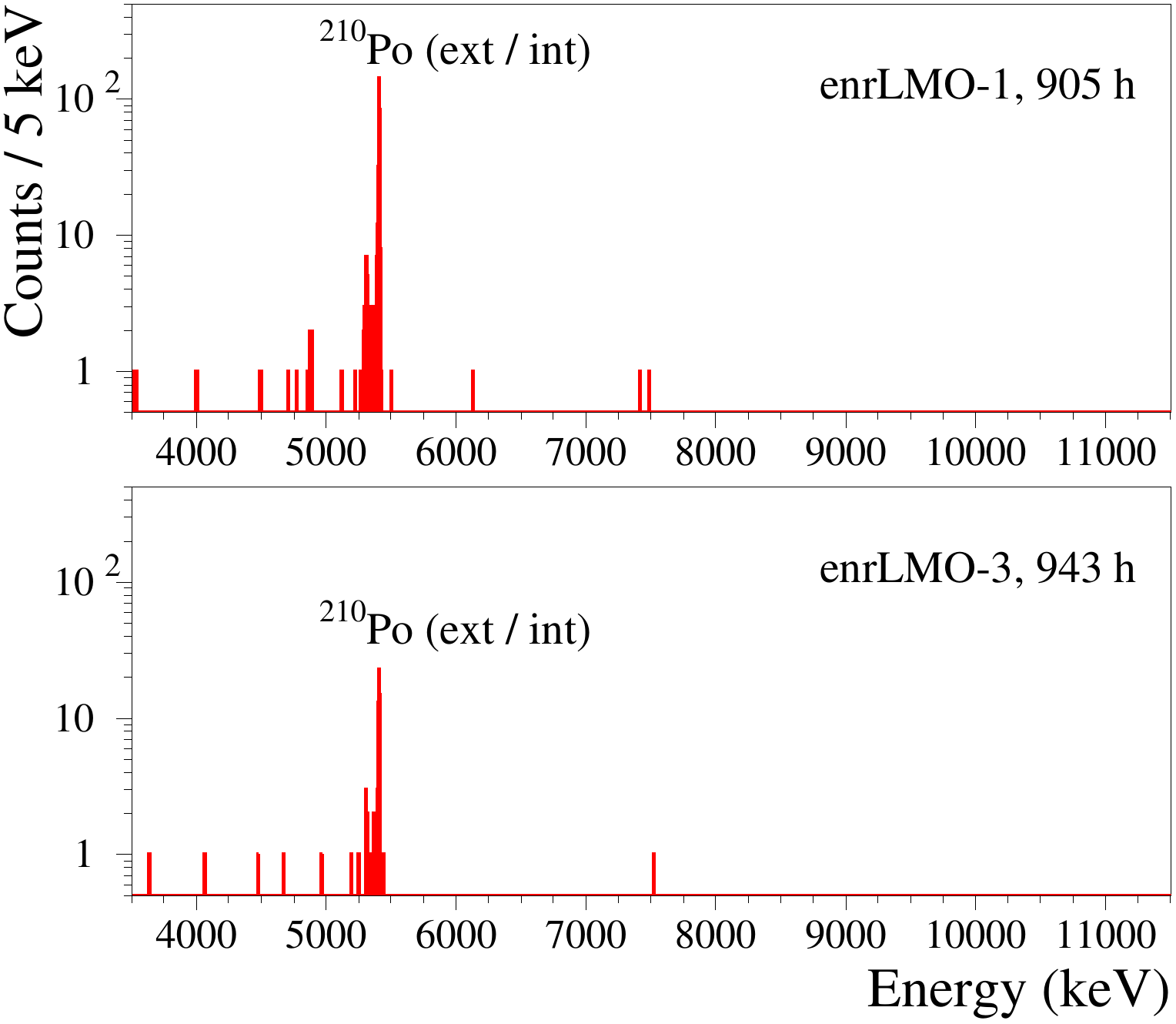}
  \caption{Energy spectrum of $\alpha$ events in the \lmo\ scintillating bolometers. }
  \label{fig:cupid_mo_alpha_spect}
\end{SCfigure}

We measured the radioactivity in the crystal bulk
using the $\alpha$ spectra obtained in a measurement campaign
with four enriched \lmo\ detectors (enrLMO-1, enrLMO-2, enrLMO-3, and enrLMO-4) and about 10~kg$\cdot$d exposure \cite{Armengaud:2017} \cite{Poda2017} \cite{CupidMo:Mo2b2n}.
Figure~\ref{fig:cupid_mo_alpha_spect} shows the energy spectra
in two enriched Li$_2$MoO$_4$ crystals 

Only the peak from $^{210}$Po is observed, therefore upper limits have been set for the other radionuclides in the $^{238}$U and $^{232}$Th decay chains.
The measured activity of  $^{210}$Po is between 20 mBq/kg and 450 mBq/kg, depending on the crystal. However, $^{210}$Po is not a potential source of background since it has no associated $\gamma$s or $\beta$s with high enough energies to produce events in the ROI. Table~\ref{LMO_contamination} gives the measured radioactivity levels in the bulk of \lmo\ crystals.



\begin{table}[htb]
  \centering
  \caption{Radiopurity in the bulk of $^{100}$Mo enriched \lmo\ crystals.  Uncertainties are given at 68$\%$ CL and upper limits at 90$\%$ CL.}
  \label{LMO_contamination}
  \begin{tabular}{cccccc}
    \toprule
    \bf{Activities}    & \bf{enrLMO-1}    &   \bf{enrLMO-2}  &  \bf{enrLMO-3} &  \bf{enrLMO-4} & \bf{Ref}     \\
    \bf {[$\mu$Bq/kg]} &       186 g           & 204 g              & 213 g          & 207 g   &  \\
    \midrule
    $^{232}$Th      & $<$3            &    $<$11           &               & &  \cite{Armengaud:2017}   \\
    $^{228}$Th      & $<$4            &    $<$6           &   $<$3        &  $<$5 &  \cite{Poda2017} \\
    \midrule
    $^{238}$U       & $<$5            &     $<$11        &                &     &  \cite{Armengaud:2017}     \\
    $^{234}$U       &   $<$7           &   $<$11          &                &     &  \cite{Armengaud:2017}   \\
    $^{230}$Th       &   $<$3          &    $<$11        &                &     &   \cite{Armengaud:2017}   \\
    $^{226}$Ra       &  $<$6           &  $<$11           &    $<$3       & $<$9       &  \cite{Poda2017}   \\
    $^{210}$Po       & 450$\pm$30      &   200$\pm$20     & 76$\pm$10     &  20$\pm$6  & \cite{Poda2017} \\
    \midrule
    $^{235}$U       &   $<$5           &   $<$6           &                &     &  \cite{Armengaud:2017}    \\
    $^{231}$Pa       &  $<$3          &     $<$6         &                &     &  \cite{Armengaud:2017}    \\
    $^{227}$Ac       &  $<$5          &    $<$6          &                &     &  \cite{Armengaud:2017}    \\
    \midrule
    $^{40}$K       &  800$\pm$300      &   800$\pm$300   &   800$\pm$300  &  800$\pm$300   &  \cite{CupidMo:Mo2b2n}  \\
    \midrule
    $^{190}$Pt      &  $<$3             &   $<$11                &                &     &  \cite{Armengaud:2017}   \\
    \bottomrule
  \end{tabular}
\end{table}


\subsection{Predicted CUPID backgrounds}\label{sec:bkg-BBcupid}

In this section we discuss the CUPID background budget assuming the following geometry for the array: 118 towers with 13 \lmo\ crystals each. The crystals are cylinders with 5~cm height and diameter, face a Ge Light Detector (LD) on their top and bottom, and are
surrounded by a 70~$\mu$m thick reflecting foil on the side. 
The array is enclosed in a copper vessel that, just as in CUORE, acts as 10 mK thermal shield. Crystals and LDs are held by ring-shaped copper frames as in CUPID-0~\cite{Azzolini:2018tum}. In the following we will refer to these copper parts as to the ``holder.'' The cryogenic infrastructure and the radiation shield system is the same as in CUORE.

\begin{SCfigure}
  \includegraphics[width=.5\textwidth]{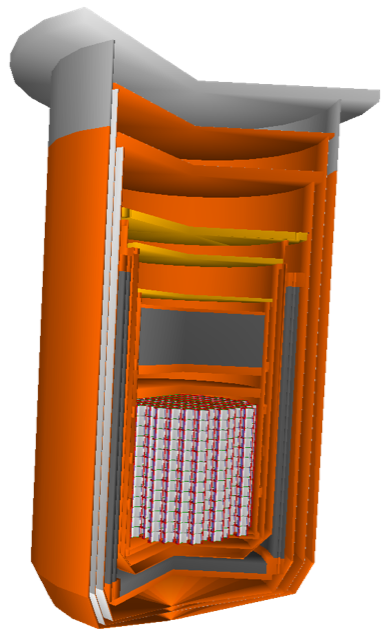}
  \caption{Geometry of the CUPID detector array with cylindrical
	crystals implemented in the CUORE Geant4 simulation software.}
  \label{fig:CUPID_simulation}
\end{SCfigure}

We implemented this detector geometry in the CUORE Geant4  simulation package, as shown in Fig.~\ref{fig:CUPID_simulation}, and used it to evaluate the expected BI in the CUPID ROI. 

We do not simulate the production and collection of scintillation light, but reconstruct the expected light signal as follows. In the simulation, we tag energy deposition in the \lmo\ crystals according to the kind of interacting particle. Therefore, for each event in a \lmo\ crystal, we know the fraction of energy deposited from $\beta/\gamma$s, from $\alpha$s, and from nuclear recoils. We use this information to evaluate the expected light signal associated to the event. In the simulations presented here, we assume a LY$_{\beta/\gamma}$ = 0.74~keV/MeV, a quenching for $\alpha$ of QF$_{\alpha}$=0.209, and a quenching for nuclear recoils of QF$_{\rm{recoil}}$=0.125. These parameters are obtained from the measurement shown in the lower panel of Fig.~\ref{fig:LightVsHeat}. From these same data, we extract the energy resolution of the light signal that we reproduce in the simulation applying a Gaussian smearing with an energy-dependent width.  Finally, we apply a cut on the LY=Light/Energy ratio (to reject $\alpha$ particles) accepting only events with LY$>$ 0.5. 

We note that the parameters adopted in the reproduction of the LY~vs.~Energy scatter plot, though extracted from a specific measurement, are representative of the average behaviour of the \lmo\ detectors. Slight changes of LY, QF, as well as of the cut adopted for $\alpha$ rejection do not imply relevant changes in the results discussed below.  The quoted background rates refer to a 70~keV width ROI centered on the $^{100}$Mo \Qbb.  They are evaluated having applied the anti-coincidence cut, a time-cut that removes pile-up events occurring in a 5~sec window, and delayed coincidences occurring on the same crystal in a 900~sec window.  This corresponds to about 5 half-lives of $^{208}$Tl. There is no gain in using larger values since the residual background becomes dominated by the inefficiency in the parent detection.


In the following sections we discuss hypotheses and results obtained for the different background sources.

\begin{table}[htb]
  \centering
  \caption{Radiopurity levels assumed in the CUPID background model for close materials.}
  \label{tab:BM_contamination}
  \begin{tabular}{lccc}
    \toprule
    \bf{Material}    					& $^{238}$U    & $^{232}$Th  & \bf{Reference} \\
    \midrule
    \lmo\ bulk [$\mu$Bq/kg]  					& 10        &    3   & Sec.~\ref{sec:bkg-BMLumineu}\\
    \lmo\ surface 10 nm [nBq/cm$^2$]  		& 3         &    2   & \cite{Q0_2nu}\\
    \lmo\ surface 10 $\mu$m [nBq/cm$^2$]  	& 0.8       &    $<$0.03  & \cite{Q0_2nu} \\
    Reflecting foil surface 10 $\mu$m [nBq/cm$^2$]	& 8.7   &    $<$0.7   &
    Sec.~\ref{sec:bkg-BMCupid0}\\
    Cu bulk [$\mu$Bq/kg]      				& $<$10     &    $<$2  & \cite{Q0_2nu}\\
    Cu surface  10 $\mu$m [nBq/cm$^2$]      & 14        &    5  & \cite{Q0_2nu}\\
    \bottomrule
  \end{tabular}
\end{table}

\begin{SCfigure}
  \centering
  \includegraphics[width=.7\textwidth]{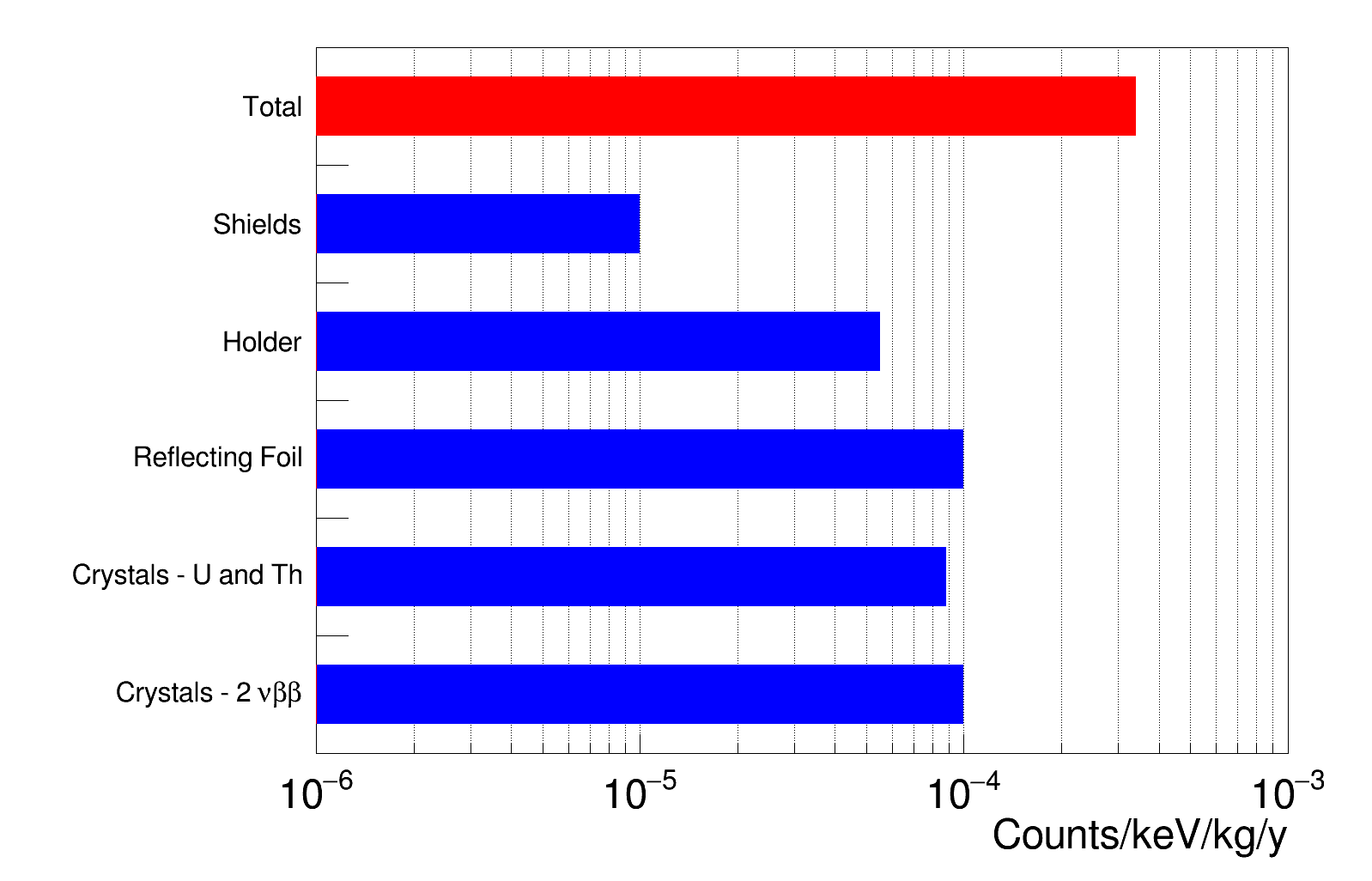}
  \caption{Breakdown of the CUPID $\beta/\gamma$ counting rate predicted by the BM in the $^{100}$Mo ROI. Here, the baseline configuration is considered. As discussed in the text, the substitution of the reflective foil with a reflective coating on \lmo\ crystals would dramatically reduce both the U and Th contributions of crystals (here dominated by surface contaminants) and that of the reflector itself.}
  \label{fig:CUPID_rate}
\end{SCfigure}

\subsubsection{$^{238}$U, $^{232}$Th from close sources}

We evaluate $^{238}$U and $^{232}$Th contribution from close sources simulating the chains in secular equilibrium and using the activities reported in Table~\ref{tab:BM_contamination}. Since the isotopes responsible of the background rate in the ROI are mainly $^{214}$Bi and $^{208}$Tl, and since secular equilibrium is often broken at the $^{226}$Ra or $^{228}$Th level (respectively for the $^{238}$U and $^{232}$Th chains), it is important to keep in mind that activities bounds must be respected, not only by the progenitors of the chains, but also by their two long-lived daughters.

Below we briefly comment on the activities used for this study while in Fig.~\ref{fig:CUPID_rate} we show the resulting BI. 
\begin{itemize}
\item For the bulk of \lmo\ crystals, based on the results reported in Sec.~\ref{sec:bkg-BMLumineu}, we have assumed  a contamination of 10 $\mu$Bq/kg in  $^{238}$U and 3 $\mu$Bq/kg in $^{232}$Th.
\item For the surface of the \lmo\ crystals, since we do not have a sensitive measurement on the \lmo\ crystals themselves, we assume the same activity measured in CUORE-0. 
  A surface contamination is characterized not only by the contaminant concentration but also by the way the contaminant is distributed in the contaminated layer. As discussed in Ref.~\cite{Q0_2nu}, a measurement of the density profile of surface contamination is not easy.  However, experimental results in CUORE-0 (as well as in CUORE and in CUPID-0) can be satisfactorily reproduced assuming two different depths: 10 nm and 10 $\mu$m~\cite{Q0_2nu}. We do the same in our CUPID background model. 
\item For the  reflecting foil, we use the results of CUPID-0 BM. Indeed, direct measurements (ICPMS and Bi-Po tracked in a scintillation detector) bound the activities of $^{238}$U and $^{232}$Th progenitors; the lowest part of the chain is bounded only in the case of $^{232}$Th. CUPID-0 data proves the existence of contamination coming from the reflecting foil. This contamination is observed from its $\alpha$ emission.  The BM attributes it to a $^{226}$Ra ($^{238}$U chain) contamination.  This is equally well represented by a bulk or by a surface contamination of the foil, as expected given the thickness of the material. The alternative hypotheses, namely that $\alpha$s are due to a $^{210}$Po contamination (therefore attributable to an excess in the $^{238}$U chain of the $^{210}$Pb isotopes and its daughters), is moderately rejected by the fit.
\item For the copper holder, we use the surface and bulk contamination measured for the same material in CUORE-0~\cite{Q0_2nu}.
\end{itemize}
It is clear from Fig.~\ref{fig:CUPID_rate} that the two most critical contribution are those coming from the reflecting foil and from the surfaces of crystals. Removing the reflecting foil will likely allow us to get rid of both these two contributions. In one case, it's because the background source is removed (the reflective foil).  In the other, it's because the efficiency of coincidence cut will be improved (since crystals will face each other without an inert material interposed in between them).


\subsubsection{$^{238}$U, $^{232}$Th in the cryostat and radiation shields}

According to the CUORE BM, the only contaminants present in the
infrastructure that may yield a background in the $^{100}$Mo ROI
are $^{214}$Bi and $^{208}$Tl in the shields and cryostat.
As previously discussed, this is a pure $\gamma$ contribution,
since a minimum thickness of few cm of copper is assumed to shield the
array from these elements. Despite the current uncertainty
on the exact location of these contaminants, a safe upper limit on
their contribution to CUPID BI can be set to about 10$^{-5}$ \ckky\ 
with the anti-coincidence cut.

\subsubsection{Cosmogenic Isotopes in the crystals and materials}\label{sec:bkg-cosmo}

The interaction of cosmic rays with the crystals and holding
structure leads to the production of radioactive nuclei
that accumulate during the production, handling, and storage
of the materials above ground~\cite{Heusser:1995,Cebrian:2017}.
The amount of cosmogenic activation is the result of the properties of the target material, the cosmic ray flux and composition, the exposure time, and the cooling time (in case an underground storage follows the production). The cosmogenic activation of \lmo\ crystals was estimated using the ACTIVIA~\cite{ACTIVIA} code, assuming a sea level exposure of 90 days an underground cooling period of one year. Among all the materials present in the CUPID setup, only molybdenum contributes with cosmogenically activated isotopes emitting radiation with energy high enough to give events in the ROI. We assume the crystals to be $95\%$ enriched in $^{100}$Mo, with the remaining isotope being only $^{98}$Mo. The only potentially dangerous activated isotopes in $^{100}$Mo are $^{82}$Rb, which is produced by the decay of the activated isotope $^{82}$Sr, $^{56}$Co, and $^{88}$Y.  These are also produced by the decay of the activated isotope $^{88}$Zr. The activation rates after the cooldown period are of the order of $10^{-5}$ counts/(kg$\cdot$day) for $^{82}$Sr, $10^{-3}$ counts/(kg$\cdot$day) for $^{56}$Co, and $10^{-1}$ counts/(kg$\cdot$day) for $^{88}$Zr and Y. Monte Carlo simulations show that the contribution of such isotopes to the CUPID background do not exceed $5\times10^{-5}$ counts/(kg$\cdot$day). Although there are different assumptions, the results are consistent with previous calculations performed with both ACTIVIA and COSMO~\cite{COSMO} codes~\cite{ChernyakPhD,Danevich:2015,Poda:2017c}.

\subsubsection{Neutron-induced backgrounds}
\label{sec:bkg_neutrons}

In a deep underground site like the LNGS, the dominant neutron flux is caused by the  radioactivity in the surrounding rock. This consists of spontaneous fission (mainly of $^{238}$U) and ($\alpha$, n) reactions, with neutrons energies below $\sim$10 MeV. The large rock overburden of the LNGS significantly reduces the cosmic ray muon flux.
Therefore the muon-induced neutron flux is several orders of magnitude weaker than that attributed to the natural radioactivity, but its neutron energy distribution extends up to a few GeV \cite{Mei:2006}. The background in the ROI for a double-beta decay search may arise mainly from (n,$\gamma$) reactions or by inelastic scattering of high-energy neutrons. The (n,$\gamma$) reactions 
can produce $\gamma$s with energies up to 10 MeV, while the neutron inelastic scattering off the elemental composition of a detector may induce several MeV, potentially populating the ROI.

CUORE features a massive external passive shield
consisting of 18 cm of polyethylene to thermalize neutrons,
2 cm of H$_3$BO$_3$ powder to capture thermal neutrons,
and at least 25 cm of lead to absorb $\gamma$ rays.
The neutron-induced background in the ROI of the \onbb\ of $^{130}$Te is estimated to be
$(8\pm6)\times10^{-6}$ \ckky\ ~\cite{Bellini:2010,Alduino:2017}.
The CUORE expectation could also be reasonably valid for the case
of CUPID and the ROI of both $^{130}$Te and $^{100}$Mo.

The difference between the CUPID baseline option and CUORE in the material properties of the detector (i.e. lower density, effective $Z$, and crystal volume of \lmo\ than those of TeO$_2$), would result in a comparable neutron-induced gamma rate. Furthermore, the particle identification capability of a Li$_2$MoO$_4$ scintillating bolometer would improve the discrimination of inelastic neutron scattering-induced events.  The latter are rejected in CUORE only by an anti-coincidence cut. 

The difference in the detector elemental composition (i.e. the presence of $^{100}$Mo and $^{\rm{nat}}$Li instead of
$^{\rm{nat}}$Te or $^{130}$Te), does not significantly impact the 
background originated from a neutron capture in a detector volume. 
The cross section energy dependence of (n,$\gamma$) reactions on 
$^{100}$Mo is rather similar to that of $^{130}$Te.
The difference is mainly in a larger number of resonances
for $^{100}$Mo in the 0.2--20~keV energy range.
These resonances can have up to a factor 10 larger
cross section \cite{Chadwick:2011}.
At the same time, the intensity of $\gamma$s with energies
above $Q_{\beta\beta}$ are about one order magnitude higher
for $^{130}$Te \cite{CapGam}.
Regarding the lithium and oxygen, the dominant isotopes are $^7$Li and $^{16}$O, respectively.
The intensity of their few broad resonances for fast neutron
capture is about four orders of magnitude lower than those of
$^{130}$Te and $^{100}$Mo. In all cases, the only potentially
harmful daughter nucleus is $^8$Li
($Q_{\beta}$ = 16 MeV, $T_{1/2}$ = 0.84 s).
The $^8$Li decay occurs through a broad ($\Gamma$ = 1.5 MeV) 3 MeV
excited state of $^8$B, which disintegrates promptly into two
$\alpha$ particles producing mixed $\beta$+$\alpha$ single events
\cite{Bhattacharya:2006}.
Considering the energy spectrum of such
events \cite{Wilkinson:1971}, and the estimation of the thermal
neutron flux even inside the CUPID R\&D cryogenic set-up
($\leq2.6\times10^{-8}$~neutrons/s/cm$^2$ \cite{Armengaud:2017}),
the expected background in the $^{100}$Mo ROI is $1.0\times10^{-5}$ and $2.6\times10^{-7}$ \ckky\ for the 100\% and 1\% transition to the $^8$B excited and ground states respectively.
It is worth stressing that this background is expected to be significantly suppressed thanks to $^6$Li(n,t)$\alpha$ reaction which is a factor of 1735 more probable than the $^7$Li(n,$\gamma$)$^8$Li reaction for natural lithium (7.6\% of $^6$Li). Furthermore, a major part of the $\beta$+$\alpha$ events occurring in the $^8$Li-$^8$B decays can be rejected with high efficiency using a scintillating bolometer approach \cite{Bekker:2016,Armengaud:2017}.

\subsubsection{Muon-induced background}

An additional background source is represented by muons interacting directly with the detectors.
Most of the muons release energy in more than one crystal, thus a strong suppression
of the muon background is obtained with the rejection of events with multiplicity $>1$ (M1).
Preliminary results from CUORE predict a background contribution of surviving M1 events
at the level of $10^{-4}$ \ckky, as depicted in Fig.~\ref{fig:CUORE_BM_contributions}.
These mainly come from muons interacting with peripheral crystals.
We plan to further reduce this background by at least an order of magnitude with the
installation of dedicated plastic scintillator panels around, above, or below the external lead shielding (Sec. \ref{sec:MuonTagger}). We are currently investigating the possible
geometrical configurations that can fit in the CUORE setup without requiring
modifications to the infrastructure. The exact suppression factor will be evaluated
with dedicated Monte Carlo simulations that are currently under development.
Preliminary results show that the required suppression factor is achievable 
even in the conservative scenario where only the sides of the shielding are
instrumented.

\subsubsection{Neutrino-induced background}

The background induced by neutrino-electron elastic scattering (neutral current process) is estimated to be of the order of 10$^{-7}$ \ckky\ for all experiments investigating $\beta\beta$ isotope with $Q_{\beta\beta}$ above 2~MeV \cite{Barros:2011}. The charged current neutrino-nucleus interaction for the case of $^{100}$Mo, characterized by the low energy threshold and the large neutrino capture cross-section \cite{Ejiri:2000}, is expected to be on the level of 2$\times$10$^{-5}$~counts/(kg$\cdot$yr) \cite{Ejiri:2017}, assuming 0.4\% energy resolution. This background can also be suppressed by tagging the charged current process thanks to a short half-life ($T_{1/2}$ = 16 s) of the intermediate nucleus $^{100}$Tc and/or multiple-hit events.


\newpage


\section{Phased Deployment}\label{sec:phases}

Bolometric experiments with a large number of individual crystals and
towers are well suited for phased deployment, starting data taking
with a fraction of the total isotopic mass. We foresee an option of
deploying the first few towers assembled for CUPID early, while the
rest of the crystals are being procured and the detectors
assembled. Such phase ``CUPID-I'' could be deployed in a separate
cryostat at LNGS, or in the \cuore\ cryostat if it is available. Similar
to \cuoreo, which was developed as part of the \cuore\ project R\&D and
was invaluable in vetting the \cuore\ assembly procedures and the
background model, CUPID-I phase would be an important step in the
project execution plan. At the same time, a detector with mass of
50-70 kg and nearly zero background, deployed in the early 2020s,
would be a leading NLDBD experiment in its own right. 

The main phase of CUPID is a detector deployed in the \cuore\
cryostat. Its size is determined by the volume of the existing 10 mK
vessel and is not limited by the available cooling power. The baseline
CUPID design parameters are listed in Table~\ref{tab:baseline}. We
assume a conservative background goal of $10^{-4}$\ckky; as discussed
in Section~\ref{sec:bkg}, this goal is readily achievable with the
existing technology.

At the same time, it is instructive to consider the ultimate
sensitivity of a bolometric detector not limited by the current
technology and infrastructure. We consider two additional scenarios
for the purposes of computing the ultimate sensitivity of the CUPID
program. One is the detector deployed in the \cuore\ cryostat, but
operating in the nearly zero-background mode, which corresponds to the
background index of $2\times10^{-5}$\ckky. This is an optimistic
``reach'' goal for CUPID. As Section~\ref{sec:bkg}
demonstrates, reaching this background goal is possible with
additional R\&D: eliminating
the reflective foil backgrounds, reducing the bulk and surface
backgrounds from the crystals (e.g. by additional purification of
\lmo\ crystals and elimination of surface backgrounds by pulse shape
discrimination), and reducing the  \nnbb\ pileup background to below
$10^{-5}$\ckky\ level through the use of higher-bandwidth sensors. 

Finally, we consider a strawman ultimate bolometric detector
``CUPID-1T'', consisting of 1.8 tons of \lmo, or 1000~kg of \Mo. Such
detector could be accommodated in a new cryostat approximately
4 times larger than \cuore, within the capabilities of the \cuore\
cryogenic systems. For optimal sensitivity, we assume that
the backgrounds could be further reduced to the level of
$5\times10^{-6}$\ckky. Such background levels are within the realm of
possibility for the 
transition energy of 3034~keV, especially considering advances in
material screening and radiopurity, and advanced high-speed,
high-resolution sensors being developed within CUPID
(Section~\ref{sec:tes}). Such a detector would be the ultimate Phase
III of the CUPID program, a bolometric detector with the sensitivity
in the Normal Hierarchy of neutrino masses. In case of a discovery,
such a detector could also explore other isotopes, e.g. \Te. 

We summarize the parameters of the possible CUPID detector phases in
Table~\ref{tab:phases}. The sensitivity to various models of NLDBD are
discussed in Section~\ref{sec:sensitivity}. 

\begin{table}[b!]
  \caption{Parameters of the CUPID detector in the baseline scenario,
    in the optimistic background scenario, and for a large bolometric
    detector with 1 metric ton of \Mo\ isotope. 
}
  \label{tab:phases}
  \centering
  \begin{tabular}{lrrr}
    \toprule
    Parameter & CUPID Baseline & CUPID-reach & CUPID-1T \\
    \midrule
    Crystal & \enrlmo & \enrlmo & \enrlmo \\
    Detector mass (kg) & 472 & 472 & 1871 \\ 
    $^{100}$Mo mass (kg) & 253 & 253 & 1000 \\ 
    Energy resolution FWHM (keV) & 5 & 5 & 5 \\ 
    Background index (\ckky) & $10^{-4}$ & $2\times10^{-5}$ & $5\times10^{-6}$ \\ 
    Containment efficiency & 79\% & 79\% & 79\% \\ 
    Selection efficiency  & 90\% & 90\% & 90\% \\ 
    Livetime (years) & 10 & 10 & 10 \\
    Half-life exclusion sensitivity (90\% C.L.) & $1.5\times10^{27}$ y & $2.3\times10^{27}$ y & $9.2\times10^{27}$ y \\
    Half-life discovery sensitivity ($3\sigma$) & $1.1\times10^{27}$ y & $2\times10^{27}$ y & $8\times10^{27}$ y\\
    \mbb\ exclusion sensitivity (90\% C.L.)     & 10--17~meV           & 8.2--14~meV          & 4.1--6.8~MeV \\
    \mbb\ discovery sensitivity ($3\sigma$)     & 12--20~meV           & 8.8--15~meV          & 4.4--7.3~meV \\
    \bottomrule
  \end{tabular}
\end{table}

\section{Physics sensitivity and other measurements}
\label{sec:sensitivity}

\subsection{Sensitivity to \onbb\ decay}

\begin{figure}[htbp]
     \centering
     \includegraphics[height=.309\textheight]{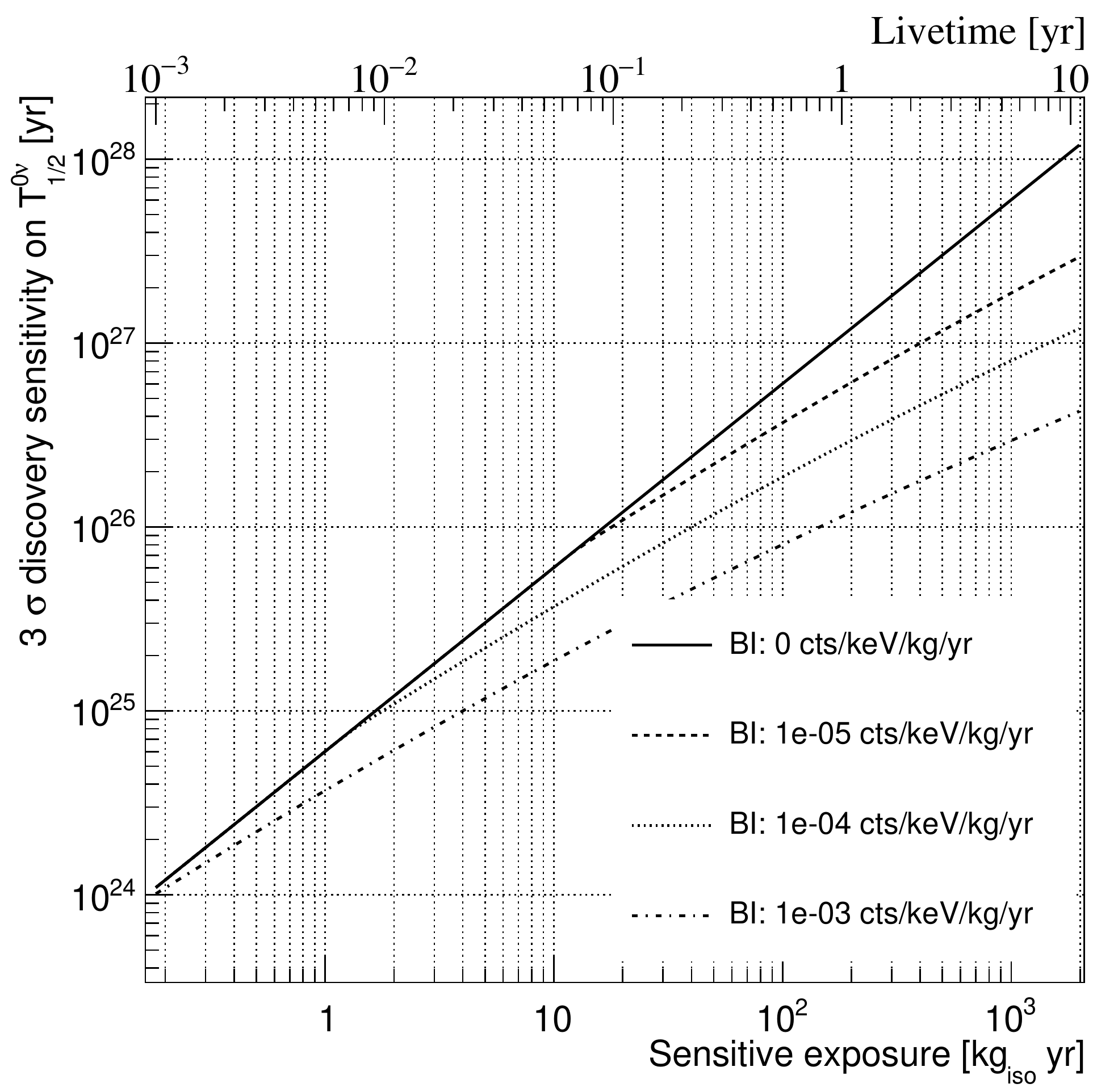}\label{fig:DiscSensBkg}\quad
     \includegraphics[height=.309\textheight]{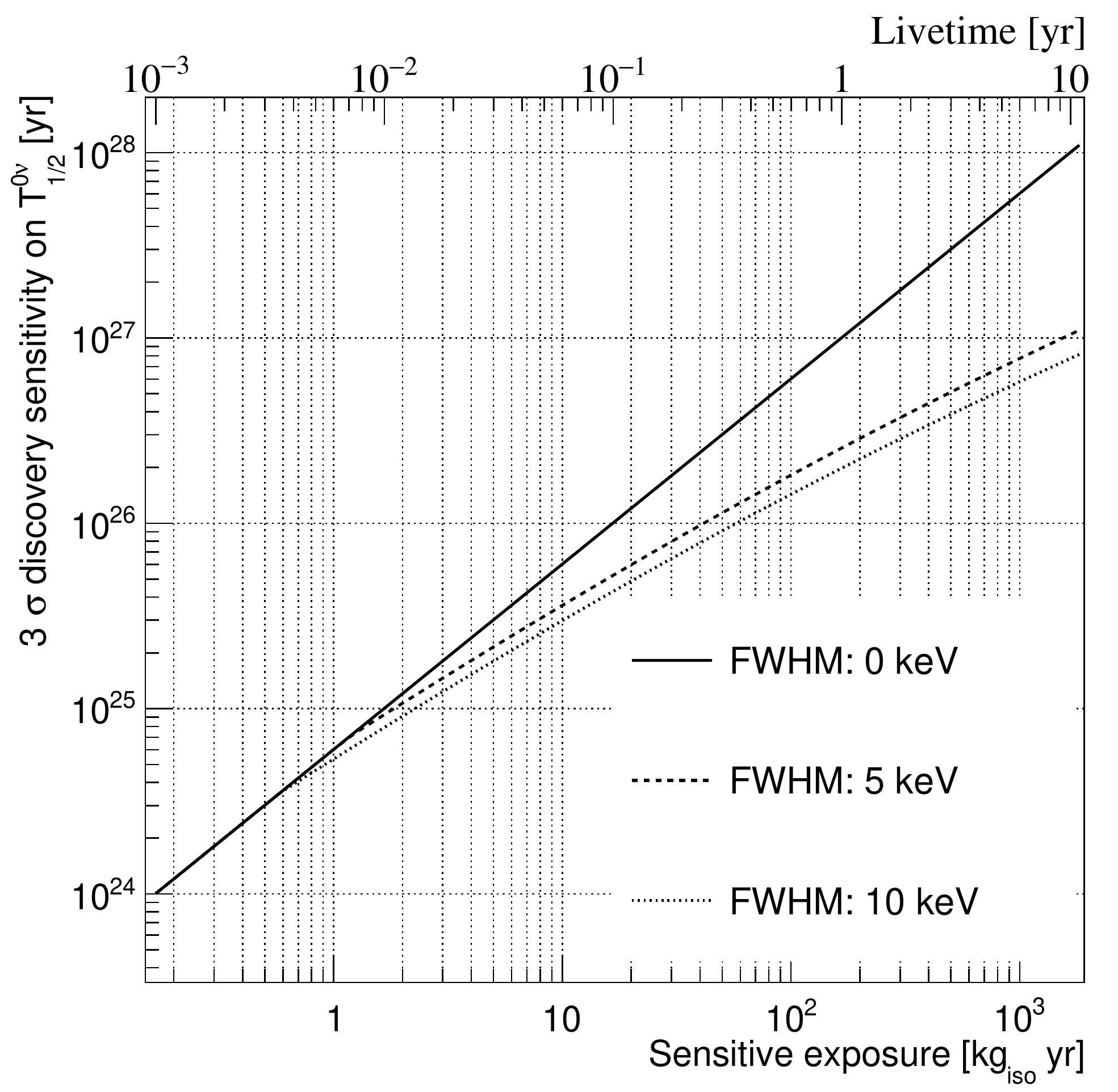}\label{fig:DiscSensFWHM}
     \caption{Left -- Discovery sensitivity on the \onbb\ decay half-life
     for an energy resolution of 5~keV FWHM and different background levels.  Right -- Discovery sensitivity for a BI of $10^{-4}$~\ckky\ and different energy resolutions.}
     \label{fig:DiscSensHalflife}
\end{figure}

The main physics goal of \cupid\ is searching for \onbb\ decay
with a discovery sensitivity greater than $10^{27}$~yr in the half-life
of the process, \thalf. The sensitivity scales linearly with the livetime
of the measurement for an experiment with zero background, or as the
square root of the live-time for a background-dominated experiment.
In practice, the relevant parameters are 1) the sensitive exposure,
defined as the product of the isotope mass, live-time, and total efficiency,
and 2) the sensitive background, defined as the expected number of background
events in the ROI normalized by the sensitive exposure.
The sensitive background depends on the BI at \Qbb,
measured in \ckky, and on energy resolution, which defines the ROI width.
Fig.~\ref{fig:DiscSensHalflife} shows the $3 \sigma$ discovery sensitivity
of \cupid\ for different values of the BI and energy resolutions.
In order to maximize the sensitivity and make the most efficient use
of the active material, the measurement has to be performed as close as
possible to the zero background approximation.
Fig.~\ref{fig:DiscSensHalflife} also shows that the energy
resolution is a somewhat less relevant parameter once the background
is low enough.

\begin{figure}
    \centering
    \includegraphics[width=\textwidth]{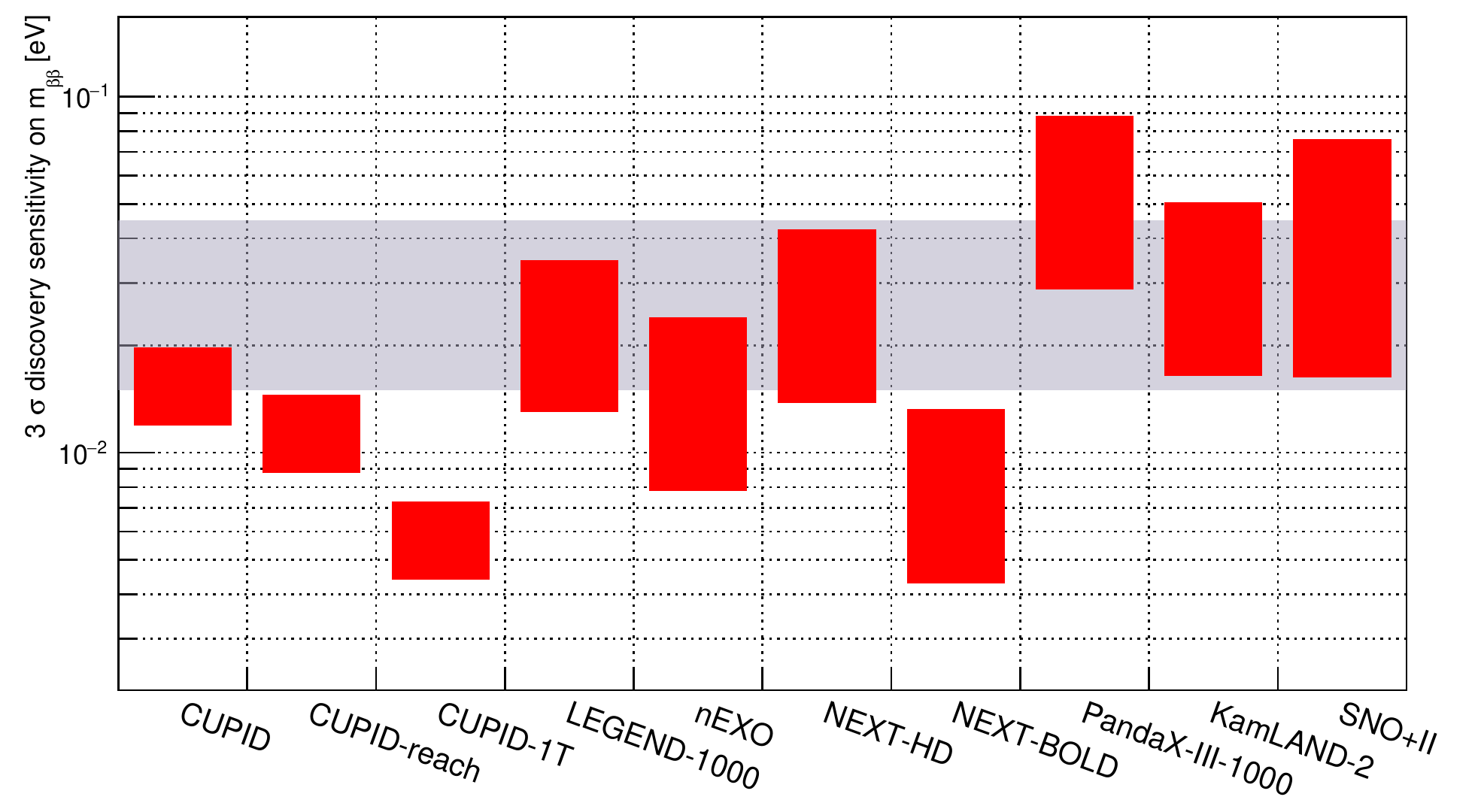}
    \caption{Discovery sensitivity for a selected set of next-generation ton-scale experiments.
    The grey shaded region corresponds to the parameter region
    allowed in the Inverted Hierarchy of the neutrino mass. The red error bars show the $m_{\beta\beta}$ values such that an experiment can make at least a $3\sigma$ discovery, within the range of the nuclear matrix elements for a given isotope.
    Parameters of the CUPID detector are listed in Table~\ref{tab:phases}.
    The parameters of the other experiments are taken from
    Refs.~\cite{Agostini:2017jim,Gomez-Cadenas:2019sfa,Galan:2019ake}.
    }
    \label{fig:DiscSensMbb}
\end{figure}

The \onbb\ decay can be induced by several mechanisms~\cite{DellOro:2016tmg},
with the following relation between \thalf\ and the parameter describing
the new physics, $|f|$:
\begin{equation}\label{eq:newphysics}
\frac{1}{T_{1/2}^{0\nu}} = G_{0\nu} g_A^4 \mathcal{M}^2 |f|^2\qquad,    
\end{equation}
where $G_{0\nu}$ is the phase space, $g_A$ is the axial vector coupling
constant, and $\mathcal{M}$ the nuclear matrix element (NME) relevant
for the process. Under the minimal assumption that only the three known
neutrinos are exchanged in the reaction, $|f|$ corresponds to the
effective Majorana mass:
\begin{equation}\label{eq:majorana-mass}
    |f| = m_{\beta\beta} = \frac{\left| \sum_{i=1}^3 U_{ei}^2 m_i \right|}{m_e}\qquad,
\end{equation}
where $U$ is the PMNS mixing matrix, $m_i$ the neutrino mass eigenvalues,
and $m_e$ the electron mass. The discovery sensitivities on \mbb\ 
for several proposed next-generation experiments measuring \onbb\ decay, including the CUPID phases described in Section~\ref{sec:phases}, 
are shown in Fig.~\ref{fig:DiscSensMbb}, assuming a livetime of 10~yr for each experiment.
The final sensitive exposure of each experiment depends on its isotope mass
and total efficiency. Experiments with a large isotope mass and/or high
efficiency (e.g. nEXO and SNO+) have a larger sensitive exposure
than experiments with a lower isotope mass, e.g. CUPID.
For a given sensitive exposure, the sensitivity on \mbb\ of a given
experiment has a width that depends on the NME for the corresponding isotope,
which is represented by the two lines in the figure.
For each isotope, we used the phase space factors from
Ref.~\cite{Kotila:2012zza}, and all NMEs available in
literature~\cite{Barea:2015kwa,Simkovic:2013qiy,Hyvarinen:2015bda,Neacsu:2014bia,Menendez:2008jp,Rath:2013fma,Rodriguez:2010mn,Mustonen:2013zu,Meroni:2012qf,Vaquero:2014dna,Yao:2014uta,Horoi:2015tkc,Senkov:2014wtz}.
The band for \Mo\ is narrower than for the other isotopes due to the lack
of NMEs computed with the interacting shell model.
A dedicated calculation was requested by the CUPID collaboration
and is ongoing~\cite{Menendez}.
Fig.~\ref{fig:DiscSensMbb} immediately shows a viable strategy for CUPID:
even with a relatively small isotope mass, we can cover most of the region
allowed in the inverted ordering (IO) even for the largest NME values.

Alternatively, we can consider the case in which the \onbb\ decay is mediated
by currently unknown heavy neutrino(s) predicted by several extensions
of the Standard Model (SM)~\cite{DellOro:2016tmg}.
These possibilities have an increasing appeal because they offer more
natural explanations for the mass of the known neutrinos,
which can be obtained within the SM only assuming an extremely
small coupling to the Higgs boson. The common drawback of these models
is the dependence on various assumptions and the presence of
additional parameters in the theory.
An additional motivation for the study of \onbb\ decay mediated by heavy
neutrinos is given by several independent measurements that favor
the normal ordering. In parallel, the cosmological measurements
are putting increasingly stronger bounds on the sum of neutrino masses,
making the degenerate region unlikely.
If this scenario is confirmed, the expected value
of \mbb\  can be considerably smaller than in the IO.
On the other hand, if next generation experiments find \onbb\ decay
with half-lives shorter than predicted by light neutrino exchange and normal ordering, the exchange of heavy neutrinos must be involved.
The new physics term of Eq.~\ref{eq:newphysics} is:
\begin{equation}\label{eq:heavyneutrino}
    |f| = M_{\beta\beta} = \left| \frac{ \sum_i V_{ei}^2 M_i }{ p^2 - M_i^2 } \right|\qquad,
\end{equation}
where the sum runs over the number of heavy neutrinos introduced by the 
SM extension under consideration, $V$ is an extended version of the PMNS mixing matrix
that comprises the heavy neutrino(s), $M_i$ are the heavy neutrino mass
eigenvalues, and $p$ is the momentum exchanged in the reaction ($p\sim200$~MeV).

\begin{SCfigure}
    \centering
    \includegraphics[width=0.5\textwidth]{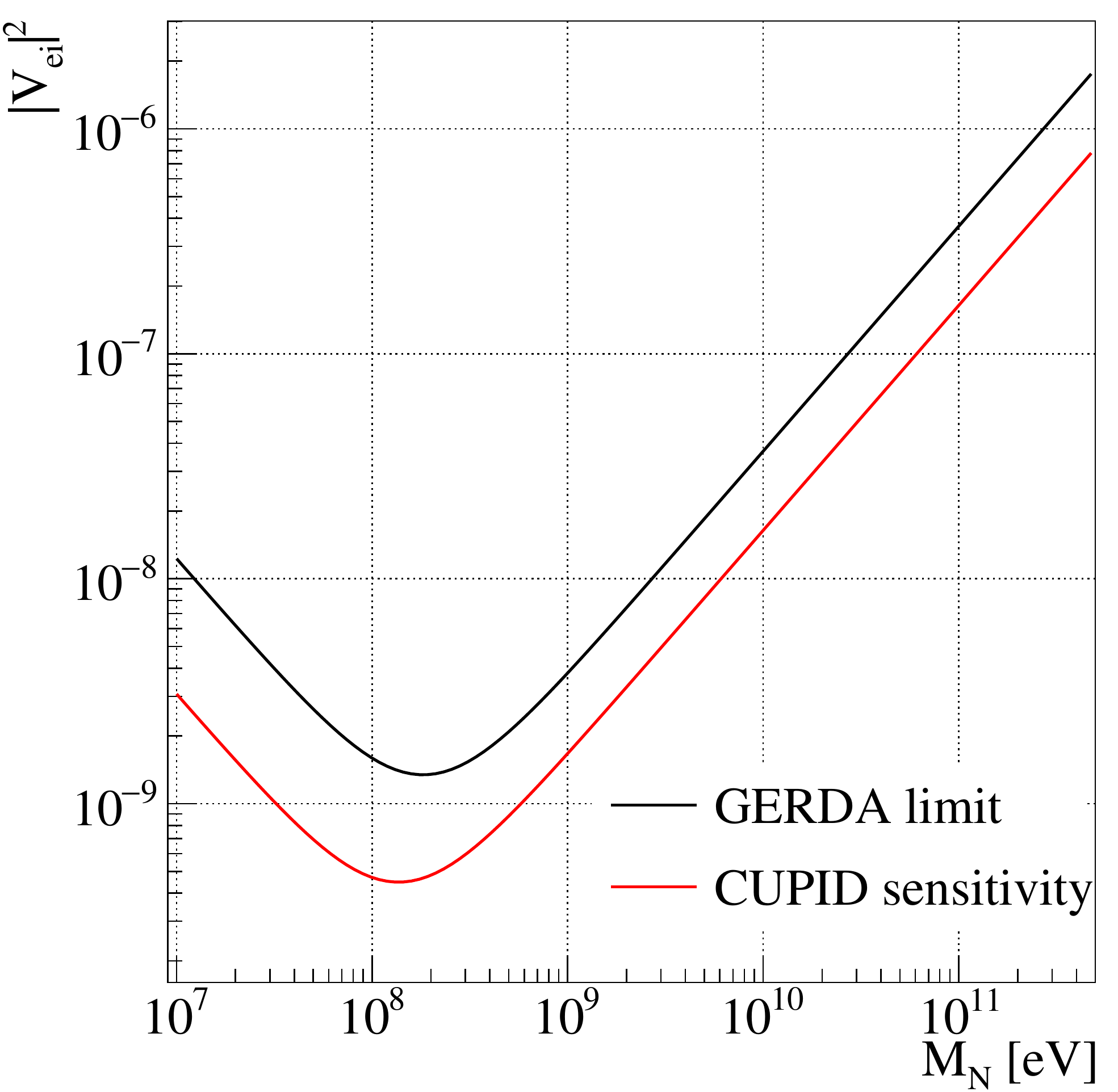}
    \caption{Sensitivity to \onbb\ mediated by a single heavy neutrino exchange.
    The red line corresponds to a CUPID sensitivity of $10^{28}$~yr, the black is
    for the currently exclusion limit set by GERDA on
    $^{76}$Ge~\cite{Agostini:2018tnm}.}
    \label{fig:heavyneutrino}
\end{SCfigure}

We consider one of the several possible models as an example,
which involves the exchange
of an additional single heavy neutrino~\cite{Mitra:2011qr}.
The discovery of \onbb\ decay would give a measurement of the new
term in Eq.~\ref{eq:heavyneutrino}, corresponding to a line
in the $(M_i,V_{ei})$ plane shown in Fig.~\ref{fig:heavyneutrino}.
In this hypothesis, CUPID has a discovery sensitivity superior
by an order of magnitude to the current limit by GERDA.

\subsection{Exotic processes}
Although \onbb\ is the main objective of CUPID, other processes are open to experimental investigation.  The anticipated low background rate promises  competitive sensitivities for many of them. 

The accessible processes include alternative modes of double beta decay as well as more exotic processes predicted by some extensions of the Standard Model.  The experimental investigation of the validity limits of fundamental physics principles like charge conservation or CPT/Lorentz invariance deserves particular attention since most of our theoretical framework is based on them.  Finally, the search for dark matter candidates, a project pursued by the first experiments on \onbb, is still one of the most appealing objectives of CUPID.

\subsubsection{Other double beta decays}
The best double beta decay experimental sensitivity is generally for the transition to the ground state of the daughter nucleus. However, double beta decay (in both 2$\nu$ and 0$\nu$ modes) may also occur to an excited state of the daughter nucleus. 
In the case of \onbb, these transitions can disclose the exotic mechanisms (e.g. right-handed currents) which mediate the decay~\cite{TOMODA2000245,SIMKOVIC2002201}, while for \nnbb\ they can provide unique insight to the details of the mechanisms responsible for the nuclear transition~\cite{AUNOLA1996133,SUHONEN1998124}.
From the experimental point of view, most of the interest is motivated by the fact that in a close-packed array like CUPID, the strong signature  provided by the simultaneous detection of one or two gammas can lead to an almost background-free search. 
In this respect, the transitions to 0+ states are favoured while states with larger spin (e.g. 2+) are generally suppressed by angular momentum conservation.
The adopted strategy exploits multiple coincidence patterns to select topological configurations characterized by a lower background contribution~\cite{CUORE02012503}.
A number of 0+ and 2+ excited states of $^{100}$Mo are accessible to CUPID with unprecedented sensitivity~\cite{BARABASH20078}.

Exotic modes such as $\beta^+\beta^+$, $\epsilon\beta^+$ and $\epsilon\epsilon$ are generally less appealing because of the lower available energy. However, they can only be mediated by peculiar mechanisms and can therefore provide unique information on the decay details~\cite{VERGADOS1983109,IACHELLO2014064319}. 
With a transition energy of 1649.1 keV, $^{92}$Mo is an excellent candidate, with sensitivities close to theoretical expectations for the \nnbb\ channel.

$^{100}$Mo is certainly the most favourable isotope for \onbb\ searches. However, another possible \onbb\ candidate is $^{98}$Mo.
Its large natural isotopic abundance (24.13 \%) could be appealing, but the very low transition energy (112 keV), and the fact that we will work with material isotopically enriched in $^{100}$Mo, ultimately leads to a very poor sensitivity for CUPID.

Exotic \onbb\ decays characterized by the emission of a massless Goldstone boson, called a Majoron, are predicted by some theoretical models~\cite{GELMINI1981411}. 
The precise measurements of the invisible Z width at LEP has greatly disfavoured the original Majoron triplet and pure doublet.
However, several new  models have been developed~\cite{BURGESS19945925,BAMERT199525}. 
All these models predict different (continuous) spectral shapes for the sum energies of the emitted electrons, which extend from zero to the transition energy Q$_{\beta\beta}$:
\begin{equation}
\frac{dN}{dT} \sim (Q_{\beta\beta}-T)^n,
\end{equation}
where $T$ is the electron summed kinetic energy and the spectral index, $n$, depends on the decay details. Single Majoron emissions are characterized by $n$ values between 1 and 3, while double Majoron decays can have either $n=3$ or $n=7$. The precise measurement of n allows one to discriminate between the Majoron processes and \nnbb\ ($n=5$).
As for any process characterized by continuous spectra, the experimental sensitivity is mainly limited by the background contributions and the detector mass~\cite{ELLIOTT19871649,ARNOLD2006483,KZENMAJORON2012}. Because of this, we anticipate excellent CUPID sensitivity to these processes. 

\subsubsection{Violations of fundamental principles}
The decay of an atomic electron is probably the most sensitive test of electric charge conservation. 
Charge non-conservation (CNC) can be obtained by including additional interactions of leptons and photons which lead to the decay of the electron: $e \to \gamma \nu$ or $e \to \nu_e\nu_X\bar{\nu}_X$. These modes conserve all known quantities apart from electric charge.
In addition, CNC can also involve interactions with nucleons. Discussions of CNC in the context of gauge theories can be found in a number of BSM gauge models~\cite{VOLOSHIN1978145,OKUN1978597,IGNATIEV1979315}.

While the signature of the neutrino mode for CNC is quite poor, the coincidence between the  decayed gamma and the X-rays from atomic de-excitations can give rise to interesting topological configurations that can help to lower the background contributions. 
The most stringent limits on CNC have been obtained as secondary results in experiments characterized by large masses and very low backgrounds~\cite{DAMA2000117301,BACK200229}.
Indeed, the large detection efficiency, low threshold, and excellent energy resolution expected for CUPID are crucial to detect the low energy de-ectitation X-rays or Auger electrons.  This, in combination with the ton-size scale of the experiment, makes us believe we can anticipate competitive results for these processes. 

Lorentz invariance and CPT violations arising from the spontaneous breaking of the underlying space-time symmetry are interesting theoretical features that can be parmeterized within the so-called Standard Model Extension (SME)~\cite{SME19976760,SME1998116002,SME2004105009}. 
Lorentz-violating effects in the neutrino sector can appear both in the two-neutrino and in the neutrino-less decay mode~\cite{JORGE2014036002}. Indeed, a distortion of the two-electron summed energy is expected for \nnbb\ due to an extra term in the phase space factor, while \onbb\ could be directly induced by a Lorentz violating term.
The signature is therefore very similar to the one expected for Majoron searches with a deformation of the upper part of the \nnbb\ spectrum.
The dominant contribution from \nnbb\ of $^{100}$Mo, and the extremely low background expected above 2 MeV, make CUPID particularly sensitive to these effects.

The Pauli exclusion principle (PEP) is one of the basic principles of physics upon which modern atomic and nuclear physics are built. 
Despite its well known success, the exact validity of PEP is still an open question and experimental verification is therefore extremely important~\cite{GREENBERG198983}. 
Indeed, a number of experimental investigations have been carried out both in the nuclear and atomic sector~\cite{BXINO2010034317,VIP2018319,BELLI1999236}. 
In all the cases, the signature is a  transition between already-occupied (atomic or nuclear) levels, which is clearly prohibited by PEP. Most of the low activity experiments exploit large masses and/or low background rates to search for the emission of specific electromagnetic or nuclear radiation from atoms or nuclei. In contrast, dedicated PEP-violation searches aim at improving the sensitivity by filling already complete atomic levels with fresh electrons and measuring the corresponding X-ray transitions. Unfortunately, a model linking the two experimental observations is still missing and a comparison of the sensitivities is impossible. 
CUPID belongs to the first category and will exploit the excellent energy resolution and the very low background to search for the emission of X-rays, $\gamma$s, or nucleons from the detector atoms and/or nuclei. 
PEP-prohibited nuclear decays are also possible, although a lower experimental sensitivity is anticipated for CUPID. 

\subsubsection{Tri-nucleon decays}
Baryon number (B) conservation is an empirical symmetry of the SM.
Its violation is predicted by a number of SM extensions. Furthermore, it is expected that quantum gravity theories violate B and that theories with extra dimensions permit nucleon decay via interactions with dark matter~\cite{BABU20135285}. 
In particular, some SM extensions, which allow for small neutrino masses, anticipate $\Delta$B=3 transitions in which three baryons can simultaneously disappear from the nucleus, frequently leaving an unstable isotope~\cite{BABU200332}. The coincidence between the tri-nucleon decay and the radioactive decay of the daughter nuclei is then a robust signature which can help mitigate the backgrounds. The dominant $\Delta$B=3 decay modes are $ppp \to e^+ \pi^+ \pi^+$, $ppn \to e^+ \pi^+$, $pnn \to e^+ \pi^0$, and $nnn \to \bar{\nu} \pi^0$. The decay-mode-specific signatures (charged fragments) include an initial saturated event followed by one or more radioactive decays. The invisible decay-mode signatures are composed of two successive decays and hence have two energy constraints and one time constraint.

\subsubsection{Dark Matter}
Although its existence is strongly suggested by several gravitational effects~\cite{DELPOPOLO20131548}, dark matter (DM) represents one of the deepest mysteries in modern physics and its nature still remains unknown. 
The WIMP hypothesis is the  most attractive and simplest scenario to explain DM.  However, the absence of an observed signal has led to very stringent constraints on WIMP properties, encouraging physicists to look for alternatives and stronger signatures~\cite{ROSZKOWSKI2018066201,ARCADI2018203}. 
In this framework, the possibility of observing a time-varying signal related to the relative motion of the earth with respect to the DM halo of the galaxy has attracted considerable interest~\cite{MAYET20161}.
A low energy threshold and a large mass are the most appealing features for experiments hoping to observe these effects.  With the possibility of separating nuclear recoils (characterized by a lower light yield) from the dominant electron-based background, CUPID's technology is particularly well-suited for searches of this kind.  We anticipate excellent sensitivity for observing the WIMP seasonal modulation generated by the motion of the Earth. 

Non-WIMP DM candidates have also been explored extensively in the literature~\cite{BAER20151}. 
Among them, ultralight axions, which emerge as solutions to the strong CP problem in QCD, are a valid candidate for (cold) DM~\cite{IPSER1983925}. 
They are expected to couple with the electromagnetic field and directly to leptons and quarks.
A comparatively large production yield is therefore expected in dense, hot matter like the sun’s core.
In particular, the $^{57}$Fe 14.4 keV nuclear transition, which can be populated in the inner region of the sun via thermal excitation, is a good candidate for the production of a monochromatic axion, whose observation in CUPID will be based on the axio-electric effect~\cite{CUORE2013007}.
Here, the most relevant experimental parameters are the energy threshold and the background level at very low energies. CUPID will contribute to significantly restricting the axion parameter space. 
A particularly promising process for the observation of solar axions in \lmo\ is the resonant absorption on $^7$Li nuclei~\cite{KRVCMAR2001115016}, which is accompanied by the excitation of the first nuclear level of lithium with the corresponding emission of a 477.6 keV $\gamma$-ray: $a + ^7$Li$ \to ^7$Li$^* \to ^7$Li$ + \gamma$.

\newpage


\section{Alternative configuration: enriched TeO$_2$ detectors} \label{sec:Te}

Tellurium ($^{130}$Te isotope) has been chosen as an alternative \onbb\ emitter along with $^{100}$Mo for the CUPID experiment. 
The selection of TeO$_2$ has been made for two main reasons:
\begin{itemize}
\item TeO$_2$ bolometers are very well known detectors, thanks to the overall CUORE experience (material procurement, crystal handling, and detector optimization);
\item The high isotopic abundance of $^{130}$Te (about 33$\%$) significantly reduces enrichment costs.
\end{itemize}

The main disadvantage of this choice relies in the fact that tellurium-based bolometers are not scintillating. Nevertheless \al/\be\ discrimination can be obtained exploiting the emission of Cherenkov light in TeO$_2$ crystals \cite{TabarellideFatis:2009zz}. In TeO$_2$ crystals, the threshold for Cherenkov light emission is about 50 keV for electrons and about 400 MeV for \al\ particles. Therefore, in the \onbb\ region of interest, the \al\ particle energies are below the Cherenkov emission threshold, thus emitting no light signal.
It is noteworthy that the Cherenkov light emitted by the crystals steeply decreases with increasing crystal size \cite{Casali2015}, corresponding to about 100 eV for \be\ particles in the ROI in TeO$_2$ crystals with CUORE's size (5$\times$5$\times$5 cm$^{3}$). Therefore, the detection of such low energy light requires light detectors with a RMS noise of about 20 eV.

During past few years, the light detector technology has improved to the level that “standard” Ge-based light detectors with NTD readout that can reach baseline values of few tens of eV (see Section~\ref{sec:Ge_LD_NTD}) are now available. Even better results have been recently achieved using NTD Ge-based light detectors with Neganov-Luke amplification (see Section ~\ref{sec:NL-NTD}). This demonstrates that an event-by-event active particle identification is possible for full-size CUORE TeO$_2$ bolometers \cite{Berge2018}.
The latter light detection technique has been recently applied to two $^{130}$TeO$_2$ enriched crystals of 92$\%$ $^{130}$Te, having a mass of 435 g compared to the 750 g CUORE crystals \cite{Artusa2017}. In this case, excellent event-by-event particle discrimination was obtained (95$\%$ acceptance level). Moreover, the two crystals showed performances in terms of baseline noise and energy resolution of 4.3 and 6.5 keV FWHM respectively, fully compatible with the CUPID baseline.

The R\&D on enriched TeO$_2$ thus far clearly demonstrates two main issues. 
First, significant raw material purification will be required to eliminate
radioactive Th and U, as well as the non-radioactive elements, mainly Na, Si, Fe, Cu, Sn, and Sb, which can degrade the quality of the crystals as bolometers.
The second issue involves the altered isotopic concentration of enriched TeO$_2$ crystals if the Bridgman crystal growth technique is used, as it was for the CUORE crystals.
The other problem, general for all enriched crystals, consists in finding methods for the highest possible raw material utilization coefficient (possibly 100$\%$). In the following, we describe proposed research and development to address these issues.

\subsection{Enriched $^{130}$TeO$_2$ raw material for crystal growth}

The synthesis of TeO$_2$ powder and general chemical processing of raw material for crystal growth is well mastered.  The crystal growth process at the Shanghai Institute of Ceramics of the Chinese Academy of Science (SICCAS) was proven to be acceptable as far as radio-contamination is concerned \cite{Arnaboldi2010}.
Nevertheless, our experiments with the small crystals \cite{Artusa2017} demonstrated that the enriched detectors had excessive radioactive backgrounds of Th and U coming from the isotopic enrichment process of natural tellurium.
Accordingly, the enriched Te or \teo\ must be purified to reduce the Th and U content. To achieve the purification of the input material, we propose an experimental effort to develop a technique to zone-refine TeO$_2$. If successful, this could also be used to recycle the scrap material from the crystal fabrication. In a first attempt, our team has explored chemical purification with the help of the Brookhaven National Laboratory, and we concluded that it was cost-prohibitive to reach our required levels. For a deeper, cost-effective study of possible TeO$_2$ purification schemes by zone-refinement, the University of South Carolina purchased a 1250 $^\circ$C, computer-controlled zone-furnace from MELLEN, Inc., in Concord, NH. At the time of this writing, the zone refining facility is nearing completion.  Funds are available to complete the installation, including: the platinum zone boats, natural TeO$_2$ powder for the trials, and some evaluation of the results by ICP-MS. 

The ICP-MS tests are being explored in cooperation with the Pacific Northwest National Laboratory. A possible side effect of this activity is that the equipment used for the purification of TeO$_2$ (with a melting point 733 $^\circ$C) could be used for the purification of \lmo\ (melting point 705 $^\circ$C) or of MoO$_3$ (melting point 795 $^\circ$C).  The latter is a principal precursor of the raw material used for \lmo crystal growth. Because all these materials have close melting points, the same type of crucible (e.g. platinum) can be used for all.
Accordingly, the R$\&$D to explore the purification of input and scrap TeO$_2$ can also serve as R$\&$D for the \lmo\ option for CUPID.
In any case, a strong case can be made to develop the zone refining technique for TeO$_2$ and to attempt a similar technique for \lmo.
To start in this direction, MELLEN Inc. was contacted  to learn about zone refining TeO$_2$ with the MELLEN Series EDG12.5-Sunfire, 1250 $^\circ$C Programmable Multi-Zone Furnace. It has a 6-inch bore, with 24 one-inch zones. Delivery of the apparatus was in late August 2018 and commissioning ended in Spring 2019.
The purification of input and scrap TeO$_2$  will be 
possible in this furnace.  Note, the CUPID group from the University of South 
Carolina has past experience with these methods and were able to achieve 70\% overall yield in 
detector mass from Ge enriched to 88\% in $^{76}$Ge 
for the MAJORANA DEMONSTRATOR.
This required continuous zone refining of scrap material, where the yield is 98$\%$, rather than chemical reprocessing. The cost of the TeO$_2$ enriched in $^{130}$Te will be significantly less expensive than enriched Ge (15-20 K\$/kg).


\subsection{Enriched $^{130}$TeO$_2$ crystal growth}
Preliminary crystal growth tests have shown that the method used for the production of CUORE crystals at SICCAS is not applicable when creating enriched $^{130}$TeO$_2$ crystals. Two trial enriched detectors were produced for feasibility tests and the $^{130}$Te fraction was reduced from 93\% in the starting material (metallic powder) to 73\% in the final crystal due to the migration of the material from the natural seed throughout the crystal ingot during the growth process. This is because the growth method applied (modified Bridgman technique) requires the use of a large seed, which is partially melted in the initial phase of the crystal growth with consequent alteration of the isotopic concentration of the resulting crystal. There are two solutions to this problem: (i) maintain the same method of growth, but use an enriched seed, possibly obtained through the Czochralski method; (ii) use the Czochralski technique for the entire crystal production process.

Note that in the Czochralski method, a very tiny seed is used with a mass negligible compared to that of the crystal.  Therefore, diluting the isotopic concentration is not an issue in this case. If we apply the Czochralski method, this would most likely have to be done in the US since this technology is not well mastered at SICCAS. Early communications have begun with Gooch and Housego Inc. in Cleveland, Ohio, to determine if producing the enriched TeO$_2$ CUORE crystals by the Czochralski technique would be cost-competitive. Moreover, since we still want to do purification in the US, one advantage of growing the crystals with Gooch and Housego Inc. would be to avoid the expense and complications of shipping scrap back to the US and then returning reprocessed material back to China. While the cost per crystal could be higher, the savings in the complex import/export logistics could be substantial.  Moreover, the additional cost and complication of providing enriched seeds might well make the US production of the crystals, as well as the US purchase, purification, and recycling of the TeO$_2$, possibly cost-competitive. 

\subsection{Recovery and recycling}
Our experience with CUORE demonstrated that only 30\% to 35\% of the input material ended up in crystals. This implies that a large amount of tellurium oxide must be re-processed for each crystal pulled. If the crystals were produced at SICCAS, the many shipments of scrap material for reprocessing, and its return to SICCAS, would be very complicated, costly and time consuming. In addition, SICCAS does not grow crystals by the Czochralski method, so large crystal
seeds would have to be grown in the US and shipped to China periodically. 

As described above, we have 
identified one possible scenario for enriched crystal production, which includes an optimized efficiency in the purification of the input TeO$_2$, as well as the recycling of the scrap material from the crystal fabrication. This option would have the TeO$_2$ crystals grown by the Czochralski method in the U.S. where zone-refinement would be used for both primary and recycling purification.

\section{Implementation and Timeline}
The most time-sensitive, critical path task is the procurement of the enriched \Mo\ isotope and production of the crystals. Given the relatively conservative baseline design of the CUPID detector, and the strategy of deploying CUPID in the existing CUORE cryostat as soon as CUORE finishes data taking, the commissioning and the scientific operations of the detector could start approximately five years after the start of the project. 




\newpage

\section{CUPID Group of Interest}
The CUPID Interest group is a robust international collaboration that brings together an array of experts that will contribute to a successful program.\\


\noindent {\bf Argonne National Laboratory, Argonne, IL, USA}

\noindent W.R. Armstrong, C. Chang, K. Hafidi, M. Lisovenko, V. Novosad,  J. Pearson, T. Polakovic, G. Wang, V. Yefremenko, J. Zhang\\

\noindent {\bf INFN Laboratori Nazionali del Gran Sasso, Assergi (AQ) Italy}

\noindent C. Bucci, L. Canonica, L. Cappelli, V. Caracciolo, S. Copello, A. D'Addabbo, P. Gorla, S. Nisi, D. Orlandi, C. E. Pagliarone, L. Pattavina, S. Pirro, C. Rusconi, K. Schaffner \\

\noindent {\bf Lawrence Berkeley National Laboratory and University of California, Berkeley, CA, USA}

\noindent G. Benato, A. Drobizhev , B. K. Fujikawa, R. Huang, Yu. G. Kolomensky, L. Marini, E. Norman, M. Sakai, B. Schmidt, V. Singh, K. Vetter, S. Wagaarachchi, J. Wallig, B. Welliver\\

\noindent {\bf Virginia Tech, Blacksburg, VA, USA}

\noindent S. Dell'Oro,  T. O'Donnell\\

\noindent {\bf INFN Sezione di Bologna and University of Bologna, Bologna, Italy}

\noindent S. Zucchelli, N. Moggi\\

\noindent {\bf INFN Sezione di Bologna and CNR-IMM, Bologna, Italy}

\noindent V.Boldrini, F. Mancarella, R.Rizzi \\

\noindent {\bf Massachusetts Institute of Technology, Cambrige, MA, USA}

\noindent J. Johnston, J. Ouellet, J. Formaggio, L. Winslow\\

\noindent {\bf University of South Carolina, Columbia, SC, USA}

\noindent F. Avignone, C. Rusconi, R. Creswick and K. Wilson \\

\noindent {\bf INFN Laboratori Nazionali di Frascati, Frascati, Italy}

\noindent A. Franceschi, T. Napolitano\\

\noindent {\bf INFN Sezione di Genova and University of Genova, Genova, Italy}

\noindent A. Caminata, S. Di Domizio, M. Pallavicni\\

\noindent{\bf SIMAP Grenoble, France}

\noindent M. Velazquez\\

\noindent {\bf University of Science and Technology of China, Hefei, China}

\noindent H. Peng, M. Xue\\

\noindent {\bf KINR Kiev, Ukraine}

\noindent F. Danevich, V. Kobychev, O. Polischuk, V. Tretyak\\

\noindent {\bf INFN Laboratori Nazionali di Legnaro, Italy}

\noindent O. Azzolini, G. Keppel, C. Pira\\

\noindent {\bf INFN Sezione di Roma and Gran Sasso Science Institute, L'Aquila, Italy}

\noindent F. Ferroni\\

\noindent {\bf INFN Laboratori Nazionali del Gran Sasso and Gran Sasso Science Institute, L'Aquila, Italy}

\noindent V. Domp\`e, G. Fantini\\

\newpage 
\noindent {\bf University of California Los Angeles, Los Angeles, CA, USA}

\noindent K. Alfonso, H.Z. Huang \\

\noindent {\bf IPNL Lyon, France}

\noindent Q. Arnaud, C. Augier, J. Billard, A. Cazes, F. Charlieux, E. Elkhoury, J. Gascon, M. De Jesus, A. Juillard, D. Misiak, V. Sanglard, L. Vagneron\\

\noindent {\bf INFN Sezione di Milano Bicocca and University of Milano  Bicocca, Milano, Italy}

\noindent M. Beretta, M. Biassoni, C. Brofferio, S. Capelli, P. Carniti, D. Chiesa, M. Clemenza, O. Cremonesi, M. Faverzani, E. Ferri, A. Giachero, L. Gironi, C. Gotti, M. Nastasi, I. Nutini, L. Pagnanini, M. Pavan, G. Pessina, S. Pozzi, E. Previtali, A. Puiu, M.	Sisti\\

\noindent {\bf ITEP Moscow, Russia}

\noindent A. Barabash, S. Konovalov, V. Yumatov\\

\noindent {\bf Yale University, New Haven, CT, USA}

\noindent K. Heeger, R. Maruyama, J. Nikkel, D. Speller, P. T. Surukuchi\\

\noindent {\bf NIIC Novosibirsk, Novosibirk, Russia}

\noindent V. Shlegel\\

\noindent {\bf CSNSM Orsay, France} 

\noindent L. Berg\'e, M. Chapellier, L. Dumoulin, A. Giuliani, H. Khalife, P. de Marcillac, S. Marnieros, E. Olivieri, D. Poda, T. Redon, A. Zolotarova\\

\noindent {\bf LAL Orsay, France}

\noindent M. Bri\`ere, C. Bourgeois, E. Guerard, P. Loaiza\\

\noindent {\bf INFN Sezione di Padova, Padova, Italy}

\noindent L. Taffarello\\

\noindent {\bf Drexel University, Philadelphia, PA, USA}

\noindent G. Karapetrov \\

\noindent {\bf INFN Sezione di Roma and Sapienza University of Rome, Rome, Italy}

\noindent F. Bellini, L. Cardani, N. Casali, A. Cruciani, I. Dafinei, V. Pettinacci, G. D' Imperio, C. Tomei, M. Vignati\\ 

\noindent {\bf INFN Sezione di Roma and CNR-NANOTEC, Rome, Italy}

\noindent I. Colantoni\\

\noindent {\bf CEA Saclay, France}

\noindent E. Armengaud, A. Charrier, M. de Combarieu, F. Ferri, Ph. Gras, M. Gros, D. Helis, X.F. Navick, C. Nones, P. Pari, B. Paul\\

\noindent {\bf Cal Poly, San Luis Obispo, CA, USA}

\noindent T. Gutierrez\\

\noindent {\bf Shanghai Jiao Tong University, Shangai, China}

\noindent K. Han\\

\noindent {\bf Fudan University, Shangai, China}

\noindent L. Ma, Y. Shen, W. He\\

\noindent {\bf Universidad de Zaragoza, Zaragoza, Spain} 

\noindent M. Martinez\\



\newpage

\bibliographystyle{apsrev4-1}
\bibliography{main}
\end{document}